\documentclass[11pt]{article}
\usepackage{latexsym,amssymb,lscape,graphics,setspace}
\usepackage{amsmath}
\usepackage{graphicx}        % pacchetto grafico standard LaTeX
\usepackage{longtable}
\usepackage{multicol}
\usepackage{slashed,epsfig}
\usepackage{amsfonts}
\usepackage{hyperref}
\newcommand{\mathsym}[1]{{}}

\newtheorem{definizione}{Definition}[section]
\newtheorem{teorema}{Theorem}[section]

\newtheorem{lemma}{Lemma}[section]

\newtheorem{proofteo}{Proof}[teorema]

\newcommand{\bd}{\begin{definizione}}
\newcommand{\ed}{\end{definizione}}
\def\bfzero{\relax{\rm I\kern-.18em 0}}
\def\bfone{\relax{\rm 1\kern-.35em 1}}

\DeclareFontFamily{U}{rsf}{} \DeclareFontShape{U}{rsf}{m}{n}{
  <5> <6> rsfs5 <7> <8> <9> rsfs7 <10-> rsfs10}{}
\DeclareMathAlphabet\Scr{U}{rsf}{m}{n}

\newcommand{\ft}[2]{{\textstyle\frac{#1}{#2}}}
\def\tilde{\widetilde}

\def\1bar{1\hskip -.275cm -}
\def\2bar{2\hskip -.275cm -}
\def\3bar{3\hskip -.275cm -}

\newsavebox{\uuunit}
\sbox{\uuunit}
                 {\setlength{\unitlength}{0.825em}
                      \begin{picture}(0.6,0.7)
                                      \thinlines
                                      \put(0,0){\line(1,0){0.5}}
                                      \put(0.15,0){\line(0,1){0.7}}
                                      \put(0.35,0){\line(0,1){0.8}}
                                     \multiput(0.3,0.8)(-0.04,-0.02){10}{\rule{0.5pt}{0.5pt}}
                      \end {picture}}

\makeatletter \@addtoreset{equation}{section} \makeatother
\def\bfone{\relax{\rm 1\kern-.35em 1}}

\def\bfone{\relax{\rm 1\kern-.35em 1}}
 
\begin{document}
\begin{titlepage}
\begin{center}
%%%%%%%%%%%%%%%%%%%%%%%%%%%%%%%%%%%%%%%%%%%%%%%%%%%%%%%%%%%%%%%%%%%%
%%%%%%%%%%%%%%%%%%%%%%%%%%%%%%%%%%%%%%%%%%%%%%%%%%%%%%%%%%%%%%%%%%%%
%%%%%%%%%%%%%%%%%%%%%%%%%%%%%%%%%%%%%%%%%%%%%%%%%%%%%%%%%%%%%%%%%%%%
{ {\Large \sc Chaos from Symmetry }  \\
$\null$ \\
{\Large \it Navier Stokes equations, Beltrami fields  and \\
the Universal Classifying Crystallographic Group}\footnote{\large
This article presents the new original results of an investigation
performed within the framework of the Project ALMA FLUIDA,
cofinanced by the Regione Toscana, in connection with the
Consultancy Contract signed between the Company ITALMATIC Presse e
Stampi and the DISAT of Torino Politecnico. It focuses on the
theoretical aspects. The calculational \textbf{AlmafluidaNSPsystem}
written in Wolfram MATHEMATICA language finalized to the explicit
construction of Beltrami fields and to the analysis  of their group
theoretical structure is posted on the Wolfram Community site and
can be downloaded from there. }}
\\[0.5cm]
%%%%%%%%%%%%%%%%%%%%%%%%%%%%%%%%%%%%%%%%%%%%%%%%%%%%%%%%%%%%%%%%%%%%
Pietro~G.~Fr\'e${}^{\; a,c,d}$  and
Mario Trigiante${}^{\; b,c,d}$ \\[8pt]
\vspace{2pt}  {${}^a$\sl\small Dipartimento di Fisica, Universit\'a
di Torino\\
via P. Giuria 1,  10125 Torino,  Italy}\\
\vspace{2pt}
{${}^b${{\sl\small  DISAT Politecnico di Torino,}}\\
{\em C.so Duca degli Abruzzi, 24, I-10129 Torino, Italy}\\
\vspace{2pt} {{\em $^{c}$\sl\small INFN --
 Sezione di Torino
}}\\
\vspace{2pt}
 \centerline{$^{(d)}$ \it Arnold-Regge Center}}
%%%%%%%%%%%%%%%%%%%%%%%%%%%%%%%%%%%%%%%%%%%%%%%
\end{center}
\vspace{3pt}
\begin{abstract}
%\begin{left}
The core of this paper is the group-theoretical approach, initiated
in 2015 by one of the present authors in collaboration with
Alexander Sorin brings into the classical field of mathematical
fluid-mechanics a brand new vision, allowing for a more systematic
classification and  algorithmic construction of  Beltrami flows on
torii $\mathbb{R}^3/\Lambda$ where $\Lambda$ is a crystallographic
lattice. Here this new hydro-theory is based on the focal idea of a
Universal Classifying Group $\mathfrak{UG}_\Lambda$  is revised,
reorganized, improved and extended.  In particular, we construct the
so far missing $\mathfrak{UG}_{\Lambda_{Hex}}$ for the hexagonal
lattice and we advocate that, mastering the cubic and hexagonal
instances of this group, we can cover all cases. The relation
between the classification of Beltrami Flows with that of contact
structures is enlightened. The recent developments about the
framework of $\mathfrak{b}$-manifolds  are considered and it is
shown that the choice of the allowed critical surfaces for the
$\mathfrak{b}$-deformation of a Beltrami field seems to be strongly
related with the group-theoretical structure of the latter. This
opens new directions of investigation about a group theoretical
classification of critical surfaces. Apart from that the most
promising research direction opened by the present work streams from
the fact that the Fourier series expansion of a generic Navier-Stokes solution can be regrouped into  an infinite sum of
contributions $\mathbf{W}_r$, each associated with a spherical layer
of quantized radius $r$ in the momentum lattice. Each $\mathbf{W}_r$
 is the superposition of a Beltrami field $\mathbf{W}_r^+$ plus
an anti-Beltrami field $\mathbf{W}_r^-$. These latter have a priori
exactly the same decomposition into irreps of the group
$\mathfrak{UG}_\Lambda$ that  are variously repeated on higher
layers. This crucial property enables the construction of generic
Fourier series with prescribed hidden symmetries as candidate
solutions of the NS equations. Alternatively the Fourier series
representation of known solutions can be analyzed from the point of
view of such symmetries. As a further result of this research
programme a complete and versatile system of MATHEMATICA Codes named
\textbf{AlmafluidaNSPsystem} has been constructed and is now
available through the site of the Wolfram Community. The exact
solutions presented in this paper have to be considered as an
illustration of the new conceptions and ideas that have emerged and
of what can be further done utilizing the computer codes as an
instrument. The main message streaming from our constructions is
that the more symmetric the Beltrami Flow the highest is the
probability of the onset of chaotic trajectories.
%\end{left}
\end{abstract}
%\end{center}
\end{titlepage}
\tableofcontents \noindent {}
\newpage
\section{Introduction}
The present paper is at the same time a research paper and partially
a review one since its basic aim is the development of an entirely
new and original theory of periodic incompressible hydro-flows that
was initiated seven years ago by one of the present authors in
collaboration with Aleksander S. Sorin \cite{Fre:2015mla}. The core
of this new theory consists of the introduction in the context of
three-dimensional crystallographic groups, of a new general and well
defined concept,  namely that of the \textbf{Universal Classifying
Group} $\mathfrak{UG}_\Lambda$. As it will become clear in the
course of our exposition (see in particular section
\ref{gruppafunda}) $\mathfrak{UG}_\Lambda$ is essentially defined as
the smallest finite group that contains, as possible subgroups, all
the \textbf{space groups} $\mathfrak{G}^{space}_\Lambda$ associated
with the \textbf{point group} $\mathfrak{P}_\Lambda$ of a specific
lattice $\Lambda$. The reason why this group is relevant for
hydro-flows is that solutions of Beltrami equation can be
systematically constructed on the three torus
$\mathbb{R}^3/\Lambda$, utilizing the decomposition into
$\mathfrak{P}_\Lambda$-orbits of the dual momentum lattice
$\Lambda^\star$. The obtained solutions are the appropriate
generalization to an exhaustive  finite family of building blocks of
the ABC models that pertain only to the smallest point group orbit
of the cubic lattice (the six dimensional one) and are moreover a
special truncation thereof. The fascinating discovery pointed out in
\cite{Fre:2015mla} is that the classification of these building
block Beltrami fields is naturally organized into irreducible
representations of the group $\mathfrak{UG}_\Lambda$ (see in
particular section \ref{triplettoni}). Although it was already clear
in 2015 that the unveiled hidden symmetry structure of Beltrami
fields was a prominent new weapon in the study of hydro-flows, the
development of this new group-based theory of hydrodynamics stopped
at the level of the systematic construction of Beltrami fields for
the cubic lattice, the other maximal lattice, the hexagonal one,
being only touched upon, and, most significantly, the general scheme
of utilization of the group theoretical weapon in the context of
true solutions of Navier-Stokes equations being not envisaged.
\par
The development of such a new theory is re-addressed in the present
paper from scratch in view of three critical observations that open
the road to a wealth of systematic new studies:
\begin{description}
  \item[A)] On each spherical layer (or energy shell) $\mathrm{SL}_r$  of the momentum
  lattice (see in particular section \ref{furetto})
  the contribution $\mathbf{U}_r$ to the Fourier development
  of a generic Navier-Stokes velocity field $\mathbf{U}$ is always partitioned
  into the sum of Beltrami field $\mathbf{U}_r^+$ (eigenstate of the Beltrami operator with positive eigenvalue $\pi r$) and an
  anti-Beltrami field $\mathbf{U}_r^-$ (eigenstate of the Beltrami operator with negative eigenvalue $- \pi r$)
  \item[B)] The group theoretical structure of Beltrami and anti
  Beltrami fields is identical so that the decomposition into
  irreducible representations of the Universal Classifying group can
  be applied to the complete Fourier development of a generic
  velocity field.
  \item[C)] It appears from old results \cite{beltraspectra} which made no reference to
  the, at that time unknown, $\mathfrak{UG}_\Lambda$ group structure, that the Beltrami modes
  are weakly interacting and that the so named Beltrami spectrum is
  almost conserved (see section \ref{furetto})
\end{description}
Since paper \cite{Fre:2015mla} was mostly disregarded by the
reference scientific community  and, more relevantly, since the
conceptual perspective is now deeply changed and the framing of
derivations has been substantially improved and clarified, we
decided to present the new results together with the old ones of
\cite{Fre:2015mla} in a unified exposition, logically organized
around the pivot of the new ideas presented above. Specifically the
new results are:
\begin{enumerate}
  \item The derivation of the Universal Class Group $\mathfrak{U}_{72}$
  for the hexagonal lattice case, the construction of its irreps and
  character table.
  \item The development of the construction algorithm of Beltrami/anti-Beltrami
  fields for all the orbits of $\Lambda^\star_{hexag}$.
  \item The development of the decomposition algorithm of
  Beltrami/antiBeltrami fields into irreps of $\mathfrak{U}_{72}$
  and subgroups thereof.
  \item The inspection of some examples of Beltrami flows, both in the
  cubic and hexagonal lattice case, that are endowed with  large
  groups of hidden symmetries. These examples confirm the general
  idea of \textit{Chaos from Symmetry} since the streamlines appear
  more chaotic wider is the hidden symmetry.
\end{enumerate}
In view of the three critical observations exposed above, since the
hidden symmetries of Beltrami fields can be extrapolated to generic
Fourier expansions, the search for quasi chaotic trajectories or
streamlines is now liable to be established in much more general
setups.
\par
Just only in order to further emphasize the role of the hidden
symmetries, we have also considered the recent interesting results
concerning the extension of Beltrami equation to so named
$b$-manifolds. Our attention was captured by the result of
\cite{cardone2019} where $b$-deformation of the classical ABC-models
was studied and it was shown that the deformation, by means of the
introduction of two critical surfaces on the torus, can be done only
if the parameter that is named $C$ in our  normalizations, vanishes.
The previous group-theoretical analysis of the ABC models presented
in \cite{Fre:2015mla} and repeated here for completeness was invoked
in order to unveil the group theoretical meaning of the condition
$C=0$. We refer the reader to section \ref{dippo} for all the
details. The message is very clear and worth of systematic
investigation. The ABC models, that for several decades have been
the focus of a lot of attention, are just only, in their own
functional definition, the tip of an iceberg. They encode half of
the 6-parameter Beltrami field obtained from the lowest lying
6-dimensional orbit of the cubic lattice point group
$\mathrm{O_{24}}$ which provides a precise irreducible
representation of the relevant Universal Classifying Group. The
splitting into two halves is obtained by decomposing this
6-dimensional irrep with respect to a proper subgroup that admits a
three-dimensional irrep corresponding to the A,B,C parameters. The
nullification of the C-parameter corresponds to choosing a further
subgroup with respect to which C is a singlet. This reveals that
$b$-deformations are in correspondence with subgroups of
$\mathfrak{UG}_\Lambda$. Clarifying in a full-fledged manner such a
correspondence is a research plan that some-one should address. From
the point of view of the present article this is just an interesting
side issue which by no means constitutes the goal and the core of
the paper. Notwithstanding this, in order to present this lateral
issue while keeping the paper readable for a mixed community of
readers that hopefully, besides symplectic geometers includes also
general relativists, supergravity/string theorists (as we are),
group theorists and also fluid-mechanics experts working with CFD
simulations, we had to provide some schematic but essential
background on contact structures, symplectic manifolds and all that,
in order to introduce the notion of $\mathfrak{b}$-deformations.
This is done in section \ref{geomfund}. Finally we should also
mention that one of the main connected achievement of the present
investigation project has been the construction in Wolfram
MATHEMATICA language of the \textbf{AlmafluidaNSPsystem} published
on the Wolfram community site at
\url{https://community.wolfram.com/groups/-/m/t/2555905}. This code
system is able to construct Beltrami and anti-Beltrami fields on any
spherical momentum space layer, decompose them into irreps of any
chosen hidden symmetry group, integrate the corresponding stream
lines and plot them graphically. Hence this system provides  the
building blocks for further studies in the various theoretical
directions emerged from the development of this new group-based
theory of incompressible fluid-dynamics.

\section{The Navier Stokes Equations and their elaboration} \label{genteoria}
\label{NavStok} Our primary object of study  is the fundamental
equation of classical hydrodynamics of an \textit{ideal,
incompressible, viscous fluid} subject to some external forces,
namely the Navier Stokes equation in three dimensional Euclidian
space $\mathbb{R}^3$, which, in our adopted notation, reads as
follows:
\begin{equation}\label{EulerusEqua}
    \frac{\partial}{\partial t}\mathbf{u}\, + \, \mathbf{u}\cdot \nabla \mathbf{u} \, =
     \, - \, \nabla p  \, + \, \nu \, \Delta \,\mathbf{u} \, + \, \mathbf{f} \quad ;
     \quad \nabla \cdot \mathbf{u} \, = \, 0
\end{equation}
In equation (\ref{EulerusEqua}), $\mathbf{u} \, = \,
\mathbf{u}\left( x \, , \,t\right)$ denotes the local velocity
field, $p(\mathbf{x},t)$ denotes the local pressure field, $\nu$ is
viscosity and $\mathbf{f}$ is the external force, if it is
introduced. The symbol $\Delta\, = \,\sum_{i=1}^3
\,\frac{\partial^2}{\partial x_i^2}$ stands for the standard
laplacian. In vector notation eq. (\ref{EulerusEqua}) takes the
following form:
\begin{equation}\label{EulerusEquaVec}
    \frac{\partial}{\partial t} u^i\, + \, u^j \, \partial_j \, u^i
    \, = \, - \, \partial^i \, p  \, + \, \nu \,\Delta u^i\, + \, f^i \quad ;
    \quad \partial^\ell  \, u_\ell\, = \, 0
\end{equation} and admits some straightforward rewriting that, notwithstanding
the kinder-garden arithmetic involved in its derivation, is at the
basis of several profound and momentous theoretical  developments
which have kept  the community of  dynamical system theorists busy
for already fifty years
\cite{arnoldus,Childress0,Childress,Henon,Dombre,ArnoldBook,Bogoyav0,Bogoyav,
arnoldorussopapero,FFMF,Dynamo,Gilbert,Etnyre2000,Ghrist2007}.
\par
Here we aim at extending to the case where $\nu \neq 0$ previous
results applying to the case of null viscosity, namely to Euler
equation. The scope, however, is more ample since, as we already
anticipated, we introduce a more direct reference to contact
structures and to the recent developments occurred in this field of
mathematics, where the notion of \textit{singular contact
structures} has been introduced
\cite{cardone2019,miranda2021,trasversozero,unochiuseinteg,guglielminoconmiranda,pollosingolare,singularwinstein,scappascappa}
to account, in particular for boundaries of a certain type
(cylindrical ends), which are potentially momentous for some
applications.
\par
The core of our paper is the group-theoretical approach, initiated
in \cite{Fre:2015mla} that brings into the classical field of
mathematical fluid-mechanics a brand new vision allowing for a more
systematic classification and  algorithmic construction of the so
named Beltrami flows, providing new insight into their properties.
Combining the group theoretical classification of Beltrami
(anti-Beltrami) fields and their generalized relation with
\textit{contact structures} possibly \textit{admitting
singularities} is one of  the promising follow up of our work. The
other, as we already stressed, is the general scheme for the
construction of exact or approximate solutions of Navier Stokes
equations with prescribed hidden symmetries and calculable Beltrami
spectra.
\par
The notion of {Universal Classifying Group} introduced in
\cite{Fre:2015mla} and mentioned in the introduction  is an
intrinsic property of the considered crystallographic lattice
$\Lambda$ and of its point group $\mathfrak{P}^{max}_\Lambda$,
which, by definition, is the maximal finite subgroup of
$\mathrm{SO(3)}$ leaving the lattice $\Lambda$ invariant.
\par
The reason why lattices and crystallography are brought into the
study of fluid dynamics is that  we focus on \textit{hydro-flows}
that are confined within some bounded domain, as it happens in a
large variety of cases of interest for technological applications
like industrial autoclavs, pipelines, thermal machines of various
kind, blood vessels in physiology, liquid helium micro-tubes in
superconducting magnets, chemical reactors with mechanical agitation
sytems and so on. The argument goes as follows. Solutions of partial
differential equations (PDE.s) like the NS-equation in
(\ref{EulerusEqua}), that encode the characterizing feature of being
confined to finite regions of space can be obtained essentially by
means of two alternative strategies:
\begin{description}
  \item[A)] By brutally imposing boundary conditions that simulate the
  walls of the chamber, tube, box or whatever else contains the
  flowing fluid. This strategy is the most direct and suitable for
  numerical computer aided integration of the PDE.s but it is hardly
  viable in the search of exact analytic solutions of the same PDE.s with the
  ambition of establishing some rational taxonomy.
  \item[B)] The use of periodic boundary conditions which amounts to
  restricting one's attention to a compact space $\mathcal{M}_3$ without boundary ($\partial \mathcal{M}_3 \, = \, 0$) as
  a mathematical model of the finite volume region of interest.
\end{description}
The use of alternative B) amounts to developing in some suitable
Fourier series some functions (the velocity components) that are not
necessarily periodic but which, on a bounded support, coincide with
periodic functions admitting a Fourier series development.
\par
This being clarified a systematic way of imposing periodic boundary
conditions  is the  identification of the $\mathcal{M}_3$ manifold
with a $\mathrm{T^3}$ torus obtained by modding $\mathbb{R}^3$ with
respect to a three dimensional lattice $\Lambda \subset
\mathbb{R}^3$:
\begin{equation}\label{metricT3}
 \mathcal{M}_3 \, = \,    \mathrm{T}^3_g \, = \, \frac{\mathbb{R}^3}{\Lambda}
\end{equation}
Abstractly the  lattice $\Lambda$ is a an abelian infinite group
isomorphic to $\mathbb{Z}\times \mathbb{Z} \times \mathbb{Z}$ which
is embedded in some way into $\mathbb{R}^3$. Using
eq.(\ref{metricT3}) the topological torus
\begin{equation}\label{tritorustop}
    \mathrm{T^3} \simeq \mathbb{S}^1 \times \mathbb{S}^1 \times \mathbb{S}^1
\end{equation}
comes out automatically equipped with a flat constant metric.
Indeed, according with (\ref{metricT3}) the flat Riemaniann space
$\mathrm{T}^3_g$ is defined as the set of equivalence classes with
respect to the following equivalence relation: $ {\mathbf{r}}^\prime
\, \sim \,  {\mathbf{r}} \quad \mbox{iff} \quad  {\mathbf{r}}^\prime
\, - \,  {\mathbf{r}} \, \in \, \Lambda $. The metric $g$ defined on
$\mathbb{R}^3$ is inherited by the quotient space and therefore it
endows  the topological torus (\ref{tritorustop}) with a flat
Riemaniann structure. Seen from another point of view the space of
flat metrics on  $\mathrm{T}^3$ is just the coset manifold
$\mathrm{SL(3,\mathbb{R})}/\mathrm{SO(3)}$ encoding all possible
symmetric matrices, alternatively all possible space lattices, each
lattice being spanned by an arbitrary triplet of basis vectors.
\paragraph{\bf Lattices} To make the above statement
precise let us consider the standard $\mathbb{R}^3$ manifold and
introduce a basis of three linearly independent 3-vectors that are
not necessarily orthogonal to each other and of equal length:
\begin{equation}\label{sospirone}
 {\mathbf{w}}_\mu \, \in \, \mathbb{R}^3 \quad \mu \, = \, 1,
\dots ,\, 3
\end{equation}
Any vector in $\mathbb{R}$ can be decomposed along such a basis and
we have:
\begin{equation}\label{xvec}
 {\mathbf{r}} \, = \,  r^\mu {\mathbf{w}}_\mu
\end{equation}
The flat, constant metric on $\mathbb{R}^3$ is defined by:
\begin{equation}\label{gmunu}
g_{\mu\nu} \, = \, \langle   {\mathbf{w}}_\mu \, , \,
 {\mathbf{w}}_\nu \rangle
\end{equation}
where $\langle \, ,\, \rangle$ denotes the standard euclidian scalar
product. The space lattice $\Lambda$ consistent with the metric
(\ref{gmunu}) is the free abelian group (with respect to the sum)
generated by the three basis vectors (\ref{sospirone}), namely:
\begin{equation}\label{reticoloLa}
\mathbb{R}^3 \, \ni \,    {\mathbf{q}}  \, \in \, \Lambda \,
\Leftrightarrow \,  {\mathbf{q}} \, = \, q^\mu \,
 {\mathbf{w}}_\mu \quad \mbox{where} \quad q^\mu \, \in \,
\mathbb{Z}
\end{equation}
\paragraph{\bf Dual lattices} Any time we are given a lattice in the sense of the definition (\ref{reticoloLa}) we obtain
a dual lattice $\Lambda^\star$ defined by the property:
\begin{equation}\label{reticoloLastar}
\mathbb{R}^3 \, \ni \,    {\mathbf{p}}  \, \in \, \Lambda^\star \,
\Leftrightarrow \, \langle  {\mathbf{p}} \, , \,
 {\mathbf{q}}\rangle \, \in \, \mathbb{Z} \quad \forall \,
 {\mathbf{q}}\, \in \, \Lambda
\end{equation}
A basis for the dual lattice is provided by a set of three
\textit{dual vectors} $ {\mathbf{e}}^\mu$ defined by the
relations\footnote{In the sequel for the scalar product of two
vectors we utilize also the equivalent shorter notation $
{\mathbf{a}}\, \cdot  {\mathbf{b}} \, = \, \langle
 {\mathbf{a}}\, \cdot  {\mathbf{b}}\rangle $}:
\begin{equation}\label{dualvecti}
    \langle  {\mathbf{w}}_\mu \, , \,  {\mathbf{e}}^\nu \rangle \, = \, \delta^\nu_\mu
\end{equation}
so that
\begin{equation}\label{pcompi}
\forall \,  {\mathbf{p}} \, \in \, \Lambda^\star \quad
 {\mathbf{p}} \, = \, p_\mu \,  {\mathbf{e}}^\mu \quad
\mbox{where } \quad p_\mu \, \in \, \mathbb{Z}
\end{equation}
According with such a definition it immediately follows that the
original lattice is always a subgroup of the dual lattice and
necessarily a normal one, due to the abelian character of both the
larger and smaller group:
\begin{equation}\label{tarragonese}
    \Lambda \, \subset \, \Lambda^\star
\end{equation}
\subsection{Rewriting of the equations of  hydrodynamics in a
geometrical set up} Let us then begin with the rewriting of
eq.(\ref{EulerusEquaVec}) which is the starting point of the entire
adventure. The first step to be taken in our raising conceptual
ladder is that of promoting the fluid trajectories, defined as the
solutions of the following first order differential
system\footnote{In mathematical hydrodynamics people distinguish two
notions, that of trajectories, which are the solutions of the
differential equations (\ref{streamlines}) and that of streamlines.
Streamlines are the instantaneous curves that at any time $t=t_0$
admit the velocity field $u^{i}(x,t_0)$ as tangent vector.
Introducing a new parameter $\tau$, streamlines at time $t_0$, are
the solutions of the differential system:
\begin{equation}\label{strimotti}
    \frac{d}{d\tau}x^i(\tau) \, = \, u^i(\mathbf{x}(\tau),t_0)
\end{equation}
In the case of steady flows where the velocity field is independent
from time, trajectories and streamlines coincide.}:
\begin{equation}\label{streamlines}
    \frac{d}{dt}x^i(t) \, = \, u^i(x(t),t)
\end{equation}
to smooth maps:
\begin{equation}\label{mappini}
    \mathcal{S} \, : \, \mathbb{R}_t \, \rightarrow \, \mathcal{M}_g
\end{equation}
from the time real line $\mathbb{R}_t $ to a smooth Riemaniann
manifold $\mathcal{M}_g$ endowed with a metric $g$. The classical
case corresponds to $\mathcal{M} \, = \, \mathbb{R}^3$, $g_{ij}(x)
\, = \, \delta_{ij}$, but any other Riemaniann three-manifold might
be used and there exist also generalizations to higher dimensions.
Adopting this point of view, the velocity field $\mathbf{u}\left( x
\, , \,t\right)$ is turned into a time evolving vector field on
$\mathcal{M}$ namely into a smooth family of sections of the tangent
bundle $\mathcal{T}\mathcal{M}$:
\begin{equation}\label{Ufildo}
    \forall \, t \, \in \, \mathbb{R} \, : \, u^i(x,t) \, \partial_i \,
    \equiv \, {\mathrm{U}}(t) \, \in \, \Gamma\left(\mathcal{T}\mathcal{M}\, , \, \mathcal{M}\right)
\end{equation}
Next, using the Riemaniann metric, which allows to raise and lower
tensor indices, with any ${\mathrm{U}}(t)$ we can associate a family
of sections of the cotangent bundle $\mathcal{T}^\star\mathcal{M}$
defined by the following time evolving one-form:
\begin{equation}\label{Omfildo}
  \forall \, t \, \in \, \mathbb{R} \, : \,    \Omega^{[\mathrm{U}]}(t)
  \, \equiv \, g_{ij} \, u^i(x,t) \, dx^j \, \in \, \Gamma\left(\mathcal{T}^\star\mathcal{M}\, , \, \mathcal{M}\right)
\end{equation}
Utilizing the exterior differential and the contraction operator
acting on differential forms, we can evaluate the Lie-derivative of
the one-form $\Omega^{[\mathrm{U}]}(t)$ along the vector field
$\mathrm{U}$. Applying definitions (see for instance
\cite{Fre:2013ika}, chapter five, page 120 of volume two) we obtain:
\begin{eqnarray}\label{trivialone}
\mathcal{L}_\mathrm{U} \Omega^{[\mathrm{U}]}(t) & \equiv & {\rm
i}_\mathrm{U}  \cdot {\rm d} \Omega^{[\mathrm{U}]} \, + \, {\rm d}
\left({\rm i}_\mathrm{U} \cdot \Omega^{[\mathrm{U}]} \right) \, = \,
\left ( u^\ell \partial_\ell \, u^i  \, + \, g^{i k}
\partial_{k} \, \underbrace{\parallel \mathrm{U} \parallel^2}_{g_{mn}
\, u^m \, u^n} \right)\, g_{ij} \, dx^j
\end{eqnarray}
and the Navier Stokes  equation can be rewritten in the the
following  index-free reformulation
\begin{eqnarray}
    - \, {\rm d } \left( p \, + \, \ft 12 \,\parallel \mathrm{U} \parallel^2\right ) & = &
    \partial_t \Omega^{[\mathrm{U}]} \, + \,  {\rm i}_\mathrm{U}
    \cdot {\rm d}  \Omega^{[\mathrm{U}]} \,  \, - \,\nu \,  \Delta_g \, \Omega^{[\mathrm{U}]}
    \, - \, \mathbf{f} \label{bernullini}
\end{eqnarray}
Where $\Delta_g$ is  the Laplace-Beltrami operator on 1-forms,
written in an index free notation as it follows:
\begin{equation}\label{laplacciobeltramo}
    \Delta_g \, = \, \delta\, \mathrm{d} \, + \, \mathrm{d} \, \delta \quad ;
    \quad\delta \equiv \star_g \, \mathrm{d} \star_g
\end{equation}
where with $\star_g$ we have denoted the Hodge duality operation in
the background of the metric $g$.
\par
Eq.(\ref{bernullini}) is one of the possible formulations of
classical Bernoulli theorem. To begin with, consider inviscid fluids
($\nu \, = \, 0$) with no external forces ($\mathbf{f} = 0$). Then
equation eq.(\ref{bernullini}) becomes:
\begin{equation}\label{bernullone}
- \, {\rm d } \left( p \, + \, \ft 12 \,\parallel \mathrm{U}
\parallel^2\right ) \, = \,
\partial_t \Omega^{[\mathrm{U}]} \, + \,  {\rm i}_\mathrm{U}
\cdot {\rm d}  \Omega^{[\mathrm{U}]} \,
\end{equation}
and from eq.(\ref{bernullini}) we immediately conclude that
\begin{equation}\label{finocchionabiscotta}
    H_B \, = \, p \, + \, \ft 12 \,\parallel \mathrm{U} \parallel^2
\end{equation}
is constant along the trajectories defined by
eq.(\ref{streamlines}). Turning matters around we can say that in
\textbf{steady flows} of inviscid free fluids, where
\begin{equation}\label{stedone}
\partial_t \,\Omega^{[\mathrm{U}]} \, = \,0
\end{equation}
the fluid trajectories necessarily lay on the level surfaces
$H_B(\mathrm{x})\, = \, h \, \in \, \mathbb{R}$ of the function:
\begin{equation}\label{ignobilia}
    H \, : \, \mathcal{M} \, \rightarrow \, \mathbb{R}
\end{equation}
defined by (\ref{finocchionabiscotta}) and hereafter named, as it is
traditional in Fluid Mechanics, the \textbf{Bernoulli function}.
\par
An identical conclusion can be reached in the case of non-vanishing
viscosity if the steady flow condition (\ref{stedone}) is replaced
by:
\begin{equation}\label{genstedona}
    \partial_t \,\Omega^{[\mathrm{U}]} \, = \,\nu \, \Delta_g\, \Omega^{[\mathrm{U}]} \, + \, \mathbf{f}
\end{equation}
For instance if at time $t=t_0$, the $1$-form
$\Omega^{[\mathrm{U}]}$ is the superposition of a collection of $N$
eigenstates of the Laplace-Beltrami operator:
\begin{equation}\label{zeroide}
    \Omega^{[\mathrm{U}]}\mid_{t=t_0}\, = \, \sum_{i=1}^N \, \omega_{i} \quad ;
    \quad \Delta_g\,\omega_{i}\, = \, \lambda_i \, \omega_i
\end{equation}
choosing a subset of such forms, say those from $i=1$ to $i=M < N$,
one can solve the condition (\ref{genstedona}) by setting the
driving force as follows:
\begin{equation}\label{laterano}
    \mathbf{f} \, = \, -\, \nu \, \sum_{i=1}^{M} \lambda_i \, \omega_i
\end{equation}
and the 1-form flow as follows:
\begin{equation}\label{carnitina}
\Omega^{[\mathrm{U}]}\, = \, \sum_{i=1}^{M} \, \omega_i \, +
\sum_{i=M+1}^N \, \omega_i \, \exp\left[-\lambda_i \, t\right]
\end{equation}
For viscid fluids, flows satisfying eq.(\ref{genstedona}) will be
referred to as \textbf{generalized steady flows}. It follows that in
the case of  steady and generalized steady flows the fluid
trajectories necessarily lay on the level surfaces
$H_B(\mathrm{x})\, = \, h \, \in \, \mathbb{R}$ of the Bernoulli
function (\ref{ignobilia}) defined by (\ref{finocchionabiscotta}).
\vskip 0.2cm
\subsubsection{Foliations} Then if $H_B(\mathbf{x})$ has a non trivial
$x$-dependence, locally, in open charts $\mathcal{U}_n\subset
\mathcal{M}_n$ of the considered  $n$-dimensional manifold, it
defines a natural foliation of such charts $\mathcal{U}_n$  into a
smooth family of $(n-1)$-manifolds (all diffeomorphic among
themselves) corresponding to the level surfaces.
\par
The global topological  and analytic structure of level  surfaces of
the Bernoulli function is the object of interesting recent
mathematical studies (see for instance \cite{cardonebotto}) that we
avoid addressing since the focus of the present discussion is only
local and heuristic since the $3$-dimensional manifolds eventually
considered in this paper are just flat, non singular torii
$\mathbb{R}^3/\Lambda$ where $\Lambda$ is a lattice. Then in the
mentioned open charts, as already advocated, the trajectories,
\textit{i.e.} the solutions of eq.(\ref{streamlines}), lay on these
surfaces. In other words the dynamical system encoded in
eq.(\ref{streamlines}) is effectively $(n-1)$-dimensional admitting
$H$ as an additional conserved hamiltonian. In the classical  case
$n\, = \, 3$ this means that the differential system
(\ref{streamlines}) is actually two-dimensional, namely non-chaotic
and in some instances even integrable\footnote{Here we rely on a
general result established by the theorem of Poincar\'e-Bendixson
\cite{PoincareNonchaos,bendix}on the limiting orbits of planar
differential systems  whose corollary is generally accepted to
establish e that two-dimensional continuous systems cannot be
chaotic.}. Consequently we reach the conclusion that no chaotic
trajectories (or streamlines) can exist in those domains where the
Bernoulli function $H_B(x)$ has a non trivial $x$-dependence: the
only window open for lagrangian chaos occurs in those domains where
$H_B$ is a constant function. Looking at
eq.s(\ref{bernullini}-\ref{bernullone}) we realize that the previous
argument implies that in steady and generalized steady flows,
 chaotic trajectories can occur only if   velocity
field satisfies the following condition:
\begin{equation}\label{cinesinotulipano}
    {\rm i}_\mathrm{U}  \cdot {\rm d}  \Omega^{[\mathrm{U}]} \, = \, 0
\end{equation}
This weak condition (\ref{cinesinotulipano}) is certainly satisfied
if the velocity field $\mathrm{U}$ satisfies the following strong
condition that is named \textbf{Beltrami equation}:
\begin{equation}
    {\rm d}  \Omega^{[\mathrm{U}]} \, = \, \lambda \, \star_g \, \Omega^{[\mathrm{U}]} \quad \Leftrightarrow \quad
   \star_g \, {\rm d}  \Omega^{[\mathrm{U}]} \, = \, \lambda \,  \Omega^{[\mathrm{U}]} \label{Belatramus}
\end{equation}
where $\star_g$, as already specified, denotes the Hodge duality
operator in the metric $g$:
\begin{eqnarray}\label{gongo}
    \star_g \, \Omega^{[\mathrm{U}]} &  = & \epsilon_{\ell m n} \, g^{\ell k}\, \Omega^{[\mathrm{U}]}_k \, dx^m \, \wedge \,dx^n
    \, =\,  u^\ell \, dx^m \, \wedge \,dx^n \, \epsilon_{\ell m n}\label{gruscia1}\\
    \star_g \, \mathrm{d} \, \Omega^{[\mathrm{U}]} &  = & \epsilon_{\ell m n} \, g^{m p}\,g^{n q}
    \partial_p \left( g_{qr} u^r\right)  \, dx^\ell \label{gruscia2}
\end{eqnarray}
\vskip 0.2cm
\subsubsection{Arnold theorem} The heuristic argument
which leads to  consider velocity fields that satisfy
\textit{Beltrami condition} (\ref{Belatramus}) as the unique steady
candidates compatible with chaotic trajectories was transformed by
Arnold into a rigorous theorem \cite{ArnoldBook} which, under the
strong hypothesis that $(\mathcal{M},g)$ is a closed, compact
Riemaniann three-manifold, states the following:

%%%%%%%%%%%%%%%%%%%%%%%%%%%%%%%%%%%%%%%%%%%%%%%%%%%%%%%%%%%%%%%%%%%
\begin{teorema}
\label{fundarnoltheo} Assume that a region $D\subset \mathcal{M}$ of
the considered \textbf{three-dimensional} Riemannian manifold $\left
(\mathcal{M},g\right)$ is bounded by a compact analytic surface and
that the velocity field $\mathbf{U}$ does not satisfy Beltrami
equation everywhere in $D$, namely $\Omega^{[\mathbf{U}]}\, \neq \,
\lambda \star_g \, d\Omega^{[\mathbf{U}]}$, where $\lambda \in
\mathbb{R}$ is a real number. Then the region of the flow can be
partitioned by an analytic submanifold into a finite number of
cells, in each of which the flow is constructed in a standard way.
Namely the cells are of two types: those fibered into tori invariant
under the flow and those fibered into surfaces invariant under the
flow, diffeomorphic to the annulus $\mathbb{R}\times \mathbb{S}^1$.
On each of these tori the flow lines are either all closed or all
dense, and on each annulus all the flow lines are closed.
\end{teorema}
As one sees, in steady flows, when the velocity vector field of the
fluid  is not a Beltrami field, then streamlines either lie on
surfaces that have the topology of torii or on surfaces that are
cylindrical. In both cases chaotic streamlines are excluded.
\par
%%%%%%%%%%%%%%%%%%%%%%%%%%%%%%%%%%%%%%%%%%%%%%%%%%%%%%%%%%%%%%%%%%%
\textit{Chaotic trajectories or streamlines} are of particular
interest, both from the point of view of  theory and of
applications, since, in many scenarios, chaotic flows are  desirable
in order either to homogenize the heath exchange between the fluid
and some external objects immersed in the flow, as it happens in
\textit{autoclavs}, or to promote the mixing of two different
fluids, like it happens in \textit{chemical reactors}. The examples
are multiple and the mentioned ones are just an illustration.
\par
On the other hand the chaotic trajectories are desirable in all
these applications at \textit{small scales} while on \textit{larger
scales} the fluid should appear as moving steadily in some given
direction. The intrinsic non linearity of the NS equation forbids
the linear combination of solutions as new solutions and the
superposition of different regimes at different scales is a very
difficult mathematical problem that requires specialized analysis.
\par
The desire to investigate the on-set of chaotic trajectories in
steady (or generalized steady) flows of incompressible fluids
motivated the interest of the  dynamical system community in
Beltrami vector fields defined by  the condition (\ref{Belatramus}).
Furthermore, in view of the above powerful theorem proved by Arnold,
the focus of attention concentrated on the mathematically very
interesting case  of compact three-manifolds. Within this class, the
most easily treatable case is that of flat compact manifolds without
boundary, so that the most popular playground turned out to be the
three torus $\mathrm{T^3}$, whose possible role in applications has
already been emphasized.  Certainly many physical contexts for fluid
dynamics do not correspond to the idealized situation of a motion in
a compact manifold or, said differently, periodic boundary
conditions are not the most appropriate to be applied either in a
river, or in the atmosphere or in the charged plasmas environing a
compact star, yet the message conveyed by Arnold theorem that
Beltrami vector fields play a distinguished role in chaotic behavior
is to be taken seriously into account and gives an important hint.
 In view of what we are going to discuss in section
\ref{viacontattosa} this hint is properly developed by considering
the one-to-one relation between Beltrami fields and contact
structures on three-manifolds that is now extended to contact
structures with singularities.
\subsection{The path leading to contact geometry}
\label{viacontattosa} Beltrami vector fields are intimately related
with the mathematical notion of \textit{contact geometry and
contact topology}.   As we have seen from our sketch of Arnold
Theorem, the main obstacle to the onset of chaotic trajectories has
a distinctive geometrical flavor: trajectories are necessarily
ordered and non-chaotic if the manifold where they take place has a
foliated structure $\Sigma_h \times \mathbb{R}_h$, the two
dimensional level sets $\Sigma_h$ being invariant under the action
of the velocity vector field $U$. In this case each streamline lays
on some surface $\Sigma_h$. Equally adverse to chaotic trajectories
is the case of \textit{gradient flows} where there is  a foliation
provided by the level sets of some function $H(x)$ and the velocity
field $\mathrm{U}=\nabla H$ is just the gradient of $H$. In this
case all trajectories are orthogonal to the leaves $\Sigma_h$ of the
foliation and their well aligned tangent vectors are parallel to its
normal vector.
\par
\begin{figure}[!hbt]
\begin{center}
%\iffigs
\includegraphics[height=40mm]{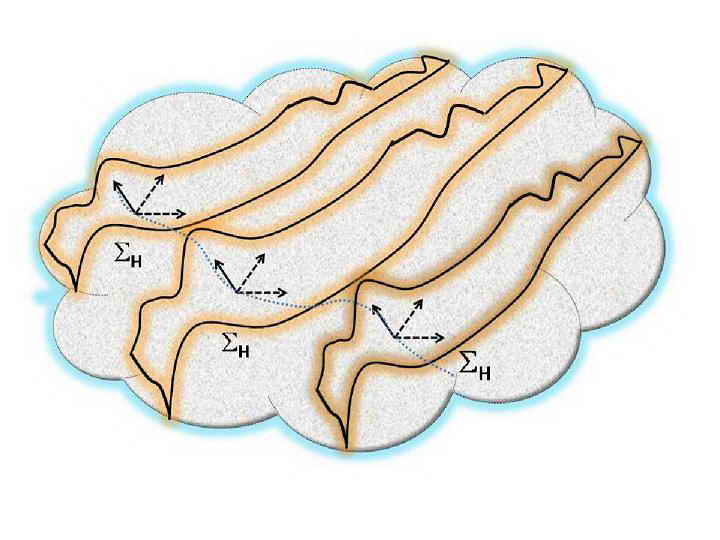}
%\else
\end{center}
 %\fi
\caption{\it  Schematic view of the foliation of a three dimensional
manifold $\mathcal{M}$. The family of two-dimensional surfaces
$\Sigma_h$ are typically the level sets $H(\mathbf{x}) =h$ of some
function $H\, : \, \mathcal{M} \, \rightarrow\, \mathbb{R}$. At each
point of $p\, \in \, \Sigma_h \subset \mathcal{M}$ the dashed
vectors span the tangent space $T_p^\parallel\Sigma_h$, while the
solid vector span the normal space to the surface
$T_p^\perp\Sigma_h$. Equally adverse to chaotic trajectories is the
case where the velocity field $\mathrm{U}$  lies in
$T_p^\perp\Sigma_h$ (gradient flow) or in $T_p^\parallel\Sigma_h$ }
\label{fogliattone}
 %\iffigs
 \hskip 1cm \unitlength=1.1mm
 %\end{center}
 % \fi
\end{figure}
In conclusion in presence of a foliation (or a local foliation) we
have the following decomposition of the tangent space to the
manifold $\mathcal{M}$ at any point $p \in \mathcal{M}$
 \begin{equation}\label{finchius}
    T_p\mathcal{M} \, = \, T_p^\perp\Sigma_h \, \oplus \, T_p^\parallel\Sigma_h
\end{equation}
and no chaotic trajectories are possible in a region $\mathfrak{S}
\subset \mathcal{M}$ where $\mathrm{U}(p) \, \in \,
T_p^\perp\Sigma_h $ or $\mathrm{U}(p) \, \in \,
T_p^\parallel\Sigma_h$ for $\forall p \, \in \, \mathfrak{S} $ (see
fig.\ref{fogliattone}).
\par
This matter of fact motivates an attempt to capture the geometry  of
the bundle of subspaces orthogonal to the lines of flow by
introducing  an intrinsic topological indicator that distinguishes
necessarily non-chaotic flows from possibly chaotic ones. Let us
first consider the extreme case of a gradient flow where
$\Omega^{[\mathrm{U}]} \, = \, \mathrm{d} H$ is an exact form. For
such flows we have:
\begin{equation}\label{canedellascala}
\Omega^{[\mathrm{U}]} \, \wedge \, \mathrm{d}\Omega^{[\mathrm{U}]}
\, = \, \Omega^{[\mathrm{U}]} \, \wedge \, \underbrace{\mathrm{d} \,
\mathrm{d}H}_{= \, 0} \, = \, 0
\end{equation}
Secondly let us consider the opposite case where the velocity field
$\mathrm{U}$ is orthogonal to a gradient vector field $\nabla H$ so
that the integral curves of $\mathrm{U}$  lay on the level surfaces
$\Sigma_h$. Furthermore let us assume that $\mathrm{U}$ is self
similar on neighboring level surfaces. We can characterize this
situation in a Riemaniann manifold $(\mathcal{M},g)$ by the
following two conditions:
\begin{equation}\label{gattoascensore}
{\rm i}_{\nabla H}\Omega^{[\mathrm{U}]} \, \Leftrightarrow \,
g\left(\mathrm{U}\, ,\, \nabla H \right) \, = \, 0 \quad ; \quad
\left[ U\, , \, \nabla H\right] \, = \, 0
\end{equation}
The first of eq.s(\ref{gattoascensore}) is obvious. To grasp the
second it is sufficient to introduce, in the neighborhood of any
point $p\, \in \, \mathcal{M}$, a local coordinate system composed
by $(h,x^\parallel )$ where $h$ is the value of the function $H$ and
$x^\parallel $ denotes some local coordinate system on the level set
$\Sigma_h$.  The situation we have described corresponds to assuming
that:
\begin{equation}\label{caruccio}
    \mathrm{U} \, \simeq \, U^{\parallel} (x^\parallel) \, \partial_\parallel  \quad ; \quad \partial_h
    \, U^{\parallel} (x^\parallel) \, = \, 0
\end{equation}
Under the conditions spelled out in eq.(\ref{gattoascensore}) we can
easily prove that:
\begin{equation}\label{suschione}
    {\rm i}_{\nabla H} \, \mathrm{d}\Omega^{[\mathrm{U}]} \, = \,0
\end{equation}
Indeed from the definition of the Lie derivative we obtain:
\begin{eqnarray}\label{cincischiando}
    {\rm i}_{\nabla H} \, \mathrm{d}\Omega^{[\mathrm{U}]} \, = \, \underbrace{\mathcal{L}_{\nabla H}
    \,\Omega^{[\mathrm{U}]}}_{= \, \Omega^{\left[[U\, ,\, \nabla H]\right]} \, = \, 0} \, - \,
    \mathrm{d}\left(\underbrace{{\rm i}_{\nabla H}\Omega^{[\mathrm{U}]}}_{= 0}\right)
\end{eqnarray}
Since we have both ${\rm i}_{\nabla H}\Omega^{[\mathrm{U}]} \, = \,
0$ and ${\rm i}_{\nabla H}\mathrm{d}\Omega^{[\mathrm{U}]} \, = \, 0$
it follows that also in this case:
\begin{equation}\label{franceschiello}
    \Omega^{[\mathrm{U}]} \, \wedge \, \mathrm{d}\Omega^{[\mathrm{U}]} \, = \, 0
\end{equation}
Therefore in order not to exclude chaotic trajectories one has to
assume that
\begin{equation}\label{rebomba}
    \Omega^{[\mathrm{U}]} \, \wedge \, \mathrm{d}\Omega^{[\mathrm{U}]} \, \neq \, 0
\end{equation}
and the above condition is what leads us to \textit{contact
geometry}.
\section{Geometrical Foundations}\label{geomfund}
In this section we just summarize some  definitions and theorems of
basic differential geometry that we shall later utilize or quote, for
their conceptual relevance in the development of our original
arguments.
\subsection{Contact Geometry}
Contact Geometry is both an old and a relatively new chapter of
Mathematics, since it springs from some classical results of
analysis that date back to Darboux, Goursat, Lie and other XIX
century maitres, yet it has been vigorously developed in the last
two decades from a relatively small community of mathematicians. To
say it in short, \textit{Contact Geometry} is a mathematical theory
aiming at providing an intrinsic geometrical-topological
characterization of \textit{non integrability}.
\par
Contact Geometry deals exclusively with real \textit{Differential
Manifolds $\mathcal{M}_{2n+1}$ of odd-dimension} and on the other
hand it has a symbiotic relation with  \textit{Symplectic Manifolds
$\mathcal{S}_{2n+2}$ and $\mathcal{S}_{2n}$} in the two adjacent
even dimensions, upper and lower.
\par
In the present concise summary  we closely follow the excellent
review \cite{contatoregeiges}.
\subsubsection{Contact structures}
We consider an odd dimensional differential manifold
$\mathcal{M}_{2n+1}$  its tangent bundle $ \mathcal{TM}_{2n+1} \,
\stackrel{\pi}{\longrightarrow} \, \mathcal{M}_{2n+1}$ whose space
of sections
$\Gamma\left[\mathcal{TM}_{2n+1},\mathcal{M}_{2n+1}\right]$ is
composed by vector fields, whose local description is in terms of
first order differential operators $\mathbf{X} \, = \, X^\mu
\left(x\right) \,\frac{\partial}{\partial x^\mu} $ and the cotangent
bundle $ \mathcal{T}^\star\mathcal{M}_{2n+1} \,
\stackrel{\pi_\star}{\longrightarrow} \, \mathcal{M}_{2n+1}$  whose
space of sections
$\Gamma\left[\mathcal{T}^\star\mathcal{M}_{2n+1},\mathcal{M}_{2n+1}\right]$
is composed by differential 1-forms $ \omega \, = \, \omega_\mu
\left(x\right)\,dx^\mu \ $. A hyperplane bundle is a reduction of
the tangent bundle where the fibres over each point constitute a
codimension one vector subspace of the tangent space in the same
point,  the transition functions being accordingly derived:
\begin{eqnarray}\label{iperpiano}
    \mathcal{HY}\,
    \stackrel{\mathcal{P}}{\longrightarrow} \, \mathcal{M} & ;
    & \forall p \in \mathcal{M} \, , \,
    \mathcal{P}^{-1}(p)
    \subset\pi^{-1}(p) \quad \text{where} \quad \mathcal{T}\mathcal{M} \, \stackrel{\pi}{\longrightarrow} \,
    \mathcal{M}\nonumber\\
    \mathrm{dim}_\mathbb{R}\mathcal{M}\, = \, m & ;&\mathrm{dim}_\mathbb{R}\pi^{-1}(p) \, = \, m \quad ; \quad
    \mathrm{dim}_\mathbb{R}\mathcal{P}^{-1}(p) \, = \, m-1
\end{eqnarray}
A simple  way of constructing a hyperplane bundle is by means of the
choice of a section of the cotangent bundle namely of some 1-form
$\omega \in
\Gamma\left[\mathcal{T}^\star\mathcal{M},\mathcal{M}\right]$. Then
the hyperplane sub-bundle $\mathcal{HY}^\omega \subset \mathcal{TM}$
of the tangent bundle is implicitly defined by stating what is the
space of its sections
$\Gamma\left[\mathcal{HY}^\omega,\mathcal{M}\right]$, namely
mentioning all the possible vector fields that are sections of
$\mathcal{HY}^\omega$. Utilizing a precise mathematical language let
$\mathbf{X}\in \Gamma\left[\mathcal{TM},\mathcal{M}\right]$ be a
vector field, we write
\begin{equation}\label{kernello}
    \mathbf{X} \, \in \, \Gamma\left[\mathcal{HY}^\omega,\mathcal{M}\right] \quad
    \text{iff} \quad \mathbf{X} \, \in \,\mathrm{ker}\, \omega \quad
    \textit{i.e.} \quad \omega\left(\mathbf{X}\right) \equiv 0 \quad
    (\text{everywhere})
\end{equation}
\begin{definizione}
Given a manifold $\mathcal{M}_{2n+1}$ of odd dimension, a
\textbf{contact structure} on $\mathcal{M}_{2n+1}$ is a rank $2n$
sub-bundle $\xi \, \stackrel{\mathcal{P}}{\longrightarrow}\,
\mathcal{M}_{2n+1}$ of the tangent bundle $\mathcal{TM}_{2n+1} \,
\stackrel{\pi}{\longrightarrow}\, \mathcal{M}_{2n+1}$ that can be
identified with the hyperplane bundle $\mathcal{HY}^\alpha$ where
the 1-form $\alpha$ satisfies the following condition:
\begin{equation}\label{latisanaditiglio}
    \alpha \, \wedge \, \underbrace{\mathrm{d}\alpha \, \wedge \,\dots \, \, \wedge
    \,\mathrm{d}\alpha}_{n-\text{times}} \, \neq \, 0 \quad
    (\text{everywhere on $\mathcal{M}_{2n+1}$})
\end{equation}
The $1$-form $\alpha$ is named a \textbf{contact form}.
\end{definizione}
\begin{definizione}
A \textbf{contact manifold} is a pair
$\left(\mathcal{M}_{2n+1},\xi\right)$ of an odd dimensional manifold
and a contact structure $\xi \,
\stackrel{\mathcal{P}}{\longrightarrow}\, \mathcal{M}_{2n+1}$.
\end{definizione}
Few relevant observations are in order in relation with the above
two definitions. The first is that the same contact structure can be
defined by several different contact forms $\alpha$,
$\alpha^\prime$,$\dots$. Indeed all multiples of a given contact
form $\alpha$ through a scalar, nowhere vanishing, function
$\lambda\, : \, \mathcal{ M}_{2n+1}\, \to \, \mathbb{R}$ define the
same contact structure. Secondly it is quite possible that the same
odd--dimensional manifold $\mathcal{ M}_{2n+1}$ can admit more that
one contact structure. The classification of these latter, modulo
trivial diffeomeorphisms, is an interesting and relevant
mathematical problem. It is therefore mandatory to single out the
notion of \textbf{contactomorphism}.
\begin{definizione}
Let $\left(\mathcal{ M},\xi\right)$ and $\left(\mathcal{
N},\chi\right)$ be two contact-manifolds and let:
\begin{equation}\label{diffeomorfus}
    \varphi \, : \, \mathcal{M} \, \longrightarrow \,\mathcal{N}
\end{equation}
be a diffeomorphism of the former on the latter manifold (obviously
$\mathcal{M}$ and $\mathcal{N}$ must have the same dimension in
order for $\varphi$ to possibly exist). Let $\alpha$ be a contact
form defining $\xi$ and let $\beta$ be a contact form defining
$\chi$. The considered diffeomorphism $\varphi $ is named a
\textbf{contactomorphism} if and only if:
\begin{equation}\label{pullabacco}
    \varphi^\star\left(\beta\right) \, = \, \lambda \, \, \alpha
\end{equation}
where $\varphi^\star$ is the pull-back map and
\begin{equation}\label{nowhere}
    \lambda \, : \, \mathcal{M} \, \longrightarrow \, \mathbb{R}
\end{equation}
is a nowhere vanishing real function on the contact manifold
$\mathcal{M}$. If a contactomorphism exists between them, the two
considered contact manifolds are named \textbf{contactomorphic}.
\end{definizione}
In the above definition the manifold $\mathcal{M}$ and $\mathcal{N}$
might be the same. In this case what we are actually considering is
the transformation by means of a diffeomorphism of a contact
structure into another one by means of a contactomorphism.
\begin{definizione}
Given two contact structures $\xi$ and $\chi$ on the same manifold
$\mathcal{M}_{2n+1}$ they are to be identified as the same if there
exists a \textit{contactomorphism} that maps one into the other.
\end{definizione}
In conclusion the relevant mathematical problem is that of
classifying contact structures on $\mathcal{M}_{2n+1}$ modulo
contactomorphisms.
\subsubsection{Integrability and Frobenius Theorem}
We refrain here from providing a detailed discussion of Frobenius theorem about
integrability. We shall limit ourselves to sketch the basic concepts underlying its
formulation. One just begins with the observation that every vector
field $\mathbf{X}$ on a manifold $\mathcal{M}$ of whatever
dimension defines its own integral curves $\mathcal{I}_\mathbf{X}$,
namely those curves that at any of their points admit the local
value of the vector field $\mathbf{X}$ as tangent vector. Since any
point $p\in \mathcal{M}$ lies on some integral curve
$\mathcal{I}_\mathbf{X}$, we are guaranteed that a single vector
field induces a \textit{foliation} of the manifold $\mathcal{M}$
into one-dimensional submanifolds. It is more tricky to establish
whether a sub-bundle of the tangent bundle $\mathcal{E}
\longrightarrow \mathcal{M}$ of rank  $r > 1$ induces or not a
foliation of $\mathcal{M}$. In this case, to say it in a
not-completely rigorous, yet intuitive and qualitatively correct
way, by \textit{foliation} we mean the covering of the manifold with
a family of \textit{leaves}, namely of  sub-manifolds diffeomorphic
among themselves,  $\mathcal{F}_{\pmb{\nu}} \subset \mathcal{M}$ of
dimension equal to the rank $r$ of the sub-bundle $\mathcal{E}$,
each of which can be thought as the level set hypersurface  for $r$
functions $u_i(p)$ ($i=1,\dots,r$) that originate from the
integration of a basis of sections $\mathbf{X}_i$ of the sub-bundle
$\mathcal{E} \longrightarrow \mathcal{M}$.
\begin{eqnarray}\label{sfogliatellanapoletana}
    \mathcal{F}_{\pmb{\nu}} & = & \left\{p \, \in \mathcal{M} \,
    \mid \, u_i(p) \, = \,\nu_i \right\} \quad ; \quad \pmb{\nu}\,
    \equiv\, \{\nu_1, \dots \nu_r\} \quad ; \quad \nu_i\, = \,
    \textit{real constants}\nonumber\\
    \nabla u_i(p) &=& \mathbf{X}_i\mid_p
\end{eqnarray}
When the above situation is realized one says that the sub-bundle
$\mathcal{E} \longrightarrow \mathcal{M}$ is \textbf{integrable}.
\par
\textbf{Frobenius theorem} establishes the necessary condition for
such integrability.
\begin{teorema} Let $\mathcal{M}$ be a manifold and $\mathcal{E}
\longrightarrow \mathcal{M}$ a sub-bundle of its tangent bundle of
rank  $r > 1$. The necessary and sufficient condition for
$\mathcal{E}$ to be integrable is that:
\begin{equation}\label{frobeniata}
    \forall\, \mathbf{X},\mathbf{Y}  \in  \Gamma[\mathcal{E},\mathcal{M}] \quad :
    \quad
    \, \left[\mathbf{X},\mathbf{Y} \right] \, \in \, \Gamma[\mathcal{E},\mathcal{M}]
\end{equation}
\end{teorema}
In the case where $\mathcal{E} \longrightarrow \mathcal{M}$ is an
hyperplane bundle defined by a 1-form $\omega$ Frobenius
integrability condition can also be formulated as:
\begin{equation}\label{frombolino}
    \omega \, \wedge \, d\omega \, = \, 0
\end{equation}
This shows that a contact structure defined by a contact form is the
exact opposite of an integrable sub-bundle. Indeed one might show
that it corresponds to maximal non-integrability.
\subsubsection{Isotropic submanifolds of a contact manifold and non integrability}
We begin with the following
\begin{definizione}
Let $\left(\mathcal{M}_{2n+1}\, , \, \xi\right)$ be a contact
manifold and $\mathcal{L}\subset\mathcal{M}_{2n+1}$ a submanifold.
Consider the tangent bundle of such a submanifold
$\mathcal{TL}\,\stackrel{\pi_{\tau}}{\longrightarrow}\, \mathcal{L}$
and the contact structure bundle $\xi
\stackrel{\pi_\xi}{\longrightarrow}\, \mathcal{M}$. The submanifold
$\mathcal{L}$ is named \textbf{isotropic} if and only if
\begin{equation}\label{isotropicamente}
    \forall p \in \mathcal{L} \quad  : \quad \pi_{\tau}^{-1}(p) \,
    \subset \, \pi_{\xi}^{-1}(p)
\end{equation}
Equivalently, if the contact structure $\xi$ is defined by the
contact-form $\alpha$, the sub-manifold $\mathcal{L}$ is
\textbf{isotropic} if any vector field $\mathbf{X}$ tangent to
$\mathcal{L}$,  is also in $\text{ker}\,\alpha$:
\begin{equation}\label{cundoiso}
\mathbf{X}\in \Gamma\left[\mathcal{TL},\mathcal{L}\right] \quad
\Rightarrow \quad \alpha(\mathbf{X}) \, = \, 0
\end{equation}
\end{definizione}
We  introduce the additional definition
\begin{definizione}
Let $\left(\mathcal{M}_{2n+1}\, , \, \xi\right)$ be a contact
manifold and $\tilde{\mathcal{M}}_{2m+1} \subset \mathcal{M}_{2n+1}$
an odd dimensional submanifold of codimension $2(n-m)\geq 0$. Let
$\alpha$ be the contact one form defining the contact structure
$\xi$ and $\iota$ the inclusion map:
\begin{equation}\label{includendoiota}
    \iota \quad : \quad \tilde{\mathcal{M}}_{2m+1} \,
    \longrightarrow \, \mathcal{M}_{2n+1}
\end{equation}
Then $\left(\tilde{\mathcal{M}}_{2m+1},\chi\right)$ is named a
\textbf{contact submanifold} of $\left(\mathcal{M}_{2n+1}\, , \,
\xi\right)$ if the contact structure $\chi$ on
$\tilde{\mathcal{M}}_{2m+1}$ is defined by the \textbf{contact-form}
$\iota^\star \, \alpha$, in other words if:
\begin{equation}\label{cariota}
    \chi \, = \, {\mathrm{ker}} \, \iota^\star\,\alpha
\end{equation}
\end{definizione}
The main reason why contact geometry is relevant for chaotic flows
in fluid dynamics streams from the following
\begin{teorema}
\label{barbagianni} Let $\left(\mathcal{M}_{2n+1},\xi\right)$ be a
contact manifold in $2n+1$-dimensions and
$\mathcal{L}\subset\mathcal{M}_{2n+1}$ an isotropic submanifold.
Then $\mathrm{dim}\, \mathcal{L} \leq n$.
\end{teorema}
In order to prove theorem \ref{barbagianni} we need first the
following
\begin{lemma}
Let  $\left(\mathcal{M}_{2n+1},\xi\right)$ be a contact manifold
whose contact structure $\xi$ is defined as  $\mathrm{ker}
\,\alpha$, in terms of the contact 1-form $\alpha$. Because of the
defining condition $\ref{latisanaditiglio}$ it follows that
$\mathrm{d}\alpha\mid_\xi \neq 0$ and for every point $p\in
\mathcal{M}_{2n+1}$ the $2n$-dimensional fibre $\xi_p \subset T_p
\,\mathcal{M}_{2n+1}$ is a vector-space equipped with a
skew-symmetric 2-form of maximal rank (no-zero eigenvalues) exactly
provided by the restriction to $\xi_p$ of $\mathrm{d}\alpha$
\textit{i.e.} $\mathrm{d}\alpha\mid_{\xi_p}$. Hence the contact
structure is a \textit{symplectic bundle} with respect to the 2-form
$\mathrm{d}\alpha\mid_\xi$.
\end{lemma}
\begin{proofteo}
{\rm In order to prove the theorem, consider the inclusion map: $
\iota \quad : \quad \mathcal{L} \, \longrightarrow \,
\mathcal{M}_{2n+1} $ and consider the pull-back of the contact form
on the isotropic manifold. By definition of isotropy
$\iota^\star\alpha = 0$. Hence we have also $\iota^\star
\,\mathrm{d}\alpha = 0$. At each point $p\in \mathcal{L}$, the
tangent space $\mathcal{T}_p \mathcal{L}$ is a subspace of the
symplectic space $\xi_p$ on which the symplectic 2-form vanishes
$\mathrm{d}\alpha\mid_{\xi_p}$. From elementary linear algebra it
follows that such a subspace has at most one-half of the dimension
of $\xi_p$. Indeed it suffices to put the skew 2-form in canonical
form:
$
    \left(\begin{array}{c|c}
    \mathbf{0}_{n\times n}& \mathbf{1}_{n\times n}\\
    \hline
    -\mathbf{1}_{n\times n} &\mathbf{0}_{n\times n} \\
    \end{array}\right)\nonumber\\
$
and the statement becomes evident. This proves the theorem
 $\blacksquare$.}
\end{proofteo}
\par
What are the consequences of this theorem? It states that if we have
a contact structure $\xi$, induced by a contact form $\alpha$, then
we can exclude a foliation of the contact manifold into
hypersurfaces $\Sigma_h \subset \mathcal{M}_{2n+1}$ of codimension
1:
\begin{equation}\label{sfoliazione}
  \mathcal{M}_{2n+1}\, \backsimeq \, \Sigma_h \times \mathbb{R}_h
\end{equation}
such that for each $h\in \mathbb{R}$ the  tangent bundle of
$\Sigma_h$ is comprised within the contact structure. Indeed if that
happened each leave $\Sigma_h$ of the foliation would be an
isotropic submanifold of dimension $2\times n$ which is what the
theorem forbids.
\begin{definizione}
An isotropic submanifold $\mathcal{L} \subset \mathcal{M}_{2n+1}$ of
maximal possible dimension, namely $n$, of a contact manifold in
dimensions $2n+1$, is named a \textbf{Legendrian submanifold}.
\end{definizione}
Furthermore
\begin{definizione}\label{ribatriestina} Associated with a contact form $\alpha$ one has
the so called \textbf{Reeb vector field}  $\mathbf{R}_\alpha$,
defined by the two conditions:
\begin{eqnarray}\label{Ribbo}
&&\alpha\left(\mathbf{R}_\alpha\right) \, = \, \lambda(\mathbf{x})
\quad =
 \quad
 \text{nowhere vanishing function on $\mathcal{M}_{2n+1}$}\nonumber\\
 &&\forall \mathbf{X}\in \Gamma\left[\mathcal{TM}_{2n+1},\mathcal{M}_{2n+1}\right] \quad :
 \quad
 \mathrm{d}\alpha\left(\mathbf{R}_\alpha,\mathbf{X}\right)\, = \,
 0
\end{eqnarray}
\end{definizione}
If the contact manifold $\mathcal{M}_{2n+1}$ is equipped with a
Riemannian metric $g$, then the contact 1-form $\alpha$ and its Reeb
field $\mathbf{R}_\alpha$ are related one to the other by the
raising and lowering of indices. Suppose that we start from the Reeb
field:
\begin{equation}\label{ribbone}
    \mathbf{R} \, = \, R^\mu \,
    \frac{\partial}{\partial x^\mu}
\end{equation}
The corresponding $\alpha$ is obtained by setting:
\begin{equation}\label{contactribbo}
    \alpha \, = \, \Omega^{[\mathbf{R}]} \,  \equiv \, g_{\mu\nu} R^\mu \,
    dx^\nu
\end{equation}
and the contact structure condition (\ref{latisanaditiglio}) is
turned into the following condition on the Reeb field components:
\begin{equation}\label{gridolino}
    \epsilon^{\lambda \mu_1\nu_1\mu_2\nu_2\dots\mu_n\nu_m} \,
    R_\lambda \, \partial_{\mu_1}\,R_{\nu_1}
    \,\partial_{\mu_2}\,R_{\nu_2}\, \dots\,
    \partial_{\mu_n}\,R_{\nu_n}\, \neq \, 0 \quad \text{nowhere vanishes}
\end{equation}
On the contrary, if one begins with the contact form $\alpha$, the
components of the Reeb field are obtained by setting:
\begin{equation}\label{crinolina}
    R_\alpha^\mu \, = \, g^{\mu\nu} \, \alpha_\nu
\end{equation}
Note that the nowhere vanishing function $\lambda$ mentioned in the
definition \ref{ribatriestina} is just the squared norm of the Reeb
field or of the contact form which coincide:
\begin{equation}\label{normaquadrata}
    \lambda \, =\, \parallel \mathbf{R}\parallel^2 \, = \, \parallel
    \Omega^{[\mathbf{R}]}\parallel^2 \, \equiv \, g_{\mu\nu} \, R^\mu \,
    R^\nu
\end{equation}
\subsubsection{Contact structures in $D=3$ and hydro-flows}
Let us now consider the case relevant for fluid dynamics, namely
that of three dimensional contact manifolds $\left(\mathcal{M}_3\,
,\,\xi_\alpha\right)$, where, in the notation $\xi_\alpha$, we
mention the contact form $\alpha$ defining the contact structure.
The consequence of theorem \ref{barbagianni} is that in such contact
manifolds, the Legendrian submanifolds are all 1-dimensional, namely
they are \textit{curves} or, as it is customary to name them in the
present context, \textit{knots}.
\par
Hence in three dimensions there are two kind of knots, the
\textbf{Legendrian knots} whose tangent vector belongs to
$\text{ker}\, \alpha$ and the \textbf{transverse knots} whose
tangent vector is parallel to the Reeb field at each point of the
trajectory.
\par
\begin{figure}[!hbt]
\begin{center}
\includegraphics[height=90mm]{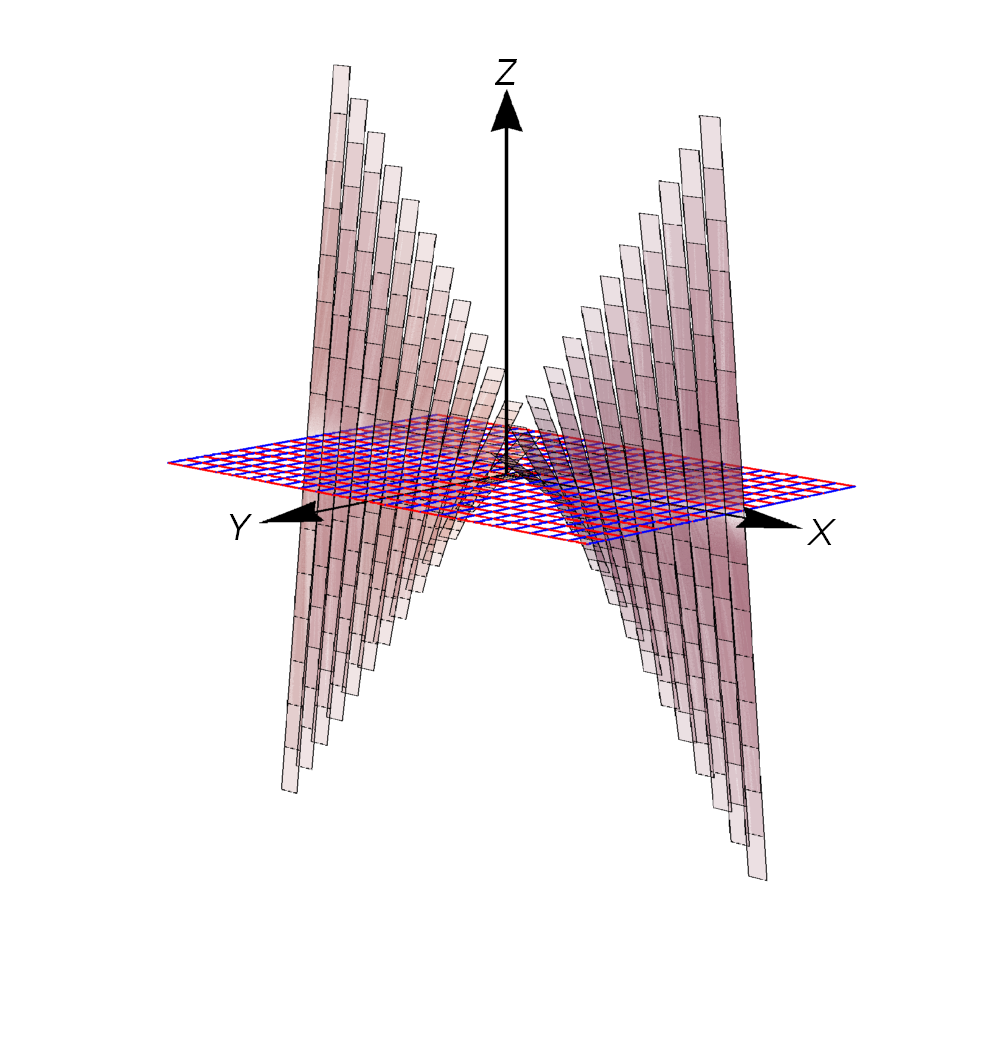}
\end{center}
\caption{\it  Schematic view of  the standard contact structure
$\mathbb{R}^3$} \label{berlucchino}
 \hskip 1cm \unitlength=1.1mm
\end{figure}

Furthermore in D=3 the condition (\ref{gridolino}) becomes:
\begin{equation}\label{urlonotturno}
    \epsilon^{\lambda\mu\nu}  \,R_\lambda \, \partial_\mu \, R_\nu
    \, \neq \, 0
\end{equation}
\paragraph{The standard contact structure on $\mathbb{R}^3$.} The
flat Euclidian space in three dimensions whose coordinates we denote
as $x,y,z$ is endowed with a standard contact structure that admits
the following contact form:
\begin{equation}\label{romildino}
    \alpha_s \, = \, \mathrm{d}z+ x \,\mathrm{d}y
\end{equation}
A picture of the local planes defining the contact structure
(\ref{romildino}) is shown in fig.\ref{berlucchino}.
\subsubsection{Relation with Beltrami vector fields}
As wee see the main reason to introduce the contact form  conception
is that, so doing one liberates the notion of a vector field capable
to generate chaotic trajectories from the use of any  metric
structure. A vector field $U$ is potentially interesting for chaotic
regimes if it is a Reeb  field for at least one contact form
$\alpha$. In this way the mathematical theorems about the
classification of contact structures modulo diffeomorphisms
(theorems that are metric-free and of topological nature) provide
new global methods to capture the topology of hydro-flows.
\par
Instead if we work in a Riemaniann manifold endowed with a metric
$(\mathcal{M},g)$ we can always invert the procedure and define the
contact form $\alpha$ that can admit $\mathrm{U}$ as a Reeb vector
field by identifying
\begin{equation}\label{gominato}
    \alpha \, = \, \Omega^{[\mathrm{U}]}
\end{equation}
In this way the first of the two conditions (\ref{Ribbo}) is
automatically satisfied: ${\rm i}_\mathrm{U}\Omega^{[\mathrm{U}]} \,
= \, \parallel \mathrm{U}\parallel^2 \, > 0$. It remains to be seen
whether $\Omega^{[\mathrm{U}]}$ is indeed a contact form, namely
whether $\Omega^{[\mathrm{U}]} \, \wedge \, \mathrm{d}
\Omega^{[\mathrm{U}]} \, \ne \,0$ and whether the second condition
${\rm i}_\mathrm{U} \, \mathrm{d} \Omega^{[\mathrm{U}]} \, = \, 0$
is also satisfied. Both conditions are automatically fulfilled if
$\mathrm{U}$ is a Beltrami field, namely if it is an eigenstate of
the operator $\star_g \, \mathrm{d}$ as advocated in
eq.(\ref{Belatramus}). Indeed the implication ${\rm i}_\mathrm{U} \,
\mathrm{d} \Omega^{[\mathrm{U}]} \, = \, 0$ of Beltrami equation was
shown  in eq. (\ref{cinesinotulipano}), while from the Beltrami
condition it also follows:
\begin{eqnarray}\label{cirimellaG}
 \Omega^{[\mathrm{U}]} \, \wedge \, \mathrm{d} \Omega^{[\mathrm{U}]}
 & = &\Omega^{[\mathrm{U}]} \, \wedge \, \star_g \Omega^{[\mathrm{U}]}
 \, = \, \parallel \mathrm{U}\parallel^2 \, \mathrm{Vol}  \, \ne \, 0
 \quad ; \quad
 \mathrm{Vol} \,  \equiv \, \frac{1}{3!} \,
 \times \, \epsilon_{ijk} \, dx^i \, \wedge \, dx^j \, \wedge \, dx^k
\end{eqnarray}
In this way the conceptual circle closes and we see that all
Beltrami vector fields can be regarded as Reeb  fields for a
bona-fide contact form. Since the same contact structure (in the
topological sense) can be described by different contact forms, once
Beltrami fields have been classified it remains the task to discover
how many inequivalent contact structures they actually describe. Yet
it is reasonable to assume that every contact structure has a
contact form representative that is derived from a Beltrami Reeb
field. Indeed a precise correspondence is established by a theorem
proved in \cite{Ghrist2007}:
\begin{teorema}
\label{ghristheo} Any rotational Beltrami vector field on a
Riemaniann $3$-manifold is a Reeb  field for some contact form.
Conversely any Reeb field associated to a contact form on a
$3$-manifold is a rotational Beltrami field for some Riemaniann
metric. Rotational Beltrami field means an eigenfunction of the
$\star_g \mathrm{d}$ operator corresponding to a non-vanishing
eigenvalue $\lambda$.
\end{teorema}
\subsubsection{Darboux's theorem}
We finally mention, without providing its proof that can be found in
\cite{contatoregeiges}, a  classical theorem named after Darboux,
which shows that the standard contact structure on $\mathbb{R}^3$
displayed in eq.(\ref{romildino}) and graphically shown in
fig.\ref{berlucchino} is not just a choice, rather it corresponds to
the  canonical local form of any contact structure on any contact
manifold.
\begin{teorema}
\label{darbuso} Let $\left(\mathcal{M}_{2n+1},\xi\right)$ be an
$(2n+1)$-dimensional contact manifold and $\alpha$ a contact
$1$-form defining $\xi = \text{ker}\,\alpha$. Let $p\in
\mathcal{M}_{2n+1}$ be any point of the manifold and $U \subset
\mathcal{M}_{2n+1}$ an open neighborhood of $p$. Then we can always
find a local homomorphism: $\varphi \, : \, U \, \to \,
\mathbb{R}^{2n+1}$ such that, naming
$\{y_0,x_i,y_i\}$,$(i=1,\dots,n)$ the coordinates on $\varphi(U)
\subset \mathbb{R}^{2n+1}$ we obtain:
\begin{equation}\label{localU}
    \alpha\mid_{U} \, = \, dy_0 \, + \,\sum_{i=1}^n \,x_i\,dy_i
\end{equation}
\end{teorema}
In the case $n=1$ eq.(\ref{localU}) reproduces eq.(\ref{romildino}).
Hence for all three-dimensional contact manifolds $\mathcal{M}_3$
that in (\ref{romildino}) is the universal local form of the contact
$1$-form $\alpha$.
\subsection{$\mathfrak{b}$-Contact Geometry and Singular Beltrami Fields}
\label{bmanstory}
%%%%%%%%%%%%%%%%%%%%%%%%%%%%%%%%%%%%%%
As we emphasized in the introduction, the main difficulty in solving
NS or Euler equations comes from the non-linearity of the transport
term which forbids the generic linear superposition of solutions, a
limited superposition being possible, within the landscape approach
to be discussed in section \ref{panoramica}, with Beltrami fields
belonging to the same spherical layer. As we stressed, Beltrami
fields are essential, via their relation with contact structures, in
order to create the possibility of chaotic streamlines at small
scales, yet they are defined on compact manifolds without boundary,
in particular on torii, while the geometry of physical systems of
relevance for applications is certainly  not that of torii, rather
that of finite portions of $\mathbb{R}^3$ delimited by boundaries,
like finite 3D cylinders. Furthermore at larger scales, the fluids
of interest for applications should present a non-chaotic behavior
similar to that of the Poiseuille flow (see for instance
\cite{wangsteady}). How could we try to reconcile the two
conflicting needs? A new window of opportunity opens up with the
relatively new set up of Beltrami fields in
$\mathfrak{b}$-manifolds, which  can be viewed as compact manifolds
with boundaries. In this section we make a short review of this new
approach which, as already stressed, we desire to combine with our
group theoretical classification of Beltrami fields. Essentially we
collect the main definitions and concepts developed in particular by
Victor Guillemin, Eva Miranda, Robert Cardona, Daniel Peralta Salas
and other collaborators in
\cite{cardone2019,miranda2021,trasversozero,unochiuseinteg,guglielminoconmiranda,pollosingolare},
having, as main goal, that of discussing the example of the
$\mathfrak{b}$-modified ABC flow\footnote{see section
\ref{abcperbambini} for the definition of ABC flows} presented in
\cite{miranda2021}. Such a discussion will be done, in view of the
underlying group theoretical structures, in section \ref{bABCfildo}.
\par
In order to introduce the $\mathfrak{b}$-generalization of contact
manifolds we have first to set the stage by recalling essential
facts and definitions about symplectic manifolds and Poisson
structures.
\subsubsection{Symplectic and Poisson Manifolds}
We begin with
\begin{definizione}
A symplectic manifold is a pair
$\left(\mathcal{SM}_{2n+2},\omega\right)$ of a smooth manifold
$\mathcal{SM}_{2n+2}$ in even dimension $2n+2$ and a $2$-form
$\omega$ which is closed and non degenerate of maximal rank:
\begin{equation}\label{simplicioformo}
    \mathrm{d}\omega \, = \, 0 \quad ; \quad
    \omega\wedge\omega\wedge \dots \wedge \omega \, \neq 0 \quad
    \text{everywhere on $\mathcal{SM}_{2n+2}$}
\end{equation}
\end{definizione}
On a symplectic manifold we have a naturally defined antisymmetric
quadratic form on the space of sections of the tangent bundle,
\textit{i.e.} on the vector fields:
\begin{eqnarray}\label{tardicchio}
    \omega &: & \Gamma\left[\mathcal{TSM}_{2n+2}\, , \,
    \mathcal{SM}_{2n+2}\right] \times \Gamma\left[\mathcal{TSM}_{2n+2}\, , \,
    \mathcal{SM}_{2n+2}\right] \, \longrightarrow \,
    C^{(\infty)}\left(\mathcal{SM}_{2n+2}\right)\nonumber\\
   \forall X,Y\, &\in & \Gamma\left[\mathcal{TSM}_{2n+2}\, , \,
    \mathcal{SM}_{2n+2}\right] \quad , \quad  \omega\left(X,Y\right) \,
    \in\,
    C^{(\infty)}\left(\mathcal{SM}_{2n+2}\right)
\end{eqnarray}
Poisson manifolds are instead defined as follows.
\begin{definizione}
\label{poissone} A Poisson manifold
$\left(\mathcal{PM}_m,\{,\}\right)$ is the pair of a smooth manifold
$\mathcal{PM}_m$ of dimension $m$ and a Poisson bracket $\{,\}$
which is a binary operation on the space of smooth functions on the
manifold:
\begin{equation}\label{pesceparente}
    \{,\} \quad : \quad C^{(\infty)} \left(\mathcal{PM}_m\right)
    \, \times \, C^{(\infty)} \left(\mathcal{PM}_m\right)\,
    \longrightarrow \, C^{(\infty)} \left(\mathcal{PM}_m\right)
\end{equation}
satisfying the following three properties:
\begin{description}
  \item[1)] Antisymmetry $\quad \{f\, ,\, g\} \, = \, - \, \{g\, ,\, f\}
  $, $\quad\forall f,g \in C^{(\infty)} \left(\mathcal{PM}_m\right)$
  \item[2)] Jacobi Identity $\quad \{f\, , \, \{g \, ,\, h\}\}+\{g\, , \, \{h \, ,\, f\}\}+\{h\, , \, \{f \, ,\, g\}\}\, = \,
  0$, $\quad \forall f,g,h \in C^{(\infty)} \left(\mathcal{PM}_m\right)$
  \item[3)] Leibniz rule $\quad \{f\, , \, g.h\} \, = \, \{f\, , \,
  g\} \, h \, + \, g \,\{f\, , \,
  h\}$,  $\quad \forall f,g,h \in C^{(\infty)} \left(\mathcal{PM}_m\right)$
\end{description}
\end{definizione}
The first two properties mentioned in the definition \ref{poissone}
guarantee that the space of functions on the Poisson manifold
becomes a Lie algebra once equipped with the Poisson bracket. On the
other hand the third property implies that to each function $f \in
C^{(\infty)}$ the Poisson bracket associates a derivation of the
commutative algebra of functions on the manifold, namely, by
definition a vector field $\mathbf{X}_f$, named the
\textbf{hamiltonian vector field} of $f$.
\par
Locally, in any coordinate patch $\{x_1,\dots,x_j\}$, the Poisson
bracket takes the following form:
\begin{equation}\label{localedelpesce}
    \{f,g\} \, = \,\pi^{ij}(x)\,\frac{\partial f}{\partial x^i}\, \frac{\partial g}{\partial
    x^j} \quad ; \quad \pi^{i,j}(x) \, = \, - \, \pi^{ji}(x)
\end{equation}
where the controvariant antisymmetric tensor $\pi^{ij}(x)$ is
usually called a \textbf{bivector}. The hamiltonian vector field
$\mathbf{X}_f$ is then easily identified:
\begin{equation}\label{carneadehamilt}
    \mathbf{X}_f \, = \, \pi^{ij}\partial_i f \, \partial_j
\end{equation}
Let us now suppose that the dimension of the Poisson manifold is
even $m=2n+2$ and that the bivector $\pi^{ji}(x)$ is an everywhere
invertible matrix. Setting: $ \omega \, = \, \pi^{-1}_{k\ell} dx^k
\wedge dx^\ell $ we obtain a symplectic 2-form of maximal rank which
is closed as a consequence of the Jacobi identities satisfied by the
bivector. In this way the Poisson manifold is recognized to be a
symplectic manifold. In particular we can set:
\begin{equation}\label{agniziono}
    \{f,g\} \, = \, \omega\left(\mathbf{X}_f,\mathbf{X}_g\right)
\end{equation}
\begin{definizione}
\label{liuvillocampillo} Let
$\left(\mathcal{SM}_{2n+2},\omega\right)$ be a symplectic manifold.
A \text{Liouville vector field} is a vector field that leaves the
symplectic form $\omega$ invariant, namely:
\begin{equation}\label{taleggio}
  \omega\, = \,  \mathcal{L}_X\omega \, \equiv \, i_X\,
  \mathrm{d}\omega \, + \, \mathrm{d} (i_X\omega)\,,
\end{equation}
where $\mathcal{L}_X$ denotes the Lie derivative along the vector
field.
\end{definizione}
\subsubsection{Relation between symplectic and contact manifolds}
Let us consider a symplectic manifold $\left(
\mathcal{SM}_{2n+2},\omega\right)$ and let us assume that it admits
at least one Liouville vector field $\mathbf{L}$. Let moreover
$\Sigma_{\mathbf{L}}\subset \mathcal{SM}_{2n+2} $ be a hypersurface
transverse to the Liouville vector field $\mathbf{L}$. Then we
realize that $\Sigma_{\mathbf{L}}$ is a contact manifold with
contact form $\alpha \, = \,i_{\mathbf{L}}\omega$. If
$\Sigma_{\mathbf{L}}$ is transverse to $\mathbf{L}$ the form
$\alpha$ vanishes on $\mathbf{L}$ and no-where vanishes on
$T\Sigma_{\mathbf{L}}$. To verify that it is indeed a contact form
we just have to compute:
\begin{eqnarray}\label{gelso}
    &\null&\alpha\wedge \underbrace{\mathrm{d}\alpha \wedge \dots \wedge
    \mathrm{d}\alpha}_{n-\text{times}} \, = \, i_\mathbf{L}\omega\wedge\underbrace{\mathrm{d} i_\mathbf{L}\omega \wedge \dots \wedge
    \mathrm{d} i_\mathbf{L}\omega}_{n-\text{times}}\,=\,i_\mathbf{L}\omega\wedge\underbrace{\omega \wedge \dots \wedge
    \omega}_{n-\text{times}}\, = \, \frac{1}{n+1} i_\mathbf{L}\left(\underbrace{\omega \wedge \dots \wedge
    \omega}_{(n+1)-\text{times}}
    \right)\, = \, \text{Vol}_{\Sigma_\mathbf{L}}\nonumber\\
\end{eqnarray}
The last equation is true because $\omega^{n+1}$ is the volume form
of the ambient symplectic manifold and the hypersurface
$\Sigma_\mathbf{L}$ is by hypothesis transverse to the Liouville
vector field.
\par
Conversely given a contact manifold
$\left(\mathcal{M}_{2n+1},\xi\right)$ with contact form $\alpha$ and
Reeb field $\mathbf{R}$, any hypersurface $\Sigma \subset
\mathcal{M}_{2n+1}$ which is transverse to the Reeb field
$\mathbf{R}$ automatically acquires the structure of a symplectic
manifold with symplectic form $\tilde{\omega}\, = \,
d\alpha\mid_\Sigma$.
\par
Hence we can have odd-dimensional contact manifolds that sit in
between two symplectic manifolds of adjacent dimensions as shown in
the following diagram:
\begin{equation}\label{diagrammus}
\begin{array}{ccccc}
    \left(\mathcal{SM}_{2n},\tilde{\omega}=d\alpha\right)
    &\stackrel{\iota}{\hookrightarrow}
    &\left(\mathcal{M}_{2n+1},\alpha=i_\mathbf{L}\omega\right)
    &\stackrel{\iota}{\hookrightarrow}
    & \left(\mathcal{SM}_{2n+2},\omega\right)\\
    \Downarrow &\null& \Downarrow &\null & \Downarrow\\
    \text{symplectic}&\null& \text{contact}& \null &
    \text{symplectic}\\
\text{transverse to Reeb field}&\null& \text{transverse to Liouville
field}& \null &
    \null\\
    \end{array}
\end{equation}
The scheme described in eq.(\ref{diagrammus}) reminds that occurring
with Sasaki manifolds that sit in between two K\"ahler manifolds
which, indeed, are special instances of symplectic manifolds, the
symplectic form being the K\"ahler 2-form.
\subsubsection{$\mathfrak{b}$-Manifolds}
Having recalled for reader's ease the above concepts and definitions
we come to our main goal that is the definition of
$\mathfrak{b}$-manifolds. Following \cite{trasversozero,cardone2019}
we set:
\begin{definizione}
\label{bmanif} A $\mathfrak{b}$-manifold is a pair
$(\mathcal{M},\Sigma)$ where $\mathcal{M}$ is a differentiable
manifold and $\Sigma \subset \mathcal{M}$ is a hypersurface, namely
a submanifold of codimension one.
\end{definizione}
Given two $\mathfrak{b}$-manifolds $(\mathcal{M},\Sigma)$ and
$(\mathcal{N},\Pi)$ one defines as follows a smooth
$\mathfrak{b}$-map between them.
\begin{definizione}
A smooth map
\begin{equation}\label{mappas}
    f \quad : \quad \mathcal{M} \longrightarrow \mathcal{N}
\end{equation}
is a $\mathfrak{b}$-map:
\begin{equation}\label{bimappas}
    \null^\mathfrak{b}f \quad : \quad (\mathcal{M},\Sigma) \,\longrightarrow\,(\mathcal{N},\Pi)
\end{equation}
if $f$ is transverse to $\Pi$ and $f^{-1}(\Pi) \, = \, \Sigma$
\end{definizione}
With this setup one can re-establish all the basic ingredients of
differential geometry in the $\mathfrak{b}$-version. We begin with
vector fields.
\begin{definizione}
A $\mathfrak{b}$-vector field on $\mathfrak{b}$-manifold
$(\mathcal{M}_{m+1},\Sigma_m)$ is a vector field
$\null^{\mathfrak{b}}\mathbf{X}$ which is tangent to the
hypersurface $\Sigma_m$ in all $p\in \Sigma_m$
\end{definizione}
In an open neighborhood $U \subset \mathcal{M}_{m+1}$  that contains
the point $p\in \Sigma$ we can choose the coordinates in the
following way. Let $\sigma(x_0,x_1,\dots,x_{m})$ be the function
whose vanishing defines the surface $\Sigma_m$ in that neighborhood.
We can trade one of the standard coordinates $x_i$, say  $x_0$, for
the value $s=\sigma(x_0,\dots,x_{m+1})$ of the function, regarding
the remaining ones  $\mathbf{x}=\{x_1,\dots,x_m\}$ as coordinates on
the hypersurface $\Sigma$. Using such coordinate frame a vector
field parallel to the surface is of the form:
\begin{equation}\label{parallelone}
    \null^{\mathfrak{b}}\mathbf{X}\, =\, s
    X_0(\mathbf{x})\frac{\partial}{\partial s} + \sum_{i=1}^m
    X^i(s,\mathbf{x}) \frac{\partial}{\partial x^i}
\end{equation}
One can easily check that under standard commutation the
${\mathfrak{b}}$-vector fields form a Lie subalgebra of the Lie
algebra of vector fields. They can be considered the sections of a
new vector-bundle on $\mathcal{M}_{m+1}$ that we name the
$\mathfrak{b}$-tangent bundle:
$\null^{\mathfrak{b}}T\mathcal{M}_{m+1}\,\stackrel{\pi}{\longrightarrow}\,{\mathcal{M}}_{m+1}$.
This being established the road easily climbs down. We obtain the
the $\mathfrak{b}$-cotangent bundle by usual duality.
\par
In practice as shown in \cite{guglielminoconmiranda} the
$\mathfrak{b}$ de-Rham complex is structured as follows. A $k$-form
$\null^{\mathfrak{b}}\omega\,\in\,\null^{\mathfrak{b}}\Omega^k(\mathcal{M})$,
namely a section of the $k$-th external power of the cotangent
bundle $\null^{\mathfrak{b}}T^\star \mathcal{M}$ can always be
written as:
\begin{equation}\label{alfabetone}
\null^{\mathfrak{b}}\omega \, = \,    \frac{ds}{s}\wedge \alpha \, +
\, \beta \quad ; \quad \alpha\,\in\,\Omega^{k-1}(\mathcal{M})\quad ;
\quad
    \beta\,\in\,\Omega^{k}(\mathcal{M})
\end{equation}
Furthermore in \cite{guglielminoconmiranda} it is stated and shown
that although $\alpha,\beta$ are not unique in the bulk of the
manifold they are unique at every point $p\, \in \, \Sigma$ on the
distinguished surface or boundary.
\par
This provides an algorithmic tool to perform the $b$-deformation of
any given Riemannian metric on a given manifold $\mathcal{M}$.
\par
Relevant to our goals is the $\mathfrak{b}$-generalization of the
definition of contact manifolds.
\begin{definizione} Let $(\mathcal{M},\Sigma)$ be a
$(2n+1)$-dimensional, $\mathfrak{b}$-manifold. A
$\mathfrak{b}$-contact structure is the kernel of a
$\mathfrak{b}$-one-form $\alpha\, \in \,
\null^{\mathfrak{b}}T^\star\mathcal{M}$ that satisfies the
condition:
\begin{equation}\label{ravanone}
    \alpha \wedge \mathrm{d}\alpha \wedge \dots \wedge \mathrm{d}\alpha\, \neq \, 0
\end{equation}
In this case $\alpha$ is a $\mathfrak{b}$-contact form and $\xi \, =
\, \text{ker}\,\alpha$ is a $\mathfrak{b}$-contact structure.
\end{definizione}
As in the un-deformed case we can introduce the $\mathfrak{b}$-Reeb
field as that particular $\mathfrak{b}$-vector field $\mathbf{R}$
which satisfies the two conditions:
\begin{equation}\label{ribbetto}
    i_\mathbf{R}\cdot \mathrm{d}\alpha \, = \, 0 \quad ; \quad
    \alpha(\mathbf{R}) \, = \, 1
\end{equation}
As we are going to see in section \ref{bABCfildo}, the use of
$\mathfrak{b}$-deformations can introduce modified Beltrami fields
that are parallel to certain boundaries. The open deep question that
is touched upon and put into evidence in chapter \ref{dippo} is that
the choice of an allowed distinguished surface $\Sigma$ seem to
depend on the group structure of the Beltrami field one wants to
$\mathfrak{b}$-deform. Up to the knowledge of the authors this
aspect was not so far discussed in the literature. It appears to be
a very momentous question worth an in depth investigation.
%%%%%%%%%%%%
\section{Harmonic Analysis and the Algorithm}
\subsection{Beltrami equation and harmonic analysis}
In the present section which is partly based on a corresponding
section of \cite{Fre:2015mla}, partly new, we stress that all the
arguments presented above have been instrumental to enlighten the
role of Beltrami vector fields from various viewpoints related with
hydrodynamics and lagrangian chaos. Let us now consider from a more
general point of view Beltrami equation (\ref{Belatramus}). The one
here at stake  is the case $p=1$ of an eigenvalue equation that can
be written in any $(2\, p + 1)$-dimensional Riemaniann manifold
$\left(\mathcal{M}_p \, , \, g\right)$, namely:
\begin{equation}\label{choucrut}
    \star_g  \mathrm{d}\omega^{(p)} \, = \, \lambda \, \omega^{(p)}
\end{equation}
The eigenfunctions of the $\star_g  \mathrm{d}$ operator are 1-forms
for $p=1$, namely in three-dimensions, but they are higher
differential forms in higher odd dimensions. Another particularly
interesting case is that of $7$-manifolds where the eigenfunctions
of $\star_g  \mathrm{d}$ are three-forms and can be related with a
$\mathrm{G_2}$-structure of the manifold
\cite{englert,miol168,Cerchiai:2018shs}. On the other hand the
relation encoded in theorem \ref{ghristheo} between
eq.(\ref{choucrut}) and contact structures, as they are defined in
current mathematical literature, is true only for $p=1$ and it is
lost for higher $p$. Indeed contact structures are always defined in
terms of a contact one-form $\alpha$ and by the condition:
\begin{equation}\label{salsicciafresca}
    \alpha \, \wedge \, \underbrace{\mathrm{d}\alpha \,
    \wedge \,\mathrm{d}\alpha \, \dots \, \mathrm{d}\alpha}_{p-\mbox{times}} \, \ne \, 0
\end{equation}
Hence the problem of determining the spectrum and the eigenfunctions
of the operator $\star_g  \mathrm{d}\omega^{(p)}$  is a general one
and can be addressed in the same way in all odd-dimensions, yet its
relation with flows and contact-structures is peculiar to $d=3$ and
has not a general significance.  In any case it is absolutely  clear
that once the correspondence of theorem \ref{ghristheo} has been
established, the classification of Beltrami fields is reduced to a
classical problem of differential geometry whose solution can be
derived within a time honored framework which makes no reference to
trajectories and contact structures.
\par
The framework we refer to is that of \textit{harmonic analysis} on
compact Riemaniann manifolds $\left( \mathcal{M},g\right)$  and its
application to the spectral analysis of Laplace-Beltrami  operators
(for reviews see the book \cite{castdauriafre} and the articles
\cite{univer,spectfer,Fabbri:1999mk,Fre1999xp,gualto}). As
thoroughly discussed in the quoted references there are, on a
Riemann manifold $\left( \mathcal{M},g\right)$, several invariant
differential operators, generically named Laplace-Beltrami some of
which are of second order, some other of first order. They act on
the sections of vector bundles $E \rightarrow \mathcal{M}$  of
different rank, for instance the tangent bundle, the bundle of
$p$-forms, the bundle of symmetric two tensors, the spinor bundle
etc. Among the first order operators the most important ones are the
Dirac operator acting on sections of the spinor bundle and the
$\star_g \mathrm{d}$-operator acting on $p$-forms in a $(2 \, p
+1)$-dimensional manifold. The spectrum of all Laplace-Beltrami
operators is sensitive both to the topology and to the metric of the
underlying manifold. Each eigenspace is organized into irreducible
representations of the isometry group $\mathrm{G}$ of the metric $g$
and the eigenfunctions assigned to a particular representation are
generically named \textit{harmonics}.
\par
Here comes an important distinction in relation with the nature of
the group $\mathrm{G}$. If $\mathrm{G}$  is a Lie group and if the
manifold $\mathcal{M}$ is homogeneous under its action, than
$\mathcal{M} \sim \mathrm{G/H}$ where $\mathrm{H}\subset \mathrm{G}$
is the stability subgroup of some reference point $p_0 \, \in \,
\mathcal{M}$. In this case harmonic analysis reduces completely to
group-theory and the spectrum of any Laplace-Beltrami operator can
be derived in pure algebraic terms without ever using any
differential operations. In the case $\mathrm{G}$ is not a Lie group
and/or $\mathcal{M}$ is not homogeneous under its action, then
matters become more complicated and \textit{ad hoc} techniques have
to be utilized case by case to analyze the spectrum of invariant
operators.

\subsubsection{Harmonic analysis on the $\mathrm{T^3}$ torus and the Universal Classifying Group}
\label{introT3} The reasons to compactify Arnold-Beltrami flows on a
$\mathrm{T}^3$ have already been discussed and we do not resume the
issue. We just observe that $\mathbb{R}^3$ is a non-compact coset
manifold so that harmonic analysis over $\mathbb{R}^3$ is a
complicated matter of functional analysis. After compactification,
namely after imposing periodic boundary conditions, things
drastically simplify.
\par
Firstly, as we explained above  the compactification is obtained by
quotienting $\mathbb{R}^3$ with respect to a discrete subgroup of
the translation group which constitutes a lattice (see
eq.(\ref{metricT3})).
\par Secondly we implement the programme of
harmonic analysis by presenting a general algorithm to construct
solutions of the Beltrami equation which utilizes as main ingredient
the orbits under the action of the point group
$\mathfrak{P}_\Lambda$ of three-vectors in the momentum lattice
$^\star\Lambda$ that is just the dual of the lattice $\Lambda$. In
the language of crystallography the point group  is just the
discrete subgroup $\mathfrak{P}_\Lambda \subset \mathrm{SO(3)}$ of
the rotation group which maps the lattice $\Lambda$ and its dual
$^\star\Lambda$ into themselves:
\begin{equation}\label{puntarellona}
    \mathfrak{P}_\Lambda \, \Lambda \, = \, \Lambda \quad ; \quad \mathfrak{P}_\Lambda \, ^\star\Lambda\, = \, ^\star\Lambda
\end{equation}
In the case of the cubic lattice, that is the main example studied
in paper \cite{Fre:2015mla} we have $\mathfrak{P}_{\mathrm{cubic}}\,
= \, \mathrm{O_{24}}$ where $\mathrm{O_{24}} \sim \mathrm{S_4}$ is
the proper octahedral group of order $|\mathrm{O_{24}}|\, = \, 24$.
In the case of the hexagonal lattice which was only briefly touched
upon in \cite{Fre:2015mla} and which  instead we analyze in depth in
the present work, the point group is the dihedral group
$\mathrm{Dih_6}$ of order $|\mathrm{Dih_{6}}|\, = \, 12$.
\par
Thirdly, as it was originally  conceived and introduced for the
first time in \cite{Fre:2015mla},  a general argument, inspired by
the logic that crystallographers used to derive and classify space
groups, leads to introduce a large finite group
$\mathfrak{UG}_\Lambda$, named by the authors of \cite{Fre:2015mla}
the \textit{Universal Classifying Group for the Lattice $\Lambda$},
made out of discretized rotations and translations that are defined
by the structure of $\Lambda$. All eigenfunctions of the $\star_g
\mathrm{d}$-operator can be organized into a finite number of
classes and each class decomposes in a specific unique way into the
irreducible representations of $\mathfrak{UG}_\Lambda$. Hence all
Arnold-Beltrami vector fields are in  correspondence with the irreps
of $\mathfrak{UG}_\Lambda$. Knowing the branching rules of such
irreps with respect to its various subgroups $\mathrm{H_i} \subset
\mathfrak{UG}_\Lambda$ and selecting the identity representation one
obtains Arnold-Beltrami vector fields invariant with respect to
those $\mathrm{H_i}$ for which we are able to find an identity irrep
$D_1$ in the branching rules. In this way we can classify all Arnold
Beltrami flows and also uncover their \textit{hidden symmetries}.
\par
Such a conclusion was already reached in \cite{Fre:2015mla}.
\par
As we recalled above, the authors of \cite{Fre:2015mla}  considered
in an extensive way the case of the cubic lattice and constructed
the corresponding Universal Classifying Group $\mathfrak{UG}_{cubic}
\, = \, \mathrm{G_{1536}}$. This latter is a finite group  of order
$|\mathrm{\mathrm{G_{1536}}}| \, = \, 1536$ which was studied in
full detail in \cite{Fre:2015mla}. All of its 37 irreducible
representations were derived and the associated character table was
also constructed. A large class of its subgroups $\mathrm{H_i
}\subset \mathrm{\mathrm{G_{1536}}}$ were also singled out and each
of them was studied systematically, by constructing their irreps and
character tables. This allowed the derivation of all the
\textit{branching rules} of the 37 irreps of
$\mathrm{\mathrm{G_{1536}}}$ with respect to the considered
subgroups which were displayed in dedicated tables in the appendices
of \cite{Fre:2015mla}. In the present paper one of the goals is that
of providing the same group theoretical lore for the case of the
hexagonal lattice which in \cite{Fre:2015mla} was only briefly
touched upon and sketched.
\par
Since the crystallographic lattices are more than two one might
think that covering these two cases is only part of the work.  It is
not so. Mastering the Universal Classifying Groups for the cubic and
hexagonal lattices is sufficient to provide the entire picture.
Indeed the crystallographic lattices in D=3 subdivide just in two
classes:
\begin{description}
  \item[A)] The lattices whose basis vectors $\mathbf{w}_\lambda$ provide an orthogonal
  basis (although not necessarily orthonormal):
  \begin{equation}\label{ortobambo}
    \left(\mathbf{w}_\lambda \, , \, \mathbf{w}_\mu\right) \, = \,
    a_\lambda^2 \, \delta_{\lambda\mu}
  \end{equation}
where $\delta_{\lambda\mu}$ is the Kronecker delta and $a_\lambda$
is the lattice spacing in direction $\lambda\, = \, 1,2,3$.
  \item[B)] The lattices whose basis vectors $\mathbf{w}_\lambda$
  are arranged as follows:
  \begin{eqnarray}\label{hexabambo}
    \left(\mathbf{w}_1 \, , \, \mathbf{w}_1\right)\, = \,\left(\mathbf{w}_2 \, , \,
    \mathbf{w}_2\right) & = & a^2 \quad ; \quad
    \left(\mathbf{w}_3 \, , \,
    \mathbf{w}_3\right) \, = \, b^2\nonumber\\
\left(\mathbf{w}_1 \, , \, \mathbf{w}_2\right)& = & a^2 \, \cos
\left[\frac{2\pi}{3} \right] \nonumber\\
\left(\mathbf{w}_1 \, , \, \mathbf{w}_3\right)\, =
\,\left(\mathbf{w}_2 \, , \,
    \mathbf{w}_3\right) & = & 0
  \end{eqnarray}
  $a$  being the lattice spacing in each horizontal plane spanned by
  $\mathbf{w}_{1,2}$
  which is endowed with a hexagonal tesselation and $b$ the lattice
  spacing in the third vertical direction.
\end{description}
The point groups pertaining to the lattices of class A) are:
\begin{equation}\label{Aclasspointgroups}
    \mathfrak{P}_{\Lambda_A} \, = \, \left ( \mathrm{C_2, \, C_4, \, Dih_2,
    \, Dih_4, \, T_{12}, \, O_{24}} \right )
\end{equation}
where $\mathrm{C_n}$ denotes the cyclic group of order $\mathrm{n}$,
$\mathrm{Dih_m}$ denotes the dihedral group of order $\mathrm{m}$
and $\mathrm{T_{12}}$ is the tetrahedral group, while
$\mathrm{O_{24}}$ is the already mentioned octahedral group. All the
point groups in the list (\ref{Aclasspointgroups}) are subgroups of
the maximal one $\mathrm{O_{24}}$.
\par
The point groups pertaining to the lattices of class B) are:
\begin{equation}\label{Bclasspointgroups}
    \mathfrak{P}_{\Lambda_B} \, = \, \left ( \mathrm{C_3, \, C_6, \, Dih_3,
    \, Dih_6} \right )
\end{equation}
All the point groups in the list (\ref{Bclasspointgroups}) are
subgroups of the maximal one $\mathrm{Dih_6}$.
\par
This fact has the important consequence that the Universal
Classifying Group for the cubic lattice contains as subgroups the
Universal Classifying Groups for all the other lattices of class A),
while the Universal Classifying group for the hexagonal lattice
contains as subgroups all the Universal Classifying Groups for the
lattices of class B). Since, as we explain below, the construction
of Beltrami fields is organized into irreps of such classifying
groups, once we have the algorithm for the largest group we have
also that for all its subgroups.
\par
In the case of the cubic lattice, the main result of
\cite{Fre:2015mla} was the proof that the $\mathrm{O_{24}}$ orbits
in the cubic lattice arrange into $48$ equivalence classes, the
parameters of the corresponding Beltrami vector fields  filling all
the 37 irreducible representations of $\mathrm{\mathrm{G_{1536}}}$.
\subsubsection{The classical ABC flows} \label{abcperbambini}
 The following vector field:
\begin{equation}\label{bagcigaluppi}
  \mathbf{u}(x,y,z) \, = \, \mathcal{V}^{(ABC)}(x,y,z) \, \equiv \, \left(
\begin{array}{l}
 C \cos (2 \pi  y)+A \sin (2 \pi  z) \\
 A \cos (2 \pi  z)+B \sin (2 \pi  x) \\
 B \cos (2 \pi  x)+C \sin (2 \pi  y)
\end{array}
\right)
\end{equation}
which satisfies the Beltrami condition with eigenvalue $\lambda \, =
\, 1$ and which contains three real parameters $A,B,C$ defines  what
is known in the literature by the name of an ABC-flow
(Arnold-Beltrami-Childress) and during the last half century it was
the target of fantastically numerous investigations.
\par
Main motivation of the paper \cite{Fre:2015mla}  was to understand
the principles underlying the construction  of the ABC-flows in
order to use systematically such principles to construct and
classify all other Arnold-like Beltrami flows, deriving also, as a
bonus, their hidden discrete symmetries. For instance symmetries of
Beltrami flows have proved to be crucial in connection with their
use in modeling \textit{magneto-hydrodynamic fast
dynamos}\cite{zeldus},\cite{Dynamo},\cite{FFMF},\cite{Gilbert}. By
this words it is understood the mechanism that in a steady flow of
charged particles generates a large scale magnetic field whose
magnitude might be exponentially increasing with time. No analytic
results do exist on fast dynamos and all studies have been so far
numerical, yet while dealing with these latter, crucial
simplifications occur and optimization algorithms become available
if the steady flow possesses a large enough group
$\mathrm{G_{sim}}\subset \mathfrak{UG}_\Lambda$ of symmetries. In
this case the magnetic field can be developed into irreducible
representations of $\mathrm{G_{sim}} $ and this facilitates the
numerical determination of growing rates of different modes. It is
important to stress that the linearized dynamo equations for the
magnetic field $\mathbf{B}$ coincide with the linearized equations
for perturbations around a steady flow. Therefore the same
development of perturbations into irreps of $\mathrm{G_{sim}}$ is of
great relevance also for the study of fluid instabilities.
\par
As already stressed a much shorter sketch of the Hexagonal Lattice
was provided in \cite{Fre:2015mla} in order to emphasize the
generality of the applied methods, yet the authors did not address
the construction of the Universal Classifying Group  which is one of
the tasks addressed in the present paper.

\subsection{The spectrum of
the $\star \mathbf{d}$ operator on $\mathrm{T}^3$}
\label{fantasmabeltrami} The main issue of paper \cite{Fre:2015mla}
was the construction of vector fields defined over the three-torus
$\mathrm{T^3}$ that are eigenstates of the $\star_g \mathrm{d}$
operator, namely  solutions of the following equation:
\begin{eqnarray}
\star_g \mathrm{d}\Omega^{(n;I)} &=& \, m_{(n)} \, \Omega^{(n;I)}
\quad ; \quad \Omega^{(n;I)}\left[V_{(m;J)} \right] \,=\,
\delta^n_m \, \delta^I_J \label{formaduale}
\end{eqnarray}
where $\mathrm{d}$ is the exterior differential, and $\star_g$ is
the Hodge-duality operator which, differently from the exterior
differential, can be defined only with reference to a given metric
$g$. By $\Omega^{(n;I)}$ we denote a one-form:
\begin{equation}\label{omegas}
    \Omega^{(n;I)} \, = \, \Omega^{(n;I)}_\mu \,dx^\mu
\end{equation}
which is declared to be dual to the vector field we are interested
in:
\begin{eqnarray}\label{vettocampo}
V_{(m;J)} & = & V_{(m;J)}^\mu \, \partial_\mu \nonumber\\
\Omega^{(n;I)}\left[V_{(m;J)} \right] & \equiv &  \Omega^{(n;I)}_\mu
\, V_{(m;J)}^\mu \, = \, \delta^n_m \, \delta^I_J
\end{eqnarray}
and by means of the composite index $(n;I)$ we make reference to the
quantized eigenvalues $m_{(n)}$ of the $\star_g \mathrm{d}$ operator
(ordered in increasing magnitude $|m_{(n)}|$) and to a basis of the
corresponding eigenspaces
\begin{equation}\label{superpongomega}
\star_g \mathrm{d}\Omega^{(n)} \, = \, \, m_{(n)} \, \Omega^{(n)}
\quad \Rightarrow \quad \Omega^{(n)}\, = \, \sum_{I=1}^{d_n} \,c_I
\, \Omega^{(n;I)}
\end{equation}
the symbol $d_n$ denoting the degeneracy of $|m_{(n)}|$ and $c_I$
being constant coefficients.
\par
Indeed, since $\mathrm{T^3}$ is a compact manifold, the eigenvalues
$m_{(n)}$ form a discrete set. Their values and their degeneracies
are a property of the metric $g$ introduced on it. Here we outline
the general procedure to construct the eigenfunctions of $\star_g
\mathrm{d}$, to calculate the eigenvalues and to determine  their
degeneracies. What follows is an elementary and straightforward
exercise in harmonic analysis.
\par
In tensor notation, equation (\ref{formaduale}) has the following
appearance:
\begin{equation}\label{tensoBeltra}
\frac{1}{2} \, g_{\mu\nu} \, \epsilon^{\nu\rho\sigma} \partial_\rho
\Omega_\sigma \, = \, m \, \Omega_\mu
\end{equation}
The equation written above was named Beltrami equation since it was
already considered by the great italian mathematician Eugenio
Beltrami in 1881 \cite{beltramus}, who presented one of its periodic
solutions previously constructed by  Gromeka in 1881. Such a
solution was inherited by Arnold and it is essentially the basis of
his Hydrodynamical Model. We will see that Arnold Model just
corresponds to the lowest eigenfunction of the $\star_g \,
\mathrm{d}$-operator in the case of the cubic lattice. Many more
similar models can be constructed choosing higher eigenvalues,
choosing irreducible representation of the point group  in their
eigenspaces or changing the lattice.
\par
Introducing the basis vectors of the dual lattice $\Lambda^\star$ we
can write:
\begin{equation}\label{bardacco1}
    \Omega \, = \, \Omega_\mu \, dr^\mu  \, = \, \Omega_\mu \, e^\mu_i \, dx^i \, = \, \Omega_i \, dx^i
\end{equation}
where $e^\mu_i$ are the components of the vectors $
{\mathbf{e}}^\mu$ in a standard orthogonal basis of $\mathbb{R}^3$
and
\begin{equation}\label{bardacco2}
    x^i \, = \, w_\mu^i \, r^\mu
\end{equation}
are a new set of euclidian coordinates obtained from the original
ones $r^\mu$ by means  of the components $w_\mu^i$ of the basis
vectors $ {\mathbf{w}}_\mu$ of the space lattice $\Lambda$.
Recalling that:
\begin{equation}\label{derivosugiu}
\partial_\mu \, =\, \frac{\partial}{\partial r^\mu} \, = \, w^\mu_i \, \partial_i \, = \, w^\mu_i \, \frac{\partial}{\partial x^i}
\end{equation}
with a little bit of straightforward algebra we can rewrite
eq.(\ref{formaduale}) in the equivalent universal way:
\begin{equation}\label{tensoBeltra2}
\frac{1}{2} \,  \epsilon_{ijk} \partial_j \Omega_k \, = \, \mu  \,
\Omega_i \quad ; \quad \mu \, = \, \frac{m}{\mbox{det} w}
\end{equation}
where by $\mbox{det} w$ we denote the determinant of the $3 \times
3$ matrix $w_\mu^i$.
\subsection{Fourier expansions and Beltrami
chirality}\label{furetto}
It is now the appropriate moment to point out that the first order
Beltrami operator is a chirality operator that splits the ordinary
Fourier spectrum of any vector field defined over the three torus in
two disjoint sectors of \textbf{positive} and \textbf{negative}
\textbf{Beltramicity}, respectively.
\par
This statement is easily understood by means of the following
elementary discussion. Given a Riemannian three-manifold
$\left(\mathcal{M},g\right)$, the Laplace--Beltrami operator on
one-forms is given by:
\begin{equation}\label{LapBelOm}
    \Delta_g \, = \, \star_g \, \mathrm{d} \, \star_g \, \mathrm{d} \,
\end{equation}
namely it is the square of the Beltrami operator $\mathfrak{B}_g \,
\equiv \, \star_g \, \mathrm{d}$. Hence any eigenstate
$\Omega^{[\mathbf{U}]}$ of the  Beltrami operator with eigenvalue
$\mu$ is automatically an eigenstate of the Laplace--Beltrami
operator $\Delta_g$ with eigenvalue $\mu^2$ :
\begin{equation}\label{roccagrimalda}
    \mathfrak{B}_g \,\Omega^{[\mathbf{U}]} \, = \, \mu \,\Omega^{[\mathbf{U}]} \,
    \quad \Rightarrow \quad \Delta_g \,\Omega^{[\mathbf{U}]} \, = \, \mu^2 \,\Omega^{[\mathbf{U}]}
\end{equation}
Inverting the argument one expects that the eigenspace of $\Delta_g$
corresponding to an eigenvalue $E = \mu^2 > 0$, that is a linear
vector space, can be partitioned in two vector subspaces,
respectively spanned by the solutions of Beltrami equation with
eigenvalue $\mu=\pm \sqrt{E}$. This is precisely what it happens for
the one-form duals of vector fields defined on torii $\mathrm{T}^3
\, = \, \mathbb{R}^3/\Lambda$ where $\Lambda$ is a lattice.
\par
Let us name $U_i( {\mathbf{x}})$  a generic vector field on
$\mathrm{T}^3$, its most general form is necessarily provided by the
standard Fourier expansion that we write as follows\footnote{Take
note that the latin indices $i,j,..$ refer to the standard euclidian
metric of $\mathbb{R}^3$ in whose basis the  components of the
momentum vectors lying in the dual lattice $\Lambda^\star$ are not
necessarily integer valued.}:
\begin{eqnarray}\label{harmogen}
U_i\left( {\mathbf{x}}\right)& = & \sum_{\mathbf{k}\in \,
\Lambda^\star}
\,Y_i\left( {\mathbf{k}}\, | \, {\mathbf{x}}\right) \nonumber\\
Y_i\left( {\mathbf{k}}\, | \, {\mathbf{x}}\right) & = & v_i\left(
{\mathbf{k}}\right) \,\cos\left( 2\,\pi \, {\mathbf{k}}\cdot
 {\mathbf{x}}\right)
\, +\, \omega_i\left( {\mathbf{k}}\right) \,\sin\left( 2\,\pi \,
 {\mathbf{k}}\cdot  {\mathbf{x}}\right)
\end{eqnarray}
The condition that the momenta $ {\mathbf{k}}$ included in the
Fourier expansion should belong to the dual lattice guarantees that
each mode $Y_i( {\mathbf{x}})$ is periodic with respect to the space
lattice $\Lambda$ and, a fortiori, such is the vector field
$U_i\left( {\mathbf{x}}\right)$. Indeed, by means of the very
definition of the dual lattice (\ref{reticoloLastar}) it follows
that:
\begin{equation}\label{periodico}
\forall \,  {\mathbf{q}} \, \in \, \Lambda \, : \quad Y_i\left(
{\mathbf{x}}\, + \,  {\mathbf{q}}\right) \, = \, Y_i\left(
{\mathbf{x}}\right)
\end{equation}
If $U_i\left( {\mathbf{x}}\right)$ is supposed to be the velocity
field of a fluid, then, in force of Navier-Stokes or Euler
equations, it must be divergenceless  $\partial^i \, U_i \, = \, 0$
and this requires that we impose such a condition on each mode,
namely $\partial^i Y_i\left( {\mathbf{k}}\, | \,
{\mathbf{x}}\right)\, = \, 0$. Imposing this constraint on  the
general ansatz (\ref{harmogen}) we obtain:
\begin{equation}\label{transversocondo}
    {\mathbf{k}}\, \cdot \,  {\mathbf{v}}\left( {\mathbf{k}}\right) \, = \, 0 \quad ; \quad  {\mathbf{k}}\,
   \cdot \,  {\mathbf{\omega}}\left( {\mathbf{k}}\right) \, = \, 0
\end{equation}
which reduces the 6 parameters per mode contained in the general
ansatz (\ref{harmogen}) to 4 per mode. At the same time we can
easily verify that the vector field:
\begin{equation}\label{vettoromodo}
    \mathbf{Y}_\mathbf{k} \left( \mathbf{x}\right)\, \equiv \, Y_i\left( {\mathbf{k}}\, | \,
    {\mathbf{x}}\right) \frac{\partial}{\partial x^i}
\end{equation}
is dual to a 1-form $\Omega^{[\mathbf{Y}_\mathbf{k}]}$ that is an
eigenstate of the Laplace-Beltrami operator with the explicit
eigenvalue displayed in the following formula:
\begin{eqnarray}
\Delta_g \, \Omega^{[\mathbf{Y}_\mathbf{k}]} & = &  \mu^2 \,
 \Omega^{[\mathbf{Y}_\mathbf{k}]}\label{laplaccio}\\
\mu^2  &= &   \pi^2 \, \langle {\mathbf{k}}\, , \,
{\mathbf{k}}\rangle\label{spectra1}
\end{eqnarray}
Let us now set:
\begin{equation}\label{radius}
    r \, = \,  \sqrt{\langle {\mathbf{k}}\, , \,
    {\mathbf{k}}\rangle} \quad \Rightarrow \quad \mu^2 = \pi^2 \,
    r^2
\end{equation}
The degeneracy of each Laplace-Beltrami eigenvalue $\pi^2 \, r^2$ is
geometrically provided by counting the number of intersection points
of the dual lattice $\Lambda^\star$ with a sphere whose center is in
the origin and whose radius is $r$. For a generic lattice the number
of solutions of equation (\ref{radius}) namely the number of
intersection points of the lattice with the sphere is either $0$
(the sphere does not intersect the lattice) or just two: $\pm
{\mathbf{k}}$ (the sphere intersects the lattice in two points), so
that the typical degeneracy of each eigenvalue is just $2$. On the
other hand, if the lattice $\Lambda$ is one of the Bravais lattices
admitting a non trivial point group $\mathfrak{P}_\Lambda$, then the
number of solutions of eq.(\ref{radius}) is larger, since all
lattice vectors ${\mathbf{k}}$ that sit  in one orbit of
$\mathfrak{P}_\Lambda$ have the same norm and therefore are located
on the same spherical layer. Hence we ought to consider   the
spherical layers of radius $r_k = \sqrt{\mathbf{k}^2}$ defined as
the intersection of a sphere of such a radius with the momentum
lattice:
\begin{equation}\label{spherlayer}
    \mathrm{SL}_{r_k}\, \equiv  \,
    \mathbb{S}_{r_k}\, \bigcap \, \Lambda^\star
\end{equation}
The set of available radii for which the corresponding spherical
layer is not an empty set is an infinite increasing sequence of
rational numbers:
\begin{equation}\label{culissardo}
    0<r_1 < r_2 <\dots <r_k < \dots \infty
\end{equation}
whose explicit form depends on the chosen lattice $\Lambda$. In each
spherical layer $\mathrm{SL}_{\mathbf{k}^2}$ we find a certain
finite number of points:
\begin{equation}\label{fluttuone}
    |\mathrm{SL}_{r_k}| \, \equiv \, \#\text{ of points in
    $\mathrm{SL}_{r_k}$}
\end{equation}
which in the average steadily increases with $r_k$, yet it strongly
fluctuates on the short range (see section \ref{panoramica} for more
details on this point). Indeed, in a rather capricious way,
depending on the choice of the primary lattice $\Lambda$ and, hence,
of the point group $\mathbb{P}_\Lambda$, each spherical layer
$\mathrm{SL}_{r_k}$ decomposes into a certain number $n_{r_k}\in
\mathbb{N}$ of orbits:
\begin{equation}\label{calendulus}
    \mathrm{SL}_{r_k} \, = \,
    \bigcup_{i=1}^{n_{r_k}} \,
    \mathcal{O}^{\ell_i}_i\left(r_k\right)
\end{equation}
where $i$ is an enumeration index and $\ell_i$ is the length of the
orbit $\mathcal{O}^{\ell_i}_i\left(r_k\right)$, namely the number of
elements it contains. Each point groups admits a finite number of
orbit types of a characteristic length, whose maximal value is the
order of the point group $\mid\mathfrak{P}_\Lambda\mid$. Actually
the orbits are in one-to-one correspondence with the possible
stability subgroups $\mathcal{H}_i \subset \mathfrak{P}_\Lambda$ of
moment vectors and the orbit lengths are just the orders of these
corresponding subgroups which, by Lagrange theorem, are divisors of
the order of the point group. Since, by definition, orbits are
disjoint sets we have:
\begin{equation}\label{cucurucu}
    \mathrm{SL}_{r_k}\, = \, \sum_{j=1}^{n_{r_k}} \, \ell_j
\end{equation}
\par
In view of this discussion the general Fourier series of
eq.(\ref{harmogen}) can be reorganized in the following way:
\begin{equation}\label{bardolinus}
    \mathbf{U}\left(\mathbf{x}\right) \, = \,\sum_{q=1}^{\infty}\,
     \underbrace{\sum_{\mathbf{k}\in
    \mathrm{SL}_{r_q}} \,  \mathbf{Y}_\mathbf{k} \left(
    \mathbf{x}\right)}_{\mathbf{W}_{r_k}\left(x\right)}
\end{equation}
The vector field $\mathbf{W}_{r_k}\left(\mathbf{x}\right)$ is the
most general divergenceless one associated with the spherical layer
$\mathrm{SL}_{r_k}$ and, according with the counting provided above,
in principle it contains a number of parameters that is $4\times
|\mathrm{SL}_{r_k}|$ since there are 4 parameters for each momentum
vector $\mathbf{k}$. There is however a subtlety. Necessarily both
$\pm \mathbf{k}$  are located one the same layer since they have the
same norm. For each of these momentum pairs the number of parameters
appearing in $\mathbf{W}_{r_k}\left(x\right)$ is not 8 rather it is
4, since $\cos(\mathbf{k}\cdot \mathbf{x})= \cos(-\mathbf{k}\cdot
\mathbf{x})$ and $\sin(\mathbf{k}\cdot \mathbf{x})= -
\sin(-\mathbf{k}\cdot \mathbf{x})$. Hence the total number of
parameters appearing in $\mathbf{W}_{r_k}\left(\mathbf{x}\right)$
is:
\begin{equation}\label{boldano}
   \mathrm{N}_{r_k} \, \equiv \, \text{\# of parameters in $\mathbf{W}_{r_k}\left(\mathbf{x}\right)$}
   \, = \, 2 |\mathrm{SL}_{r_k}|
\end{equation}
\par
Hence to each layer we can associate an eigenstate of the Laplace
Beltrami operator $\Delta_g$ of eigenvalue\footnote{Originally we
defined the Laplace-Beltrami operator on the 1-forms, but its
definition trivially extends, by lowering the indices with the
metric, to the corresponding vector field} $\mu^2_{r_k} =\pi^2 \,
r^2_k$
\begin{equation}\label{rompipalle}
    \Delta_g \, \mathbf{W}_{r_k}\left(\mathbf{x}|\mathbf{F}\right)\,
    = \, \mu^2_{r_k} \,\mathbf{W}_{r_k}\left(\mathbf{x}|\mathbf{F}\right)
\end{equation}
where by $\mathbf{F}$ we have denoted the
$\mathrm{N}_{r_k}$-component vector of free parameters appearing in
the vector field $\mathbf{W}_{r_k}$. Note that, as it appears from
eq.(\ref{boldano}), $\mathrm{N}_{r_k}$ is always a multiple of $2$.
This is relevant because the parameter space can be split into two
subspaces each of dimension $|\mathrm{SL}_{r_k}|$ by imposing the
additional Beltrami/anti-Beltrami condition, mode by mode. Indeed we
can explicitly implement equation (\ref{tensoBeltra2}) and we get
the following two conditions:
\begin{eqnarray}
  \mu \, v_i \left( {\mathbf{k}}\right) &=& \pi \, \epsilon_{ij\ell} \, k_j \,
  \omega_\ell \left( {\mathbf{k}}\right) \label{curlo1} \\
  \mu \, \omega_i \left( {\mathbf{k}}\right) &=& -\pi \,
  \epsilon_{ij\ell} \, k_j \, v_\ell \left( {\mathbf{k}}\right) \label{curlo2}
\end{eqnarray}
The two equations are self consistent if and only if the eigenvalue
$\mu$ is such that $\mu^2 = \pi^2 \, \langle \mathbf{k} ,
\mathbf{k}\rangle$. Hence we can choose either $\mu = \pi\, r_k$ or
$\mu = -\pi \, r_k$ and in each case we obtain a solution of the
algebraic equations depending on 2 parameters.  This amounts to
stating that the general contribution $\mathbf{W}_{r_k}$ of the
spherical layer $\mathrm{SL}_{r_k}$ to the general Fourier series is
split in a Beltrami plus an anti-Beltrami part:
\begin{equation}\label{romilinu}
    \mathbf{W}_{r_k}\left(\mathbf{x}|\mathbf{F}\right) \, = \,
    \mathbf{W}^+_{r_k}\left(\mathbf{x}|\mathbf{F}^+\right)\, + \,
\mathbf{W}^-_{r_k}\left(\mathbf{x}|\mathbf{F}^-\right)
\end{equation}
such that:
\begin{equation}\label{crissolo}
    \star_g \mathrm{d}\Omega^{\left[\mathbf{W}^\pm_{r_k}\right]} \, =
    \,\pm  \pi \, r_k \Omega^{\left[\mathbf{W}^\pm_{r_k}\right]}
\end{equation}
We can introduce an $L^2$ functional space on each spherical shell
$\mathrm{SL_{rk}}$ by defining the scalar product of any two
eigenvector field $A(\mathbf{x})$ and $B(\mathbf{x})$ of the Laplace
Beltrami operator $\Delta_g$ with the same eigenvalue $\pi^2 r_k^2$:
\begin{equation}\label{cromagnone}
    \left(\mathbf{A},\mathbf{B}\right) \, \equiv \, \int_{FC} \,\mathrm{d}^3\mathbf{x}\,
 \mathbf{A}(\mathbf{x}) \cdot
 \mathbf{B}(\mathbf{x}) \quad ; \quad
    \mid \mathbf{A}\mid^2 \, = \, \left( \mathbf{A},\mathbf{A}\right)
\end{equation}
where by $FC$ we denote the fundamental cell (namely the torus) of
$\mathbb{R}^3$ modulus the lattice $\Lambda$. It is easy to see that
with respect to such a product Beltrami and anti-Beltrami fields are
always orthogonal to each other. Relying on this observation the
authors of \cite{beltraspectra} introduced the Beltrami index of a
stationary Navier-Stokes  solution $\mathbf{U}$  by means of the
following formula:
\begin{equation}\label{balengus}
    \beta_{r_k}\left[\mathbf{U}\right] \, = \, \frac{\mid U_{r_k}^+\mid^2 -\mid
    U_{r_k}^-\mid^2}{\mid U_{r_k}^+\mid^2 +\mid
    U_{r_k}^-\mid^2}
\end{equation}
and partially proved, partially conjectured from the results of
computer simulations a set of properties of this chiral spectral
index. The word chiral is utilized because a space reflection
$\mathbf{x} \to - \mathbf{x}$ transforms Beltrami fields into anti
Beltrami ones and viceversa. What was not even envisaged in the very
interesting papers  \cite{beltraspectra} and  \cite{ondoso} is the
group theoretical structure underlying Beltrami (and anti-Beltrami)
fields appearing in the Fourier expansions of Navier-Stokes
solutions. Indeed that group theoretical structure, based on the new
conception of the \textit{Universal Classifying Group} was unveiled
only  in 2015 in \cite{Fre:2015mla}, starting from the observation
by Arnold of a hidden roto-translation symmetry in the AAA model,
which isomorphic, as a group, to the relevant point group
$\mathrm{O_{24}}$.
\par
As stated in the introduction, the ultimate goal of the research plan
initiated by the present paper, is that of complementing the
spectral analysis of papers \cite{beltraspectra,ondoso} with the
insights provided by a systematic use of the group theoretical
structure inherent to the \textit{Universal Classifying Group}
$\mathfrak{UG}_\Lambda$. Indeed, in view of the decomposition
(\ref{calendulus}) of the spherical layer into orbits of the point
group, the general field
$\mathbf{W}_{r_k}\left(\mathbf{x}|\mathbf{F}\right)$ can be seen as
the sum of as many vector fields as there are orbits in the layer:
\begin{equation}\label{lagaccio}
    \mathbf{W}_{r_k}\left(\mathbf{x}|\mathbf{F}\right) \, = \,
    \sum_{i=1}^{n_{r_k}} \mathbf{Y}_{[\mathcal{O}_i]}\left(\mathbf{x}|\mathbf{F}^{[i]}\right)
\end{equation}
and each vector field associated with  an orbit $\mathcal{O}_i$ can
be split into its Beltrami and anti-Beltrami part. It follows that
the construction of Beltrami (or by reflection anti-Beltrami) vector
fields provides the building blocks to represent any Navier-Stokes
flow in a compact torus $\mathrm{T}^3 = \mathbb{R}^3 / \Lambda$. The
completely new lore introduced in this paper on the basis of the
results of \cite{Fre:2015mla} is that the orbit building blocks to
be described in the next subsection can be further analyzed and
organized into irreducible representations of the Universal
Classifying Group fully discussed in section \ref{gruppafunda}. It
is in view of this powerful group theoretical weapon that the
spectral analysis of \cite{beltraspectra} has to be reconsidered.
\subsection{The algorithm to construct Arnold Beltrami
Flows} \label{algoritmo} What we described in the previous
subsection provides a well defined algorithm to construct a series
of Arnold Beltrami flows that can be summarized in a few clear-cut
steps and it is quite suitable for a systematic  computer aided
implementation.
\par
The steps are the following ones:
\begin{description}
  \item[a)] Choose a Bravais Lattice $\Lambda$ with a non trivial proper point group  $\mathfrak{P}_\Lambda$.
  \item[b)] Construct the character table and the irreducible representations of $\mathfrak{P}_\Lambda$.
  \item[c)] Analyze the structure of orbits of $\mathfrak{P}_\Lambda$ on the lattice $\Lambda$ and determine
  the number  of lattice points  contained in each spherical layer $\mathrm{SL}_{r_k}$ of the dual lattice $\Lambda^\star$
  of quantized radius $r_k$, that as we already remarked is always even
  $|\mathrm{SL}_{r_k}| \, = 2 \, P_{r_k}$
  \item[d)] Construct the most general solution of the Beltrami equation with eigenvalue $\mu_{k} \, = \, \pi \, r_k$
  by using the individual harmonics discussed in the previous
  section. The corresponding anti-Beltrami field is immediately
  determined by a reflection $\mathbf{x} \to -\mathbf{x}$
\begin{equation}\label{beltramusgen}
V_i\left( {\mathbf{x}}\right) \, = \, \sum_{ {\mathbf{x}}\, \in
\,\mathfrak{S}_n } \,Y_i\left( {\mathbf{k}}\, | \,
{\mathbf{x}}\right)
\end{equation}
Hidden in each harmonic $Y_i\left( {\mathbf{k}}\, | \,
{\mathbf{x}}\right)$ there are two parameters that are the remainder
of the six parameters $v_i\left( {\mathbf{k}}\right)$ and
$\omega_i\left( {\mathbf{k}}\right)$ after conditions
(\ref{transversocondo},\ref{curlo1},\ref{curlo2}) have been imposed.
This would amount to a total of $4\, P_{r_k}$ parameters, yet, for
the already discussed reason, the number of independent parameters
is always reduced to $2 P_{r_k}$. Hence, at the end of the
construction encoded in eq. (\ref{beltramusgen}), we have a Beltrami
vector depending on a set of $2 P_{r_k}$ parameters that we can call
$F_I$ and consider as the $2\, P_{r_k}$-components of a vector
$\mathbf{F}$. Ultimately we have an object of the following form:
\begin{equation}\label{gomez}
    \mathbf{V}\left( {\mathbf{x}}\, | \, \mathbf{F}\right)
\end{equation}
which under the point group   $\mathfrak{P}_\Lambda$ necessarily
transforms in the following way:
\begin{equation}\label{Rtrasformogen}
    \forall \, \gamma \, \in \, \mathfrak{P}_\Lambda \, : \quad \gamma^{-1} \,\cdot \,
    \mathbf{V}\left(\gamma \,\cdot \, {\mathbf{x}}\, | \, \mathbf{F}\right)
    \, = \, \mathbf{V}\left( {\mathbf{x}}\, | \, \mathfrak{R}[\gamma] \,\cdot \, \mathbf{F}\right)
\end{equation}
where $\mathfrak{R}[\gamma]$ are $2\, P_{r_k} \times 2\, P_{r_k}$
matrices that form a representation of $\mathfrak{P}_\Lambda$.
Eq.(\ref{Rtrasformogen}) is necessarily true because any rotation
$\gamma \, \in \, \mathfrak{P}_\Lambda$ permutes the elements of
$\mathrm{SL}_{r_k}$ among themselves.
\item[e)] Decompose the representation $\mathfrak{R}[\gamma]$ into irreducible representations of $\mathfrak{P}_\Lambda$.
Each irreducible subspace $\mathbf{f}_{p}$ of the $2 P_{r_k}$
parameter space $\mathbf{F}$ defines an Arnold--Beltrami Flow:
\begin{equation}\label{arnoldosim}
\frac{\mathrm{d}}{\mathrm{d}t} \,  {\mathbf{x}}(t) \, = \,
\mathbf{V}\left( {\mathbf{x}}(t)\, | \, \mathbf{f}_p\right)
\end{equation}
which is worth to analyze.
\end{description}
An obvious question which arises in connection with such a
constructive algorithm is the following: how many Arnold--Beltrami
flows are there? At first sight it seems that there is an infinite
number of such systems since we can arbitrarily increase the radius
of the spherical layer and on each new layer it seems that we have
new models. Let us however observe that if on two different
spherical layers $\mathrm{SL}_{r_1}$ and $\mathrm{SL}_{r_2}$ there
are two orbits of lattice vectors $\mathcal{O}_1$ and
$\mathcal{O}_2$ that have the same order
\begin{equation}\label{parorder}
   \ell \, = \,  \left| \mathcal{O}_1\right | \, = \, \left| \mathcal{O}_2\right |
\end{equation}
and furthermore all vectors ${\mathbf{k}}_{(n_2)} \, \in
\,\mathcal{O}_2$ are simply proportional to their analogues in orbit
$\mathcal{O}_1$:
\begin{equation}\label{proporzio}
{\mathbf{k}}_{(n_2)} \, = \, \lambda \, {\mathbf{k}}_{(n_1)} \quad ;
\quad \lambda \, \in \, \mathbb{Z}
\end{equation}
then we can conclude that:
\begin{equation}\label{riduzione}
\mathbf{V}_{(n_2)}\left( {\mathbf{x}}\, | \, \mathbf{f}_p\right) \,
= \,  \mathbf{V}_{(n_1)}\left(\, \lambda \, {\mathbf{x}}\, | \,
\mathbf{f}_p\right)
\end{equation}
By redefining  the coordinate fields $\lambda \, {\mathbf{x}} \, =
\,  {\mathbf{x}}^\prime$ and rescaling time $t$ the two differential
systems (\ref{arnoldosim}) respectively  constructed from  layer
$n_1$ and layer $n_2$ can be identified.
\par
As it was demonstrated  in \cite{Fre:2015mla} analyzing the case of
the cubic lattice and the orbits of the octahedral  group there is
always a finite number of $\mathfrak{P}_\Lambda$-orbit type on each
lattice $\Lambda$. There is a  maximal orbit $\mathcal{O}_{max}$
that has order equal to the order of the point group :
\begin{equation}\label{orbitmax}
\left| \mathcal{O}_{max}\right| \, = \, \left|
\mathfrak{P}_\Lambda\right|
\end{equation}
and there are a few shortened orbits $\mathcal{O}_{i}$ ($i=\,1,\dots
,\, s$) that have a smaller order:
\begin{equation}\label{carisma}
\ell_i \, = \, \left| \mathcal{O}_{i}\right| \, < \, \left|
\mathfrak{P}_\Lambda\right|
\end{equation}
The fascinating property is that for the shortened orbits, which
seem to play an analogue role in this context to that of BPS states
in another context, property (\ref{proporzio}) is always true. The
vectors pertaining to the same orbit $\mathcal{O}_i$ in different
spherical layers are always the same up to a multiplicative factor.
Hence from the shortened orbits it was shown in \cite{Fre:2015mla}
that one always obtains a finite number of Arnold--Beltrami flows.
It remained the case of the maximal orbit for which property
(\ref{proporzio}) is not necessarily imposed. How many independent
flows do we obtain considering all the layers? The answer to the
posed question is hidden in number theory. Indeed one has to analyze
how many different type of triplets of integer numbers satisfy
Diophantine equations of the Fermat type. In section
\ref{triplettoni} we review the answer obtained in
\cite{Fre:2015mla} providing a systematic classification of such
triplets for the cubic lattice.
\par
Actually that classification is a classification of sublattices of
the cubic lattice and each sublattice is associated with irreducible
representations of the Universal Classifying Group
$\mathfrak{UG}_\Lambda$.
\par
Such result demonstrated that there is a finite number of
Arnold--Beltrami flows and each of them can be promoted to a
definite type of exact solutions of the Navier-Stokes equations
depending on a finite number of parameters that acquire a dependence
on the momenta and are, in this way, identified with Fourier
coefficients in a Fourier series expansion of the initial
conditions.
\section{Group Theory Foundations} \label{gruppafunda}
%\subsection{Review of the necessary Group Theory ingredients}
In order to make the present paper self-consistent and better
highlight the interpretation of several of the results obtained in
\cite{Fre:2015mla}, that here are clarified in a more systematic way
and extended from the cubic to the hexagonal case, we review the
main group theoretical ingredients utilized in \cite{Fre:2015mla} to
derive the \textit{Universal Classifying Group},  whose very notion
in the present paper is made more precise in view of \textit{exact
sequences} and \textit{finite group cohomology}.
\par
Skipping generalities we just remind the reader of what was already
presented in eq.s(\ref{Aclasspointgroups},\ref{Bclasspointgroups}),
namely that in three dimensions the available \textit{Lattice Point
Groups} $\mathfrak{P}_\Lambda$ are either the cyclic groups
$\mathrm{C}_h \sim\mathbb{Z}_h$ with $h=2,3,4,6$ or the dihedral
groups $\mathrm{Dih}_h$ with $h=2,3,4,6$ or the tetrahedral group
$\mathrm{T_{12}}\sim \mathrm{A_4}$ or the octahedral group
$\mathrm{O_{24}}\sim \mathrm{S_4}$.
%%%%%%%%
\subsection{The cubic lattice and the octahedral point group $\mathrm{O_{24}}$}
\label{reticolocubico} The case of the cubic lattice was analyzed in
depth in \cite{Fre:2015mla}. We review and repeat here a good deal
of the results of that paper for three reasons:
\begin{enumerate}
  \item We need to revise the conventions and the notations in order
  to make clear how the upgrading to the complete Navier-Stokes
  equations is achieved in practice.
  \item Since a large part of the results to be obtained, classified
  and visualized necessarily depends on the use of MATHEMATICA codes
  that derive from those developed in 2014-2015 by means of a
  systematic reorganization of the routines and subroutines and by
  a transcription from MATHEMATICA 5.2 to MATHEMATICA 12, it is of
  vital importance to utilize a well defined and already established
  set of conventions and nomenclature.
  \item The cubic lattice case constitutes the paradigm for the
  development of the same lore in the case of the hexagonal lattice which is
  a  goal of the present paper.
\end{enumerate}
Hence, within the general frame presented above, let us review the
cubic lattice case.
\par
\begin{figure}[!hbt]
\begin{center}
\includegraphics[height=50mm]{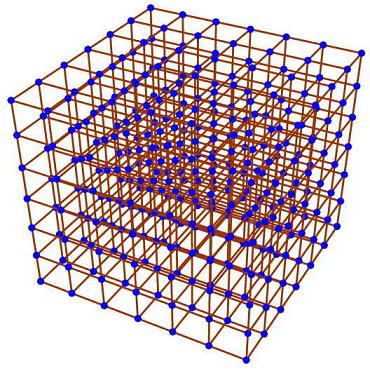}
\end{center}
\caption{\it  A view of the self-dual cubic lattice}
\label{cubicogenerale}
\end{figure}
The self-dual cubic lattice (momentum and space lattice at the same
time) is displayed in fig.\ref{cubicogenerale}.
\par
The basis vectors of the cubic lattice $\Lambda_{cubic}$ are :
\begin{equation}\label{cubicobase}
{\mathbf{w}}_1 \, = \, \{1,0,0\} \quad ; \quad {\mathbf{w}}_2 \, =
\, \{0,1,0\} \quad ; \quad {\mathbf{w}}_3 \, = \, \{0,0,1\}
\end{equation}
which implies that the metric is just the Kronecker delta:
\begin{equation}\label{Kronecca}
    g_{\mu\nu} \, = \, \delta_{\mu\nu}
\end{equation}
and the basis vectors ${\mathbf{e}}^\mu$ of the dual lattice
$\Lambda_{cubic}^\star$ coincide with those of the lattice
$\Lambda$. Hence the cubic lattice is self-dual:
\begin{equation}\label{autoduali}
     {\mathbf{w}}_\mu \, = \,  {\mathbf{e}}^\mu \quad \Rightarrow \quad \Lambda_{cubic} \, = \, \Lambda^\star_{cubic}
\end{equation}
The subgroup of the proper rotation group which maps the cubic
lattice into itself is the octahedral  group $\mathrm{O_{24}}$ whose
order is 24. In the next subsection we recall its structure.
\subsubsection{Structure of the  Octahedral Group $\mathrm{O_{24}}\sim \mathrm{S_4}$}
\label{ottostruttura} Abstractly the octahedral Group
$\mathrm{O_{24}}\sim \mathrm{S_{4}}$ is isomorphic to the symmetric
group of permutations of 4 objects. It is defined by the following
generators and relations:
\begin{equation}\label{octapresa}
T, \, S \quad : \quad    T^3 \, = \, \mbox{\bf e} \quad ; \quad S^2
\, = \, \mbox{\bf e} \quad ; \quad (S\,T)^4 \, = \, \mbox{\bf e}
\end{equation}
On the other hand $\mathrm{O_{24}}$ is a finite, discrete subgroup
of the three-dimensional rotation group and any $\gamma \, \in \,
\mathrm{O_{24}}\, \subset \, \mathrm{SO(3)}$ of its 24 elements can
be uniquely identified by its action on the coordinates $x,y,z$,  as
it is displayed below:
\begin{equation}\label{nomiOelemen}
\begin{array}{cc}
\begin{array}{|c|rcl|}
\hline
\mbox{\bf e} & 1_1 & = & \{x,y,z\} \\
 \hline
\null & 2_1 & = & \{-y,-z,x\} \\
\null &  2_2 & = & \{-y,z,-x\} \\
\null & 2_3 & = & \{-z,-x,y\} \\
C_3 & 2_4 & = & \{-z,x,-y\} \\
 \null &2_5 & = & \{z,-x,-y\} \\
\null & 2_6 & = & \{z,x,y\} \\
\null & 2_7 & = & \{y,-z,-x\} \\
\null & 2_8 & = & \{y,z,x\} \\
 \hline
\null & 3_1 & = & \{-x,-y,z\} \\
C_4^2 & 3_2 & = & \{-x,y,-z\} \\
\null & 3_3 & = & \{x,-y,-z\} \\
 \hline
\end{array} & \begin{array}{|c|rcl|}
\hline
\null & 4_1 & = & \{-x,-z,-y\} \\
\null & 4_2 & = & \{-x,z,y\} \\
C_2 &4_3 & = & \{-y,-x,-z\} \\
\null & 4_4 & = & \{-z,-y,-x\} \\
\null & 4_5 & = & \{z,-y,x\} \\
\null & 4_6 & = & \{y,x,-z\} \\
 \hline
\null & 5_1 & = & \{-y,x,z\} \\
\null & 5_2 & = & \{-z,y,x\} \\
C_4 & 5_3 & = & \{z,y,-x\} \\
\null & 5_4 & = & \{y,-x,z\} \\
\null & 5_5 & = & \{x,-z,y\} \\
\null & 5_6 & = & \{x,z,-y\}\\
 \hline
\end{array} \\
\end{array}
\end{equation}
As one sees from the above list the 24 elements are distributed into
5 conjugacy classes mentioned in the first column of the table,
according to a nomenclature which is standard in the chemical
literature on crystallography. The relation between the abstract and
concrete presentation of the octahedral  group is obtained by
identifying in the list (\ref{nomiOelemen}) the generators $T$ and
$S$ mentioned in eq. (\ref{octapresa}). Explicitly we have:
\begin{equation}\label{generatiTS}
    T \, = \, 2_8 \, = \, \left(
\begin{array}{lll}
 0 & 1 & 0 \\
 0 & 0 & 1 \\
 1 & 0 & 0
\end{array}
\right)\quad ; \quad S \, = \, 4_6 \, =\left(
\begin{array}{lll}
 0 & 1 & 0 \\
 1 & 0 & 0 \\
 0 & 0 & -1
\end{array}
\right)
\end{equation}
All other elements are reconstructed from the above two using the
multiplication table of the group which we omit for brevity.  This
observation is important in relation with representation theory. Any
linear representation of the group is uniquely specified by giving
the matrix representation of the two generators $T=2_8$ and $S=4_6$.
In the sequel this will be extensively utilized in the compact
codification of the reducible representations that emerge in our
calculations.
%%%%%%%%%%%%%%%%%%%%%%%%%%%%%
\subsubsection{Irreducible representations of the Octahedral Group}
\label{ottoirreppi}
%%%%%%%%%%%%%%%%%%%%%%
There are five conjugacy classes in $\mathrm{O}_{24}$ and therefore
according to theory there are five irreducible representations of
the same group, that we name $D_i$, $i\, =\, 1,\dots, 5$. Let us
briefly describe them.
\subsubsection{$D_1$ : the identity representation}
The identity representation which exists for all groups is that one
where to each element of $\mathrm{O}$ we associate the number $1$
\begin{equation}\label{identD1}
    \forall \, \gamma \, \in \, \mathrm{O}_{24} \,\, : \quad D_1(\gamma) \, = \, 1
\end{equation}
Obviously the character of such a representation is:
\begin{equation}\label{caretterusOD1}
    \chi_1 \, = \, \{1,1,1,1,1\}
\end{equation}
\subsubsection{$D_2$ : the quadratic Vandermonde representation}
The representation $D_2$ is also one-dimensional. It is constructed
as follows. Consider the following polynomial of order six in the
coordinates of a point in $\mathbb{R}^3$ or $\mathrm{T}^3$:
\begin{equation}\label{vPol}
    \mathfrak{V}(x,y,z) \, = \, (x^2 - y^2) \, (x^2 - z^2) \, (y^2 - z^2)
\end{equation}
As one can explicitly check under the transformations of the
octahedral  group listed in eq.(\ref{nomiOelemen}) the polynomial
$\mathfrak{V}(x,y,z)$ is always mapped into itself modulo an overall
sign. Keeping track of such a sign provides the form of the second
one-dimensional representation whose character is explicitly
calculated to be the following one:
\begin{equation}\label{caretterusOD2}
    \chi_1 \, = \, \{1,1,1,-1,-1\}
\end{equation}
\subsubsection{$D_3$ : the two-dimensional representation}
The representation $D_3$ is two-dimensional and it corresponds to a
homomorphism:
\begin{equation}\label{D3map1}
    D_3 \, : \quad \mathrm{O}_{24} \, \rightarrow \, \mathrm{SL(2,\mathbb{Z})}
\end{equation}
which associates to each element of the octahedral group a $2 \times
2$ integer valued matrix of determinant one. The homomorphism is
completely specified by giving the two matrices representing the two
generators:
\begin{equation}\label{D3map2}
    D_3(T) \, = \,
\left(
\begin{array}{ll}
 0 & 1 \\
 -1 & -1
\end{array}
\right) \quad ; \quad D_3(S) \, = \, \left(
\begin{array}{ll}
 0 & 1 \\
 1 & 0
\end{array}
\right)
\end{equation}
The character vector of $D_2$ is easily calculated from the above
information and we have:
\begin{equation}\label{caretterusOD3}
    \chi_3 \, = \, \{2,-1,2,0,0\}
\end{equation}
%%%%%%%%%%%%%%%%%%%%%%
\subsubsection{$D_4$ : the three-dimensional defining representation}
The three dimensional representation $D_4$ is simply the defining
representation, where the generators $T$ and $S$ are given by the
matrices in eq.(\ref{generatiTS}).
\begin{equation}\label{D4map}
    D_4(T)\, = \, T \quad ; \quad D_4(S) \, = \, S
\end{equation}
From this information the characters are immediately calculated and
we get:
\begin{equation}\label{caretterusOD4}
    \chi_3 \, = \, \{3,0,-1,-1,1\}
\end{equation}
\subsubsection{$D_5$ : the three-dimensional unoriented representation}
The three dimensional representation $D_5$ is simply that one  where
the generators $T$ and $S$ are given by the following  matrices:
\begin{equation}
D_5(T) \, =  \, \left(
\begin{array}{lll}
 0 & 1 & 0 \\
 0 & 0 & 1 \\
 1 & 0 & 0
\end{array}
\right)\quad ; \quad D_5(S) \, = \, \left(
\begin{array}{lll}
 0 & 1 & 0 \\
 1 & 0 & 0 \\
 0 & 0 & 1
\end{array}
\right)
\end{equation}
From this information the characters are immediately calculated and
we get:
\begin{equation}\label{caretterusOD4bis}
    \chi_5 \, = \, \{3,0,-1,1,-1\}
\end{equation}
%%%%%%%%%%%%%%%%
\begin{table}[!hbt]
  \centering
  \begin{eqnarray*}
   \begin{array}{||l||ccccc||}
    \hline
    \hline
{\begin{array}{cc} \null &\mbox{Class}\\
\mbox{Irrep} & \null\\
\end{array}} & \{\mbox{\bf e},1\} & \left\{C_3,8\right\} & \left\{C_4^2,3\right\} & \left\{C_2,6\right\} & \left\{C_4,6\right\} \\
\hline
 \hline
 D_1 \, , \quad \chi_1 \, = \,& 1 & 1 & 1 & 1 & 1 \\
D_2 \, , \quad \chi_2 \, = \,& 1 & 1 & 1 & -1 & -1 \\
 D_3 \, , \quad \chi_3 \, = \, & 2 & -1 & 2 & 0 & 0 \\
 D_4 \, , \quad \chi_4 \, = \, & 3 & 0 & -1 & -1 & 1 \\
D_5 \, , \quad \chi_5 \, = \, & 3 & 0 & -1 & 1 & -1\\
 \hline
 \hline
\end{array}
\end{eqnarray*}
  \caption{Character Table of the proper Octahedral Group}\label{caratteriO}
\end{table}
The table of characters is summarized in eq.(\ref{caratteriO}).
\subsection{The hexagonal lattice and the dihedral group $\mathrm{Dih_6}$}
\label{hexareticolo} We come next to a  discussion of the hexagonal
lattice. Since in this section all considered representations are
relative to the point group  we simplify the notation mentioning the
irreps only as $D_1, \dots, D_6$ without writing in square brackets
the group.
\subsubsection{The hexagonal lattice}
The basis vectors of the hexagonal space  lattice $\Lambda_{Hex}$
are the following ones :
\begin{equation}\label{Hexagbase}
    {\mathbf{w}}_1 \, = \, \left\{\sqrt{2},0,0\right\} \quad ; \quad {\mathbf{w}}_2 \, = \,
    \left\{-\frac{1}{\sqrt{2}},\sqrt{\frac{3}{2}},0\right\} \quad ; \quad {\mathbf{w}}_3 \, = \, \left\{0,0,\sqrt{2}\right\}
\end{equation}
which implies that the metric is the following non diagonal one:
\begin{equation}\label{nonkronecca}
    g_{\mu\nu} \, = \, \left(
\begin{array}{ccc}
 2 & -1 & 0 \\
 -1 & 2 & 0 \\
 0 & 0 & 2 \\
\end{array}
\right)
\end{equation}
The basis vectors $ {\mathbf{e}}^\mu$ of the dual momentum lattice
$\Lambda_{Hex}^\star$ do not coincide with those of the lattice
$\Lambda_{Hex}$. They are the following ones:
\begin{equation}\label{Hexagdualbas}
    {\mathbf{e}}^1 \, = \, \left\{\frac{1}{\sqrt{2}},\frac{1}{\sqrt{6}},0\right\} \quad ;
    \quad {\mathbf{e}}^2 \, = \,\left\{0,\sqrt{\frac{2}{3}},0\right\} \quad ;
    \quad {\mathbf{e}}^3 \, = \,\left\{0,0,\frac{1}{\sqrt{2}}\right\}
\end{equation}
so that the space lattice is now a proper subgroup of its dual
$\Lambda_{Hex}^\star$, named also the \textit{momentum-lattice}. In
order to understand the structure of the hexagonal lattice one ought
to consider first the hexagonal tesselation of a plane that is
generated by the first two basis vectors ${\mathbf{w}}_{1,2}$.
\par
\begin{figure}[!hbt]
\begin{center}
\includegraphics[width=150mm]{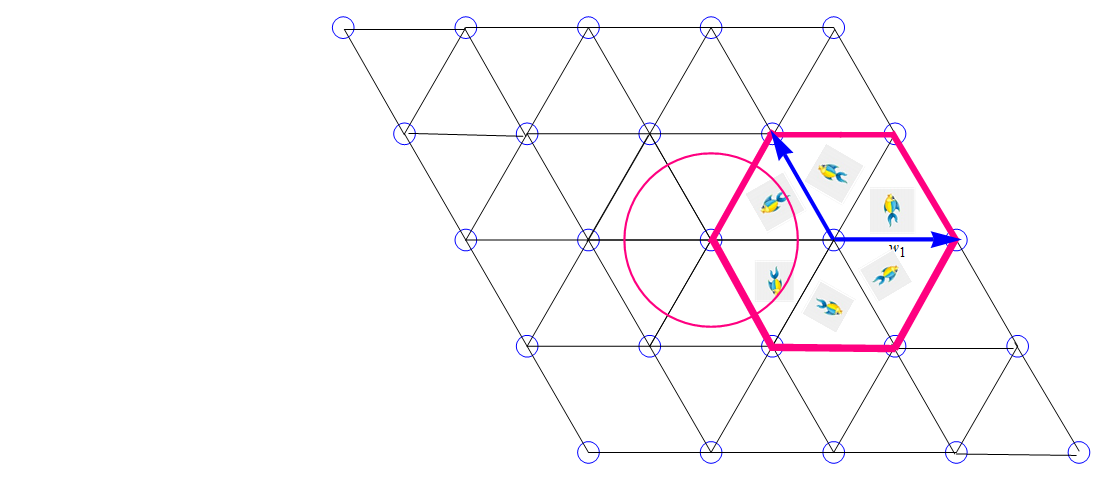}
\caption{{\it  A view of the hexagonal tesselation of the plane. The
hexagonal two dimensional lattice coincides with the $A_2$ root
lattice. Indeed the projection on the plane of the two basis vectors
$\mathbf{w}_1$ and $\mathbf{w}_2$ (the two blue vectors) are the two
simple roots of the $A_2$ Lie algebra. Each point of the lattice can
be regarded as the center of a regular hexagon whose vertices are
the first nearest neighbors. These hexagons provide a tesselation of
the infinite plane. } \label{Hexagtessere}}
\end{center}
\end{figure}

To this effect it is convenient to look at fig.\ref{Hexagtessere}
The space lattice which provides a tiling of the plane by means of
regular hexagons coincides with the root lattice of the  $A_2$ Lie
algebra its generators being the two simple roots $\alpha_{1,2}$.
\par
The plane projection of the dual lattice $\Lambda^\star_{Hex}$ is
just the weight lattice of $A_2$ the plane projection of the basis
vectors $\mathbf{e}_{1,2}$ being just the fundamental weights
$\lambda_{1,2}$. This is illustrated in the next fig.\ref{pesorete}.
\begin{figure}[!hbt]
\begin{center}
\includegraphics[width=90mm]{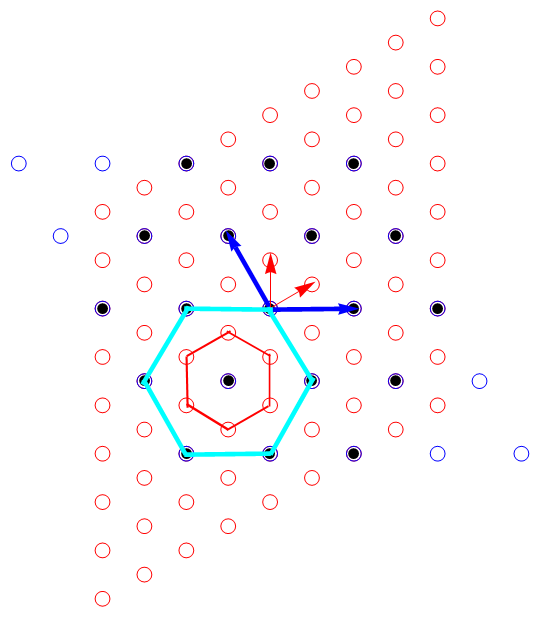}
\caption{{\it  Illustration of the dual momentum lattice of the
hexagonal lattice in the plane. The red circles are the points of
the momentum lattice, while the blue ones are the points of the
space lattice. In the finite portions of the two lattices that we
show in this picture the black points are the common ones. As we see
each point of the space--lattice is surrounded by two hexagons; the
vertices of the smaller hexagon are moment-lattice points that do
not belong to space-lattice, while the vertices of the bigger
hexagon are the space-lattice nearest neighbors, as already remarked
in the caption of fig.\ref{Hexagtessere}.} \label{pesorete}}
\end{center}
\end{figure}
There it is clearly shown that the space lattice is a sublattice of
the dual momentum lattice.
\par
The three-dimensional hexagonal lattice is obtained by adjoining an
infinite number of equally spaced planes each tiled in the way shown
in fig.s \ref{Hexagtessere} and \ref{pesorete}.
%%%%%%%%%%%%%%%%%
\begin{figure}[!hbt]
\begin{center}
\includegraphics[height=70mm]{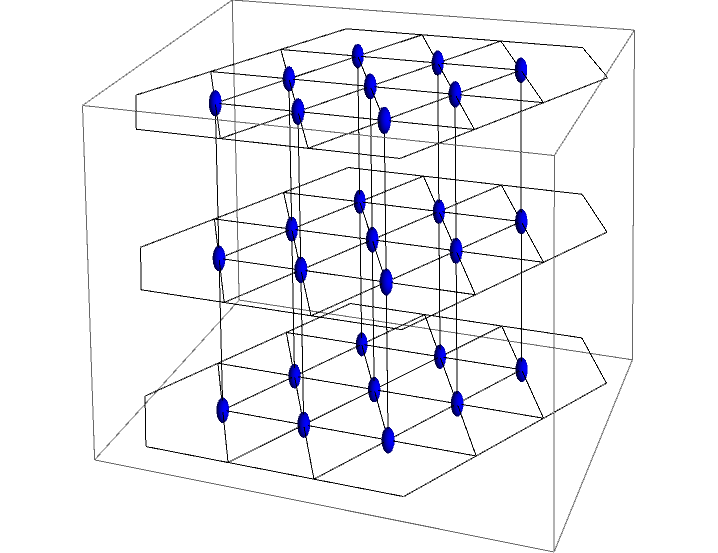}
\includegraphics[height=70mm]{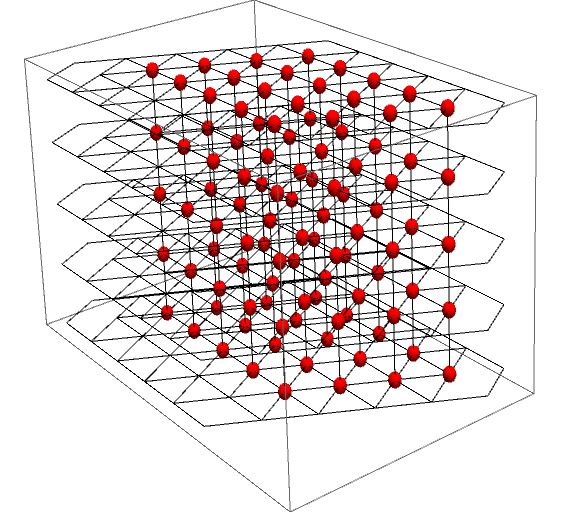}
\caption{\it  A view of the hexagonal space lattice $\Lambda_{Hex}$
(blue points on the left) and momentum momentum lattice
$\Lambda_{Hex}^\star$ (red points on the right)} \label{HexagLatPS}
\end{center}
\end{figure}
A view of the resulting three dimensional lattices is provided in
fig.\ref{HexagLatPS}.
\subsubsection{The point group $\mathrm{Dih_6}$}
The subgroup of the proper rotation group which maps the cubic
lattice into itself is the dihedral group $\mathrm{Dih}_6$ whose
order is 12. In the next lines we recall its structure.
\par
Abstractly the dihedral $\mathrm{Dih_6}$ group is defined by the
following generators and relations:
\begin{equation}\label{presentaD6}
    A \; , \; B \quad : \quad A^6 \, = \, \mbox{\bf e} \quad ; \quad B^2 \, = \, \mbox{\bf e} \quad ;
    \quad \left( BA\right)^2 \, = \, \mbox{\bf e}
\end{equation}
Explicitly in three dimensions we can take the following
matrix-representation for the generators of
$\mathrm{\mathrm{Dih_6}}$:
\begin{equation}\label{d3repD6}
    A \, = \, \left(
\begin{array}{lll}
 \frac{1}{2} & \frac{\sqrt{3}}{2} & 0 \\
 -\frac{\sqrt{3}}{2} & \frac{1}{2} & 0 \\
 0 & 0 & 1
\end{array}
\right) \quad ; \quad B \, = \, \left(
\begin{array}{lll}
 -1 & 0 & 0 \\
 0 & 1 & 0 \\
 0 & 0 & -1
\end{array}
\right)
\end{equation}
The group generated by the above generators has 12 elements that can
be arranged into 6 conjugacy classes, as it it is displayed in table
\ref{nomiD6elem}:
\begin{table}[!hbt]
  \centering
  \begin{eqnarray*}
  \begin{array}{|c|rcl|}
\hline
\mbox{\bf e} & 1_1 &  = & \{x,y,z\} \\
\hline \null & 2_1 &  = & \left\{\frac{1}{2} \left(x+\sqrt{3}
   y\right),\frac{1}{2} \left(y-\sqrt{3} x\right),z\right\} \\
A & 2_2 &  = & \left\{\frac{1}{2} \left(x-\sqrt{3}
   y\right),\frac{1}{2} \left(\sqrt{3} x+y\right),z\right\} \\
\hline \null & 3_1 &  = & \left\{\frac{1}{2} \left(\sqrt{3}
   y-x\right),\frac{1}{2} \left(-\sqrt{3} x-y\right),z\right\}
   \\
A^2 & 3_2 &  = & \left\{\frac{1}{2} \left(-x-\sqrt{3}
   y\right),\frac{1}{2} \left(\sqrt{3} x-y\right),z\right\} \\
 \hline
A^3 & 4_1 &  = & \{-x,-y,z\} \\
 \hline
\null & 5_1 &  = & \{-x,y,-z\} \\
B & 5_2 &  = & \left\{\frac{1}{2} \left(x-\sqrt{3}
   y\right),\frac{1}{2} \left(-\sqrt{3} x-y\right),-z\right\}
   \\
\null & 5_3 &  = & \left\{\frac{1}{2} \left(x+\sqrt{3}
   y\right),\frac{1}{2} \left(\sqrt{3} x-y\right),-z\right\} \\
 \hline
\null & 6_1 &  = & \left\{\frac{1}{2} \left(-x-\sqrt{3}
   y\right),\frac{1}{2} \left(y-\sqrt{3} x\right),-z\right\} \\
BA & 6_2 &  = & \{x,-y,-z\} \\
\null & 6_3 &  = & \left\{\frac{1}{2} \left(\sqrt{3}
   y-x\right),\frac{1}{2} \left(\sqrt{3} x+y\right),-z\right\}\\
   \hline
\end{array}
\end{eqnarray*}
  \caption{Conjugacy Classes of the Dihedral Group $\mathrm{Dih_6}$}\label{nomiD6elem}
\end{table}
In such a table every group element is uniquely identified by its
action on the three-dimensional vector $\left\{x,y,z\right\}$. The
multiplication table of the group $\mathrm{Dih_6}$ is also omitted
for brevity.
%%%%%%%%%%%%%%%%%%%%%%%
\subsubsection{Irreducible representations of the dihedral group $\mathrm{Dih_6}$
and the character table} The group $\mathrm{Dih_6}$ has six
conjugacy classes. Therefore according to theory we expect six
irreducible representations that we name $D_i$, $i\, =\, 1,\dots,
6$. Let us briefly describe them. The first four representations are
one-dimensional.
\subsubsection{$D_1$ : the identity representation}
The identity representation which exists for all groups is that one
where to each element of $\mathrm{Dih_6}$ we associate the number
$1$
\begin{equation}\label{D6identD1}
    \forall \, \gamma \, \in \, \mathrm{O} \,\, : \quad D_1(\gamma) \, = \, 1
\end{equation}
Obviously the character of such a representation is:
\begin{equation}\label{caretterusD6D1}
    \chi_1 \, = \, \{1,1,1,1,1\}
\end{equation}
\subsubsection{$D_2$ : the second one-dimensional representation}
The representation $D_2$ is also one-dimensional. It is constructed
as follows.
\begin{equation}
\begin{array}{ccccccc}
 \forall \, \gamma & \in & \{\mbox{\bf e}\} & : & D_2(\gamma) & = & 1 \\
  \forall \, \gamma & \in & \{A\} & : & D_2(\gamma) & = & -1 \\
   \forall \, \gamma & \in &\left\{A^2\right\}& : & D_2(\gamma) & = & 1 \\
 \forall \, \gamma & \in  & \left\{A^3\right\}& : & D_2(\gamma) & = & -1 \\
   \forall \, \gamma & \in & \{B\}& : & D_2(\gamma) & = & 1 \\
  \forall \, \gamma & \in  & \{{BA}\} & : & D_2(\gamma) & = & -1 \\
\end{array}
\end{equation}
Clearly the corresponding character vector is the following one.
\begin{equation}\label{caretterusD6D2}
    \chi_2 \, = \, \{1,-1,1,-1,1,-1\}
\end{equation}
Said in another way, this is the representation where $A \, = \, -1$
and $B\, = \,1$.
\subsubsection{$D_3$ : the third one-dimensional representation}
The representation $D_3$ is also one-dimensional. It is constructed
as follows.
\begin{equation}
\begin{array}{ccccccc}
 \forall \, \gamma & \in & \{\mbox{\bf e}\} & : & D_2(\gamma) & = & 1 \\
  \forall \, \gamma & \in & \{A\} & : & D_2(\gamma) & = & -1 \\
   \forall \, \gamma & \in &\left\{A^2\right\}& : & D_2(\gamma) & = & 1 \\
 \forall \, \gamma & \in  & \left\{A^3\right\}& : & D_2(\gamma) & = & -1 \\
   \forall \, \gamma & \in & \{B\}& : & D_2(\gamma) & = & -1 \\
  \forall \, \gamma & \in  & \{{BA}\} & : & D_2(\gamma) & = & 1 \\
\end{array}
\end{equation}
Clearly the corresponding character vector is the following one.
\begin{equation}\label{caretterusD6D3}
    \chi_3 \, = \, \{1,-1,1,-1,-1,1\}
\end{equation}
Said in another way, this is the representation where $A \, = \, -1$
and $B\, = \, -1$.
\subsubsection{$D_4$ : the fourth one-dimensional representation}
The representation $D_4$ is also one-dimensional. It is constructed
as follows.
\begin{equation}
\begin{array}{ccccccc}
 \forall \, \gamma & \in & \{\mbox{\bf e}\} & : & D_2(\gamma) & = & 1 \\
  \forall \, \gamma & \in & \{A\} & : & D_2(\gamma) & = & 1 \\
   \forall \, \gamma & \in &\left\{A^2\right\}& : & D_2(\gamma) & = & 1 \\
 \forall \, \gamma & \in  & \left\{A^3\right\}& : & D_2(\gamma) & = & 1 \\
   \forall \, \gamma & \in & \{B\}& : & D_2(\gamma) & = & -1 \\
  \forall \, \gamma & \in  & \{{BA}\} & : & D_2(\gamma) & = & -1 \\
\end{array}
\end{equation}
Clearly the corresponding character vector is the following one.
\begin{equation}\label{caretterusD6D2bis}
    \chi_4 \, = \, \{1,1,1,1,-1,-1\}
\end{equation}
Said in another way, this is the representation where $A \, = \, 1$
and $B\, = \, -1$.
\subsubsection{$D_5$ : the first two-dimensional representation}
The representation $D_5$ is two-dimensional and it corresponds to a
homomorphism:
\begin{equation}\label{D6mapD5}
    D_5 \, : \quad \mathrm{Dih_6} \, \rightarrow \, \mathrm{SL(2,\mathbb{C})}
\end{equation}
which associates to each element of the dihedral group a $2 \times
2$ complex valued matrix of determinant one. The homomorphism is
completely specified by giving the two matrices representing the two
generators:
\begin{equation}\label{D6mapD5bis}
    D_5(A) \, = \,
\left(
\begin{array}{ll}
 e^{\frac{i \pi }{3}} & 0 \\
 0 & e^{-\frac{i \pi }{3}}
\end{array}
\right)\quad ; \quad D_5(B) \, = \, \left(
\begin{array}{ll}
 0 & 1 \\
 1 & 0
\end{array}
\right)
\end{equation}
The character vector of $D_5$ is easily calculated from the above
information and we have:
\begin{equation}\label{caretterusD6D5}
    \chi_5 \, = \, \{2 , \, 1 ,\, -1 ,\, -2,\, 0,\,  0\}
\end{equation}
%%%%%%%%%%%%%%%%%%%%%%
\subsubsection{$D_6$ :  the second two-dimensional representation}
The representation $D_6$ is also two-dimensional and it corresponds
to a homomorphism:
\begin{equation}\label{D6mapD5tris}
    D_6 \, : \quad \mathrm{Dih_6} \, \rightarrow \, \mathrm{SL(2,\mathbb{C})}
\end{equation}
which associates to each element of the dihedral group a $2 \times
2$ complex valued matrix of determinant one. The homomorphism is
completely specified by giving the two matrices representing the two
generators:
\begin{equation}\label{D6mapD6}
    D_6(A) \, = \,
\left(
\begin{array}{ll}
 e^{\frac{2 i \pi }{3}} & 0 \\
 0 & e^{-\frac{2 i \pi }{3}}
\end{array}
\right)\quad ; \quad D_6(B) \, = \, \left(
\begin{array}{ll}
 0 & 1 \\
 1 & 0
\end{array}
\right)
\end{equation}
The character vector of $D_6$ is easily calculated from the above
information and we have:
\begin{equation}\label{caretterusD6D6}
    \chi_6 \, = \, \{2 , \, -1 ,\, -1 ,\, 2,\, 0,\,  0\}
\end{equation}
%%%%%%%%%%%%%%%%%%%%%%%%%%%%
\begin{table}[!hbt]
  \centering
  \begin{eqnarray*}
    \begin{array}{||l||cccccc||}
    \hline
    \hline
{\begin{array}{cc} \null &\mbox{Class}\\
\mbox{Irrep} & \null\\
\end{array}}& \{\mbox{\bf e},1\} & \{A,2\} & \left\{A^2,2\right\} & \left\{A^3,1\right\} & \{B,3\} &\{{BA},3\} \\
\hline \hline
 D_1 \, , \quad \chi_1 \, = \,& 1 & 1 & 1 & 1 & 1 & 1 \\
 \hline
 D_2 \, , \quad \chi_2 \, = \,& 1 & -1 & 1 & -1 & 1 & -1 \\
 \hline
 D_3 \, , \quad \chi_3 \, = \,& 1 & -1 & 1 & -1 & -1 & 1 \\
 \hline
 D_4 \, , \quad \chi_4 \, = \,& 1 & 1 & 1 & 1 & -1 & -1 \\
 \hline
 D_5 \, , \quad \chi_5 \, = \,& 2 & 1 & -1 & -2 & 0 & 0 \\
 \hline
 D_6\, , \quad \chi_6 \, = \,& 2 & -1 & -1 & 2 & 0 & 0\\
 \hline
\end{array}
\end{eqnarray*}
  \caption{The character table of the dihedral group $\mathrm{Dih_6}$}\label{D6caratter}
\end{table}
The character table of the $\mathrm{Dih_6}$ group is summarized in
table \ref{D6caratter}.
\subsection{Extensions of the Point Group with translations and the Universal Classifying Group}
\label{maingruppo} We come now to what constitutes the main
mathematical point of \cite{Fre:2015mla}, namely the extension of
the point group with appropriate discrete subgroups of the
compactified translation group $\mathrm{U(1)}^3$. This issue bears
on a classical topic dating back to the XIX century, which was
developed by crystallographers and in particular by the great
russian mathematician Fyodorov \cite{fyodorovcryst}. We refer here
to the issue of space groups which historically resulted into the
classification of the $230$ crystallographic groups, well known in
the chemical literature, for which an international system of
notations and conventions was established that is available in
numerous  encyclopedic tables and books. Although in
\cite{Fre:2015mla}  one key-point of the logic that leads to  the
classification of space groups, was utilized, yet the pursued goal
happened to be slightly different. Indeed what was aimed at was not
 the identification of the various space groups, rather the construction
of what was christened in \cite{Fre:2015mla} the \textit{Universal
Classifying Group}, namely  a single large group which contains all
the existing \textit{space groups} as subgroups. It was advocated in
\cite{Fre:2015mla} that such   \textit{Universal Classifying Group}
is the one appropriate to organize the eigenfunctions of the
$\star_g \mathrm{d}$-operator into irreducible representations  and
eventually to uncover the available hidden symmetries of all
Arnold-Beltrami flows.
%%%%%%%%%%%%%%%%%%
\subsubsection{Group extensions}
\label{spaziogruppi} The idea of space groups is naturally related
with the notion of group-extensions. Here we analyze how it arises.
The covering manifold of the $\mathrm{T^3}$ torus is $\mathbb{R}^3$
which can be regarded as the following coset manifold:
\begin{equation}\label{fantacosetto}
\mathbb{R}^3 \, \simeq \, \frac{\mathbb{E}^3}{\mathrm{SO(3)}} \quad
; \quad  \, \mathbb{E}^3 \, \equiv \, \mathrm{ISO(3)} \, \doteq \,
\mathcal{T}^3 \ltimes \mathrm{SO(3)}
\end{equation}
where $\mathcal{T}^3$  is the three dimensional translation group
acting on $\mathbb{R}^3$ in the standard way:
\begin{equation}\label{trasluco}
\forall \mathbf{t} \, \in \, \mathcal{T}^3 \, ,\, \forall \mathbf{x}
\, \in \, \mathbb{R}^3\quad | \quad\mathbf{t} \, : \,   \mathbf{x}
\, \rightarrow \, \mathbf{x}\, + \, \mathbf{t}
\end{equation}
and the Euclidian group $\mathbb{E}^3$ is the semi-direct product of
the  translation group $\mathcal{T}^3$ with the proper rotation
group $\mathrm{SO(3)}$.
\par
In an abstract notation the semi-direct product of  two groups
$\mathrm{T}$ and $\mathrm{G_0}$, where $\mathrm{T}$ is abelian and
supports an action of $\mathrm{G_0}$ which is not necessarily
abelian:
\begin{equation}\label{astrattone}
    \forall \gamma \in \mathrm{G_0}\quad \text{and} \quad  \forall t \in \mathrm{T}
    \quad\quad\quad  \gamma \quad : \quad \mathrm{T} \, \longrightarrow \,\mathrm{T}
    \, \quad ; \quad
    \gamma\circ t \,\in \, T
\end{equation}
can be presented as it follows. As a set the semidirect product:
\begin{equation}\label{strggoG}
   \mathrm{G} \, = \,\mathrm{T} \, \ltimes \, \mathrm{G_0}
\end{equation}
is the cartesian product $\mathrm{T} \times \mathrm{G_0}$ and the
product law $\bullet$ on the set of pairs of elements $\left(t\in
\mathrm{T}\, , \, \gamma \in \mathrm{G_0} \right)$ is the following
one:
\begin{equation}\label{diretprodut}
    \left(t\, , \, \alpha\right)\,\bullet\,\left(w\, , \,
    \beta\right)\, = \, \left( t+\alpha\circ w \, , \, \alpha \cdot
    \beta\right)
\end{equation}
where the product operation for the abelian group $\mathrm{T}$ has
been denoted with $+$, (the inverse is $-$) and the neutral element
is $0$, while for the group $\mathrm{\mathrm{G_0}}$ the product
operation is denoted by $\cdot$ and the neutral element is denoted
by $\mathbf{1}$. As a consequence of the definition of direct
product the original abelian group $\mathrm{T}$ is a normal subgroup
of $\mathrm{G}$:
\begin{equation}\label{fradiacono}
    \mathrm{T} \, \lhd \,\mathrm{G}
\end{equation}
The direct product construction is an example of the realization of
the following exact sequence of four maps $\mu_i$:
\begin{equation}\label{carnevalepisano}
    0 \,\underbrace{\stackrel{\iota}{\longrightarrow}}_{\mu_1} \, \mathrm{T}
    \,\underbrace{\stackrel{\iota}{\longrightarrow}}_{\mu_2}
    \, \mathrm{G} \, \,\underbrace{\stackrel{\pi}{\longrightarrow}}_{\mu_3} \,
    \mathrm{\mathrm{G_0}} \,\underbrace{\stackrel{\pi}{\longrightarrow}}_{\mu_4} \,\mathbf{ 1}
\end{equation}
The first map is the injection map of the neutral element of
$\mathrm{T}$ into the group it pertains to. The second map is the
injection map of the abstract group $\mathrm{T}$ as a normal
subgroup in some group $\mathrm{G}$, the third map is the projection
onto the quotient $\mathrm{\mathrm{G_0}} \, \equiv \,
\frac{\mathrm{G}}{\mathrm{T}}$, the fourth map is the projection of
the entire $\mathrm{\mathrm{G_0}}$ onto its neutral element
$\mathbf{1}$. The exactness property of the sequence:
\begin{equation}\label{crinaldo}
    \text{ker}(\mu_i) \, = \, \text{Im}(\mu_{i+1})
\end{equation}
is evident from the description. Any time we succeed in realizing
the middle term $\mathrm{G}$ in such an exact sequence as that in
eq. (\ref{carnevalepisano}) we say that $\mathrm{G}$ is a
\textbf{group extension} of $\mathrm{T}$ by means of the group
$\mathrm{G_0}$ which is supposed to have an automorphic action on
$\mathrm{T}$. The direct product is just one example of the
realizations of such group extensions but it is not the only one.
\subsubsection{The exact sequence for space groups and the inhomogeneous group
$\mathfrak{Ip}_\Lambda$} In modern mathematical language the space
groups of crystallography emerge just in the  way described above.
We choose a crystallographic lattice $\Lambda$ and a finite point
group $\mathfrak{P}\subset \mathrm{SO(3)}$ that is the maximal one
$\mathfrak{P}_{\Lambda}^{max}$ leaving $\Lambda$ invariant or one of
its subgroups and we write the exact sequence
\begin{equation}\label{carnevalelivorno}
    0 \,\stackrel{\iota}{\longrightarrow} \, \mathrm{\Lambda}
    \,\stackrel{\iota}{\longrightarrow}
    \, \mathfrak{S} \, \,\stackrel{\pi}{\longrightarrow} \,
    \mathrm{\mathfrak{P}} \,\stackrel{\pi}{\longrightarrow}\,\mathbf{ 1}
\end{equation}
where $\mathfrak{S}$ is the space group. One possible construction
of the exact sequence is the already mentioned semi-direct product:
\begin{equation}\label{ltimo}
    \mathfrak{S}_\times \, = \, \Lambda \,\ltimes \,\mathfrak{P}
\end{equation}
which we can reproduce quite conveniently through the use of
$4\times 4$ matrices of the following type:
\begin{equation}\label{quattroruote}
\forall \left(\mathfrak{t} , \, \gamma\right) \, \in \,
\mathfrak{S}_{\times} \, \rightarrow\, {D}_\times\left[
\left(\mathfrak{t} , \, \gamma\right) \right] \, = \, \left(
\begin{array}{c|c}
\hat{\gamma} & {\mathfrak{t}} \\
\hline 0& 1
\end{array}
\right)
\end{equation}
where by $\hat{\gamma}$ we mean the $3\times 3$ matrix realization
of the abstract group element $\gamma$ in the defining
representation of $\mathfrak{P}\subset \mathrm{SO(3)}$. Let us see
how we can realize the exact sequence (\ref{carnevalelivorno}) in a
more general way.
\par
We begin by observing that harmonic analysis on $\mathbb{R}^3$ is a
complicated matter of functional analysis since $\mathcal{T}^3$ is a
non-compact group and its unitary irreducible representations are
infinite-dimensional. The landscape changes drastically when we
compactify our manifold from $\mathbb{R}^3$ to the three torus
$\mathrm{T^3}$. Compactification is obtained taking the  quotient of
$\mathbb{R}^3$ with respect to the lattice $\Lambda \, \subset \,
\mathcal{T}^3$. As a result of this quotient the manifold becomes
$\mathbb{S}^1 \times \mathbb{S}^1 \times \mathbb{S}^1$ but also the
isometry group is reduced. Instead of $\mathrm{SO(3)}$ as rotation
group we are left with its discrete subgroup
$\mathfrak{P}^{max}_\Lambda \, \subset \, \mathrm{SO(3)}$  which
maps the lattice $\Lambda$ into itself (the maximal point group or a
subgroup thereof $\mathfrak{P}_\Lambda \subset
\mathfrak{P}^{max}_\Lambda \, \subset \, \mathrm{SO(3)}$) and
instead of the translation subgroup $\mathcal{T}^3$ we are left with
the quotient group:
\begin{equation}\label{quozienTra}
    \mathfrak{T}^3_\Lambda \, \equiv \, \frac{\mathcal{T}^3}{\Lambda} \, \simeq \,
    \mathrm{U(1)} \times \mathrm{U(1) }\times \mathrm{U(1)}
\end{equation}
In this way we obtain a new group which replaces the Euclidian group
and which is the semidirect product of $\mathfrak{T}^3_\Lambda $
with the point group $\mathfrak{P}_\Lambda$  :
\begin{equation}\label{goticoIG}
    \mathfrak{Ip}_{\Lambda}  \, \equiv \,   \mathfrak{T}^3_\Lambda \,\ltimes
    \, \mathfrak{P}_\Lambda
\end{equation}
The group $ \mathfrak{Ip}_{\Lambda}$ that can be named the
\textbf{Inhomogeneous Point Group} is an exact symmetry of Beltrami
equation (\ref{formaduale}) and its action is naturally defined on
the parameter space of any of its solutions $\mathbf{V}\left(
{\mathbf{x}} | \mathbf{F}\right)$ that we can obtain by means of the
algorithm  described in section \ref{algoritmo}.  To appreciate this
point let us state that every component of the vector field
$\mathbf{V}\left( {\mathbf{x}} | \mathbf{F}\right)$ associated with
a $\mathfrak{P}_\Lambda$ point--orbit $\mathcal{O}$ is a linear
combinations of the functions $\cos\left[2\pi \, \mathbf{k}_i \cdot
{\mathbf{x}}\right]$ and $\sin\left[2\pi \, \mathbf{k}_i \cdot
{\mathbf{x}}\right]$, where $\mathbf{k}_i \, \in \, \mathcal{O}$ are
all the momentum vectors contained in the orbit. Consider next the
same functions in a translated point of the three torus
${\mathbf{x}}^\prime \, = \,{\mathbf{x}}\,  + \, \mathbf{c}$ where
$\mathbf{c} \, = \, \{\xi_1\,,\,\xi_2\, , \, \xi_3\,\}$ is a
representative of an equivalence class $\mathfrak{c}$ of constant
vectors defined modulo the lattice:
\begin{equation}\label{cribulla}
\mathfrak{c} \, = \, \mathbf{c} \, + \, \mathfrak{t} \quad ; \quad
\forall \mathfrak{t}\, \in \, \Lambda
\end{equation}
The above equivalence classes are the elements  of the quotient
group $\mathfrak{T}^3_\Lambda $. Using standard trigonometric
identities $\cos\left[2\pi \, \mathbf{k}_i \cdot {\mathbf{x}}\, + \,
2\pi \, \mathbf{k}_i \cdot \mathbf{c} \right]$  can be reexpressed
as a linear combination of the $\cos\left[2\pi \, \mathbf{k}_i \cdot
{\mathbf{x}}\right]$ and $\sin\left[2\pi \, \mathbf{k}_i \cdot
{\mathbf{x}}\right]$ functions with coefficients that depend on
trigonometric functions of $\mathfrak{c}$. The same is true of
$\sin\left[2\pi \, \mathbf{k}_i \cdot {\mathbf{x}}\, + \, 2\pi \,
\mathbf{k}_i \cdot \mathbf{c} \right]$. Note also that because of
the periodicity of the trigonometric functions, the shift in their
argument by a lattice translation is not-effective so that one deals
only with the equivalence classes (\ref{cribulla}).  It follows that
for each element $\mathfrak{c}\in \mathfrak{T}^3_\Lambda$  we obtain
a matrix representation $\mathcal{M}_\mathfrak{c}$ realized on the
$F$ parameters and defined by the following equation:
\begin{eqnarray}\label{traslazionaBuh}
   & \mathbf{V}\left({\mathbf{x}}+\mathbf{c} |\mathbf{F}\right)\, =\,
   \mathbf{V}\left({\mathbf{x}} |\mathcal{M}_\mathfrak{c} \mathbf{F}\right) &
\end{eqnarray}
As we already noted in eq.(\ref{Rtrasformogen}), for any group
element $\gamma \, \in \, \mathfrak{P}_\Lambda$ we also have a
matrix representation induced on the parameter space by the same
mechanism:
\begin{equation}\label{RotazioneBuh}
\forall \, \gamma \, \in \, \mathfrak{P}_\Lambda \, : \quad
\gamma^{-1} \,\cdot \, \mathbf{V}\left(\gamma \,\cdot \,
{\mathbf{x}}\, | \, \mathbf{F}\right) \, = \, \mathbf{V}\left(
{\mathbf{x}}\, | \, \mathfrak{R}[\gamma] \,\cdot \,
\mathbf{F}\right)
\end{equation}
Combining eq.s(\ref{traslazionaBuh}) and (\ref{RotazioneBuh}) we
obtain a matrix realization of the entire group
$\mathfrak{G}_{\Lambda} $ in the following way:
\begin{eqnarray}
\mathbf{V}\left(\gamma\, \cdot \,  {\mathbf{x}}+\mathbf{c}
|\mathbf{F}\right)&=& \gamma \, \cdot \, \mathbf{V}\left(
{\mathbf{x}}
\,  |\, \mathfrak{R}[\gamma] \cdot \mathcal{M}_\mathfrak{c} \,\cdot \,\mathbf{F}\right) \\
&\Downarrow&\nonumber\\
\forall \left(\gamma \, , \, \mathfrak{c}\right) \, \in \,
\mathfrak{Ip}_{\Lambda} &\rightarrow& D\left[ \left(\gamma \, , \,
\mathfrak{c}\right) \right] \, = \, \mathfrak{R}[\gamma] \cdot
\mathcal{M}_\mathfrak{c}\label{direttocoriandolo}
\end{eqnarray}
Actually the construction of Beltrami vector fields in the lowest
lying point-orbit, which usually yields a faithful matrix
representation of all group elements, can be regarded as an
automatic way of taking  the quotient (\ref{quozienTra}) and the
resulting representation can be considered the defining
representation of the group $\mathfrak{Ip}_{\Lambda}$.
\par
The next point in the logic which leads to space groups is the
following observation. $\mathfrak{Ip}_{\Lambda}$ is an unusual
mixture of a discrete group (the point group ) with a continuous one
(the translation subgroup $\mathfrak{T}^3_\Lambda $). This latter is
rather trivial, since its action corresponds to shifting the origin
of coordinates in three-dimensional space and, from the point of
view of the first order differential system that defines
trajectories (see eq.(\ref{streamlines})), it simply corresponds to
varying the integration constants. Yet there are in
$\mathfrak{Ip}_{\Lambda}$ some discrete subgroups which can be
isomorphic to the point group  $\mathfrak{P}_{\Lambda}$, or to one
of its subgroups $\mathfrak{H}_\Lambda \, \subset \,
\mathfrak{P}_{\Lambda}$, without being their conjugate in
$\mathfrak{Ip}_{\Lambda}$. Such groups cannot be disposed of by
shifting the origin of coordinates and consequently they can encode
non-trivial hidden symmetries of the dynamical system
(\ref{streamlines}).
\par
The precise mathematical way of thinking is encoded in the already
presented exact sequence (\ref{carnevalelivorno}). Given the point
group $\mathfrak{P}_\Lambda$ and its semidirect product extension
with translations reduced to the unit cell $\mathcal{T}^3_{unit}
\simeq \mathrm{U(1)\times U(1) \times U(1)} $, namely
$\mathfrak{G}_{\Lambda} $, the original point group can be
identified as the quotient group:
\begin{equation}\label{colbacco}
    \mathfrak{P}_\Lambda \, \simeq \,
    \frac{\mathfrak{Ip}_{\Lambda}}{\mathcal{T}^3_{unit}}
\end{equation}
since $\mathcal{T}^3_{unit}$ is a normal subgroup:
\begin{equation}\label{cappelloconorecchie}
    \mathcal{T}^3_{unit} \lhd \mathfrak{Ip}_{\Lambda}
\end{equation}
We would like to construct the entire equivalence class of elements
in $ \mathfrak{Ip}_{\Lambda}$ for each element $\gamma \in
\mathfrak{P}_\Lambda$. Choosing representatives in these classes we
can realize the various group extensions $\mathfrak{S}$ that can
occupy the middle point in the exact sequence
(\ref{carnevalelivorno}).
\par
This is the mission accomplished by crystallographers the result of
the mission being the classification of space groups. It suffices to
realize a generalized copy of each generator of the point group and
by means of multiplication we obtain the equivalence classes of each
point group element.
\par
This leads to the so named Frobenius congruences
\cite{Aroyo}\cite{Souvignier}. Let us outline this construction.
\subsubsection{Frobenius congruences} Following classical
approaches we use the already introduced $4\times 4$ matrix
representation of the group $\mathfrak{Ip}_{\Lambda}$. Performing
the matrix product of two elements, in the translation block one has
to take into account equivalence modulo lattice $\Lambda$, namely
\begin{equation}\label{moduloLatte}
\left( \begin{array}{c|c}
\gamma_1 & \mathfrak{c_1} \\
\hline 0& 1
\end{array}
\right) \, \cdot \, \left( \begin{array}{c|c}
\gamma_2 & \mathfrak{c_2} \\
\hline 0& 1
\end{array}
\right) \, = \,  \left( \begin{array}{c|c}
\gamma_1\, \cdot \, \gamma_2 & \gamma_1\,  \mathfrak{c_2} \,
+ \, \mathfrak{c_1}\, + \, \Lambda \\
\hline 0& 1
\end{array}
\right)
\end{equation}
Utilizing this notation the next step consists of introducing
translation deformations of the generators of the point group
$\mathfrak{P}_\Lambda$ searching for deformations that cannot be
eliminated by conjugation with elements of the normal subgroup
$\mathfrak{T}^3 \lhd \mathfrak{Ip}_\Lambda$. We go through the steps
of such a construction both in the case of the maximal point group
for the cubic lattice $\mathfrak{P}_{cubic}^{max}\,= \,
\mathrm{O_{24}}$, denoting with $\mathrm{O_{24}}$ the octahedral
group, and in the case of the maximal point group for the
alternative hexagonal lattice $\mathfrak{P}_{hexag}^{max}\,= \,
\mathrm{Dih_6}$.
\subsubsection{Frobenius congruences for the
Octahedral Group $\mathrm{O_{24}}$} The octahedral  group is
abstractly defined by the presentation displayed in
eq.(\ref{octapresa}). As a first step we parameterize the candidate
deformations of the two generators $T$ and $S$ in the following way:
\begin{equation}\label{deformuccia}
\hat{T} \, = \, \left(
\begin{array}{lll|l}
 0 & 1 & 0 & \tau _1 \\
 0 & 0 & 1 & \tau _2 \\
 1 & 0 & 0 & \tau _3 \\
 \hline
 0 & 0 & 0 & 1
\end{array}
\right) \quad ; \quad \hat{S} \, = \, \left(
\begin{array}{lll|l}
 0 & 0 & 1 & \sigma _1 \\
 0 & -1 & 0 & \sigma _2 \\
 1 & 0 & 0 & \sigma _3 \\
 \hline
 0 & 0 & 0 & 1
\end{array}
\right)
\end{equation}
which should be compared with eq.(\ref{generatiTS}). Next we try
impose on the deformed generators the defining relations of
$\mathrm{O}_{24}$. By explicit calculation we find:
\begin{eqnarray*}\label{finocchiona}
 \hat{T}^3 & = &    \left(
\begin{array}{lll|l}
 1 & 0 & 0 & \tau _1+\tau _2+\tau _3 \\
 0 & 1 & 0 & \tau _1+\tau _2+\tau _3 \\
 0 & 0 & 1 & \tau _1+\tau _2+\tau _3 \\
 \hline
 0 & 0 & 0 & 1
\end{array}
\right) \quad ; \quad \hat{S}^2 \, = \, \left(
\begin{array}{lll|l}
 1 & 0 & 0 & \sigma _1+\sigma _3 \\
 0 & 1 & 0 & 0 \\
 0 & 0 & 1 & \sigma _1+\sigma _3 \\
 \hline
 0 & 0 & 0 & 1
\end{array}
\right) \; ; \; \left(\hat{S}\hat{T}\right)^4 \, = \, \left(
\begin{array}{lll|l}
 1 & 0 & 0 & 4 \sigma _1+4 \tau _3 \\
 0 & 1 & 0 & 0 \\
 0 & 0 & 1 & 0 \\
 \hline
 0 & 0 & 0 & 1
\end{array}
\right)
\end{eqnarray*}
so that we obtain the conditions:
\begin{equation}\label{frobeniale}
\tau _1+\tau _2+\tau _3  \, \in \, \mathbb{Z}\quad ; \quad \sigma
_1+\sigma _3 \, \in \, \mathbb{Z} \quad ; \quad 4 \sigma _1+4 \tau
_3 \, \in \, \mathbb{Z}
\end{equation}
which are the Frobenius congruences for the present case. Next we
consider the effect of conjugation with the most general translation
element of the group $\mathfrak{T}^3\lhd \mathfrak{Ip}_{cubic}$.
Just for convenience we parameterize the translation subgroup as
follows:
\begin{equation}\label{tmatto}
    \mathfrak{t} \, = \, \left(
\begin{array}{lll|l}
 1 & 0 & 0 & a+c \\
 0 & 1 & 0 & b \\
 0 & 0 & 1 & a-c \\
 \hline
 0 & 0 & 0 & 1
\end{array}
\right)
\end{equation}
and we get:
\begin{equation}\label{giambatto}
    \mathfrak{t} \, \hat{T} \, \mathfrak{t}^{-1} \, = \, \left(
\begin{array}{lll|l}
 0 & 1 & 0 & a-b+c+\tau _1 \\
 0 & 0 & 1 & -a+b+c+\tau _2 \\
 1 & 0 & 0 & \tau _3-2 c \\
 \hline
 0 & 0 & 0 & 1
\end{array}
\right)\quad ; \quad \mathfrak{t} \, \hat{S} \, \mathfrak{t}^{-1} \,
= \,\left(
\begin{array}{lll|l}
 0 & 0 & 1 & 2 c+\sigma _1 \\
 0 & -1 & 0 & 2 b+\sigma _2 \\
 1 & 0 & 0 & \sigma _3-2 c \\
 \hline
 0 & 0 & 0 & 1
\end{array}
\right)
\end{equation}
This shows that by using the parameters $b,c$ we can always put
$\sigma_1 \, = \, \sigma_2 \, = \,0$, while using the parameter $a$
we can put $\tau_1 \, = \,0$ (this is obviously only one possible
gauge choice, yet it is the most convenient) so that Frobenius
congruences reduce to:
\begin{equation}\label{calugone}
     \tau _2+\tau _3  \, \in \, \mathbb{Z}\quad ; \quad \sigma _3 \,
      \in \, \mathbb{Z} \quad ; \quad 4 \tau _3 \, \in \, \mathbb{Z}
\end{equation}
Eq.(\ref{calugone}) is of great momentum. It tells us that any non
trivial subgroup $\hat{\mathfrak{P}}\subset \mathfrak{Ip}_{cubic}$
which is isomorphic to the point group $\mathrm{O_{24}}$, but not
conjugate to it contains point group elements extended with rational
translations of the form $\mathfrak{c} \, = \, \left\{ \ft{n_1}{4}\,
, \, \ft{n_2}{4}\,  , \, \ft{n_3}{4}\right\}$ with $n_i \in
\mathbb{Z}$.
\paragraph{The example of the group $\mathrm{GS_{24}}$} An example
is provided by the group later named $\mathrm{GS_{24}}$ which will
repeatedly appear in our later discussions of Beltrami solutions. In
the direct product realization of the point group
$\mathfrak{P}=\mathrm{O_{24}}$ the generators $T$ and $S$ were
specified in eq.s (\ref{octapresa}) and (\ref{generatiTS}). In view
of the Frobenius congruences let us set:
\begin{equation}\label{TScap}
    \hat{T} \, = \,\left(
\begin{array}{cccc}
 0 & 0 & 1 & 0 \\
 1 & 0 & 0 & \frac{1}{2} \\
 0 & 1 & 0 & \frac{1}{2} \\
 0 & 0 & 0 & 1 \\
\end{array}
\right) \quad ; \quad \hat{S} \, = \,\left(
\begin{array}{cccc}
 0 & 0 & 1 & \frac{3}{2} \\
 0 & -1 & 0 & \frac{1}{2} \\
 1 & 0 & 0 & \frac{1}{2} \\
 0 & 0 & 0 & 1 \\
\end{array}
\right)
\end{equation}
By an immediate calculation we obtain:
\begin{eqnarray}\label{governolo}
    &&\hat{T}^3 \, = \,\left(
\begin{array}{cccc}
 1 & 0 & 0 & 1 \\
 0 & 1 & 0 & 1 \\
 0 & 0 & 1 & 1 \\
 0 & 0 & 0 & 1 \\
\end{array}
\right) \quad ; \quad \hat{S}^2 \, = \,\left(
\begin{array}{cccc}
 1 & 0 & 0 & 2 \\
 0 & 1 & 0 & 0 \\
 0 & 0 & 1 & 2 \\
 0 & 0 & 0 & 1 \\
\end{array}
\right)\quad ; \quad
    \left(\hat{S}\cdot\hat{T}\right)^4 \, = \,\left(
\begin{array}{cccc}
 1 & 0 & 0 & 0 \\
 0 & 1 & 0 & 0 \\
 0 & 0 & 1 & 2 \\
 0 & 0 & 0 & 1 \\
\end{array}
\right)\nonumber\\
\end{eqnarray}
The above equation is interpreted by stating that:
\begin{equation}\label{interpretone}
    \hat{T}^3 \, \in \, \Lambda \, \subset \, \mathfrak{S}_{\mathrm{GS}} \quad ;
    \quad \hat{S}^2 \in \, \Lambda \, \subset \, \mathfrak{S}_{\mathrm{GS}}
    \quad ; \quad \left(\hat{S}\cdot\hat{T}\right)^4 \, \in \, \Lambda
    \, \subset \, \mathfrak{S}_{\mathrm{GS}}
\end{equation}
where $\mathfrak{S}_{GS}$ is the space group in the exact sequence:
\begin{equation}\label{carnevaleorbetello}
    0 \,\stackrel{\iota}{\longrightarrow} \, \mathrm{\Lambda}
    \,\stackrel{\iota}{\longrightarrow}
    \, \mathfrak{S}_{\mathrm{GS}} \, \,\stackrel{\pi}{\longrightarrow} \,
    \mathrm{GS_{24}} \,\stackrel{\pi}{\longrightarrow}\,\mathbf{ 1}
\end{equation}
and the lattice normal subgroup is realized within
$\mathfrak{S}_{\mathrm{GS}}$ by all the matrices of the form:
\begin{equation}\label{gringus}
    \mathfrak{S}_{\mathrm{GS}} \, \rhd \, \Lambda \, \ni
    \, \left(
         \begin{array}{ccc|c}
           1 & 0 & 0 & n_1 \\
           0 & 1 & 0 & n_2 \\
           0 & 0 & 1 & n_3 \\
           \hline
           0 & 0 & 0 & 1 \\
         \end{array}
       \right) \quad ; \quad n_i \, \in \, \mathbb{Z}
\end{equation}
The group $\mathrm{GS_{24}}$ is defined as the quotient group:
\begin{equation}\label{divisionescolastica}
    \mathrm{GS_{24}} \, = \,
    \frac{\mathfrak{S}_{\mathrm{GS}}}{\Lambda}\, \sim \mathrm{O_{24}} \, \sim
    \mathrm{S_4}
\end{equation}
and $\mathfrak{S}_{\mathrm{GS}}$ is a group extension of the lattice
group $\Lambda$ by means of the abstract group octahedral point
group $\mathrm{O_{24}} $, yet it is not a semidirect product of the
normal subgroup $\Lambda$ with $\mathrm{O_{24}}$. Indeed the space
group $\mathfrak{S}_{\mathrm{GS}} $ contains translations that do
not belong to the cubic lattice.
\paragraph{A conceptual bifurcation} Up to this point our way and that of crystallographers
was the same: hereafter our paths separate. The crystallographers
classify all possible non trivial groups that extend the point group
with such translation deformations: indeed looking at the
crystallographic tables one realizes that  all known space groups
for the cubic lattice have translation components of the form
$\mathfrak{c} \, = \, \left\{ \ft{n_1}{4}\,  , \, \ft{n_2}{4}\,  ,
\, \ft{n_3}{4}\right\}$. On the other hand, we do something much
simpler which leads to a quite big group containing all possible
Space-Groups as subgroups, together with other subgroups that are
not space groups in the crystallographic sense.
\subsubsection{The Universal Classifying Group for the cubic lattice:
 $\mathrm{\mathrm{G_{1536}}}$}
\label{universalone} Inspired by the space group  construction and
by Frobenius congruences we just consider the subgroup of
$\mathfrak{G}_{cubic}$ where translations are quantized in units of
$\frac{1}{4}$. In each direction and modulo integers there are just
four translations $0, \, \ft 14, \, \ft 12, \, \ft 34$ so that the
translation subgroup reduces to $\mathbb{Z}_4 \,
\otimes\,\mathbb{Z}_4\, \otimes \, \mathbb{Z}_4$  that has a total
of $64$ elements. In this way we single out a discrete subgroup
$\mathrm{G_{1536}} \, \subset \, \mathfrak{G}_{cubic}$ of order $24
\times 64 \,  =  \, 1536$,  which is simply the semidirect product
of the point group $\mathrm{O_{24}}$ with $\mathbb{Z}_4 \,
\otimes\,\mathbb{Z}_4\, \otimes \, \mathbb{Z}_4$:
\begin{equation}\label{1536defino}
\mathfrak{G}_{cubic} \, \supset \, \mathrm{\mathrm{G_{1536}}} \,
\simeq \, \mathrm{O_{24}} \, \ltimes \, \left (\mathbb{Z}_4 \,
\otimes\,\mathbb{Z}_4\, \otimes \, \mathbb{Z}_4\right)
\end{equation}
We name $\mathrm{\mathrm{G_{1536}}}$ the universal classifying group
of the cubic lattice, and its elements can be labeled as follows:
\begin{equation}\label{elementando1536}
\mathrm{\mathrm{G_{1536}}} \, \in \, \left\{ p_q \, , \, \ft{2
n_1}{4} \, , \, \ft{2 n_2}{4} \, , \, \ft{2 n_3}{4}\right\} \quad
\Rightarrow\quad \left\{ \begin{array}{rcl}
p_q & \in & \mathrm{O_{24}}\\
\left\{ \ft{n_1}{4}\,  , \, \ft{n_2}{4}\,  , \, \ft{n_3}{4}\right\}
& \in & \mathbb{Z}_4 \, \otimes\,\mathbb{Z}_4\, \otimes \,
\mathbb{Z}_4\end{array}\right.
\end{equation}
where for the elements of the point group we use the labels $p_q$
established in eq.(\ref{nomiOelemen}) while for the translation part
our notation encodes an equivalence class of  translation vectors
$\mathfrak{c} \, = \,  \left\{ \ft{n_1}{4}\,  , \, \ft{n_2}{4}\,  ,
\, \ft{n_3}{4}\right\}$. The reason why we use $\left \{\ft{2
n_1}{4} \, , \, \ft{2 n_2}{4} \, , \, \ft{2 n_3}{4}\right\}$ is
simply due to computer convenience. In the quite elaborate
MATHEMATICA codes that were  utilized in \cite{Fre:2015mla} to
derive all the results such a notation was internally used  and the
automatic LaTeX Export of the outputs was  provided in this way. In
view of eq.(\ref{direttocoriandolo}) one can associate an explicit
matrix to each group element of $\mathrm{\mathrm{G_{1536}}}$,
starting from the construction of the Beltrami vector field
associated with one point orbit of the octahedral  group. Then one
can consider such matrices the defining representation of the group
if the representation is faithful. In \cite{Fre:2015mla} the lowest
lying $6$-dimensional orbit  was used which is indeed faithful.
Three matrices are sufficient to characterize completely the
defining representation just as any other representation: the matrix
representing the generator $T$, the matrix representing the
generator $S$ and the matrix representing the translation $\left\{
\ft{n_1}{4}\,  , \, \ft{n_2}{4}\,  , \, \ft{n_3}{4}\right\}$. In
\cite{Fre:2015mla} it was found:
\begin{equation}\label{TSindefi1536}
    \mathfrak{R}^{\mbox{defi}}[T] \, = \, \left(\begin{array}{llllll}
 0 & 0 & 0 & 0 & 1 & 0 \\
 0 & 0 & 0 & 0 & 0 & 1 \\
 0 & 1 & 0 & 0 & 0 & 0 \\
 1 & 0 & 0 & 0 & 0 & 0 \\
 0 & 0 & 0 & 1 & 0 & 0 \\
 0 & 0 & 1 & 0 & 0 & 0
\end{array}\right)\quad ; \quad \mathfrak{R}^{\mbox{defi}}[S]\, = \, \left(
\begin{array}{llllll}
 0 & 0 & 0 & 0 & 1 & 0 \\
 0 & 0 & 0 & 0 & 0 & 1 \\
 0 & 1 & 0 & 0 & 0 & 0 \\
 1 & 0 & 0 & 0 & 0 & 0 \\
 0 & 0 & 0 & 1 & 0 & 0 \\
 0 & 0 & 1 & 0 & 0 & 0
\end{array}
\right)
\end{equation}
\begin{eqnarray}
\mathcal{M}_{\{\ft{2 n_1}{2},\ft{2 n_2}{2},\ft{2
n_3}{2}\}}^{\mbox{defi}} &= & \left(
\begin{array}{llllll}
\cos \left(\ft{\pi}{2}  n _3\right) & 0 &
\sin \left(\ft{\pi}{2}  n _3\right) & 0 & 0 & 0 \\
0 & \cos \left(\ft{\pi}{2}  n _2\right) & 0 & 0 &
   -\sin \left(\ft{\pi}{2}  n _2\right) & 0 \\
 -\sin \left(\ft{\pi}{2}  n _3\right) & 0 & \cos
   \left(\ft{\pi}{2}  n _3\right) & 0 & 0 & 0 \\
 0 & 0 & 0 & \cos \left(\ft{\pi}{2}  n _1\right) & 0 &
   \sin \left(\ft{\pi}{2}  n _1\right) \\
 0 & \sin \left(\ft{\pi}{2}  n _2\right) & 0 & 0 & \cos
   \left(\ft{\pi}{2}  n _2\right) & 0 \\
 0 & 0 & 0 & -\sin \left(\ft{\pi}{2}  n _1\right) & 0 &
   \cos \left(\ft{\pi}{2}  n _1\right)
\end{array}
\right)\nonumber\\
\label{trasladefi1536}
\end{eqnarray}
Relying on the above matrices, any of the 1536 group elements
obtains an explicit $6\times 6$ matrix representation upon use of
formula (\ref{direttocoriandolo}). As already stressed one can
regard that above as the actual definition of the group
$\mathrm{\mathrm{G_{1536}}}$ which from this point on can be studied
intrinsically in terms of pure group theory without any further
reference to lattices, Beltrami flows or dynamical systems.
\subsubsection{Structure of the $\mathrm{\mathrm{G_{1536}}}$ group
and derivation of its irreps} The identity card of a finite group is
given by the organization of its elements into conjugacy classes,
the list of its irreducible representation and finally its character
table. Since ours is not any of the crystallographic groups, no
explicit information is available in the literature about its
conjugacy classes, its irreps and its character table. Hence the
authors of \cite{Fre:2015mla} were forced to do everything from
scratch by themselves and they could accomplish the task by means of
purposely written MATHEMATICA codes. Most of their results were
presented in the form of tables in the appendices of
\cite{Fre:2015mla}. We will reproduce here those that are most
relevant to the purposes of the present paper referring the reader
to \cite{Fre:2015mla} for additional details.
\paragraph{Conjugacy Classes}
The conjugacy classes of $\mathrm{\mathrm{G_{1536}}}$ are explicitly
presented in appendix A.1 of \cite{Fre:2015mla}. There are 37
conjugacy classes whose populations is distributed as follows:
\begin{multicols}{3}
\begin{description}
  \item[1)]  2 classes of length 1
  \item[2)]  2  classes of length 3
  \item[3)]  2 classes of  length 6
  \item[4)] 1 class of length 8
  \item[5)] 7 classes of length 12
  \item[6)] 4 classes of length 24
  \item[7)] 13 classes of length 48
  \item[8)] 2 classes of length 96
  \item[9)] 4 classes of length 128
\end{description}
\end{multicols}
It follows that there must be $37$  irreducible representations
whose construction is a task which  was accomplished  in
\cite{Fre:2015mla} utilizing an  iterative strategy algorithm
available for solvable groups. We refer the reader  to
\cite{Fre:2015mla} for a description of that algorithm.
\subsubsection{Derivation of $\mathrm{\mathrm{G_{1536}}}$ irreps}
\label{ciurlacca} Utilizing the above described algorithm,
implemented by means of purposely written MATHEMATICA codes, the
authors of \cite{Fre:2015mla} were able to derive the explicit form
of the $37$ irreducible representations of
$\mathrm{\mathrm{G_{1536}}}$ and its character table. The essential
tool is the following chain of normal subgroups:
\begin{equation}\label{pernicinormaliText}
    \mathrm{\mathrm{G_{1536}} }\, \rhd \, \mathrm{G_{768}} \,
    \rhd \, \mathrm{G_{256}} \, \rhd \, \mathrm{G_{128}} \, \rhd \, \mathrm{G_{64}}
  \end{equation}
where $\mathrm{G_{64}} \,  \sim \, \mathbb{Z}_4 \, \times \,
\mathbb{Z}_4 \, \times \, \mathbb{Z}_4$ is abelian and corresponds
to the compactified translation group. The above chain leads to the
following quotient groups:
\begin{equation}\label{fagianirossiText}
\frac{\mathrm{\mathrm{G_{1536}} }}{\mathrm{G_{768}}} \, \sim \,
\mathbb{Z}_2 \quad ; \quad \frac{\mathrm{G_{768}
}}{\mathrm{G_{256}}} \, \sim \, \mathbb{Z}_3 \quad ; \quad
\frac{\mathrm{G_{256} }}{\mathrm{G_{128}}} \, \sim \, \mathbb{Z}_2
\quad ; \quad \frac{\mathrm{G_{128} }}{\mathrm{G_{64}}} \, \sim \,
\mathbb{Z}_2
\end{equation}
The description of the normal subgroups is given in various sections
of the appendix of \cite{Fre:2015mla}.  The result for the
irreducible representations, thoroughly described also in the
appendix of \cite{Fre:2015mla} is summarized here. The $37$ irreps
are distributed according to the following pattern:
\begin{description}
  \item[a)] 4 irreps of dimension $1$, namely $D_1,\dots,D_4$
  \item[b)] 2 irreps of dimension $2$, namely $D_5,\dots,D_6$
  \item[c)] 12 irreps of dimension $3$, namely $D_6,\dots,D_{18}$
  \item[d)] 10 irreps of dimension $6$, namely $D_7,\dots,D_{28}$
  \item[e)] 3 irreps of dimension $8$, namely $D_{29},\dots,D_{31}$
  \item[f)] 6 irreps of dimension $12$, namely $D_{32},\dots,D_{37}$
\end{description}
The  character table calculated in \cite{Fre:2015mla} is  displayed
in that paper and we omit it here. We just stress that all such
results are incorporated into the \textbf{AlmafluidaNSPsystem} of
MATHEMATICA codes available through the Wolfram Community site
\url{https://community.wolfram.com/groups/-/m/t/2555905}.
\par
The irreducible representations of the universal classifying group
are a fundamental tool in the classification of Arnold-Beltrami
vector fields. Indeed by choosing the various  point group  orbits
of momentum vectors in the cubic lattice, according to their
classification presented in the next section \ref{triplettoni}, and
constructing the corresponding Arnold-Beltrami fields one obtains
all of the $37$ irreducible representations of
$\mathrm{\mathrm{G_{1536}}}$. Each representation appears at least
once and some of them appear several times. Considering next the
possible subgroups $\mathrm{H}_i \subset \mathrm{\mathrm{G_{1536}}}$
and the branching rules of  $\mathrm{\mathrm{G_{1536}}}$ irreps with
respect to $\mathrm{H}_i$ one obtains an explicit algorithm to
construct Arnold-Beltrami vector fields with prescribed invariance
space groups $H_i$. It suffices to select the identity
representation of the subgroup in the branching rules. \textit{These
are the hidden symmetries of the Beltrami flows}. As we have
discussed in the introduction these hidden symmetries extend to the
exact Navier-Stokes time dependent solutions.
%%%%%%%%%%%%%%%%
\subsection{Classification of the 48 sublattices of the momentum
lattice and the irreps of $\mathrm{G_{1536}}$} \label{triplettoni}
Let us now analyze the action of the octahedral  group on the cubic
lattice. We define the orbits as the sets of vectors $
{\mathbf{k}}\, \in \, \Lambda$ that can be mapped one into the other
by the action of some element of the point group , namely of
$\mathrm{O_{24}}$:
\begin{equation}\label{orbitadefi}
     {\mathbf{k}}_1 \, \in \, \mathcal{O} \quad \mbox{and}
    \quad  {\mathbf{k}}_2 \, \in \, \mathcal{O} \quad \Rightarrow
    \quad \exists \, \gamma \, \in \, \mathrm{O_{24}} \; / \; \gamma\,\cdot \,
     {\mathbf{k}}_1 \, = \,  {\mathbf{k}}_2
\end{equation}
In the case of the cubic lattice there are four type of orbits
\subsubsection{Orbits of length 6} 
\begin{figure}[!hbt]
\begin{center}
\includegraphics[height=50mm]{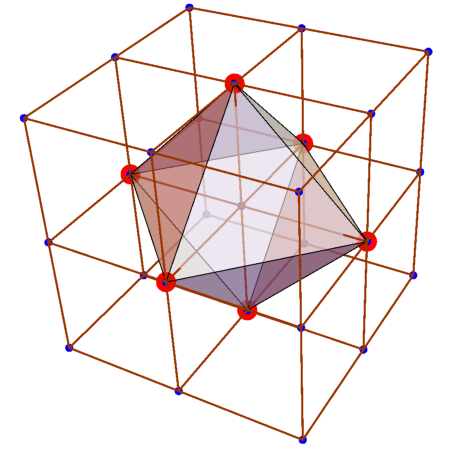}
\end{center}
\caption{\label{orbit6cub}{\it  The length 6 orbit of the octahedral
group acting on the cubic lattice corresponds to the lattice points
marked in red that can be viewed as the six vertices of a regular
octahedron.}}
\end{figure}

Each of these orbits is of the
following form:
\begin{equation}\label{orbita6}
\mathcal{O}_6 \, = \,     \left\{
\begin{array}{llllll}
 \{0,0,-n\}, & \{0,0,n\}, & \{0,-n,0\}, & \{0,n,0\}, &
   \{-n,0,0\}, & \{n,0,0\}
\end{array}
\right\}
\end{equation}
where $n \, \in \, \mathbb{Z}$ is any integer number. The six
vectors belonging to this orbit can be seen as the vertices of a
regular octahedron (see fig.\ref{orbit6cub})
\subsubsection{Orbits of length 8}
\begin{figure}[!hbt]
\begin{center}
\includegraphics[height=50mm]{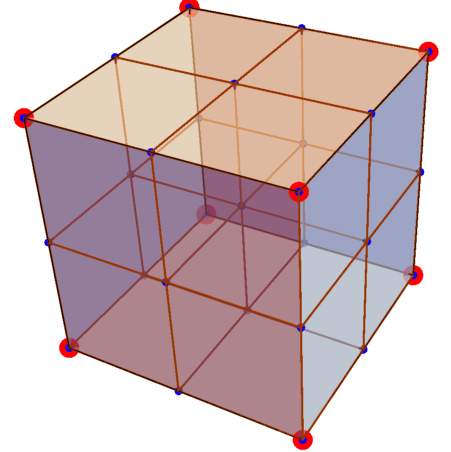}
\end{center}
\caption{\label{orbit8cub}{\it  The length 8 orbit of the octahedral
group acting on the cubic lattice corresponds to the lattice points
marked in red that can be viewed as the 8 vertices of a cube.}}
\end{figure}

Each of these orbits is of the following form
\begin{equation}\label{orbita8}
\mathcal{O}_8 \, = \,     \left\{
\begin{array}{llll}
 \{-n,-n,-n\}, & \{-n,-n,n\}, & \{-n,n,-n\}, & \{-n,n,n\}, \\
   \{n,-n,-n\}, & \{n,-n,n\}, & \{n,n,-n\}, & \{n,n,n\}\\
\end{array}
\right\}
\end{equation}
where $n \, \in \, \mathbb{Z}$ is any integer number.The 8 vectors
belonging to this orbit can be seen as the vertices of a cube (see
fig.\ref{orbit8cub})
\subsubsection{Orbits of length 12}
\begin{figure}[!hbt]
\begin{center}
\includegraphics[height=50mm]{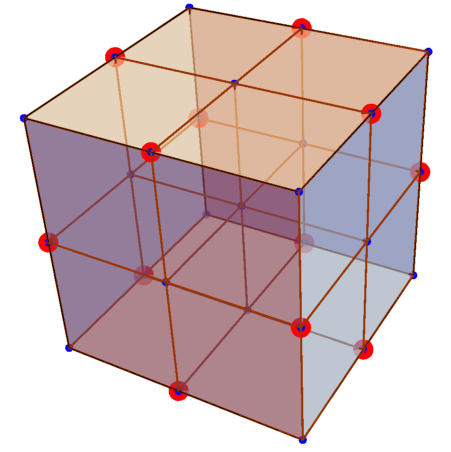}
\end{center}
\caption{\label{orbit12cub}{\it  The length 12 orbit of the
octahedral group acting on the cubic lattice corresponds to the
lattice points marked in red that can be viewed as the middle points
of the edges of a cube.}}
\end{figure}

Each of these orbits is of the following form:
\begin{equation}\label{orbita12}
\mathcal{O}_{12} \, = \,     \left\{
\begin{array}{llll}
 \{0,-n,-n\}, & \{0,-n,n\}, & \{0,n,-n\}, & \{0,n,n\}, \\
   \{-n,0,-n\}, & \{-n,0,n\}, & \{-n,-n,0\}, & \{-n,n,0\}, \\
   \{n,0,-n\}, & \{n,0,n\}, & \{n,-n,0\}, & \{n,n,0\}
\end{array}
\right\}
\end{equation}
where $n \, \in \, \mathbb{Z}$ is any integer number. The 12 vectors
belonging to this orbit can be seen as the middle points  of the
edges of a cube (see fig.\ref{orbit12cub})
\subsubsection{Orbits of length 24}
Each of these orbits is of the following form:
\begin{equation}\label{orbita24}
\mathcal{O}_{24} \, = \,     \left\{
\begin{array}{llll}
 \{-p,-q,r\}, & \{-p,q,-r\}, & \{-p,-r,-q\}, & \{-p,r,q\}, \\
   \{p,-q,-r\}, & \{p,q,r\}, & \{p,-r,q\}, & \{p,r,-q\}, \\
   \{-q,-p,-r\}, & \{-q,p,r\}, & \{-q,-r,p\}, & \{-q,r,-p\},\\
   \{q,-p,r\}, & \{q,p,-r\}, & \{q,-r,-p\}, & \{q,r,p\}, \\
   \{-r,-p,q\}, & \{-r,p,-q\}, & \{-r,-q,-p\}, & \{-r,q,p\},\\
    \{r,-p,-q\}, & \{r,p,q\}, & \{r,-q,p\}, & \{r,q,-p\},\\
\end{array}
\right\}
\end{equation}
where $\{p, \, q, \, r \}\, \in \, \mathbb{Z}$ is any triplet of
integer numbers that are not all three equal in absolute value.
\begin{figure}[!hbt]
\begin{center}
\includegraphics[height=50mm]{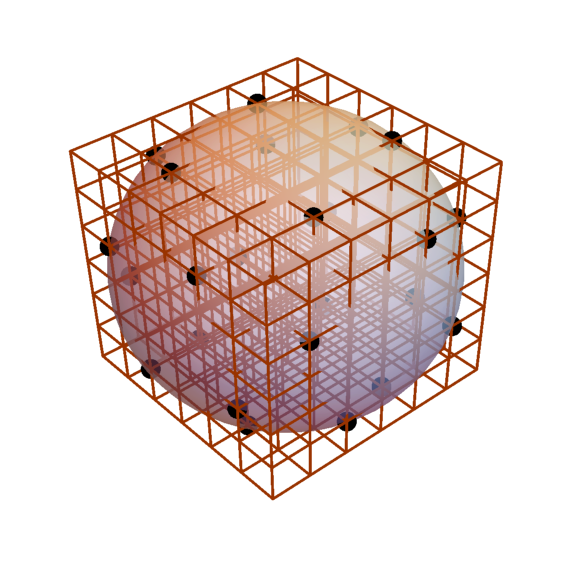}
\end{center}
\caption{\label{orbit24cub}{\it  A possible length 24 orbit of the
octahedral group acting on the cubic lattice corresponds to the
lattice points marked in black. In this case the orbit is generated
by the vector $\{1,2,3\}$ and all the orbit points belong to the
intersection of the cubic lattice $\Lambda_{cubic}$ with a sphere of
radius $r=\sqrt{14}$.}}
\end{figure}
\par
\subsubsection{Classification of the $48$ types of orbits}
The first observation is that the  group
$\mathrm{\mathrm{G_{1536}}}$ has a finite number of irreducible
representations so that the number of different types of
Arnold-Beltrami vector fields has also got to be finite,  namely as
many as the 37 irreps, times the number of different ways to obtain
them from orbits of length 6,8,12 or 24. The second observation is
the key role of the number $4$ introduced by Frobenius congruences
which was already the clue to the definition of
$\mathrm{\mathrm{G_{1536}}}$. What we should expect is that the
various orbits should be defined with integers modulo $4$ in other
words that we should just consider the possible octahedral orbits on
a lattice with coefficients in $\mathbb{Z}_4$ rather than
$\mathbb{Z}$. The easy guess, which is confirmed by computer
calculations, is that the pattern of $\mathrm{\mathrm{G_{1536}}}$
representations obtained from the construction of Arnold-Beltrami
vector fields according to the algorithm of section \ref{algoritmo}
depends only on the equivalence classes of momentum orbits modulo
$4$. Hence we have a finite number of such orbits and a finite
number of Arnold-Beltrami vector fields which we presently describe.
Let us stress that an embryo of the exhaustive classification of
orbits we are going to present was introduced by Arnold in his paper
\cite{arnoldorussopapero}. Arnold's one  was only an embryo of the
complete classification for the following two reasons:
\begin{enumerate}
\item The type of momenta orbits were partitioned  according to $odd$ and $even$
(namely according to $\mathbb{Z}_2$, rather than $\mathbb{Z}_4$)
\item The classifying group was taken to be the crystallographic $\mathrm{GS_{24}}$,
as defined  in the appendices of \cite{Fre:2015mla} and already
discussed in eq.(\ref{TScap}) and following lines.
\end{enumerate}
Let us then present the complete classification of point orbits in
the momentum lattice. First we subdivide the momenta into five
groups:
\begin{description}
  \item[A)] Momenta of type $\{a,0,0\}$ which generate $\mathrm{O_{24}}$
  orbits of length 6 and representations of the universal
  group $\mathrm{G}_{1536}$ also of dimensions 6.
  \item[B)]Momenta of type $\{a,a,a\}$ which generate $\mathrm{O_{24}}$ orbits
  of length 8 and representations of the universal group $\mathrm{G}_{1536}$
  also of dimensions 8.
  \item[C)] Momenta of type $\{a,a,0\}$ which generate $\mathrm{O_{24}}$
  orbits of length 12 and representations of the universal
  group $\mathrm{G}_{1536}$ also of dimensions 12.
  \item[D)] Momenta of type $\{a,a,b\}$ which generate $\mathrm{O_{24}}$
  orbits of length 24 and representations
  of the universal group $\mathrm{G}_{1536}$ also of dimensions 24.
  \item[E)] Momenta of type $\{a,b,c\}$ which
  generate $\mathrm{O_{24}}$ orbits of length 24 and
  representations of the universal group $\mathrm{G}_{1536}$ of dimensions 48.
\end{description}
The reason why in the cases A)\dots D) the dimension of the
representation $\mathfrak{R}\left(\mathrm{\mathrm{G_{1536}}}\right)$
coincides with the dimension $|\mathcal{O}|$ of the orbit  is
simple. For each momentum in the orbit ($\forall\mathbf{k}_i\, \in
\, \mathcal{O}$) also its negative appears in the same orbit ($- \,
\mathbf{k}_i\, \in \, \mathcal{O}$), hence the number of arguments
$\Theta_i \, \equiv \, 2\pi \,\mathbf{k}_i \cdot \mathbf{x}$ of the
independent trigonometric functions $\sin \left( \Theta_i\right)$
and $\cos \left(\Theta_i\right)$ is $\ft 12 \times 2 |\mathcal{O}|
\, = \, |\mathcal{O}|$ since $\sin \left( \pm\Theta_i\right) \, = \,
\pm  \sin \left( \Theta_i\right)$ and $\cos \left(
\pm\Theta_i\right) \, = \, \cos \left( \Theta_i\right)$.
\par In
case E), instead, the negatives of all the members of the orbit
$\mathcal{O}$ are not in $\mathcal{O}$. The number of independent
trigonometric functions is therefore 48 and such is the dimension of
the representation
$\mathfrak{R}\left(\mathrm{\mathrm{G_{1536}}}\right)$.
\par
In each of the five groups one still has  to reduce the entries to
$\mathbb{Z}_4$, namely to consider their equivalence class
$\mathrm{mod}\,4$. Each different choice of the pattern of
$\mathbb{Z}_4$ classes appearing in an orbit leads to a different
decomposition of the representation into irreducible representation
of $\mathrm{G}_{1536}$. A simple consideration of the combinatorics
leads to the conclusion that there are in total $48$ cases to be
considered. The very significant result is that all of the $37$
irreducible representations of $\mathrm{G}_{1536}$ appear at least
once in the list of these decompositions. Hence for all the
\textit{irreps} of this group one can find a corresponding Beltrami
field and for some \textit{irreps} such a Beltrami field admits a
few inequivalent realizations. The list of the $48$ distinct types
of momenta is the following one:
\begin{multicols}{3}
\begin{enumerate}
\item $\mathbf{k} \, = \,  \{0,0, 1+4 \rho \}$
\item $\mathbf{k} \, = \,  \{0,0, 2+4 \rho \}$
\item $\mathbf{k} \, = \,  \{0,0, 3+4 \rho \}$
\item $\mathbf{k} \, = \,  \{0,0, 4+4 \rho \}$
\item $\mathbf{k} \, = \,  \left\{1+4\mu ,1+4 \mu ,1+4 \mu \right\}$
\item $\mathbf{k} \, = \,  \left\{2+4 \mu ,2+4 \mu ,2+4 \mu \right\}$
\item $\mathbf{k} \, = \,  \left\{3+4 \mu ,3+4 \mu ,3+4 \mu \right\}$
\item $\mathbf{k} \, = \,  \left\{4+4 \mu ,4+4 \mu ,4+4 \mu \right\}$
\item $\mathbf{k} \, = \,  \left\{0 ,1+4 \nu ,1+4 \nu \right\}$
\item $\mathbf{k} \, = \,  \left\{0 ,2+4 \nu ,2+4 \nu \right\}$
\item $\mathbf{k} \, = \,  \left\{0 ,3+4 \nu ,3+4 \nu \right\}$
\item $\mathbf{k} \, = \,  \left\{0 ,4+4 \nu ,4+4 \nu \right\}$
\item $\mathbf{k} \, = \,  \left\{1+4  \mu ,1+4 \mu ,2+4 \rho \right\}$
\item $\mathbf{k} \, = \,  \left\{1+4 \mu ,1+4 \mu ,3+4 \rho \right\}$
\item $\mathbf{k} \, = \,  \left\{1+4 \mu ,1+4 \mu ,4+4 \rho \right\}$
\item $\mathbf{k} \, = \,  \left\{1+4\mu ,1+4 \mu ,5+4 \rho \right\}$
\item $\mathbf{k} \, = \,  \{1+4 \mu ,2+4 \mu ,2+4 \rho \}$
\item $\mathbf{k} \, = \,  \left\{2+4\mu ,2+4 \mu ,6+4 \rho \right\}$
\item $\mathbf{k} \, = \,  \left\{2+4  \mu ,2+4 \mu ,3+4 \rho \right\}$
\item $\mathbf{k} \, = \,  \left\{2+4\mu ,2+4 \mu ,4+4 \rho \right\}$
\item $\mathbf{k} \, = \,  \left\{1+4 \mu ,3+4 \mu ,3+4 \rho \right\}$
\item $\mathbf{k} \, = \,  \left\{2+4 \mu ,3+4 \mu ,3+4 \rho \right\}$
\item $\mathbf{k} \, = \,  \left\{3+4  \mu ,3+4 \mu ,7+4 \rho \right\}$
\item $\mathbf{k} \, = \,  \left\{1+4  \mu ,4+4 \mu ,4+4 \rho \right\}$
\item $\mathbf{k} \, = \,  \{2+4 \mu ,4+4 \mu ,4+4 \rho \}$
\item $\mathbf{k} \, = \,  \left\{3+4  \mu ,4+4 \mu ,4+4 \rho \right\}$
\item $\mathbf{k} \, = \,  \left\{4+4 \mu ,4+4 \mu ,8+4 \rho \right\}$
\item $\mathbf{k} \, = \,  \left\{3+4  \mu ,3+4 \mu ,4+4 \rho \right\}$
\item $\mathbf{k} \, = \,  \left\{4+4 \mu ,8+4 \nu ,12+4 \rho \right\}$
\item $\mathbf{k} \, = \,  \{1+4 \mu ,4+4 \nu ,8+4 \rho \}$
\item $\mathbf{k} \, = \,  \left\{2+4 \mu ,4+4 \nu ,8+4 \rho \right\}$
\item $\mathbf{k} \, = \,  \left\{3+4  \mu ,4+4 \nu ,8+4 \rho \right\}$
\item $\mathbf{k} \, = \,  \left\{1+4 \mu ,2+4 \nu ,4+4 \rho \right\}$
\item $\mathbf{k} \, = \,  \left\{1+4 \mu ,3+4 \nu ,4+4 \rho \right\}$
\item $\mathbf{k} \, = \,  \left\{2+4 \mu ,4+4 \nu ,6+4 \rho \right\}$
\item $\mathbf{k} \, = \,  \left\{2+4  \mu ,3+4 \nu ,4+4 \rho \right\}$
\item $\mathbf{k} \, = \,  \left\{1+4 \mu ,5+4 \nu ,9+4 \rho \right\}$
\item $\mathbf{k} \, = \,  \left\{1+4  \mu ,2+4 \nu ,5+4 \rho \right\}$
\item $\mathbf{k} \, = \,  \left\{1+4  \mu ,3+4 \nu ,5+4 \rho \right\}$
\item $\mathbf{k} \, = \,  \left\{1+4  \mu ,2+4 \nu ,6+4 \rho \right\}$
\item $\mathbf{k} \, = \,  \left\{1+4 \mu ,2+4 \nu ,3+4 \rho \right\}$
\item $\mathbf{k} \, = \,  \left\{1+4 \mu ,3+4 \nu ,7+4 \rho \right\}$
\item $\mathbf{k} \, = \,  \left\{2+4 \mu ,6+4 \nu ,10+4 \rho \right\}$
\item $\mathbf{k} \, = \,  \{2+4 \mu ,3+4 \nu ,6+4 \rho \}$
\item $\mathbf{k} \, = \,  \left\{2+4 \mu ,3+4 \nu ,7+4 \rho \right\}$
\item $\mathbf{k} \, = \,  \left\{3+4  \mu ,7+4 \nu ,11+4 \rho \right\}$
\item $\mathbf{k} \, = \,  \left\{1+4 \mu ,4+4 \nu ,5+4 \rho \right\}$
\item $\mathbf{k} \, = \,  \left\{3+4  \mu ,4+4 \nu ,7+4 \rho \right\}$
\end{enumerate}
\end{multicols}
 where $\mu,\nu,\rho \, \in \, \mathbb{Z}$. The simplest and
lowest lying representative of each of the $48$ classes of
equivalent momenta is obtained choosing $\mu \, = \, \nu \, = \, 0$.
\subsubsection{The 48 orbits type and the
irreps of the Universal Classifying Group} In this subsection,
quoting the results obtained in \cite{Fre:2015mla} for each of the
$48$ classes enumerated above we provide the decomposition of the
corresponding Beltrami vector field parameter space into
$\mathrm{\mathrm{G_{1536}}}$ irreducible representations. These
results are the outcome of extensive MATHEMATICA calculations that
were performed with purposely written codes. As already stressed the
most relevant point is that all the $37$ irreps of the Classifying
Group are reproduced: this is the main reason for its name.
\subsubsection{Classes
of momentum vectors yielding orbits of length 6: $\{$a,0,0$\}$}
\label{mortadella6} \noindent\(
\pmb{}\\
%\pmb{\text{ Eigenvalue  = } 1}\\
% \vskip 0.1 cm
\pmb{\text{ Class of the momentum vector  = } \{0,0, 1+4 \rho \}}\\
\pmb{\text{ Dimension of the $\mathrm{G}_{1536}$  representation  = } 6}\\
% \vskip 0.1 cm
\pmb{\text{ Orbit   = } D_{23}[\mathrm{G}_{1536},6]}\\
%\pmb{\text{ Eigenvalue  = } 2}\\
% \vskip 0.1 cm
\pmb{\text{ Class of the momentum vector  = } \{0,0, 2+4 \rho \}}\\
\pmb{\text{ Dimension of the $\mathrm{G}_{1536}$  representation  = } 6}\\
% \vskip 0.1 cm
\pmb{\text{ Orbit   = } D_{19}[\mathrm{G}_{1536},6]}\\
%\pmb{\text{ Eigenvalue  = } 3}\\
% \vskip 0.1 cm
\pmb{\text{ Class of the momentum vector  = } \{0,0, 3+4 \rho \}}\\
\pmb{\text{ Dimension of the $\mathrm{G}_{1536}$  representation  = } 6}\\
% \vskip 0.1 cm
\pmb{\text{ Orbit   = } D_{24}[\mathrm{G}_{1536},6]}\\
%\pmb{\text{ Eigenvalue  = } 4}\\
% \vskip 0.1 cm
\pmb{\text{ Class of the momentum vector  = } \{0,0, 4+4 \rho \}}\\
\pmb{\text{ Dimension of the $\mathrm{G}_{1536}$  representation  = } 6}\\
% \vskip 0.1 cm
\pmb{\text{ Orbit   = } D_7[\mathrm{G}_{1536},3]+D_8[\mathrm{G}_{1536},3]}\\
\pmb{ }\)
\subsubsection{Classes of momentum vectors yielding orbits
of length 8: $\{$a,a,a$\}$}
\noindent\(
\pmb{}\\
%\pmb{\text{ Eigenvalue  = } \sqrt{3}}\\
% \vskip 0.1 cm
\pmb{\text{ Class of the momentum vector  = } \left\{1+4\mu ,1+4 \mu ,1+4 \mu \right\}}\\
\pmb{\text{ Dimension of the $\mathrm{G}_{1536}$  representation  = } 8}\\
% \vskip 0.1 cm
\pmb{\text{ Orbit   = } D_{30}[\mathrm{G}_{1536},8]}\\
%\pmb{\text{ Eigenvalue  = } 2 \sqrt{3}}\\
% \vskip 0.1 cm
\pmb{\text{ Class of the momentum vector  = } \left\{2+4 \mu ,2+4 \mu ,2+4 \mu \right\}}\\
\pmb{\text{ Dimension of the $\mathrm{G}_{1536}$  representation  = } 8}\\
% \vskip 0.1 cm
\pmb{\text{ Orbit   = } D_6[\mathrm{G}_{1536},2]+D_{17}[\mathrm{G}_{1536},3]+D_{18}[\mathrm{G}_{1536},3]}\\
%\pmb{\text{ Eigenvalue  = } 3 \sqrt{3}}\\
% \vskip 0.1 cm
\pmb{\text{ Class of the momentum vector  = } \left\{3+4 \mu ,3+4 \mu ,3+4 \mu \right\}}\\
\pmb{\text{ Dimension of the $\mathrm{G}_{1536}$  representation  = } 8}\\
% \vskip 0.1 cm
\pmb{\text{ Orbit   = } D_{31}[\mathrm{G}_{1536},8]}\\
%\pmb{\text{ Eigenvalue  = } 4 \sqrt{3}}\\
% \vskip 0.1 cm
\pmb{\text{ Class of the momentum vector  = } \left\{4+4 \mu ,4+4 \mu ,4+4 \mu \right\}}\\
\pmb{\text{ Dimension of the $\mathrm{G}_{1536}$  representation  = } 8}\\
% \vskip 0.1 cm
\pmb{\text{ Orbit   = } D_5[\mathrm{G}_{1536},2]+D_7[\mathrm{G}_{1536},3]+D_8[\mathrm{G}_{1536},3]}\\
\pmb{ }\)
\subsubsection{Classes of momentum vectors yielding orbits
of length 12: $\{$0,a,a$\}$} \noindent\(
\pmb{}\\
%\pmb{\text{ Eigenvalue  = } \sqrt{2}}\\
% \vskip 0.1 cm
\pmb{\text{ Class of the momentum vector  = } \left\{0 ,1+4 \nu ,1+4 \nu \right\}}\\
\pmb{\text{ Dimension of the $\mathrm{G}_{1536}$  representation  = } 12}\\
% \vskip 0.1 cm
\pmb{\text{ Orbit   = } D_{32}[\mathrm{G}_{1536},12]}\\
%\pmb{\text{ Eigenvalue  = } 2 \sqrt{2}}\\
% \vskip 0.1 cm
\pmb{\text{ Class of the momentum vector  = } \left\{0 ,2+4 \nu ,2+4 \nu \right\}}\\
\pmb{\text{ Dimension of the $\mathrm{G}_{1536}$  representation  = } 12}\\
% \vskip 0.1 cm
\pmb{\text{ Orbit   = } D_{13}[\mathrm{G}_{1536},3]+D_{15}[\mathrm{G}_{1536},3]+D_{20}[\mathrm{G}_{1536},6]}\\
%\pmb{\text{ Eigenvalue  = } 3 \sqrt{2}}\\
% \vskip 0.1 cm
\pmb{\text{ Class of the momentum vector  = } \left\{0 ,3+4 \nu ,3+4 \nu \right\}}\\
\pmb{\text{ Dimension of the $\mathrm{G}_{1536}$  representation  = } 12}\\
% \vskip 0.1 cm
\pmb{\text{ Orbit   = } D_{32}[\mathrm{G}_{1536},12]}\\
%\pmb{\text{ Eigenvalue  = } 4 \sqrt{2}}\\
% \vskip 0.1 cm
\pmb{\text{ Class of the momentum vector  = } \left\{0 ,4+4 \nu ,4+4 \nu \right\}}\\
\pmb{\text{ Dimension of the $\mathrm{G}_{1536}$  representation  = } 12}\\
% \vskip 0.1 cm
\pmb{\text{ Orbit   = }
D_2[\mathrm{G}_{1536},1]+D_5[\mathrm{G}_{1536},2]+D_7[\mathrm{G}_{1536},3]+2
D_8[\mathrm{G}_{1536},3]}\\
\pmb{ }\)
\subsubsection{Classes of momentum vectors yielding orbits
of length 24: $\{$a,a,b$\}$} \noindent\(
\pmb{}\\
%\pmb{\text{ Eigenvalue  = } \sqrt{6}}\\
% \vskip 0.1 cm
\pmb{\text{ Class of the momentum vector  = } \left\{1+4  \mu ,1+4 \mu ,2+4 \rho \right\}}\\
\pmb{\text{ Dimension of the $\mathrm{G}_{1536}$  representation  = } 24}\\
% \vskip 0.1 cm
\pmb{\text{ Orbit   = } D_{34}[\mathrm{G}_{1536},12]+D_{35}[\mathrm{G}_{1536},12]}\\
%\pmb{\text{ Eigenvalue  = } \sqrt{11}}\\
% \vskip 0.1 cm
\pmb{\text{ Class of the momentum vector  = } \left\{1+4 \mu ,1+4 \mu ,3+4 \rho \right\}}\\
\pmb{\text{ Dimension of the $\mathrm{G}_{1536}$  representation  = } 24}\\
% \vskip 0.1 cm
\pmb{\text{ Orbit   = } D_{29}[\mathrm{G}_{1536},8]+D_{30}[\mathrm{G}_{1536},8]+D_{31}[\mathrm{G}_{1536},8]}\\
%\pmb{\text{ Eigenvalue  = } 3 \sqrt{2}}\\
% \vskip 0.1 cm
\pmb{\text{ Class of the momentum vector  = } \left\{1+4 \mu ,1+4 \mu ,4+4 \rho \right\}}\\
\pmb{\text{ Dimension of the $\mathrm{G}_{1536}$  representation  = } 24}\\
% \vskip 0.1 cm
\pmb{\text{ Orbit   = } D_{32}[\mathrm{G}_{1536},12]+D_{33}[\mathrm{G}_{1536},12]}\\
%\pmb{\text{ Eigenvalue  = } 3 \sqrt{3}}\\
% \vskip 0.1 cm
\pmb{\text{ Class of the momentum vector  = } \left\{1+4\mu ,1+4 \mu ,5+4 \rho \right\}}\\
\pmb{\text{ Dimension of the $\mathrm{G}_{1536}$  representation  = } 24}\\
% \vskip 0.1 cm
\pmb{\text{ Orbit   = } D_{29}[\mathrm{G}_{1536},8]+D_{30}[\mathrm{G}_{1536},8]+D_{31}[\mathrm{G}_{1536},8]}\\
%\pmb{\text{ Eigenvalue  = } 3}\\
% \vskip 0.1 cm
\pmb{\text{ Class of the momentum vector  = } \{1+4 \mu ,2+4 \mu ,2+4 \rho \}}\\
\pmb{\text{ Dimension of the $\mathrm{G}_{1536}$  representation  = } 24}\\
% \vskip 0.1 cm
\pmb{\text{ Orbit   = } D_{25}[\mathrm{G}_{1536},6]+D_{26}[\mathrm{G}_{1536},6]+D_{27}[\mathrm{G}_{1536},6]+D_{28}[\mathrm{G}_{1536},6]}\\
%\pmb{\text{ Eigenvalue  = } 2 \sqrt{11}}\\
% \vskip 0.1 cm
\pmb{\text{ Class of the momentum vector  = } \left\{2+4\mu ,2+4 \mu ,6+4 \rho \right\}}\\
\pmb{\text{ Dimension of the $\mathrm{G}_{1536}$  representation  = } 24}\\
% \vskip 0.1 cm
\pmb{\text{ Orbit   = }
D_3[\mathrm{G}_{1536},1]+D_4[\mathrm{G}_{1536},1]+2
D_6[\mathrm{G}_{1536},2]+3
D_{17}[\mathrm{G}_{1536},3]+3 D_{18}[\mathrm{G}_{1536},3]}\\
%\pmb{\text{ Eigenvalue  = } \sqrt{17}}\\
% \vskip 0.1 cm
\pmb{\text{ Class of the momentum vector  = } \left\{2+4  \mu ,2+4 \mu ,3+4 \rho \right\}}\\
\pmb{\text{ Dimension of the $\mathrm{G}_{1536}$  representation  = } 24}\\
% \vskip 0.1 cm
\pmb{\text{ Orbit   = } D_{25}[\mathrm{G}_{1536},6]+D_{26}[\mathrm{G}_{1536},6]+D_{27}[\mathrm{G}_{1536},6]+D_{28}[\mathrm{G}_{1536},6]}\\
%\pmb{\text{ Eigenvalue  = } 2 \sqrt{6}}\\
% \vskip 0.1 cm
\pmb{\text{ Class of the momentum vector  = } \left\{2+4\mu ,2+4 \mu ,4+4 \rho \right\}}\\
\pmb{\text{ Dimension of the $\mathrm{G}_{1536}$  representation  = } 24}\\
% \vskip 0.1 cm
\pmb{\text{ Orbit   = }
D_{13}[\mathrm{G}_{1536},3]+D_{14}[\mathrm{G}_{1536},3]+D_{15}[\mathrm{G}_{1536},3]+D_{16}[\mathrm{G}_{1536},3]+2
D_{20}[\mathrm{G}_{1536},6]}\\
%\pmb{\text{ Eigenvalue  = } \sqrt{19}}\\
% \vskip 0.1 cm
\pmb{\text{ Class of the momentum vector  = } \left\{1+4 \mu ,3+4 \mu ,3+4 \rho \right\}}\\
\pmb{\text{ Dimension of the $\mathrm{G}_{1536}$  representation  = } 24}\\
% \vskip 0.1 cm
\pmb{\text{ Orbit   = } D_{29}[\mathrm{G}_{1536},8]+D_{30}[\mathrm{G}_{1536},8]+D_{31}[\mathrm{G}_{1536},8]}\\
%\pmb{\text{ Eigenvalue  = } \sqrt{22}}\\
% \vskip 0.1 cm
\pmb{\text{ Class of the momentum vector  = } \left\{2+4 \mu ,3+4 \mu ,3+4 \rho \right\}}\\
\pmb{\text{ Dimension of the $\mathrm{G}_{1536}$  representation  = } 24}\\
% \vskip 0.1 cm
\pmb{\text{ Orbit   = } D_{34}[\mathrm{G}_{1536},12]+D_{35}[\mathrm{G}_{1536},12]}\\
%\pmb{\text{ Eigenvalue  = } \sqrt{67}}\\
% \vskip 0.1 cm
\pmb{\text{ Class of the momentum vector  = } \left\{3+4  \mu ,3+4 \mu ,7+4 \rho \right\}}\\
\pmb{\text{ Dimension of the $\mathrm{G}_{1536}$  representation  = } 24}\\
% \vskip 0.1 cm
\pmb{\text{ Orbit   = } D_{29}[\mathrm{G}_{1536},8]+D_{30}[\mathrm{G}_{1536},8]+D_{31}[\mathrm{G}_{1536},8]}\\
%\pmb{\text{ Eigenvalue  = } \sqrt{33}}\\
% \vskip 0.1 cm
\pmb{\text{ Class of the momentum vector  = } \left\{1+4  \mu ,4+4 \mu ,4+4 \rho \right\}}\\
\pmb{\text{ Dimension of the $\mathrm{G}_{1536}$  representation  = } 24}\\
% \vskip 0.1 cm
\pmb{\text{ Orbit   = } D_{21}[\mathrm{G}_{1536},6]+D_{22}[\mathrm{G}_{1536},6]+D_{23}[\mathrm{G}_{1536},6]+D_{24}[\mathrm{G}_{1536},6]}\\
%\pmb{\text{ Eigenvalue  = } 6}\\
% \vskip 0.1 cm
\pmb{\text{ Class of the momentum vector  = } \{2+4 \mu ,4+4 \mu ,4+4 \rho \}}\\
\pmb{\text{ Dimension of the $\mathrm{G}_{1536}$  representation  = } 24}\\
% \vskip 0.1 cm
\pmb{\text{ Orbit   = }
D_9[\mathrm{G}_{1536},3]+D_{10}[\mathrm{G}_{1536},3]+D_{11}[\mathrm{G}_{1536},3]+D_{12}[\mathrm{G}_{1536},3]+2
D_{19}[\mathrm{G}_{1536},6]}\\
%\pmb{\text{ Eigenvalue  = } \sqrt{41}}\\
% \vskip 0.1 cm
\pmb{\text{ Class of the momentum vector  = } \left\{3+4  \mu ,4+4 \mu ,4+4 \rho \right\}}\\
\pmb{\text{ Dimension of the $\mathrm{G}_{1536}$  representation  = } 24}\\
% \vskip 0.1 cm
\pmb{\text{ Orbit   = } D_{21}[\mathrm{G}_{1536},6]+D_{22}[\mathrm{G}_{1536},6]+D_{23}[\mathrm{G}_{1536},6]+D_{24}[\mathrm{G}_{1536},6]}\\
%\pmb{\text{ Eigenvalue  = } 4 \sqrt{6}}\\
% \vskip 0.1 cm
\pmb{\text{ Class of the momentum vector  = } \left\{4+4 \mu ,4+4 \mu ,8+4 \rho \right\}}\\
\pmb{\text{ Dimension of the $\mathrm{G}_{1536}$  representation  = } 24}\\
% \vskip 0.1 cm
\pmb{\text{ Orbit   = }
D_1[\mathrm{G}_{1536},1]+D_2[\mathrm{G}_{1536},1]+2
D_5[\mathrm{G}_{1536},2]+3
D_7[\mathrm{G}_{1536},3]+3 D_8[\mathrm{G}_{1536},3]}\\
%\pmb{\text{ Eigenvalue  = } \sqrt{34}}\\
% \vskip 0.1 cm
\pmb{\text{ Class of the momentum vector  = } \left\{3+4  \mu ,3+4 \mu ,4+4 \rho \right\}}\\
\pmb{\text{ Dimension of the $\mathrm{G}_{1536}$  representation  = } 24}\\
% \vskip 0.1 cm
\pmb{\text{ Orbit   = } D_{32}[\mathrm{G}_{1536},12]+D_{33}[\mathrm{G}_{1536},12]}\\
\pmb{ }\)
\subsubsection{Classes of momentum vectors yielding point
orbits of length 24 and $\mathrm{G_{1536}}$ representations of
dimensions 48: $\{$a,b,c$\}$} \noindent\(
\pmb{}\\
%\pmb{\text{ Eigenvalue  = } 4 \sqrt{14}}\\
% \vskip 0.1 cm
\pmb{\text{ Class of the momentum vector  = } \left\{4+4 \mu ,8+4 \nu ,12+4 \rho \right\}}\\
\pmb{\text{ Dimension of the $\mathrm{G}_{1536}$  representation  = } 48}\\
% \vskip 0.1 cm
\pmb{\text{ Orbit   = } 2 D_1[\mathrm{G}_{1536},1]+2
D_2[\mathrm{G}_{1536},1]+4 D_5[\mathrm{G}_{1536},2]+6
D_7[\mathrm{G}_{1536},3]+6 D_8[\mathrm{G}_{1536},3]}\\
%\pmb{\text{ Eigenvalue  = } 9}\\
% \vskip 0.1 cm
\pmb{\text{ Class of the momentum vector  = } \{1+4 \mu ,4+4 \nu ,8+4 \rho \}}\\
\pmb{\text{ Dimension of the $\mathrm{G}_{1536}$  representation  = } 48}\\
% \vskip 0.1 cm
\pmb{\text{ Orbit   = } 2 D_{21}[\mathrm{G}_{1536},6]+2
D_{22}[\mathrm{G}_{1536},6]+2
D_{23}[\mathrm{G}_{1536},6]+2 D_{24}[\mathrm{G}_{1536},6]}\\
%\pmb{\text{ Eigenvalue  = } 2 \sqrt{21}}\\
% \vskip 0.1 cm
\pmb{\text{ Class of the momentum vector  = } \left\{2+4 \mu ,4+4 \nu ,8+4 \rho \right\}}\\
\pmb{\text{ Dimension of the $\mathrm{G}_{1536}$  representation  = } 48}\\
% \vskip 0.1 cm
\pmb{\text{ Orbit   = } 2 D_9[\mathrm{G}_{1536},3]+2
D_{10}[\mathrm{G}_{1536},3]+2 D_{11}[\mathrm{G}_{1536},3]+2
D_{12}[\mathrm{G}_{1536},3]+4 D_{19}[\mathrm{G}_{1536},6]}\\
%\pmb{\text{ Eigenvalue  = } \sqrt{89}}\\
% \vskip 0.1 cm
\pmb{\text{ Class of the momentum vector  = } \left\{3+4  \mu ,4+4 \nu ,8+4 \rho \right\}}\\
\pmb{\text{ Dimension of the $\mathrm{G}_{1536}$  representation  = } 48}\\
% \vskip 0.1 cm
\pmb{\text{ Orbit   = } 2 D_{21}[\mathrm{G}_{1536},6]+2
D_{22}[\mathrm{G}_{1536},6]+2
D_{23}[\mathrm{G}_{1536},6]+2 D_{24}[\mathrm{G}_{1536},6]}\\
%\pmb{\text{ Eigenvalue  = } \sqrt{21}}\\
% \vskip 0.1 cm
\pmb{\text{ Class of the momentum vector  = } \left\{1+4 \mu ,2+4 \nu ,4+4 \rho \right\}}\\
\pmb{\text{ Dimension of the $\mathrm{G}_{1536}$  representation  = } 48}\\
% \vskip 0.1 cm
\pmb{\text{ Orbit   = } 2 D_{36}[\mathrm{G}_{1536},12]+2 D_{37}[\mathrm{G}_{1536},12]}\\
%\pmb{\text{ Eigenvalue  = } \sqrt{26}}\\
% \vskip 0.1 cm
\pmb{\text{ Class of the momentum vector  = } \left\{1+4 \mu ,3+4 \nu ,4+4 \rho \right\}}\\
\pmb{\text{ Dimension of the $\mathrm{G}_{1536}$  representation  = } 48}\\
% \vskip 0.1 cm
\pmb{\text{ Orbit   = } 2 D_{32}[\mathrm{G}_{1536},12]+2 D_{33}[\mathrm{G}_{1536},12]}\\
%\pmb{\text{ Eigenvalue  = } 2 \sqrt{14}}\\
% \vskip 0.1 cm
\pmb{\text{ Class of the momentum vector  = } \left\{2+4 \mu ,4+4 \nu ,6+4 \rho \right\}}\\
\pmb{\text{ Dimension of the $\mathrm{G}_{1536}$  representation  = } 48}\\
% \vskip 0.1 cm
\pmb{\text{ Orbit   = } 2 D_{13}[\mathrm{G}_{1536},3]+2
D_{14}[\mathrm{G}_{1536},3]+2
D_{15}[\mathrm{G}_{1536},3]+2 D_{16}[\mathrm{G}_{1536},3]+4 D_{20}[\mathrm{G}_{1536},6]}\\
%\pmb{\text{ Eigenvalue  = } \sqrt{29}}\\
% \vskip 0.1 cm
\pmb{\text{ Class of the momentum vector  = } \left\{2+4  \mu ,3+4 \nu ,4+4 \rho \right\}}\\
\pmb{\text{ Dimension of the $\mathrm{G}_{1536}$  representation  = } 48}\\
% \vskip 0.1 cm
\pmb{\text{ Orbit   = } 2 D_{36}[\mathrm{G}_{1536},12]+2 D_{37}[\mathrm{G}_{1536},12]}\\
%\pmb{\text{ Eigenvalue  = } \sqrt{107}}\\
% \vskip 0.1 cm
\pmb{\text{ Class of the momentum vector  = } \left\{1+4 \mu ,5+4 \nu ,9+4 \rho \right\}}\\
\pmb{\text{ Dimension of the $\mathrm{G}_{1536}$  representation  = } 48}\\
% \vskip 0.1 cm
\pmb{\text{ Orbit   = } 2 D_{29}[\mathrm{G}_{1536},8]+2
D_{30}[\mathrm{G}_{1536},8]+2
D_{31}[\mathrm{G}_{1536},8]}\\
%\pmb{\text{ Eigenvalue  = } \sqrt{30}}\\
% \vskip 0.1 cm
\pmb{\text{ Class of the momentum vector  = } \left\{1+4  \mu ,2+4 \nu ,5+4 \rho \right\}}\\
\pmb{\text{ Dimension of the $\mathrm{G}_{1536}$  representation  = } 48}\\
% \vskip 0.1 cm
\pmb{\text{ Orbit   = } 2 D_{34}[\mathrm{G}_{1536},12]+2 D_{35}[\mathrm{G}_{1536},12]}\\
%\pmb{\text{ Eigenvalue  = } \sqrt{35}}\\
% \vskip 0.1 cm
\pmb{\text{ Class of the momentum vector  = } \left\{1+4  \mu ,3+4 \nu ,5+4 \rho \right\}}\\
\pmb{\text{ Dimension of the $\mathrm{G}_{1536}$  representation  = } 48}\\
% \vskip 0.1 cm
\pmb{\text{ Orbit   = } 2 D_{29}[\mathrm{G}_{1536},8]+2
D_{30}[\mathrm{G}_{1536},8]+2
D_{31}[\mathrm{G}_{1536},8]}\\
%\pmb{\text{ Eigenvalue  = } \sqrt{41}}\\
% \vskip 0.1 cm
\pmb{\text{ Class of the momentum vector  = } \left\{1+4  \mu ,2+4 \nu ,6+4 \rho \right\}}\\
\pmb{\text{ Dimension of the $\mathrm{G}_{1536}$  representation  = } 48}\\
% \vskip 0.1 cm
\pmb{\text{ Orbit   = } 2 D_{25}[\mathrm{G}_{1536},6]+2
D_{26}[\mathrm{G}_{1536},6]+2
D_{27}[\mathrm{G}_{1536},6]+2 D_{28}[\mathrm{G}_{1536},6]}\\
%\pmb{\text{ Eigenvalue  = } \sqrt{14}}\\
% \vskip 0.1 cm
\pmb{\text{ Class of the momentum vector  = } \left\{1+4 \mu ,2+4 \nu ,3+4 \rho \right\}}\\
\pmb{\text{ Dimension of the $\mathrm{G}_{1536}$  representation  = } 48}\\
% \vskip 0.1 cm
\pmb{\text{ Orbit   = } 2 D_{34}[\mathrm{G}_{1536},12]+2 D_{35}[\mathrm{G}_{1536},12]}\\
%\pmb{\text{ Eigenvalue  = } \sqrt{59}}\\
% \vskip 0.1 cm
\pmb{\text{ Class of the momentum vector  = } \left\{1+4 \mu ,3+4 \nu ,7+4 \rho \right\}}\\
\pmb{\text{ Dimension of the $\mathrm{G}_{1536}$  representation  = } 48}\\
% \vskip 0.1 cm
\pmb{\text{ Orbit   = } 2 D_{29}[\mathrm{G}_{1536},8]+2
D_{30}[\mathrm{G}_{1536},8]+2
D_{31}[\mathrm{G}_{1536},8]}\\
%\pmb{\text{ Eigenvalue  = } 2 \sqrt{35}}\\
% \vskip 0.1 cm
\pmb{\text{ Class of the momentum vector  = } \left\{2+4 \mu ,6+4 \nu ,10+4 \rho \right\}}\\
\pmb{\text{ Dimension of the $\mathrm{G}_{1536}$  representation  = } 48}\\
% \vskip 0.1 cm
\pmb{\text{ Orbit   = } 2 D_3[\mathrm{G}_{1536},1]+2
D_4[\mathrm{G}_{1536},1]+4 D_6[\mathrm{G}_{1536},2]+6
D_{17}[\mathrm{G}_{1536},3]+6 D_{18}[\mathrm{G}_{1536},3]}\\
%\pmb{\text{ Eigenvalue  = } 7}\\
% \vskip 0.1 cm
\pmb{\text{ Class of the momentum vector  = } \{2+4 \mu ,3+4 \nu ,6+4 \rho \}}\\
\pmb{\text{ Dimension of the $\mathrm{G}_{1536}$  representation  = } 48}\\
% \vskip 0.1 cm
\pmb{\text{ Orbit   = } 2 D_{25}[\mathrm{G}_{1536},6]+2
D_{26}[\mathrm{G}_{1536},6]+2
D_{27}[\mathrm{G}_{1536},6]+2 D_{28}[\mathrm{G}_{1536},6]}\\
%\pmb{\text{ Eigenvalue  = } \sqrt{62}}\\
% \vskip 0.1 cm
\pmb{\text{ Class of the momentum vector  = } \left\{2+4 \mu ,3+4 \nu ,7+4 \rho \right\}}\\
\pmb{\text{ Dimension of the $\mathrm{G}_{1536}$  representation  = } 48}\\
% \vskip 0.1 cm
\pmb{\text{ Orbit   = } 2 D_{34}[\mathrm{G}_{1536},12]+2 D_{35}[\mathrm{G}_{1536},12]}\\
%\pmb{\text{ Eigenvalue  = } \sqrt{179}}\\
% \vskip 0.1 cm
\pmb{\text{ Class of the momentum vector  = } \left\{3+4  \mu ,7+4 \nu ,11+4 \rho \right\}}\\
\pmb{\text{ Dimension of the $\mathrm{G}_{1536}$  representation  = } 48}\\
% \vskip 0.1 cm
\pmb{\text{ Orbit   = } 2 D_{29}[\mathrm{G}_{1536},8]+2
D_{30}[\mathrm{G}_{1536},8]+2
D_{31}[\mathrm{G}_{1536},8]}\\
%\pmb{\text{ Eigenvalue  = } \sqrt{42}}\\
% \vskip 0.1 cm
\pmb{\text{ Class of the momentum vector  = } \left\{1+4 \mu ,4+4 \nu ,5+4 \rho \right\}}\\
\pmb{\text{ Dimension of the $\mathrm{G}_{1536}$  representation  = } 48}\\
% \vskip 0.1 cm
\pmb{\text{ Orbit   = } 2 D_{32}[\mathrm{G}_{1536},12]+2 D_{33}[\mathrm{G}_{1536},12]}\\
%\pmb{\text{ Eigenvalue  = } \sqrt{74}}\\
% \vskip 0.1 cm
\pmb{\text{ Class of the momentum vector  = } \left\{3+4  \mu ,4+4 \nu ,7+4 \rho \right\}}\\
\pmb{\text{ Dimension of the $\mathrm{G}_{1536}$  representation  = } 48}\\
% \vskip 0.1 cm
\pmb{\text{ Orbit   = } 2 D_{32}[\mathrm{G}_{1536},12]+2 D_{33}[\mathrm{G}_{1536},12]}\\
\pmb{ }\)
\subsubsection{The interpretation of the 48 momentum
classes as sublattices of the cubic lattice}  The union of the
orbits of the 48 vector classes for all values of the integer
parameters $\mu,\nu,\rho$ constitute infinite sublattices of the
momentum lattice.
\par
Given the class
$\mathbf{k}^{p,q,r}=\left\{p+4\mu,q+4\nu,r+4\rho\right\}$ and the
corresponding orbit of each vector in the class
$\mathcal{O}^{(p,q,r)}\left(\mu,\nu,\rho\right)$ considering all
$\mu,\nu,\rho \in \mathbb{Z}$ we obtain a sublattice of the original
lattice:
\begin{equation}\label{sublattice}
    \Lambda^{p,q,r} \, \equiv \, \bigoplus_{\mu,\nu,\rho}^\infty
    \,\mathcal{O}^{(p,q,r)} \, \subset \Lambda_{cubic}
\end{equation}
Most of these sublattices are $1$ dimensional
$\mathbb{Z}\hookrightarrow \mathbb{Z}\times \mathbb{Z}\times
\mathbb{Z}$ some are two dimensional $\mathbb{Z}\times
\mathbb{Z}\hookrightarrow \mathbb{Z}\times \mathbb{Z}\times
\mathbb{Z}$  and some are three dimensional $\mathbb{Z}\times
\mathbb{Z}\times \mathbb{Z} \hookrightarrow \mathbb{Z}\times
\mathbb{Z}\times \mathbb{Z}$.
\par
\begin{figure}[!hbt]
\begin{center}
\includegraphics[height=50mm]{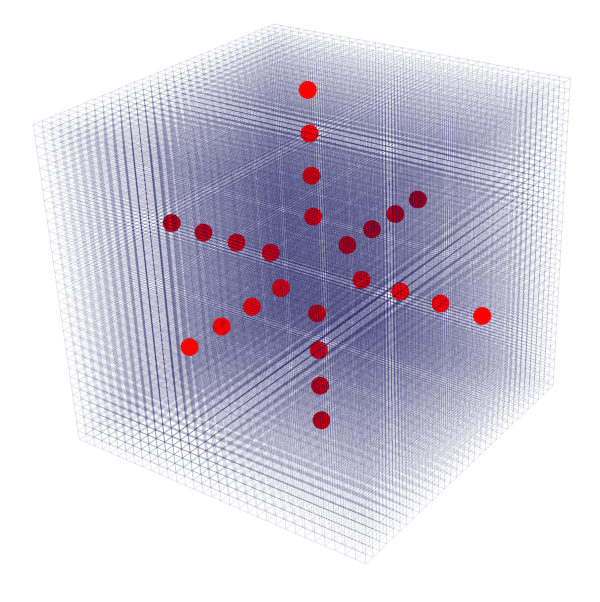}
\end{center}
\caption{\label{ceronetto1}{\it   A picture of the immersion of the
sublattice $\Lambda^{0,0,1}$ into the full cubic lattice. The points
of $\Lambda^{0,0,1}$ are painted in red on the background of the
grid of the cubic lattice.}}
\end{figure}
\begin{figure}[!hbt]
\begin{center}
\includegraphics[height=50mm]{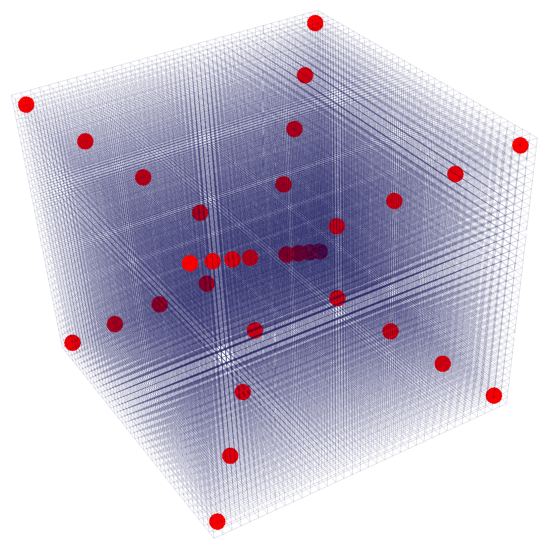}
\end{center}
\caption{\label{ceronetto2}{\it   A picture of the immersion of the
sublattice $\Lambda^{1,1,1}$ into the full cubic lattice. The points
of $\Lambda^{1,1,1}$ are painted in red on the background of the
grid of the cubic lattice.}}
\end{figure}
\begin{figure}[!hbt]
\begin{center}
\includegraphics[height=50mm]{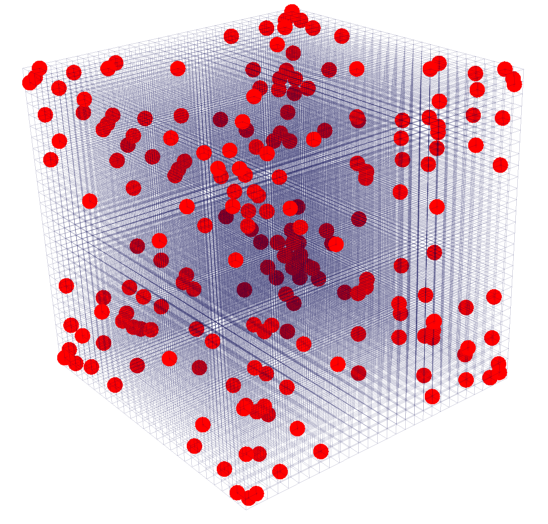}
\end{center}
\caption{\label{ceronetto3}{\it   A picture of the immersion of the
sublattice $\Lambda^{1,1,2}$ into the full cubic lattice. The points
of $\Lambda^{1,1,2}$ are painted in red on the background of the
grid of the cubic lattice.}}
\end{figure}

For instance the sublattice $\Lambda^{0,0,1}$ is formed by all the
six dimensional orbits of the vectors of type $\{0,0,1+4\mu\}$ with
$\mu\in \mathbb{Z}$. A picture of the immersion of the points of
this sublattice in the full cubic lattice is provided in
fig.\ref{ceronetto1}.
Similarly the sublattice $\Lambda^{1,1,1}$ is formed by all the
eight dimensional orbits of the vectors of type
$\{1+4\mu,1+4\mu,1+4\mu\}$ with $\mu\in \mathbb{Z}$. A picture of
the immersion of the points of this sublattice in the full cubic
lattice is provided in fig.\ref{ceronetto2}.
Finally we display sublattice $\Lambda^{1,1,2}$ which is formed by
all the eight dimensional orbits of the vectors of type
$\{1+4\mu,1+4\mu,2+4\nu\}$ with $\mu,\nu\in \mathbb{Z}$. A picture
of the immersion of the points of this sublattice in the full cubic
lattice is provided in fig.\ref{ceronetto3}.
%%%%%%%%%%%%%%%%%%%%%%%%%%%%%%%%%%%%%%%%%%%%%%%%%%%%%%%%%%%%%%%%
\subsection{The universal classifying group $\mathfrak{U}_{72}$ for the Hexagonal Lattice $\Lambda _{Hex}$}
In this section following  the same procedure utilized in the cubic
case, namely Frobenius congruences, we identify the Universal
Classifying Group for the hexagonal lattice with point group
$\mathfrak{P}_{\Lambda_{Hex}}\, = \, \mathrm{Dih_{6}}$ and we
discover that it is a group with $72$ elements.
\subsubsection{Frobenius congruences for $\mathrm{Dih_6}$}
Utilizing the block triangular representation for the semidirect
product we introduce the two candidate generators
$\hat{\mathcal{A}}$ and $\hat{\mathcal{B}}$ as it follows:
\begin{equation}\label{laperonista}
    \hat{\mathcal{A}}\, = \,\left(
\begin{array}{ccc|c}
 \frac{1}{2} & \frac{\sqrt{3}}{2} & 0 & \alpha _1 \\
 -\frac{\sqrt{3}}{2} & \frac{1}{2} & 0 & \alpha _2 \\
 0 & 0 & 1 & \alpha _3 \\
 \hline
 0 & 0 & 0 & 1 \\
\end{array}
\right) \quad ; \quad \hat{\mathcal{B}}\, = \,\left(
\begin{array}{ccc|c}
 -1 & 0 & 0 & \beta _1 \\
 0 & 1 & 0 & \beta _2 \\
 0 & 0 & -1 & \beta _3 \\
 \hline
 0 & 0 & 0 & 1 \\
\end{array}
\right)
\end{equation}
and we impose the three conditions:
\begin{equation}\label{solstizio}
    \hat{\mathcal{A}}^6 \, \in \, \hat{\Lambda} \quad ; \quad \hat{\mathcal{B}}^2 \, \in \,\hat{\Lambda}\quad ; \quad
  \left( \hat{\mathcal{B}}\cdot\hat{\mathcal{A}} \right)^2 \, \in \,  \hat{\Lambda}
\end{equation}
where the lattice subgroup is embedded in the inhomogeneous point
group $\mathfrak{Ip}_{\Lambda_{Hex}}$ as it is specified below:
\begin{equation}\label{cardenson}
    \mathfrak{Ip}_{\Lambda_{Hex}} \, \supset \, \hat{\Lambda} \, =
    \,\left\{ \left(
                \begin{array}{ccc|c}
                  1 & 0 & 0 & \sqrt{2} m_1-\frac{m_2}{\sqrt{2}} \\
                  0 & 1 & 0 & \sqrt{\frac{3}{2}} m_2 \\
                  0 & 0 & 1 & \sqrt{\frac{3}{2}} m_3 \\
                  \hline
                  0 & 0 & 0 & 0 \\
                \end{array}
              \right) \, \parallel \, m_{1,2,3} \, \in \,
              \mathbb{Z}\right \}
\end{equation}
By explicit calculation we find:
\begin{eqnarray}
\label{esplicitone}
   \hat{\mathcal{A}}^6 &=&  \left(
\begin{array}{cccc}
 1 & 0 & 0 & 0 \\
 0 & 1 & 0 & 0 \\
 0 & 0 & 1 & 6 \alpha _3 \\
 0 & 0 & 0 & 1 \\
\end{array}
\right) \quad ; \quad \hat{\mathcal{B}}^2 \, = \, \left(
\begin{array}{ccc|c}
 1 & 0 & 0 & 0 \\
 0 & 1 & 0 & 2 \beta _2 \\
 0 & 0 & 1 & 0 \\
 \hline
 0 & 0 & 0 & 1 \\
\end{array}
\right)\nonumber \\
  \left( \hat{\mathcal{B}}\cdot\hat{\mathcal{A}} \right)^2 &=& \left(
\begin{array}{ccc|c}
 1 & 0 & 0 & \frac{1}{2} \left(-\alpha _1-\sqrt{3} \alpha _2+\beta _1-\sqrt{3} \beta
   _2\right) \\
 0 & 1 & 0 & \frac{1}{2} \left(\sqrt{3} \alpha _1+3 \alpha _2-\sqrt{3} \beta _1+3
   \beta _2\right) \\
 0 & 0 & 1 & 0 \\
 \hline
 0 & 0 & 0 & 1 \\
\end{array}
\right)
\end{eqnarray}
Next introducing the generic translation group element
\begin{equation}\label{genericT}
    T \, = \, \left(
\begin{array}{ccc|c}
 1 & 0 & 0 & \text{t}_{1} \\
 0 & 1 & 0 & \text{t}_{2} \\
 0 & 0 & 1 & \text{t}_{3} \\
 \hline
 0 & 0 & 0 & 1 \\
\end{array}
\right) \quad ; \quad \text{t}_{1,2,3} \, \in \, \mathbb{R}
\end{equation}
by conjugating the two generators
$\hat{\mathcal{A}},\hat{\mathcal{B}}$ we find:
\begin{eqnarray}\label{sposalizio}
    T^{-1}\,\hat{\mathcal{A}}\, T & = & \left(
\begin{array}{ccc|c}
 \frac{1}{2} & \frac{\sqrt{3}}{2} & 0 & \frac{1}{2} \left(2 \alpha
   _1+\text{t}_{1}-\sqrt{3} \text{t}_{2}\right) \\
 -\frac{\sqrt{3}}{2} & \frac{1}{2} & 0 & \frac{1}{2} \left(2 \alpha _2+\sqrt{3}
   \text{t}_{1}+\text{t}_{2}\right) \\
 0 & 0 & 1 & \alpha_3 \\
 \hline
 0 & 0 & 0 & 1 \\
\end{array}
\right) \; ; \; T^{-1}\,\hat{\mathcal{B}}\, T \, = \,\left(
\begin{array}{ccc|c}
 -1 & 0 & 0 & \beta _1+2 \text{t}_{1} \\
 0 & 1 & 0 & \beta _2\\
 0 & 0 & -1 & \beta _3+2 \text{t}_{3} \\
 \hline
 0 & 0 & 0 & 1 \\
\end{array}
\right)
\end{eqnarray}
Hence the parameters $\mathrm{t}_{1,2}$ can be used to set
$\alpha_{1,2} \, = \, 0$, while the parameter $\mathrm{t}_3$ can be
utilized to set $\beta_3 \, = 0$. Inserting such a gauge chocie in
the conditions (\ref{solstizio}), in view of eq.(\ref{esplicitone})
and (\ref{cardenson}) we finally get:
\begin{equation}\label{argentina}
    \hat{\mathcal{A}}\, = \,\left(
\begin{array}{ccc|c}
 \frac{1}{2} & \frac{\sqrt{3}}{2} & 0 & 0 \\
 -\frac{\sqrt{3}}{2} & \frac{1}{2} & 0 & 0 \\
 0 & 0 & 1 & \frac{1}{2 \sqrt{6}} \\
 \hline
 0 & 0 & 0 & 1 \\
\end{array}
\right) \quad ; \quad \hat{\mathcal{B}}\, = \,\left(
\begin{array}{ccc|c}
 -1 & 0 & 0 & 0 \\
 0 & 1 & 0 & 0 \\
 0 & 0 & -1 & 0 \\
 \hline
 0 & 0 & 0 & 1 \\
\end{array}
\right)
\end{equation}
Following the same logic utilized in the cubic case the result that
we obtain from eq.(\ref{argentina}) is that the only fractional
translations to be considered are in the $z$-direction and that they
are of length $1/6$ of the lattice spacing. Indeed the column vector
appearing in the $\hat{\mathcal{A}}$-generator, \textit{i.e.}
\begin{equation}\label{corinzio}
    \left \{ 0,0,\frac{1}{2 \sqrt{6}} \right \} \, = \, \frac{1}{6}
\mathbf{w}_3
\end{equation}
is the generator of translational $\mathbb{Z}_6$ subgroup and the
Universal Classifying Group for the hexagonal lattice turns out to
be:
\begin{equation}\label{U72definio}
    \mathfrak{U}_{72} \, \equiv \, \mathbb{Z}_6 \, \ltimes \,\mathrm{Dih_6}
\end{equation}

\subsubsection{Structure and irreps of $\mathfrak{U}_{72}$}
Utilizing the information obtained from Frobenius congruences we
know that the abstract structure of the group that we name
\(\mathfrak{U}_{72}\) is the following one:
\begin{equation}\label{structU72}
\mathfrak{U}_{72}\,
=\,\mathbb{Z}_2\underset{\text{semidirect}}{\ltimes}\left(\mathbb{Z}_6\otimes
\mathbb{Z}_6\right)
\end{equation}
The generators and relations defining this group are as follows. We
have just three generators named
$\{\mathcal{A}$,$\mathcal{B}$,$\mathcal{T}\}$ that obey the
relations:
\begin{equation}\label{generDih6Rela}
\mathcal{A}^6=\mathcal{B}^2=\mathcal{T}^6=(\mathcal{B}\mathcal{A})^2=(\mathcal{B}\mathcal{T})^2=\mathcal{E}\text{
};\text{      }\mathcal{A}\mathcal{T}=\mathcal{T}\mathcal{A}
\end{equation}
In the case of the hexagonal lattice $\mathcal{A}$,$\mathcal{B}$ are
realized as proper rotations belonging to $\mathrm{SO(3)}$ and they
generate the dihedral group $\text{Dih}_6$. The generator
$\mathcal{T}$ is a translation (modulo lattice). However if we
suppress the generator $\mathcal{A}$ we obtain another dihedral
group $\text{Dih}_6\,\subset \,\mathfrak{U}_{72}$ realized partially
by rotations, partially by translations. The group
$\mathfrak{U}_{72}$ contains a maximal normal abelian subgroup that
we name $\mathfrak{N}_{36}\simeq \mathbb{Z}_6\otimes \mathbb{Z}_6$
which is generated by $\mathcal{A}$ and $\mathcal{T}$:
\begin{equation}\label{normalosottogruppo36}
   \mathfrak{U}_{72} \, \rhd \, \mathfrak{N}_{36}
\end{equation}
This fact is fundamental in order to construct all the irreducible
representations of \(\mathfrak{U}_{72}\) with the iterative
procedure that can be applied to solvable groups (see section
\cite{Fre:2015mla}).
%%%%%%%%%%%%%%%%%
\subsubsection{The auxiliary four dimensional representation of
$\mathfrak{U}_{72}$} \label{U72coniugato}
As we are going to see
below, none of the irreducible representation of $\mathfrak{U}_{72}$
is faithful. In order to study the algebraic structure of
$\mathfrak{U}_{72}$ and its organization in conjugacy classes, we
need a faithful representation. The smallest we found is in four
dimension.
\par
The auxiliary four dimensional representation is generated as it
follows :
\begin{eqnarray}
  \mathcal{A} &=& \left(
\begin{array}{cccc}
 \frac{1}{2} & \frac{\sqrt{3}}{2} & 0 & 0 \\
 -\frac{\sqrt{3}}{2} & \frac{1}{2} & 0 & 0 \\
 0 & 0 & 1 & 0 \\
 0 & 0 & 0 & 1 \\
\end{array}
\right) \quad ; \quad \mathcal{T}\, = \, \left(
\begin{array}{cccc}
 1 & 0 & 0 & 0 \\
 0 & 1 & 0 & 0 \\
 0 & 0 & \frac{1}{2} & \frac{\sqrt{3}}{2} \\
 0 & 0 & -\frac{\sqrt{3}}{2} & \frac{1}{2} \\
\end{array}
\right) \quad ; \quad
  \mathcal{B} \,=\, \left(
\begin{array}{cccc}
 1 & 0 & 0 & 0 \\
 0 & -1 & 0 & 0 \\
 0 & 0 & 1 & 0 \\
 0 & 0 & 0 & -1 \\
\end{array}
\right)
\end{eqnarray}
From the above generators we obtain an explicit form of all the 72
elements that are organized in 24 conjugacy classes as it is
displayed in the table below:\\
%\paragraph{The list of conjugacy classes for $\mathfrak{U}_{72}$}
\begin{doublespace}
\noindent\(\text{Class }1\mid\text{ order of elements = }1\mid\text{ $\#$ of elem in class = }1\mid\text{ representative = }\mathcal{E}\\
\text{Class }2\mid\text{ order of elements = }2\mid\text{ $\#$ of elem in class = }1\mid\text{ representative = }\mathcal{A}^3.\mathcal{T}^3\\
\text{Class }3\mid\text{ order of elements = }2\mid\text{ $\#$ of elem in class = }1\mid\text{ representative = }\mathcal{A}^3\\
\text{Class }4\mid\text{ order of elements = }2\mid\text{ $\#$ of elem in class = }1\mid\text{ representative = }\mathcal{T}^3\\
\text{Class }5\mid\text{ order of elements = }2\mid\text{ $\#$ of
elem in
class = }9\mid\text{ representative = }\mathcal{B}.\mathcal{A}.\mathcal{T}\\
\text{Class }6\mid\text{ order of elements = }2\mid\text{ $\#$ of elem in class = }9\mid\text{ representative = }\mathcal{B}.\mathcal{A}\\
\text{Class }7\mid\text{ order of elements = }2\mid\text{ $\#$ of elem in class = }9\mid\text{ representative = }\mathcal{B}.\mathcal{T}\\
\text{Class }8\mid\text{ order of elements = }2\mid\text{ $\#$ of elem in class = }9\mid\text{ representative = }\mathcal{B}\\
\text{Class }9\mid\text{ order of elements = }3\mid\text{ $\#$ of elem in class = }2\mid\text{ representative = }\mathcal{A}^2.\mathcal{T}^2\\
\text{Class }10\mid\text{ order of elements = }3\mid\text{ $\#$ of elem in class = }2\mid\text{ representative = }\mathcal{A}^2.\mathcal{T}^4\\
\text{Class }11\mid\text{ order of elements = }3\mid\text{ $\#$ of elem in class = }2\mid\text{ representative = }\mathcal{A}^2\\
\text{Class }12\mid\text{ order of elements = }3\mid\text{ $\#$ of elem in class = }2\mid\text{ representative = }\mathcal{T}^2\\
\text{Class }13\mid\text{ order of elements = }6\mid\text{ $\#$ of elem in class = }2\mid\text{ representative = }\mathcal{A}^3.\mathcal{T}^2\\
\text{Class }14\mid\text{ order of elements = }6\mid\text{ $\#$ of elem in class = }2\mid\text{ representative = }\mathcal{A}^3.\mathcal{T}\\
\text{Class }15\mid\text{ order of elements = }6\mid\text{ $\#$ of elem in class = }2\mid\text{ representative = }\mathcal{A}^2.\mathcal{T}^3\\
\text{Class }16\mid\text{ order of elements = }6\mid\text{ $\#$ of elem in class = }2\mid\text{ representative = }\mathcal{A}^2.\mathcal{T}\\
\text{Class }17\mid\text{ order of elements = }6\mid\text{ $\#$ of elem in class = }2\mid\text{ representative = }\mathcal{A}^2.\mathcal{T}^5\\
\text{Class }18\mid\text{ order of elements = }6\mid\text{ $\#$ of elem in class = }2\mid\text{ representative = }\mathcal{A}.\mathcal{T}^3\\
\text{Class }19\mid\text{ order of elements = }6\mid\text{ $\#$ of elem in class = }2\mid\text{ representative = }\mathcal{A}.\mathcal{T}^2\\
\text{Class }20\mid\text{ order of elements = }6\mid\text{ $\#$ of elem in class = }2\mid\text{ representative = }\mathcal{A}.\mathcal{T}^4\\
\text{Class }21\mid\text{ order of elements = }6\mid\text{ $\#$ of elem in class = }2\mid\text{ representative = }\mathcal{A}.\mathcal{T}\\
\text{Class }22\mid\text{ order of elements = }6\mid\text{ $\#$ of elem in class = }2\mid\text{ representative = }\mathcal{A}.\mathcal{T}^5\\
\text{Class }23\mid\text{ order of elements = }6\mid\text{ $\#$ of elem in class = }2\mid\text{ representative = }\mathcal{A}\\
\text{Class }24\mid\text{ order of elements = }6\mid\text{ $\#$ of
elem in class = }2\mid\text{ representative = }\mathcal{T}\)
\end{doublespace}
\subsubsection{Irreducible representations and the character table of
$\mathfrak{U}_{72}$} According with general theorems and with the
fact that $\mathfrak{U}_{72}$ is a solvable group we arrive at the
conclusion that it  has 24 irreps of which 8 are 1-dimensional and
16 are 2-dimensional. These representations were explicitly computed
once for all and they are incorporated in the
\textbf{AlmaFluidaNSPsytem} of MATHEMATICA codes.  The character
table of 24 irreps is also incorporated in that system and we omit
it here.
\subsection{Classification of orbits of the point group
$\mathrm{Dih_6}$ in the momentum lattice}
In complete analogy with what it was done for the cubic lattice also
in the case of the hexagonal lattice we need to classify the orbits
of the point group $\mathrm{Dih_6}$ in the lattice
$\Lambda^\star_{Hex}$. Here we have six different types of orbits:
\subsubsection{Orbits of length 2}
These are the simplest orbits and are formed by vectors of the
following type:
\begin{eqnarray}\label{grongo2}
    \mathcal{O}_2 & = & \left\{p \, \lambda_3 \, , \, -p \,\lambda_3
    \right\} \quad ; \quad p \, \in \, \mathbb{Z} \nonumber\\
    &\Downarrow & \nonumber\\
    & = &\left\{\{0,0, \frac{p}{\sqrt{2}}\} \, , \,\{0,0,
    -\frac{p}{\sqrt{2}}\}
    \right\} \quad \text{in the orthonormal basis}
\end{eqnarray}
that are arranged along the $z$-axis. The action of the
$\mathcal{A}$ generator of the dihedral group vanishes on such
vectors and they are sensitive only to the $\mathcal{B}$ generators
that flips their orientation.
\subsubsection{Orbits of length 6}
\begin{figure}[!hbt]
\begin{center}
\includegraphics[height=50mm]{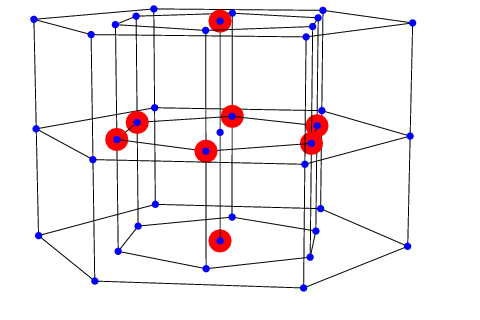}
\includegraphics[height=50mm]{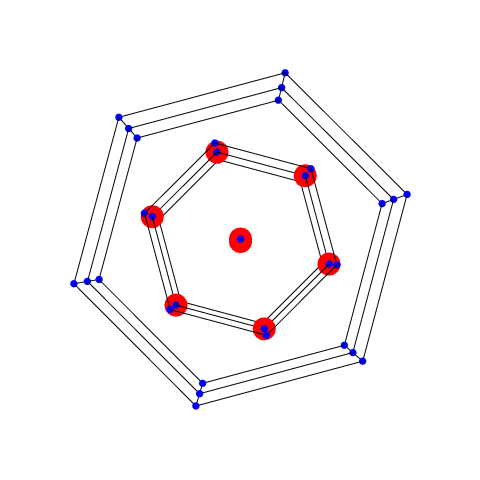}
\end{center}
\caption{\it  In this picture we mark with a big circle the points
in the hexagonal momentum lattice $\Lambda^\star_{Hex}$ that
constitute the lowest lying orbits of length 2 and 6. The orbit of
length 2 is given by the two marked antipodal points along the
vertical $z$-axis, while the orbit of length 6 is given by six
vertices of the regular hexagon lying in the horizontal plane $z=0$.
In the picture on the left we have a view of the lattice from the
front, in the picture on the right we have a view from above. The
blue points belong to the space lattice, the red points belong to
the dual momentum lattice.} \label{duepisei}
\end{figure}

The orbit of length 6 lies in the plane $z=0$ and are made by vectors
of the following type:
\begin{eqnarray}\label{grongo6}
    \mathcal{O}_6 & = & \underbrace{\left\{
\begin{array}{lcr}
 \{0, & -p, & 0 \}\\
\{ 0, & p, & 0\} \\
\{ -p, & 0, & 0 \}\\
 \{-p, & p, & 0\} \\
\{ p, & -p, & 0\} \\
\{ p, & 0, & 0\} \\
\end{array}
\right\}}_{\text{in the $\lambda_i$ basis}} \, = \, \underbrace{\left\{
\begin{array}{lcr}
 \{0, & -\sqrt{\frac{2}{3}} p, & 0 \}\\
 \{0, & \sqrt{\frac{2}{3}} p, & 0 \}\\
 \{-\frac{p}{\sqrt{2}}, & -\frac{p}{\sqrt{6}}, & 0\} \\
 \{-\frac{p}{\sqrt{2}}, & \frac{p}{\sqrt{6}}, & 0 \} \\
 \{\frac{p}{\sqrt{2}}, & -\frac{p}{\sqrt{6}}, & 0 \} \\
 \{\frac{p}{\sqrt{2}}, & \frac{p}{\sqrt{6}}, & 0\} \\
\end{array}
\right\}}_{\text{in the orthonormal basis}}\quad ; \quad p\, \in \,
\mathbb{Z}
\end{eqnarray}
See fig.\ref{duepisei}.
\subsubsection{Orbits of length 12 of type 1}
The orbits of length 12 and type 1 lie in the $z=0$ plane and are of
the following form:
\begin{eqnarray}
\label{grongo12t1}
  \mathcal{O}_{12|1} &=& \underbrace{\left\{
\begin{array}{lcr}
 \{-p, & -q, & 0 \}\\
 \{-p, & p+q, & 0 \}\\
 \{p, & -p-q, & 0\} \\
\{ p, & q, & 0 \}\\
 \{-q, & -p, & 0\} \\
\{ -q, & p+q, & 0 \} \\
\{ q, & -p-q, & 0 \}\\
\{ q, & p, & 0 \}\\
\{ -p-q, & p, & 0 \}\\
\{ -p-q, & q, & 0 \}\\
\{ p+q, & -q, & 0 \} \\
\{ p+q, & -p, & 0 \} \\
\end{array}
\right\}}_{\text{in the $\lambda_i$ basis}} \, = \,\underbrace{\left\{
\begin{array}{lcr}
 \{-\frac{p}{\sqrt{2}}, & -\frac{p+2 q}{\sqrt{6}}, & 0 \}\\
\{ -\frac{p}{\sqrt{2}}, & \frac{p+2 q}{\sqrt{6}}, & 0 \}\\
\{ \frac{p}{\sqrt{2}}, & -\frac{p+2 q}{\sqrt{6}}, & 0 \}\\
\{ \frac{p}{\sqrt{2}} & \frac{p+2 q}{\sqrt{6}} & 0 \}\\
\{ -\frac{q}{\sqrt{2}}, & -\frac{2 p+q}{\sqrt{6}}, & 0 \}\\
\{ -\frac{q}{\sqrt{2}}, & \frac{2 p+q}{\sqrt{6}}, & 0 \}\\
\{ \frac{q}{\sqrt{2}}, & -\frac{2 p+q}{\sqrt{6}}, & 0 \}\\
\{ \frac{q}{\sqrt{2}}, & \frac{2 p+q}{\sqrt{6}}, & 0 \}\\
\{ -\frac{p+q}{\sqrt{2}}, & \frac{p-q}{\sqrt{6}}, & 0 \}\\
\{ -\frac{p+q}{\sqrt{2}}, & \frac{q-p}{\sqrt{6}}, & 0 \}\\
\{ \frac{p+q}{\sqrt{2}}, & \frac{p-q}{\sqrt{6}}, & 0 \}\\
\{ \frac{p+q}{\sqrt{2}}, & \frac{q-p}{\sqrt{6}}, & 0 \}\\
\end{array}
\right\}}_{\text{in the orthonormal basis}} \quad ; \quad p,q \, \in
\, \mathbb{Z}
\end{eqnarray}
\subsubsection{Orbits of length 12 of type 2}
The orbits of length 12 and type 2 are the most generic ones that
depend on three integers $p,q,r$ with no relation among them capable
of nullify some of the orthonormal components of the vectors
belonging to the orbit. Explicitly we find:
\begin{eqnarray}
\label{grongo12t2}
  \mathcal{O}_{12|2} &=& \underbrace{\left(
\begin{array}{lll}
\{-p , & -q , & r\} \\
\{-p , & p+q , & -r\} \\
\{ p , & -p-q , & -r\} \\
\{ p , & q , & r\} \\
\{-q , & -p , & -r\} \\
\{-q , & p+q , & r\} \\
\{q , & -p-q , & r\} \\
\{ q , & p , & -r\} \\
\{ -p-q , & p , & r\} \\
\{ -p-q , & q , & -r\} \\
\{ p+q , & -q , & -r\} \\
\{ p+q , & -p , & r\} \\
\end{array}
\right)}_{\text{in the $\lambda_i$ basis}} \, =
\,\underbrace{\left\{
\begin{array}{lll}
 \{-\frac{p}{\sqrt{2}} , & -\frac{p+2 q}{\sqrt{6}} , & \frac{r}{\sqrt{2}}\} \\
 \{-\frac{p}{\sqrt{2}} , & \frac{p+2 q}{\sqrt{6}} , & -\frac{r}{\sqrt{2}}\} \\
 \{\frac{p}{\sqrt{2}} , & -\frac{p+2 q}{\sqrt{6}} , & -\frac{r}{\sqrt{2}}\} \\
 \{\frac{p}{\sqrt{2}} , & \frac{p+2 q}{\sqrt{6}} , & \frac{r}{\sqrt{2}}\} \\
 \{-\frac{q}{\sqrt{2}} , & -\frac{2 p+q}{\sqrt{6}} , & -\frac{r}{\sqrt{2}}\} \\
 \{-\frac{q}{\sqrt{2}} , & \frac{2 p+q}{\sqrt{6}} , & \frac{r}{\sqrt{2}}\} \\
 \{\frac{q}{\sqrt{2}} , & -\frac{2 p+q}{\sqrt{6}} , & \frac{r}{\sqrt{2}}\} \\
 \{\frac{q}{\sqrt{2}} , & \frac{2 p+q}{\sqrt{6}} , & -\frac{r}{\sqrt{2}}\} \\
 \{-\frac{p+q}{\sqrt{2}} , & \frac{p-q}{\sqrt{6}} , & \frac{r}{\sqrt{2}}\} \\
 \{-\frac{p+q}{\sqrt{2}} , & \frac{q-p}{\sqrt{6}} , & -\frac{r}{\sqrt{2}}\} \\
 \{\frac{p+q}{\sqrt{2}} , & \frac{p-q}{\sqrt{6}} , & -\frac{r}{\sqrt{2}}\} \\
 \{\frac{p+q}{\sqrt{2}} , & \frac{q-p}{\sqrt{6}} , & \frac{r}{\sqrt{2}}\} \\
\end{array}
\right\}}_{\text{in the orthonormal basis}}
\end{eqnarray}
The parameters in the above orbit must satisfy the following
conditions:
\begin{equation}\label{condorbita2}
 p,q,r \,
\in \, \mathbb{Z} \quad \text{and} \quad
\left \{\begin{array}{ccc}q&\neq &- 2p \\ q &\neq & \pm p\\
\end{array}\right .
\end{equation}
\begin{figure}[!hbt]
\begin{center}
\includegraphics[height=50mm]{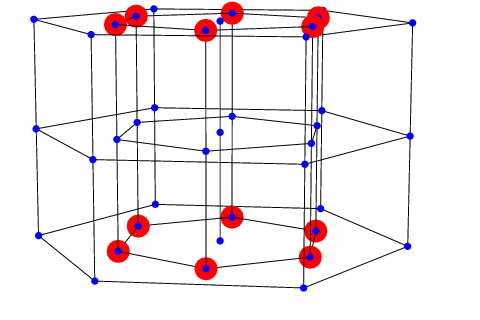}
\includegraphics[height=50mm]{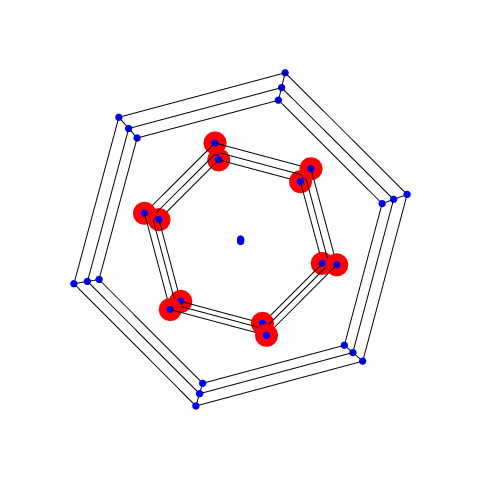}
\end{center}
\caption{\it  In this picture we mark with a big circle the points
in the hexagonal momentum lattice $\Lambda^\star_{Hex}$ that
constitute  an orbit of length 12 of type 2,3 or 4. The orbit of
length 12 is given by 12 vertices of a polyhedron which has two
opposite faces (upper and lower) corresponding to regular hexagons
on horizontal planes symmetrical under $z$-reflection and 6
rectangular vertical faces. In the picture on the left we have a
view of the lattice from the front, in the picture on the right we
have a view from above. The blue points belong to the space lattice,
the red points belong to the dual momentum lattice. The distinction
among type 2,3,4 depends only on the orientation of the hexagonal
faces in the lattice.} \label{dozzina}
\end{figure}
\subsubsection{Orbits of length 12 of type 3}
The orbits of length 12 and type 3 correspond to the degeneration of
the orbits of type 1 when $q=-p$. Explicitly we find:
\begin{eqnarray}
\label{grongo12t3}
  \mathcal{O}_{12|3} &=& \underbrace{\left\{
\begin{array}{ccc}
\{ 0 , & -p , & -r\} \\
\{ 0 , & -p , & r\} \\
\{ 0 , & p , & -r\} \\
\{ 0 , & p , & r\} \\
\{ -p , & 0 , & -r\} \\
\{ -p , & 0 , & r\} \\
\{ -p , & p , & -r\} \\
\{ -p , & p , & r\} \\
\{ p , & -p , & -r\} \\
\{ p , & -p , & r\} \\
\{ p , & 0 , & -r\} \\
\{ p , & 0 , & r\} \\
\end{array}
\right\}}_{\text{in the $\lambda_i$ basis}} \, =
\,\underbrace{\left\{
\begin{array}{lll}
 \{0 , & -\sqrt{\frac{2}{3}} p , & -\frac{r}{\sqrt{2}}\} \\
 \{0 , & -\sqrt{\frac{2}{3}} p , & \frac{r}{\sqrt{2}}\} \\
 \{0 , & \sqrt{\frac{2}{3}} p , & -\frac{r}{\sqrt{2}}\} \\
 \{0 , & \sqrt{\frac{2}{3}} p , & \frac{r}{\sqrt{2}}\} \\
 \{-\frac{p}{\sqrt{2}} , & -\frac{p}{\sqrt{6}} , & -\frac{r}{\sqrt{2}}\} \\
 \{-\frac{p}{\sqrt{2}} , & -\frac{p}{\sqrt{6}} , & \frac{r}{\sqrt{2}}\} \\
 \{-\frac{p}{\sqrt{2}} , & \frac{p}{\sqrt{6}} , & -\frac{r}{\sqrt{2}}\} \\
 \{-\frac{p}{\sqrt{2}} , & \frac{p}{\sqrt{6}} , & \frac{r}{\sqrt{2}}\} \\
 \{\frac{p}{\sqrt{2}} , & -\frac{p}{\sqrt{6}} , & -\frac{r}{\sqrt{2}}\} \\
 \{\frac{p}{\sqrt{2}} , & -\frac{p}{\sqrt{6}} , & \frac{r}{\sqrt{2}}\} \\
 \{\frac{p}{\sqrt{2}} , & \frac{p}{\sqrt{6}} , & -\frac{r}{\sqrt{2}}\} \\
 \{\frac{p}{\sqrt{2}} , & \frac{p}{\sqrt{6}} , & \frac{r}{\sqrt{2}}\} \\
\end{array}
\right\}}_{\text{in the orthonormal basis}} \quad ; \quad  p,r \,
\in \, \mathbb{Z}
\end{eqnarray}
\subsubsection{Orbits of length 12 of type 4}
The orbits of length 12 and type 4 correspond to the degeneration of
the orbits of type 1 when $q=-2p$. Explicitly we find:
\begin{eqnarray}
\label{grongo12t4}
  \mathcal{O}_{12|4} , &=, & \underbrace{\left\{
\begin{array}{lll}
\{ -p , & -p , & -r \}\\
\{ -p , & -p , & r \}\\
\{ -p , & 2 p , & -r \}\\
\{ -p , & 2 p , & r \}\\
\{ p , & -2 p , & -r\} \\
\{ p , & -2 p , & r \}\\
\{ p , & p , & -r \} \\
\{ p , & p , & r\} \\
\{ -2 p , & p , & -r\} \\
\{ -2 p , & p , & r\} \\
\{ 2 p , & -p , & -r\} \\
\{ 2 p , & -p , & r\} \\
\end{array}
\right\}}_{\text{in the $\lambda_i$ basis}} \, =
\,\underbrace{\left\{
\begin{array}{lll}
\{ -\frac{p}{\sqrt{2}} , & -\sqrt{\frac{3}{2}} p , & -\frac{r}{\sqrt{2}}\} \\
\{ -\frac{p}{\sqrt{2}} , & -\sqrt{\frac{3}{2}} p , & \frac{r}{\sqrt{2}}\} \\
\{ -\frac{p}{\sqrt{2}} , & \sqrt{\frac{3}{2}} p , & -\frac{r}{\sqrt{2}} \}\\
\{ -\frac{p}{\sqrt{2}} , & \sqrt{\frac{3}{2}} p , & \frac{r}{\sqrt{2}} \}\\
\{ \frac{p}{\sqrt{2}} , & -\sqrt{\frac{3}{2}} p , & -\frac{r}{\sqrt{2}}\} \\
\{ \frac{p}{\sqrt{2}} , & -\sqrt{\frac{3}{2}} p , & \frac{r}{\sqrt{2}}\} \\
\{ \frac{p}{\sqrt{2}} , & \sqrt{\frac{3}{2}} p , & -\frac{r}{\sqrt{2}}\} \\
\{ \frac{p}{\sqrt{2}} , & \sqrt{\frac{3}{2}} p , & \frac{r}{\sqrt{2}} \}\\
\{ -\sqrt{2} p , & 0 , & -\frac{r}{\sqrt{2}}\} \\
\{ -\sqrt{2} p , & 0 , & \frac{r}{\sqrt{2}}\} \\
\{ \sqrt{2} p , & 0 , & -\frac{r}{\sqrt{2}}\} \\
\{ \sqrt{2} p , & 0 , & \frac{r}{\sqrt{2}}\} \\
\end{array}
\right\}}_{\text{in the orthonormal basis}} \quad ; \quad p,r \, \in
\, \mathbb{Z}
\end{eqnarray}
See a picture of orbits of length 12 of type 2,3,4 in
fig.\ref{dozzina}
%%%%%%%%%%%%%%%%%%%%%%%%%%%%%%%%%%%%%%%%%%%%%%%%%%%%%%%%%%%%%%%%%%%%%%%%%%%%%%%%%%
As we see the shortest orbit of length $2$ is actually vertical,
namely the associated Beltrami Flows correspond to decoupled systems
where only the coordinate $z(t)$ obeys a non linear differential
equation. The other two coordinates form a free system. Similarly
the orbits of length $6$ and the first orbit of length $12$ are all
planar. In the corresponding Beltrami Flows there is no dependence
on the coordinate $z$ which forms a free system. Presumably all the
Beltrami Flows of this type are integrable. Only the maximal orbits
of length $12$  of type two, three and four are truly
three-dimensional and give rise to systems that might develop chaos.
%%%%%%%%%%%%%%%%%%%%%%%%%%%%%%%%%%%%%%%%%%%%%%%%%%%%%%%
\section{Group Theory and $\mathfrak{b}$-Beltrami fields}
\label{dippo}
\subsection{The Euler equations in a $\mathfrak{b}$-three-manifold and
the ABC model as a test ground}\label{bABCfildo} In view of the
geometrical setup discussed in chapter \ref{bmanstory}, in the
present one we reconsider Euler equations and Beltrami fields in
$\mathfrak{b}$-manifolds, following the approach of
\cite{cardone2019} and focusing in particular on the example of the
$ABC$-flows that they used there. Our aim is to bring up to evidence
the relation existing between the necessary condition found in
\cite{cardone2019} for the consistency of Beltrami equation in a
particular $b$-manifold with a particular boundary surface $\Sigma$
and the group theoretical structure of the ABC-model that was
exhaustively presented in \cite{Fre:2015mla}. We will argue that
such a relation is most likely general and that the possible types
of boundary surfaces $\Sigma$ which can be introduced in Beltrami
fields have to be classified in group theoretical terms also in the
case of the much more complicated Beltrami flows originated by
higher orbits of the point-group in the momentum lattice.
\subsubsection{The appropriate geometrical rewriting of Euler equations
on general three-manifolds} In order to implement our programme we
come once again back to Euler equation as written in eq.
(\ref{bernullone}) which, in view of the definition of the Bernoulli
function given in eq.(\ref{finocchionabiscotta}) and for steady
flows can be stated as follows:
\begin{equation}\label{grammomole}
    i_{\mathbf{U}}\cdot \mathrm{d}\Omega^{\mathbf{U}} \, =\, - \mathrm{d}H_B
\end{equation}
We remind the reader that, geometrically, the one-form
$\Omega^{\mathbf{U}}$ is the \textit{contact form}, the velocity
field $\mathbf{U}$ is its \textit{Reeb-field} and $H_B$ is indeed
the Bernoulli-function. In addition to eq. (\ref{grammomole}) the
dynamical system requires, in order to be complete, the
divergenceless condition:
\begin{equation}\label{nonscappo}
   \nabla\cdot \mathbf{U} \, \, \equiv \, \frac{1}{\sqrt{\text{det} g}}
  \, \partial_\ell \left(\sqrt{\text{det} g}\, U^\ell\right) \, = \, 0
\end{equation}
where $g_{ij}$ is the metric tensor of the three-manifold. Also
equation (\ref{nonscappo}) admits an index-free totally geometrical
rewriting in terms of the volume three-form defined belove:
\begin{equation}\label{lopezdevega}
    \text{Vol}_g \, \equiv \, \frac{1}{3!} \, \sqrt{\text{det} g} \,
    \epsilon_{ijk} \, dx^i \wedge dx^j \wedge dx^k
\end{equation}
An easy straightforward calculations shows that:
\begin{equation}\label{speusippo}
    \mathrm{d} \left( i_{\mathbf{U}}\cdot\text{Vol}_g \right) \, = \, \nabla\cdot
    \mathbf{U} \, \times \, \text{Vol}_g
\end{equation}
Hence Euler equations reduce to:
\begin{eqnarray}
\label{eulerazna}
  i_{\mathbf{U}}\cdot \mathrm{d}\Omega^{\mathbf{U}} & =& - \mathrm{d}H_B \nonumber\\
  \mathrm{d} \left( i_{\mathbf{U}}\cdot\text{Vol}_g \right) &=& 0
\end{eqnarray}
\subsubsection{The $b$-deformation of the ABC-model}
\label{bABCsekzia} Next we consider the ABC model vector field as
defined in the next section in eq.(\ref{ABCnostro}) and we try to
convert the $\mathrm{T^3}$ torus, obtained by quotiening
$\mathbb{R}^3$ with respect to the cubic lattice, into a
$\mathfrak{b}$-manifold by choosing, in the covering space
$\mathbb{R}^3$, the surface $\Sigma_{x=0}\subset \mathbb{R}^3$
identified by the equation $x=0$. The Beltrami vector becomes
parallel to the surface $\Sigma_{x=0}$ by means of the substitution:
\begin{equation}\label{bigiazione}
    \partial_x \, \longrightarrow \, x\, \partial_x
\end{equation}
Hence we have:
\begin{eqnarray}\label{pariola}
    \null^{\mathfrak{b}}\mathbf{V}_{ABC}&=&
    \left(2 A \cos [2 \pi  y]+2 B \cos [2 \pi  z]\right)\,x
    \partial_x\nonumber\\
   &&(2 C \cos [2 \pi  x]-2 B \sin [2 \pi  z])\,\partial_y \,
    +\,\left(2 A \sin [2 \pi  y]-2 C
   \sin [2 \pi  x]\right)\,\partial_z
\end{eqnarray}
According to the principles of $\mathfrak{b}$-manifolds summarized
in in section \ref{bmanstory} the metric and the differential forms
are accordingly modified. We have:
\begin{eqnarray}
\label{perlafiorita}
\null^{\mathfrak{b}}ds^2 &=& \null^{\mathfrak{b}}g_{ij}dx^i \times dx^j
\, = \, \left(\frac{dx}{x}\right)^2 + dy^2 +dz^2\nonumber \\
\null^{\mathfrak{b}}\text{Vol}_{g} &=& \frac{dx}{x}\wedge dy \wedge
dz
\end{eqnarray}
so we easily compute:
\begin{eqnarray}\label{girdiniere}
    i_{\null^{\mathfrak{b}}\mathbf{V}_{ABC}}\cdot\null^{\mathfrak{b}}\text{Vol}_{g}&
    = & 2 dy\wedge dz \left(A \cos [2 \pi  y]+B \cos [(2 \pi  z]\right)\nonumber\\
    &&-\frac{2 dx\wedge
   dy \left(C \sin [2 \pi  x]-A \sin [2 \pi  y]\right)}{x}
   -\frac{2 dx\wedge dz \left(C
   \cos [2 \pi  x]-B \sin [2 \pi  z]\right)}{x}\nonumber\\
\end{eqnarray}
and we immediately verify the second of eq.s (\ref{eulerazna})
\begin{equation}\label{bnonscappo}
    \mathrm{d}\left(i_{\null^{\mathfrak{b}}\mathbf{V}_{ABC}}\cdot\null^{\mathfrak{b}}\text{Vol}_{g}\right)\,
    = \, 0
\end{equation}
As we know from the discussion in the introduction, the first of
eq.s(\ref{eulerazna}) is certainly satisfied if the stronger
Beltrami equation (\ref{Belatramus}) is enforced and before the
$\mathfrak{b}$-deformation the contact form
$\Omega^{\left[\mathbf{V}_{ABC}\right]}$ certainly satisfies it by
construction. It is to be seen whether the new
$\mathfrak{b}$-contact form
$\null^{\mathfrak{b}}\Omega^{\left[\mathbf{V}_{ABC}\right]}$ still
satisfies it. We easily compute:
\begin{eqnarray}\label{bicontattoOmABC}
   \null^{\mathfrak{b}}\Omega^{\left[\mathbf{V}_{ABC}\right]}\,
   =\,\null^{\mathfrak{b}}g_{ij}\null^{\mathfrak{b}}\mathbf{V}_{ABC}^i
   \, dx^j & = & \frac{dx \left(2 A \cos [2 \pi  y]+2 B \cos [2 \pi  z]\right)}{x}\nonumber\\
   &&+dy \left(2 C \cos [2 \pi  x]-2 B \sin [2 \pi
   z]\right)+dz \left(2 A \sin [2
   \pi  y]-2 C \sin [2 \pi  x]\right)\nonumber\\
\end{eqnarray}
Taking the Hodge dual of Beltrami equation (\ref{Belatramus}) we can
equivalently rewrite it as follows:
\begin{equation}\label{machiavellico}
    \mathrm{d}\null^{\mathfrak{b}}\Omega^{\left[\mathbf{V}_{ABC}\right]}\,
    = \, \lambda \, \left(\star_{\,\null^{\mathfrak{b}}g}\,
    \null^{\mathfrak{b}}\Omega^{\left[\mathbf{V}_{ABC}\right]} \right) \,
    \equiv
    \, \lambda \, \ft 12 \left(
    i_{\null^{\mathfrak{b}}\mathbf{V}_{ABC}}\cdot\null^{\mathfrak{b}}\text{Vol}_{g}\right)
\end{equation}
The second member was already calculated, the first is immediately
calculated and we find that setting $\lambda\, = \, 2\pi$ which is
its original value prior to the deformation, we have:
\begin{equation}\label{kunago}
    \mathrm{d}\null^{\mathfrak{b}}\Omega^{\left[\mathbf{V}_{ABC}\right]}\,-\,
    \pi \,\left(
    i_{\null^{\mathfrak{b}}\mathbf{V}_{ABC}}\cdot\null^{\mathfrak{b}}\text{Vol}_{g}\right)
    \, = \,-\frac{4 \pi  C (x-1) \left(\sin [2 \pi  x] dx\wedge dy+\cos [2 \pi
    x]
   dx\wedge dz\right)}{x}
\end{equation}
As it was done in \cite{cardone2019} we have no other way out then
choosing $C=0$. Hence we conclude that the complete ABC-model cannot
be $\mathfrak{b}$-deformed but the AB0-model can. In
\cite{cardone2019} the boundary surface was posed at $z=0$ and the
authors reached the same conclusion in the form of the constraint
$A=0$. As we argue in the next section these two choices are
perfectly equivalent since we can interchange $A,B,C$ with
transformations of the subgroup $\mathrm{GF_{192}}\subset
\mathrm{G_{1536}}$ of which the ABC model constitutes an irreducible
three dimensional representation. The important thing is that
setting one of the three parameters ABC, equal to zero one obtains a
two-parameter model which constitutes an irreducible representation
of a subsgroup $\mathrm{G_{128}^{(AB0)}}\subset
\mathrm{G_{1536}}$\footnote{The subgroups $\mathrm{G_{128}^{(0BC)}}$
and $\mathrm{G_{128}^{(0BC)}}$ are obviously conjugate in $G_{1536}$
to $\mathrm{G_{128}^{(AB0)}}$ and therefore isomorphic  to this
latter and among themselves}. Inside $\mathrm{G_{128}^{(AB0)}}$ the
stabilizer of the two vector $(A,B)$ is a group of order 16,
$G^{(AB0)}_{16}$ which contains a purely translational subgroup
$G^{(AB0)}_4 \sim \mathbb{Z}_4$ made by the quantized translation of
$1/4$ in the $y$-direction. This makes the dynamical system actually
two-dimensional. It is a remarkable fact that the
$\mathfrak{b}$-deformation of the chosen type is possible only in
presence of this particular hidden symmetry. We come back to this
question at the end of the chapter. First we recall from
\cite{Fre:2015mla} the group-theory behind the ABC models.
\subsection{Group theoretical interpretation of the
ABC flows} From the  analysis \cite{Fre:2015mla}  the following
pattern emerged. The Universal Classifying Group $\mathrm{G_{1536}}$
contains at least two\footnote{It is known that there are 4
different Space-Groups $\Gamma^{I}_{24}$ ($I=1,\dots,4$) of order
$24$, isomorphic to the point group $\mathrm{O_{24}}$ but not
conjugate one to the other under the action of the continuous
translation group. One of them is the point group  itself
$\Gamma^{1}_{24}\, = \, \mathrm{O_{24}}$ which is a subgroup of the
first of the two groups of order 192 identified in
\cite{Fre:2015mla}: $\mathrm{O_{24}} \subset \mathrm{G_{192}}$.
Another of the four mentioned groups is $\Gamma^{2}_{24}\, = \,
\mathrm{GS_{24}}$ which is a subgroup of the second group of order
192 identified by the authors of \cite{Fre:2015mla}:
$\mathrm{GS_{24}} \subset \mathrm{GF_{192}}$. It remains to see
whether $\Gamma^{3}_{24} $ and $ \Gamma^{4}_{24}$ are contained in
the two already identified subgroups $\mathrm{G_{192}}$ and
$\mathrm{GF_{192}}$ or if there exists other two such non conjugate
subgroups of order 192 that respectively contain $\Gamma^{3}_{24} $
and $\Gamma^{3}_{24} $. The answer was not worked out in
\cite{Fre:2015mla}.  Extensive but lengthy calculations could
resolve the issue.} isomorphic but not conjugate subgroups of order
192, namely $\mathrm{G_{192}}$ and $\mathrm{GF_{192}}$ in the
adopted nomenclature. The classical ABC-flows are obtained from the
lowest lying momentum orbit of length 6 which produces an
irreducible $6$-dimensional representation of the Universal
Classifying Group: $D_{23}\left[\mathrm{\mathrm{G_{1536}}},6\right]
$. The vector field is the following one:
\begin{equation}\label{orbit6vector}
{\mathbf{V}^{(6)}}({\mathbf{r}}|\mathbf{F})\,  = \,\left(
\begin{array}{l}
 2 \cos \left(2 \pi  z\right) F_1+2 \cos \left(2 \pi  y\right)
   F_2+2 \sin \left(2 \pi  z\right) F_3-2 \sin \left(2 \pi  y
   \right) F_5 \\
 -2 \sin \left(2 \pi  z\right) F_1+2 \cos \left(2 \pi  z\right)
   F_3+2 \cos \left(2 \pi  x\right) F_4+2 \sin \left(2 \pi  x\right) F_6 \\
 2 \sin \left(2 \pi  y\right) F_2-2 \sin \left(2 \pi  x\right)
   F_4+2 \cos \left(2 \pi  y\right) F_5+2 \cos \left(2 \pi  x\right) F_6
\end{array}
\right)
\end{equation}
where $F_i$ ($i\,=\,1,\dots\, ,6$) are real numbers. The three
parameter ABC-flow is just the irreducible $3$-dimensional
representation $D_{12}\left[\mathrm{GF_{192}},3\right]$ in the split
\begin{equation}\label{splittus}
    D_{23}\left[\mathrm{\mathrm{G_{1536}}},6\right] \, = \,
D_{12}\left[\mathrm{GF_{192}},3\right]\oplus
D_{15}\left[\mathrm{GF_{192}},3\right]
\end{equation}
 With respect to the
isomorphic but not conjugate subgroup $\mathrm{G_{192}}$ the
representation $D_{23}\left[\mathrm{\mathrm{G_{1536}}},6\right] $
remains instead irreducible:
\begin{equation}\label{nosplittus}
    D_{23}\left[\mathrm{\mathrm{G_{1536}}},6\right] \, = \,
D_{20}\left[\mathrm{G_{192}},6\right]
\end{equation}
so that there is no proper way of reducing the six parameters to
three.
\par
Indeed, as shown in \cite{Fre:2015mla}, we have the following chain
of inclusions:
\begin{equation}\label{baldacchinoA}
    \mathrm{\mathrm{G_{1536}}} \, \supset \, \mathrm{GF_{192}} \, \supset \, \mathrm{GS_{24}}
\end{equation}
that is parallel to the other one:
\begin{equation}\label{baldacchinoB}
    \mathrm{\mathrm{G_{1536}}} \, \supset \, \mathrm{G_{192}} \, \supset \, \mathrm{O_{24}}
\end{equation}
$\mathrm{G_{192}}$ being another subgroup, isomorphic to
$\mathrm{GF_{192}}$, but not conjugate to it in
$\mathrm{\mathrm{G_{1536}}}$. (see appendices A.6 and A.7 of
\cite{Fre:2015mla} for the detailed description of these two
subgroups of the Universal Classifying Group $\mathrm{G_{1536}}$ of
the cubic lattice). Since $\mathrm{G_{192}}$ and $\mathrm{GF_{192}}$
are isomorphic they have the same irreps and the same character
table. Yet, since they are not conjugate, the branching rules of the
same $\mathrm{\mathrm{G_{1536}}}$ irrep with respect to the former
or the latter subgroup can be different. In the case of the
representation $D_{23}\left[\mathrm{\mathrm{G_{1536}}},6\right]$,
which is that produced by the fundamental orbit of order six, we
have (see appendix D of \cite{Fre:2015mla}):
\begin{equation}\label{passerotto}
D_{23}\left[\mathrm{\mathrm{G_{1536}}},6\right] \, = \,
\left\{\begin{array}{rcl}
D_{20}\left[\mathrm{G_{192}},6 \right] &=&D_4\left[\mathrm{O_{24}},3\right]  \oplus D_5\left[\mathrm{O_{24}},3\right]\\
D_{12}\left[\mathrm{GF_{192}},3 \right]  \oplus
D_{15}\left[\mathrm{GF_{192}},3
\right]&=&D_1\left[\mathrm{GS_{24}},1\right]
\oplus  D_3\left[\mathrm{GS_{24}},2\right] \oplus  D_4\left[\mathrm{GS_{24}},3\right]\\
\end{array}
\right.
\end{equation}
where in the second line we have used the branching rules:
\begin{eqnarray}
D_{12}\left[\mathrm{GF_{192}},3 \right]  &=& D_1\left[\mathrm{GS_{24}},1\right] \oplus  D_3\left[\mathrm{GS_{24}},2\right]  \\
D_{15}\left[\mathrm{GF_{192}},3 \right] &=&
D_4\left[\mathrm{GS_{24}},3\right] \label{passamontagna}
\end{eqnarray}
that, in view of the isomorphism, are identical with:
\begin{eqnarray}
 D_{12}\left[\mathrm{G_{192}},3 \right]  &=& D_1\left[\mathrm{O_{24}},1\right] \oplus  D_3\left[\mathrm{O_{24}},2\right]  \\
  D_{15}\left[\mathrm{G_{192}},3 \right] &=&  D_4\left[\mathrm{O_{24}},3\right] \label{passamontagnabis}
\end{eqnarray}
Eq.(\ref{passerotto}) has far reaching consequences. While there are
no Beltrami vector fields obtained from this orbit that are
invariant with respect to the octahedral point group
$\mathrm{O_{24}}$, there exists such an invariant Beltrami flow with
respect to the isomorphic $\mathrm{GS_{24}}$: it corresponds to the
$D_1\left[\mathrm{GS_{24}},1\right]$ irrep in the second line of
(\ref{passerotto}). Furthermore while the six parameter space
$\mathbf{F}$ is irreducible with respect to the action of the group
$\mathrm{G_{192}}$ (the irrep $D_{20}\left[\mathrm{G_{192}},6
\right]$) it splits into two three-dimensional subspaces with
respect to $\mathrm{GF_{192}}$. This is the origin of the ABC-flows.
Indeed the ABC Beltrami flows can be identified with the irreducible
representation $ D_{12}\left[\mathrm{GF_{192}},3 \right] $. Let us
see how. Explicitly we have the following projection operators on
the two irreducible representations,
 $D_{12}$ and $D_{15}$:
\begin{eqnarray}
  \Pi^{(12)}\left[\mathrm{GF_{192}},3\right]\,\mathbf{F}  &=& \left\{F_1,F_2,0,F_4,0,0\right\} \\
  \Pi^{(15)}\left[\mathrm{GF_{192}},3\right]\,\mathbf{F}&=& \left\{0,0,F_3,0,F_5,F_6\right\}
\end{eqnarray}
If we set $F_3\,=\,F_5\, = \, F_6=0$ we kill the irreducible
representation $ D_{15}\left[\mathrm{GF_{192}},3 \right] $ and the
residual Beltrami vector field, upon the following identifications:
\begin{equation}\label{agnorizo}
   A\, = \, F_1 \quad ; \quad B\, = \, F_4 \quad ; \quad C \, = \, F_2
\end{equation}
coincides with the time honored ABC flow of eq.(\ref{bagcigaluppi}).
Indeed inserting  the special parameter vector $\mathbf{F}\, = \,
\{A,C,0,B,0,0\}$ in eq.(\ref{orbit6vector}) we obtain:
\begin{eqnarray}\label{sicumerulo}
\mathbf{V}^{(6)}\left(\left\{x,y,z\right\}\,
\mid\{A,C,0,B,0,0\}\right)& \equiv \,
&\mathbf{V}_{(ABC)}(x,y,z)\nonumber\\
\mathbf{V}_{(ABC)}\left(x+\ft 34,y,z-\ft
14\right)&=&\mathcal{V}^{(ABC)}(x,y,z)
\end{eqnarray}
the vector field $\mathcal{V}^{(ABC)}(x,y,z)$ being that defined by
eq.(\ref{bagcigaluppi}).
\par
For future quick reference it is convenient to write explicitly the
ABC Beltrami field $\mathbf{V}^{(ABC)}(x,y,z)$ in the normalization
we utilize in the sequel:
\begin{equation}\label{ABCnostro}
    \mathbf{V}^{(ABC)}(x,y,z)\, = \, \left(
\begin{array}{c}
 2 A \cos (2 \pi  y)+2 B \cos (2 \pi  z) \\
 2 C \cos (2 \pi  x)-2 B \sin (2 \pi  z) \\
 2 A \sin (2 \pi  y)-2 C \sin (2 \pi  x) \\
\end{array}
\right)
\end{equation}
\par
The next step is provided by considering the explicit form of the
decomposition of the $ D_{12}\left[\mathrm{GF_{192}},3 \right] $
irrep, \textit{i.e.} the ABC flow, into irreducible representations
of the subgroup $\mathrm{GS_{24}}$. The two invariant subspaces are
immediately characterized in terms of the parameters $A,B,C$, as it
follows:
\begin{eqnarray}
 D_1\left[\mathrm{GS_{24}},1\right]  &\Leftrightarrow& A\, = \, B \, = \, C \, \ne \, 0 \\
 D_3\left[\mathrm{GS_{24}},2\right] &\Leftrightarrow& A\, + \, B \, + \, C \, = \, 0 \label{ergastolo}
\end{eqnarray}
The most symmetric case $A:A:A=1$ simply corresponds to the identity
representation of the subgroup  $\mathrm{GS_{24}}\subset
\mathrm{GF_{192}}$ which occurs in the splitting  of  the
$3$-dimensional representation:
\begin{equation}\label{GS24splitto}
   D_{12}\left[\mathrm{GF_{192}},3\right]\, = \,
D_{1}\left[\mathrm{GS_{24}},1\right]\oplus
D_{3}\left[\mathrm{GS_{24}},2\right]
\end{equation}
\subsubsection{The $\mathrm{(A,A,A)}$-flow invariant under
$\mathrm{GS_{24}}$} 
\begin{figure}[!hbt]
\begin{center}
\includegraphics[height=60mm]{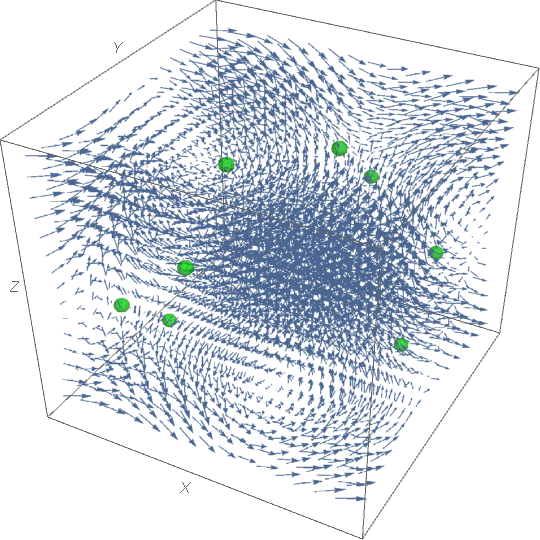}
\end{center}
\caption{\it  A plot of the $A:A:A=1$ Beltrami vector field
invariant under the group $\mathrm{GS_{24}}$ with a view of its
eight stagnation points of eq.(\ref{gommaliqua})}
\label{AAAcamporella} \hskip 1cm \unitlength=1.1mm
\end{figure}

This information suffices to understand the role
of the $A:A:A\, =\,1$ Beltrami vector field often considered in the
literature. It is the unique one invariant under the order 24 group
$\mathrm{GS_{24}}$ isomorphic to the octahedral point group.
Explicitly, in our notations, it takes the following
form\footnote{Observe that here and in the sequel we stick to our
conventions for $x,y,z$, which differ from those of
eq.(\ref{bagcigaluppi}) by the already mentioned shift $\{ \ft 34 ,
0,-\ft 14\}$)}:
\begin{equation}\label{AAAfildo}
  \mathbf{V}_{(A,A,A)}(\mathbf{r})\, = \,   \mathbf{V}_{(A,A,A)}(x,y,z)\, \equiv \, 2 A \, \left(
\begin{array}{l}
  (\cos (2 \pi  y)+\cos (2 \pi  z)) \\
  (\cos (2 \pi  x)-\sin (2 \pi  z)) \\
 (\sin (2 \pi  y)-\sin (2 \pi  x))
\end{array}
\right)
\end{equation}
This vector field $\mathbf{V}_{(A,A,A)}(x,y,z)$ is everywhere non
singular in the fundamental unit cube (the torus $\mathrm{T^3}$)
apart from eight isolated   \textit{stagnation points} where it
vanishes. They are listed below.
\begin{equation}\label{gommaliqua}
\begin{array}{rclcrcl}
s_1 &= & \left\{\frac{1}{8},\frac{1}{8},\frac{3}{8}\right\} &; &
s_2 &= &\left\{\frac{1}{8},\frac{3}{8},\frac{1}{8}\right\}\\
s_3 &= & \left\{\frac{3}{8},\frac{1}{8},\frac{5}{8}\right\} &;&
s_4 &= &\left\{\frac{3}{8},\frac{3}{8},\frac{7}{8}\right\}\\
s_5 &= & \left\{\frac{5}{8},\frac{5}{8},\frac{7}{8}\right\} &;&
s_6&=&\left\{\frac{5}{8},\frac{7}{8},\frac{5}{8}\right\}\\
s_7&= &\left\{\frac{7}{8},\frac{5}{8},\frac{1}{8}\right\} &;& s_8 &
=& \left\{\frac{7}{8},\frac{7}{8},\frac{3}{8}\right\}
\end{array}
\end{equation}
A numerical plot of this vector field is displayed in fig.
\ref{AAAcamporella}.
\par
\begin{figure}[!hbt]
\begin{center}
\includegraphics[height=60mm]{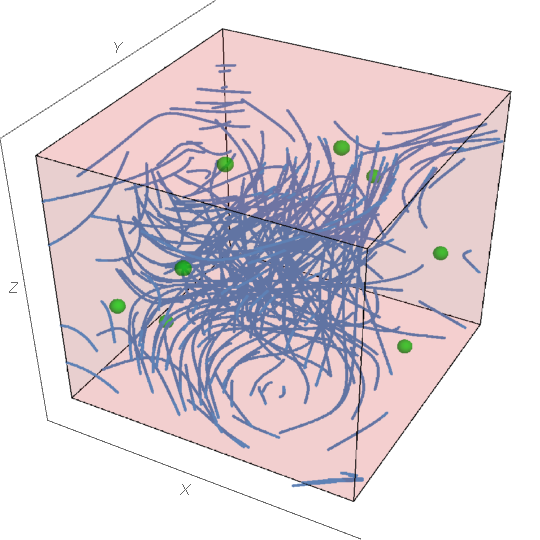}
\end{center}
\caption{\it  A plot of $216$ streamlines of the $A:A:A=1$ Beltrami
vector field with equally spaced initial conditions. The numerical
solutions are smooth in $\mathbb{R}^3$. When a branch reaches a
boundary of the unit cube it is continued with its image in the cube
modulo the appropriate lattice translation. The circles in this
figure are the eight stagnation points } \label{AAAfiliderba}
\end{figure}

In order to provide the reader with a visual impression of the
dynamics of this flow, in fig.\ref{AAAfiliderba} we display a set of
$5\times 5 \times 5 = 125$ streamlines, namely of numerical
integrations of the differential system:
\begin{equation}\label{babbione}
    \frac{dr}{dt} \, = \, \mathbf{V}_{(A,A,A)}(\mathbf{r})
\end{equation}
with initial conditions:
\begin{equation}\label{iniconda}
    \mathbf{r}(0) \, = \, \mathbf{r}_0 \, = \, \left\{\frac{n_1}{6},\, \frac{n_2}{6}, \,\frac{n_3}{6}\right\} \quad ;
    \quad n_{1,2,3} \, =\, 0,1,2,3,4,5
\end{equation}
%%%%%%%%
\subsection{Chains of subgroups and the flows $(A,B,0)$, $(A,A,0)$ and
$(A,0,0)$} In the literature a lot of attention has been given to
the special subcases of the $\mathrm{ABC}$-flow where one or two of
the parameters vanish or two are equal among themselves and one
vanishes. Also these cases can be thoroughly characterized in group
theoretical terms and their special features can be traced back to
the hidden subgroup structure associated with them.
\subsubsection{The
$(A,B,0)$ case and its associated chain of subgroups}
\label{AB0floppo} First we consider the case where we put to zero
one of the three parameters leaving the other two undetermined.
\par A preliminary important observation is the following. Each of
the three parameters is associated in eq.(\ref{ABCnostro}) with the
trigonometric functions of one of the three variables $x,y,z$. Hence
permuting the variables $x,y,z$ is equivalent to permute the $A,B,C$
coefficients. There are also some changes of sign but all these
operations are contained in the point group $\mathrm{O_{24}}$ as one
can immediately realize looking at eq.(\ref{nomiOelemen}). Hence in
the Universal Classifying Group $\mathrm{G_{1536}}$ that contains
the point group there are certainly elements that can map the
parameter vector $\{A,B,C\}$ in any other permutation of the same
letters. That means that considering the case $C=0$ is no loss of
generality. The invariance groups that we determine for this case
will just be conjugate to the invariance groups appearing in the
case $A=0$ or in the case $B=0$. So let us make the choice $C=0$
which was already done in \cite{Fre:2015mla}.
\paragraph{When we put $C=0$} we define a two dimensional subspace of the representation
$D_{12}\left[\mathrm{GF_{192}},3\right]$ which is invariant under
some proper subgroup $\mathrm{H}^{\mathrm{(A,B,0)}} \subset
\mathrm{GF_{192}}$. This group $\mathrm{H}^{\mathrm{(A,B,0)}}$ can
be calculated and found to be of order 64, yet we do not dwell on it
because the subgroup of the classifying group
$\mathrm{\mathrm{G_{1536}}}$ which leaves the subspace
$(A,0,0,B,0,0)$ invariant is larger than
$\mathrm{H}^{\mathrm{(A,B,0)}}$ and it is not contained in
$\mathrm{GF_{192}}$. It has order 128 and we name it
$\mathrm{G_{128}^{(A,B,0)}}$. This short discussion is important
because it implies the following: the flows $\mathrm{(A,B,0})$
should not be considered just as a particular case of the
$ABC$-flows rather as a different set of flows, whose properties are
encoded in the group $\mathrm{G_{128}^{(A,B,0)}}$.
\par
\begin{figure}[!hbt]
\begin{center}
%\iffigs
\includegraphics[height=60mm]{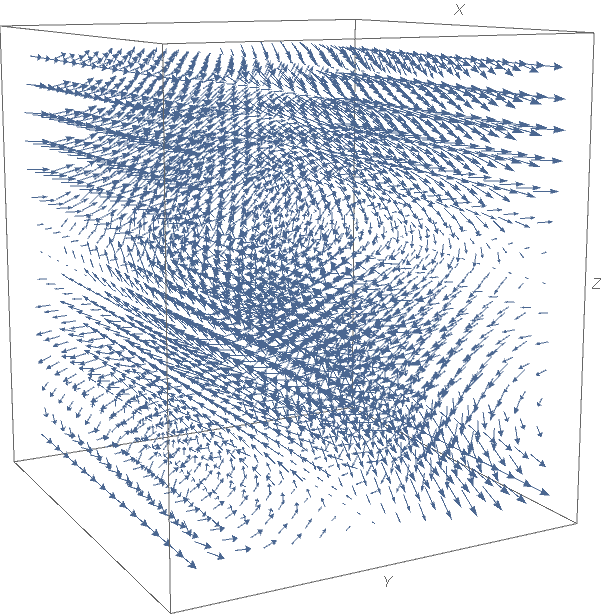}
\includegraphics[height=60mm]{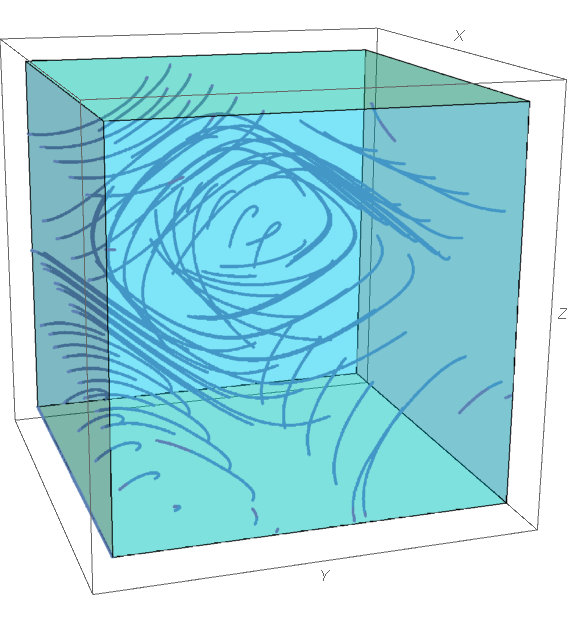}
%\else
\end{center}
%\fi
\caption{\it A plot of the Betrami vector field
$\mathbf{V}_{(A,B,0)}(\mathbf{r})$ (on the left) with $A=5$, $B=7$.
On the right a family of streamlines with equally spaced initial
conditions is displayed.} \label{ABfilotti}
%\iffigs
%\hskip 1cm \unitlength=1.1mm
%\end{center}
%\fi
\end{figure}

The group $\mathrm{G_{128}^{(A,B,0)}}$ is solvable and a chain of
normal subgroups can be found, all of index 2 which ends with the
abelian $\mathrm{G_{4}^{(A,B,0)}}$ isomorphic to $\mathbb{Z}_4$.
This latter is nothing else than the group of quantized translation
in the $y$-direction and its inclusion in the group leaving the
space $(A,0,0,B,0,0)$ invariant actually means that the differential
system must be  $y$-independent and hence two dimensional. The chain
of normal subgroups is displayed here below:
\begin{center}
\begin{picture}(200,100)
%%%%%%%%%%%%%%%%%%%%%%%%%%%%%%%%%
\put(-90,45){$\mathbb{Z}_4$} \put(-75,45){$\sim$} \put
(-60,45){$\mathrm{G_4^{(A,B,0)}}$} \put (-20,45){$\vartriangleleft
$} \put (-5,45){$\mathrm{G_8^{(A,B,0)}}$} \put
(35,45){$\vartriangleleft $} \put
(50,45){$\mathrm{G_{16}^{(A,B,0)}}$} \put (90,45){$\vartriangleleft
$} \put (103,47){\line (1,1){20}} \put (103,47){\line (1,-1){20}}
\put (127,65.5){$\vartriangleleft $} \put (127,24){$\vartriangleleft
$} \put (142,24){$\mathrm{G_{32}^{(A,A,0)}}$} \put
(142,65.5){$\mathrm{G_{32}^{(A,B,0)}}$} \put
(182,65.5){$\vartriangleleft $} \put
(197,65.5){$\mathrm{G_{64}^{(A,B,0)}}$} \put
(237,65.5){$\vartriangleleft $} \put
(252,65.5){$\mathrm{G_{128}^{(A,B,0)}}$}
%%%%%%%%%%%%%%%%%%%%%%%%%%%%%%%%%
%%%%%%%%%%%%%%%%%%%%%%%%%%%%%%%%%
\end{picture}
\end{center}
\begin{equation}
\null \label{goffo}
\end{equation}
and it allows for the construction of irreducible representations of
$\mathrm{G_{128}^{(A,B,0)}}$ and all other members of the chain, by
means of the induction algorithm. Such a construction we have not
done, but all the groups of the chain are listed, with their
conjugacy classes in appendix E of \cite{Fre:2015mla}. The group
$\mathrm{G_{128}^{(A,B,0)}}$ leaves the subspace $(A,0,0,B,0,0)$
invariant but still mixes the parameters $A$ and $B$ among
themselves. The subgroup $\mathrm{G_{16}^{(A,B,0)}}\lhd
\mathrm{G_{128}^{(A,B,0)}} $ instead stabilizes the very vector
$(A,0,0,B,0,0)$. This means that any ${\mathrm{(A,B,0)}}$-flow has a
hidden symmetry of order 16 provided by the group
$\mathrm{G_{16}^{(A,B,0)}}$. The general form of these Beltrami
fields is the following one:
\begin{equation}\label{AB0campus}
\mathbf{V}_{(A,B,0)}(\mathbf{r})\, = \,
\mathbf{V}_{(A,B,0)}(x,y,z)\, \equiv \,     \left(
\begin{array}{c}
 2 A \cos (2 \pi  y)+2 B \cos (2 \pi  z) \\
 -2 B \sin (2 \pi  z) \\
 2 A \sin (2 \pi  y) \\
\end{array}
\right)
\end{equation}
In fig.\ref{ABfilotti} we display a plot of the vector field and an
example of equally spaced streamlines.
\par
\begin{figure}[!hbt]
\begin{center}
%\iffigs
\includegraphics[height=60mm]{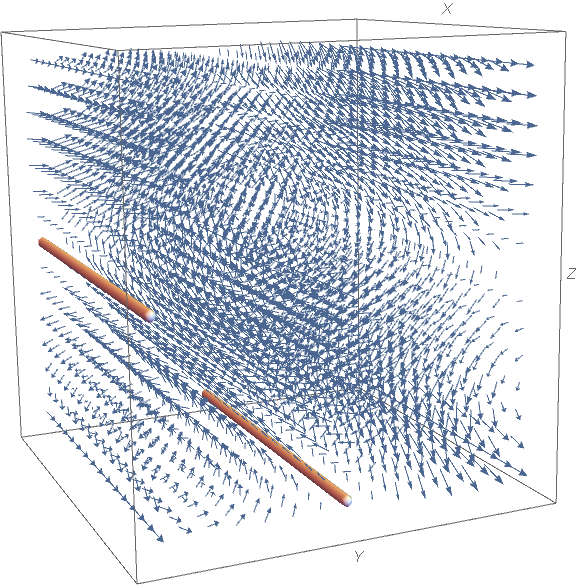}
\includegraphics[height=60mm]{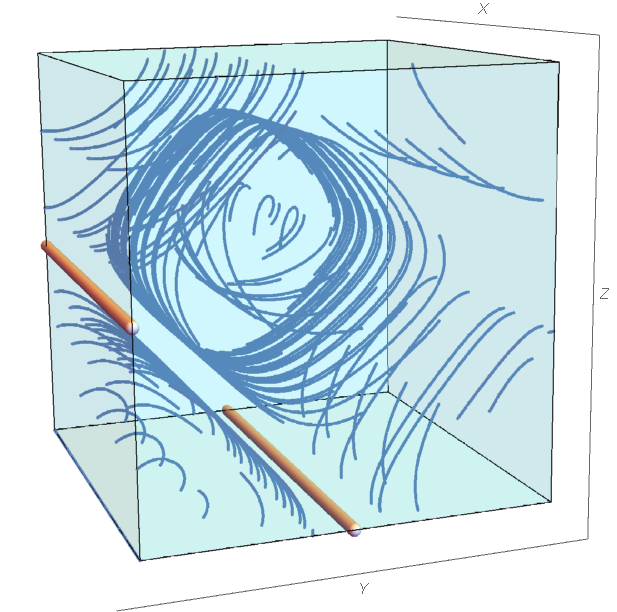}
%\else
\end{center}
%\fi
\caption{\it A plot of the Betrami vector field
$\mathbf{V}_{(A,A,0)}(\mathbf{r})$ (on the left) where are visible
(fat lines) the two stagnation lines of the flow. On the right a
family of streamlines with equally spaced initial conditions is
displayed.} \label{AAfilotti}
%\iffigs
%\hskip 1cm \unitlength=1.1mm
%\end{center}
%\fi
\end{figure}

Looking at  eq.(\ref{goffo}) we notice that there is another group
of order 32, namely  $\mathrm{G_{32}^{(A,A,0)}}$ which contains
$\mathrm{G_{16}^{(A,B,0)}}$ but it is not contained neither in
$\mathrm{G_{128}^{(A,B,0)}} $ nor in $\mathrm{GF_{192}}$. This group
is the stabilizer of the vector $(A,0,0,A,0,0)$ and hence it is the
hidden symmetry group of the flows of type $(A,A,0)$. Once again the
very fact that $\mathrm{G_{32}^{(A,A,0)}}$ is not contained in
$\mathrm{G_{128}^{(A,B,0)}}$ shows that the $(A,A,0)$ flow should
not be considered as a particular case of the $(A,B,0)$-flows rather
as a new type of its own. Let us also stress the difference with the
case of the $(A,A,A)$-flow. Here the hidden symmetry group
$\mathrm{GS_{24}}$ is contained in $\mathrm{GF_{192}}$ and the
interpretation of the $(A,A,A)$-flow as a particular case of the
$(A,B,C)$-flows is permitted. Having set:
\begin{equation}\label{AA0campus}
\mathbf{V}_{(A,A,0)}(\mathbf{r})\, = \,
\mathbf{V}_{(A,A,0)}(x,y,z)\, \equiv \,   A \,  \left(
\begin{array}{c}
 2 A \cos (2 \pi  y)+2 A \cos (2 \pi  z) \\
 -2 A \sin (2 \pi  z) \\
 2 A \sin (2 \pi  y) \\
\end{array}
\right)
\end{equation}
in fig.\ref{AAfilotti} we display a plot of the vector field
$\mathbf{V}_{(A,A,0)}(\mathbf{r})$ and a family of its streamlines.
In the case of this flow there are not isolated stagnation points,
rather, because of the $x$-independence of the Beltrami vector
field, there are two entire stagnation lines explicitly given below:
\begin{equation}\label{cannastagnante}
\mathrm{sl}_1\, = \, \left\{x,\frac{1}{2},0\right\}\quad ; \quad
\mathrm{sl}_2\  = \, \left\{x,0,\frac{1}{2}\right\}
\end{equation}
Let us finally come to the case of the flow $(A,0,0)$. The
one-dimensional subspace of vectors of the form $(A,0,0,0,0,0)$ is
left invariant by a rather big subgroup of the classifying group
which is of order $256$. We name it  $\mathrm{G_{256}^{(A,0,0)}}$
and its description is given in appendix E of \cite{Fre:2015mla}. It
is a solvable group with a chain of normal subgroups of index $2$
which ends into a subgroup of order $16$ isomorphic to $\mathbb{Z}_4
\times \mathbb{Z}_4$. This information is summarized in the equation
below:
\begin{center}
\begin{picture}(60,90)
%%%%%%%%%%%%%%%%%%%%%%%%%%%%%%%%%
\put(-120,45){$\mathbb{Z}_4 \times \mathbb{Z}_4$}
\put(-75,45){$\sim$} \put (-60,45){$\mathrm{G_{16}^{(A,0,0)}}$} \put
(-20,45){$\vartriangleleft $} \put
(-5,45){$\mathrm{G_{32}^{(A,0,0)}}$} \put (35,45){$\vartriangleleft
$} \put (50,45){$\mathrm{G_{64}^{(A,0,0)}}$} \put
(90,45){$\vartriangleleft $} \put (103,47){\line (1,-1){20}} \put
(126,27){\circle{3}} \put (135,24){$\mathrm{G_{128}^{(A,0,0)}}$}
\put (175,24){$\vartriangleleft $} \put
(190,24){$\mathrm{G_{256}^{(A,0,0)}}$} \put (103,7){\line (1,1){20}}
\put (90,6){$\subset $} \put (-60,6){$\mathrm{G_{16}^{(A,B,0)}}$}
\put (-20,6){$\vartriangleleft $} \put
(-5,6){$\mathrm{G_{32}^{(A,B,0)}}$} \put (35,6){$\vartriangleleft $}
\put (50,6){$\mathrm{G_{64}^{(A,B,0)}}$} \put
(-75,6){$\vartriangleleft $} \put
(-115,6){$\mathrm{G_{8}^{(A,B,0)}}$} \put (-130,6){$\vartriangleleft
$} \put (-170,6){$\mathrm{G_{4}^{(A,B,0)}}$}
\put(-200,6){$\mathbb{Z}_4 $} \put(-185,6){$\sim$}
%%%%%%%%%%%%%%%%%%%%%%%%%%%%%%%%%
%%%%%%%%%%%%%%%%%%%%%%%%%%%%%%%%%
\end{picture}
\begin{equation}
\label{goffo2} \null
\end{equation}
\end{center}
The group $\mathrm{G_{256}^{(A,0,0)}}$ leaves the subspace
$(A,0,0,0,0,0)$ invariant but occasionally changes the sign of $A$.
The subgroup $\mathrm{G_{128}^{(A,0,0)}} \, \subset \,
\mathrm{G_{256}^{(A,0,0)}}$ stabilizes the very vector
$(A,0,0,0,0,0)$ and therefore it is the hidden symmetry of the
$(A,0,0)$ flows encoded in the planar vector field:
\begin{equation}\label{A00campus}
\mathbf{V}^{(A,0,0)}(\mathbf{r})\, = \,
\mathbf{V}^{(A,0,0)}(x,y,z)\, \equiv \,   A \, \left(
\begin{array}{l}
 \cos (2 \pi  z) \\
 -\sin (2 \pi  z) \\
 0
\end{array}
\right)
\end{equation}
Looking back at equation (\ref{goffo2}) it is important to note that
the group $\mathrm{G_{128}^{(A,0,0)}} \ne
\mathrm{G_{128}^{(A,B,0)}}$ is different from the homologous group
appearing in the group-chain of the $\mathrm{(A,B,0)}$-flows. So
once again the $\mathrm{(A,0,0)}$-flows cannot be regarded as
particular cases of the $\mathrm{(A,B,0)}$-flows. Yet the group
$\mathrm{G_{128}^{(A,0,0)}}$ contains the entire chain of normal
subgroups $\mathrm{G_{128}^{(A,B,0)}}$ starting from
$\mathrm{G_{64}^{(A,B,0)}}$. There is however a very relevant
proviso $\mathrm{G_{64}^{(A,B,0)}}$ is a subgroup of
$\mathrm{G_{128}^{(A,0,0)}}$ but it is not normal.
\begin{figure}[!hbt]
\begin{center}
\includegraphics[height=50mm]{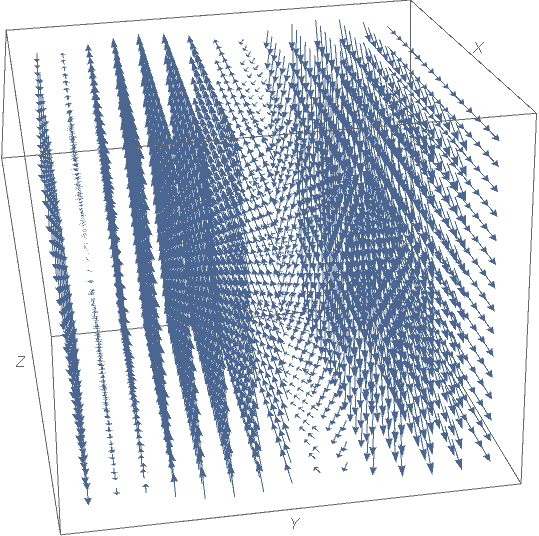}
\includegraphics[height=50mm]{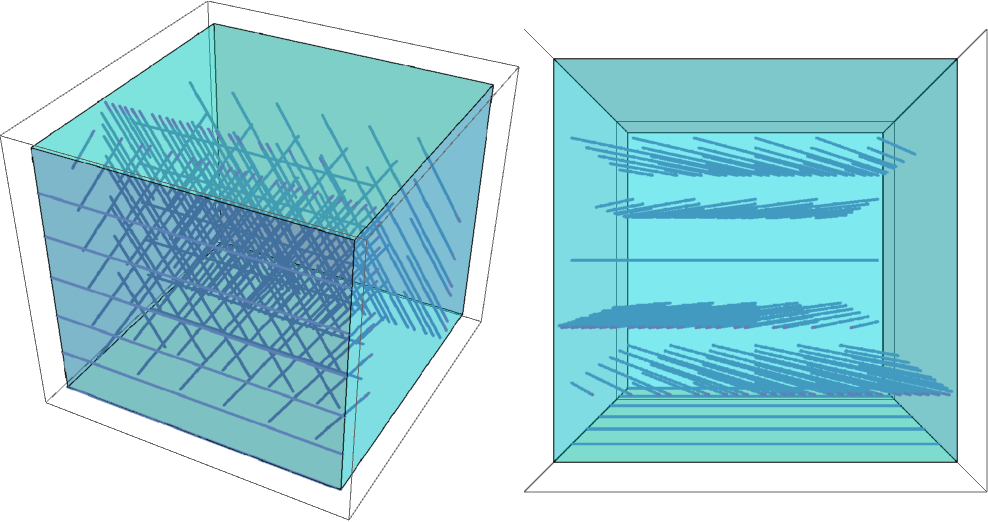}
\end{center}
\caption{\it A plot of the Beltrami vector field
$\mathbf{V}^{(A,0,0)}(\mathbf{r})$ (on the left). On the right a
family of streamlines with equally spaced initial conditions is
displayed. The planar structure of the streamlines, that are all
straight lines, is quite visible. In the center a standard viewpoint
shot of the streamlines, on the right a view from above. }
\label{A00filucci}
\end{figure}
In fig. \ref{A00filucci} we show a plot of the vector field
$\mathbf{V}^{(A,0,0)}(\mathbf{r})$ and a family of its streamlines.
\subsection{Temporary Conclusion}
Comparing the group theoretical analysis of the ABC models with
$b$-deformations of the simple considered type it becomes obvious
that there is a link between the symmetry group of a Beltrami-flow
and the surfaces $\Sigma$ that can be utilized to introduce a
$b$-deformed manifold able to host the $b$-deformation of that
Beltrami-flow. At the moment the precise relation between the
boundary surface $\Sigma$ and the symmetry group is by no means
clear yet it is evident that it exists and it should be explored.
Such exploration requires a study of the possible
$\mathfrak{b}$-deformations in the Beltrami flows associated with
higher point group orbits in the momentum lattice. It is obviously a
research direction that should be pursued. Indeed all other Beltrami
flows arising from different instances of the $48$ classes of
momentum vectors have similar structures. The result of the
construction algorithm produces a representation of the Universal
Classifying Group that can be either reducible or irreducible. This
latter can be split into irreps of either $\mathrm{G_{192}}$ or
$\mathrm{GF_{192}}$ and apparently all cases of invariant Beltrami
vector fields have invariance groups that are subgroups of one of
the two groups $\mathrm{G_{192}}$ or $\mathrm{GF_{192}}$. It would
be interesting to transform this observation into a theorem. At the
moment we have not found an obvious proof.
\subsubsection{A look at the streamlines of the $b$-deformed $AB0$-model} In order to see what the
$b$-deformations might be good for, we consider plotting the
$\mathfrak{b}$-deformed field and some of its trajectories. For the
sake of possible applications it is much better to work on compact
spaces rather than on non compact $\mathbb{R}^3$, preserving the
periodicity. As it was already remarked in \cite{cardone2019} as
equation of the boundary, instead of $x=0$, one can choose
$\sin[2\pi x]=0$. So instead of eq.s
(\ref{bigiazione},\ref{pariola}), we get
\begin{equation}\label{sinbigiazione}
    \partial_x \, \longrightarrow \,\sin[2\pi x]\, \partial_x
\end{equation}
and:
\begin{eqnarray}\label{sinpariola}
    \null^{\mathfrak{b}}\mathbf{V}_{ABC}&=&\left(2 A \cos [2 \pi  y]+2 B \cos [2 \pi  z]\right)\,\sin[2\pi x]\, \partial_x \nonumber\\
   &&(2 C \cos [2 \pi  x]-2 B \sin [2 \pi  z])\,\partial_y \,  +\,\left(2 A \sin [2 \pi  y]-2 C
   \sin [2 \pi  x]\right)\,\partial_z
\end{eqnarray}
and all the other formulae in section \ref{bABCsekzia} hold true
upon the substitution of the denominators $1/x$ with $1/\sin[2\pi
x]$. The conclusion remains the same. The $\mathfrak{b}$-deformed
Beltrami equation holds true if and only if $C=0$. In the next
figure \ref{bdefAB0} we present a picture of $\mathfrak{b}$-deformed
$AB0$-field and family of streamlines with the same parameters
$A=5$,$A=7$ utilized for the un-deformed case in fig.
\ref{ABfilotti}.
\begin{figure}[!hbt]
\begin{center}
\includegraphics[height=70mm]{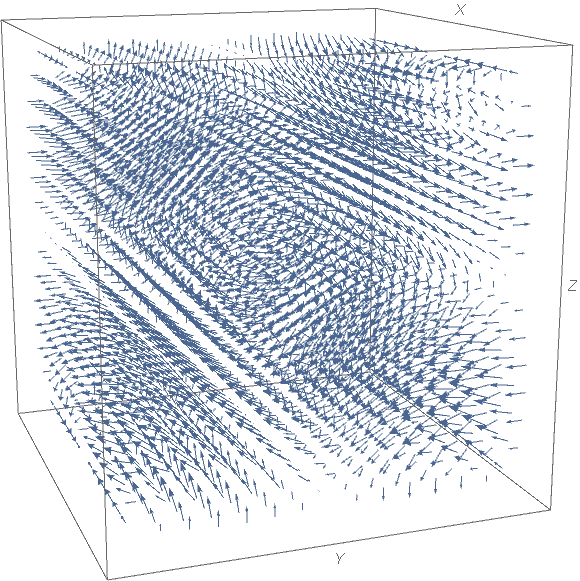}
\includegraphics[height=70mm]{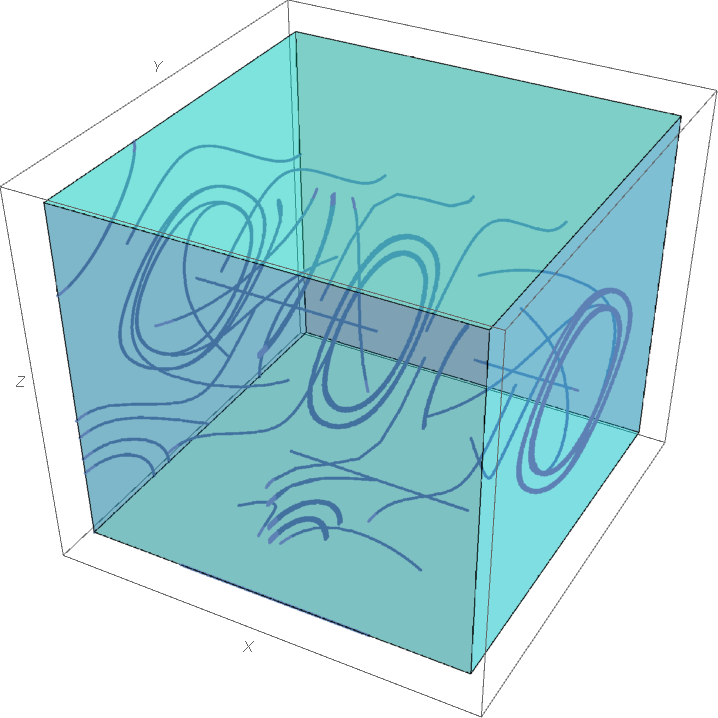}
\end{center}
\caption{\it A plot of the Betrami vector field
$\null^{\mathfrak{b}}\mathbf{V}_{ABC}$ (on the left). On the right a
family of streamlines with equally spaced initial conditions is
displayed. What it means to be parallel to the boundary becomes
clear in this picture: the trajectories that come very close to
$x=0$ or $x=1$ end up swinging on the two faces of the cubic cell.
} \label{bdefAB0}
\end{figure}
\section{The Landscape Conception with Examples} \label{panoramica} Having clarified the
group theoretical foundations of Arnold--Beltrami Flows we come back
to the issue of producing exact solutions of the Navier-Stokes
equations based on velocity fields that satisfy Beltrami equation.
In presence of non-vanishing viscosity we can name such solution
\textbf{NS-Beltrami generalized steady flows}
\subsection{Beltrami equation and generalized steady flows}
The two pillars on which the solutions we consider reside are
provided by the following:
\begin{description}
  \item[A)] Implementation of the \textbf{generalized steady flow} condition  displayed in eq.
  (\ref{genstedona})
  \item[B)] Constancy of the Bernoulli hamiltonian function $H_B$ defined in
  eq.(\ref{finocchionabiscotta})
\end{description}
The pillar B) is easily implemented by setting the pressure field
equal to a constant $h$ minus the squared norm of velocity field:
\begin{equation}\label{tarielka}
    p\left(\mathbf{x},t\right) \, = \, h \, -\, \ft 12 \parallel
    U(\mathbf{x},t)\parallel^2 \, = \,\, h \, - \, \text{const}
    \,\times\,\frac{
    \Omega^{[\mathrm{U}]} \wedge \star_{g} \Omega^{[\mathrm{U}]}}{\text{Vol}}
\end{equation}
where
\begin{equation}\label{volume3forma}
    \text{Vol} \, \equiv \, \ft{1}{3!} \, \mathrm{det}\left(g\right)
    \, \mathrm{d}x\wedge \mathrm{d}y \wedge \mathrm{d}z
\end{equation}
is the volume $3$-form. If the velocity field satisfies Beltrami
equation with eigenvalue $\mu$
\begin{equation}\label{carnenonvale}
    \star_{g} \,\mathrm{d }\Omega^{[\mathrm{U}]} \, = \,  \mu \, \Omega^{[\mathrm{U}]}
\end{equation}
then $\Omega^{[\mathrm{U}]}$ is a \textbf{contact form} and we get:
\begin{equation}\label{pagnuflone}
    \Omega^{[\mathrm{U}]} \wedge \star_{g} \Omega^{[\mathrm{U}]} \,
    = \, \frac{1}{\mu} \, \Omega^{[\mathrm{U}]} \wedge
    \mathrm{d}\Omega^{[\mathrm{U}]} \, = \, \lambda(\mathbf{x},t) \,
    \text{Vol}
\end{equation}
So that the physical pressure field (apart from the additive
constant $h$) obtains an inspiring geometrical interpretation:
indeed it is the nowhere vanishing function $\lambda(\mathbf{x},t)$
mentioned in the definition \ref{ribatriestina} of the Reeb field.
\par
As for pillar A) it is sufficient to recall eq.s
(\ref{laterano},\ref{carnitina}). The essential point is that, as a
consequence of Beltrami equation, the contact one-form
$\Omega^{[\mathrm{U}]}$, whose normalized Reeb field is just the
velocity field $\mathrm{U}\left(\mathbf{x},t\right)$, is an
eigenstate of the Laplace-Beltrami operator $\Delta$ with eigenvalue
$\mu^2$
\begin{equation}\label{kalenkamoya}
    \Delta \,\Omega^{[\mathrm{U}]} \, = \, \mu^2 \, \Omega^{[\mathrm{U}]}
\end{equation}
Then the implementation of the generalized steady flow condition
goes as follows. Consider the finite dimensional vector space
provided by the eigenspace pertaining to the eigenvalue $\mu$:
\begin{equation}\label{ramazzato}
 \mathbf{V}_\mu \ni \Omega^{[\mathrm{u}]} \quad \Rightarrow \quad   \star_{g} \,\mathrm{d }\Omega^{[\mathrm{u}]} \, = \,  \mu \,
    \Omega^{[\mathrm{u}]} \quad ; \quad \Omega^{[\mathrm{u}]} \, =
    \, \sum_{i=1}^{N_\mu} \, F_i \, \Omega^{[\mathrm{u}_i]}
\end{equation}
where $\mathrm{u}_i(\mathbf{x})$ are the normalized Reeb fields of a
basis of solutions $\Omega^{[\mathrm{u}_i]}$ and $F_i$ the free
parameters spanning the eigenspace $\mathbf{V}_\mu$. The number
$N_\mu$ is the degeneracy of the eigenvalue $\mu$ namely the
dimension of the eigenspace.  Next subdivide the $\mathbf{V}_\mu$ in
two freely chosen subspaces:
\begin{equation}\label{rominaEalbano}
    \mathbf{V}_\mu \, = \, \mathbf{V}_\mu^0 \oplus \mathbf{V}_\mu^t
    \quad ; \quad \mathrm{dim}\,\mathbf{V}_\mu^0\, = \, M_0 \, < \, N_\mu \quad ;
    \quad \mathrm{dim}\,\mathbf{V}_\mu^t \, = \, N_\mu\, -\, M_0
\end{equation}
Correspondingly the contact form $\Omega^{[\mathrm{u}]}$ and its
normalized Reeb field $\mathrm{u}(\mathbf{x})$ will split in two
parts:
\begin{equation}\label{strappone}
    \Omega^{[\mathrm{u}]} \, = \, \Omega^{[\mathrm{u}^0]} \, + \,
    \Omega^{[\mathrm{u}^t]} \quad ; \quad \Omega^{[\mathrm{u}^0]}
    \in \mathbf{V}_\mu^0 \quad ; \quad \quad \Omega^{[\mathrm{u}^t]}
    \in \mathbf{V}_\mu^t
\end{equation}
Then setting the driving force as follows:
\begin{equation}\label{patentediguida}
    \mathbf{f}\, = \, -\, \nu \,\mu \,\Omega^{[\mathrm{u}^0]}
\end{equation}
and the contact form (Reeb field) as follows
\begin{equation}\label{NStanda}
    \Omega^{[\mathrm{U}]}\, = \, \Omega^{[\mathrm{u}^0]}+
    \exp\left[- \mu^2 \,t\right]\,\Omega^{[\mathrm{u}^t]}
\end{equation}
the generalized steady flow condition (\ref{genstedona}) is
satisfied and the velocity field
\begin{equation}\label{corricorri}
    \mathrm{U}(\mathbf{x},t) \, = \, \mathrm{u}^0(\mathbf{x}) \, +
    \,\exp\left[- \mu^2 \,t\right] \,\mathrm{u}^t(\mathbf{x})
\end{equation}
fulfils the Navier-Stokes equation (\ref{EulerusEqua}).
\subsection{The landscape conception}
It follows from the above discussion that the main issue in order to
construct the NS-Beltrami generalized steady flows is the
construction of the eigenspaces $\mathbf{V}_\mu$ and their
organization in subspaces according with symmetry principles. This
is what leads to the \textit{landscape conception}.
\par
When the manifold $\mathcal{M}_3$ is the torus defined by
eq.(\ref{metricT3}), the construction of the eigenspace
$\mathbf{V}_\mu$ can be performed geometrically, relying on the
algorithm explained in section \ref{algoritmo} and on the orbits of
the point group $\mathfrak{P}_\Lambda$ in the momentum lattice
$\Lambda^\star$. We just need to consider all those orbits for which
the squared norm of the momentum vectors $\mathbf{k}$ is the same.
Geometrically this amounts to consider the spherical layers of
radius $r = \sqrt{\mathbf{k}^2}$ defined in section \ref{furetto}.
This solution of Beltrami equation constitutes a reducible
representation of the Universal Classifying Group
$\mathfrak{UG}_\Lambda$ of dimension $N_{\mathbf{k}^2}$
\begin{equation}\label{ramirezsancho}
    \mathrm{SL}_{\mathbf{k}^2} \, \stackrel{\text{Beltrami
    field}}{\Longrightarrow} \,
    D\left[\mathfrak{UG}_\Lambda,N_{\mathbf{k}^2}\right]
\end{equation}
which can be decomposed into irreps
\begin{equation}\label{sferodecompo}
    D\left[\mathfrak{UG}_\Lambda,N_{\mathbf{k}^2}\right] \, = \,
    \bigoplus_{i=1}^{\mathfrak{r}}\, a_i \,
    \,D_i\left[\mathfrak{UG}_\Lambda,n_i\right] \quad ; \quad
    \sum_{i}^{\mathfrak{r}} \,a_i \, n_i \, = \, N_{\mathbf{k}^2}
\end{equation}
having denoted by $\mathfrak{r}$ the number of conjugacy classes and
hence of irreps of $\mathfrak{UG}_\Lambda$, by $n_i$ the dimension
of the $i$-th irrep and by $a_i$ its multiplicity. Since
$N_{\mathbf{k}^2} \to \infty$ when $\mathbf{k}^2 \to \infty$ it is
obvious that enlarging the landscape the same representations will
reappear again and again with increasing multiplicity.
\par
The essential thing is that the Beltrami and anti-Beltrami solutions
associated with the same layer decompose exactly in the same way
with respect to the Universal Classifying Group
$\mathfrak{UG}_\Lambda$.
\par
The \textbf{AlmafluidaNSPsystem} posted in Wolfram Community and
available from that site  is  finalized to:
\begin{description}
  \item[a)] to the construction of a large landscape
  \item[b)] to the
construction of the Beltrami solution on each chosen spherical layer
of that landscape
  \item[c)] to the group theoretical analysis of the corresponding
representation
$D\left[\mathfrak{UG}_\Lambda,N_{\mathbf{k}^2}\right]$ including its
further decomposition with respect to subgroups of
$\mathfrak{UG}_\Lambda$.
\end{description}
\subsection{Sketches of the cubic and hexagonal landscapes}
In this section we flash through a pair of inspiring examples from
both instances of main lattice families, the cubic and hexagonal
ones.
\subsubsection{The cubic landscape}
Utilizing the background MATHEMATICA code
\textbf{UniClasGroupCubicLat} of the \textbf{AlmafluidaNSPsystem},
we have constructed a rather large portion of the self-dual cubic
lattice $\Lambda_{cubic}$ containing 117649 lattice points. In this
portion of the lattice we found 1057 spherical layers that we
analyzed with our computer code. In this way we found a maximally
large representation of dimension 792 residing on the largest radius
sphere hosted by this lattice region:
\begin{equation}\label{massimone}
     \text{Max} \text{Dim}\, = \, \dim  D\left[
G_{1536}, 792\right]\, = \, 792 \quad \Leftrightarrow\quad
\left|\mathbf{k}\right|^2 =689
\end{equation}
\subsubsection{An example of Chaos from symmetry from the cubic
lattice} 
\begin{figure}[!ht]
\begin{center}
\includegraphics[width=80mm]{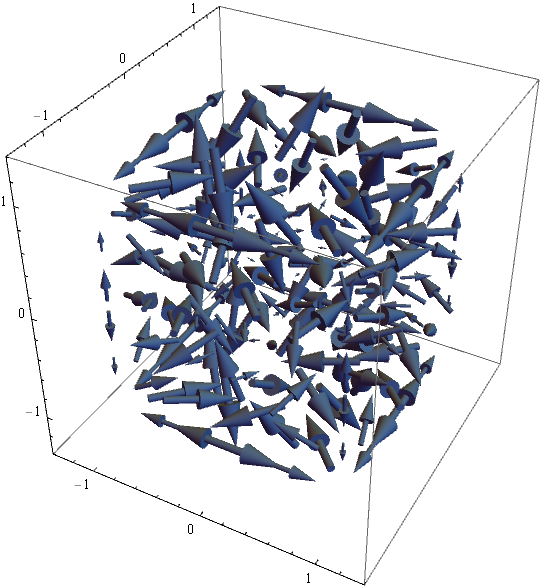}
\includegraphics[width=80mm]{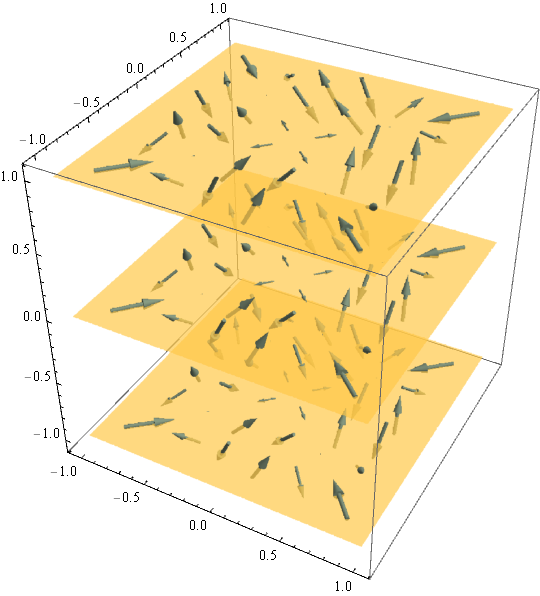}
\end{center}
\caption{\it In this figure we show the three dimensional vector
plot of the unique Beltrami vector field invariant under the largest
symmetry group $G_{1536}$ that arises in the eigenspace pertaining
to the Beltrami eigenvalue $\mu \, = \, 24$ namely on the spherical
layer $\mathbf{k}^2 \, = \, 576$. The high symmetry of the vector
field is almost evident at eye-sight. } \label{vettoronesing}
\end{figure}

\begin{figure}[!hb]
\begin{center}
\includegraphics[width=70mm]{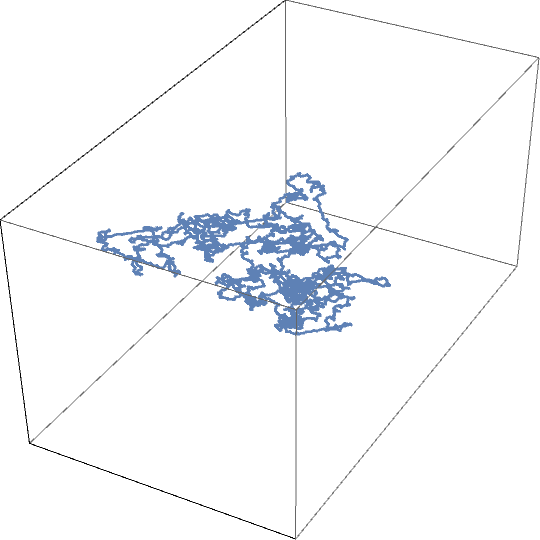}
\includegraphics[width=70mm]{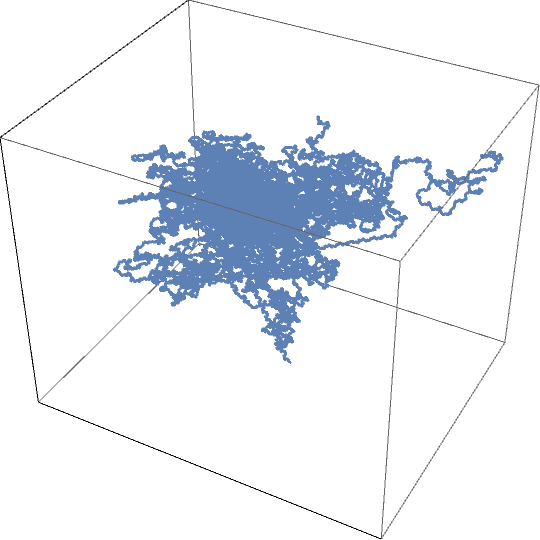}
\end{center}
\caption{\it In the picture on the left we show just one fluid
element trajectory starting in a randomly chosen point
$p=\{1/7,2/9,5/33\}$. In the picture on the right we display the
plot of 27 streamlines whose starting point are equally spaced over
the three dimensions. After 50 integration cycles they make an
inextricable pattern. This is the visual manifestation of the
contact structure. \label{gomitolinocub}}
\end{figure}

As an illustration of the Beltrami construction we
considered the Beltrami fields associated with a specific layer
namely that one where:
\begin{equation}\label{layer}
    |\mathbf{k}|^2 = 576
\end{equation}
We find that the number of points on this layer is 30 that arrange
themselves in a point group orbit $\mathcal{O}_6$ of length 6  plus
another one $\mathcal{O}_{24}$ of length 24. The Beltrami solution
corresponding to this layer has therefore eigenvalue $\mu \, = \, 24
\pi$ and the reducible representation of the Universal Classifying
Group is found to decompose into irreps as follows:
\begin{equation}\label{singolodecompo}
    D\left[G_{1536},30\right] \, = \, D_1\left[G_{1536},1\right]
    +D_2\left[G_{1536},1\right]+2 D_5\left[G_{1536},2\right]
    +4 D_7\left[G_{1536},3\right]+4
   D_8\left[G_{1536},3\right]
\end{equation}
As one sees the considered layer contains one singlet of the maximal
possible symmetry group. It is interesting to visualize both the
plot of this vector field and some of its trajectories. In
fig.\ref{vettoronesing}
Next we show the example of just one trajectory and of 27 equally
spaced streamlines of this symmetric vector field that we have
followed for 50 iterations of numerical integrations. The plots are
displayed in fig.\ref{gomitolinocub}
\subsubsection{The hexagonal landscape}
\begin{figure}[!htb]
\begin{center}
\includegraphics[width=50mm]{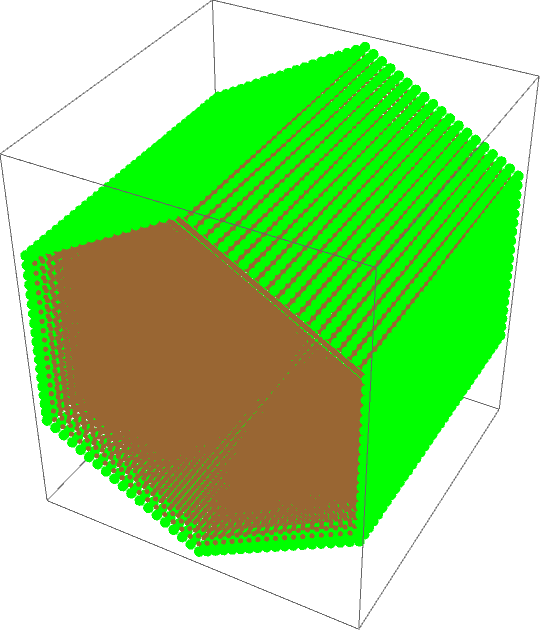}
\includegraphics[width=50mm]{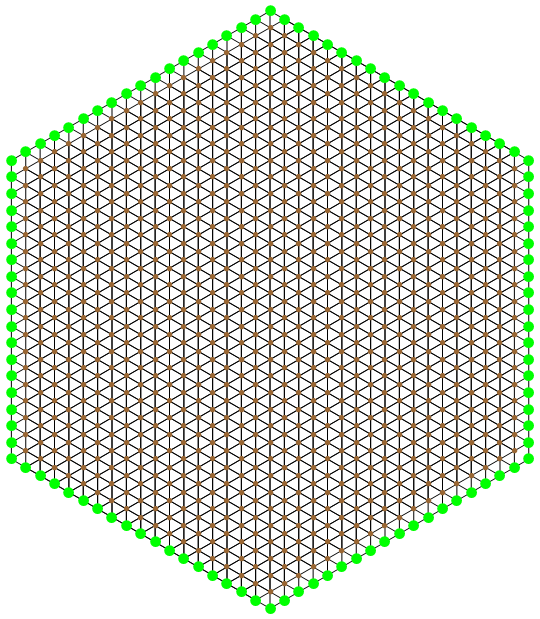}
\end{center}
\caption{\it The portion of considered hexagonal momentum lattice is
taken to be a polyhedron with an hexagonal basis that is shown on
the right, extended in the $z$-direction just as much as it extends
in the $xy$-plane. The lattice points on the 6 lateral faces of this
polyhedron have been displayed in green while the lattice  points
that are inside the polyhedron have been displayed in brown. There
are 3888 points inside the polyhedron and 4800 points on the 6
faces. Note that for visual convenience we have aligned the $z$-axis
horizontally and the $y$-axis vertically.
\label{bottigliaesagonale}}
\end{figure}

As for the hexagonal lattice we have so far constructed a landscape
portion portion of the infinite momentum lattice that is shaped as a
polyhedron with an hexagonal basis and it is displayed
fig.\ref{bottigliaesagonale}. This landscape contains 33084
\textit{interior points} and 3888 \textit{points on its boundary}.
This distinction has no intrinsic meaning and it simply corresponds
to the geometrical shape of the considered lattice portion.
We have intersected this polyhedron shaped portion of the lattice
with spheres and we have found 544 spherical layers.
\subsection{An example of chaos from
symmetry in the hexagonal landscape}
Among the records of this
landscape we have considered the spherical layer defined by:
\begin{equation}\label{ciusone}
    \mathbf{k}^2 \, = \, \frac{128}{3}
\end{equation}
which contains 90 lattice point. These 90 lattice points intercepted
by the sphere of radius $\sqrt{\frac{128}{3}}$ are organized in the
following orbits of the point group $\mathrm{Dih_6}$:
\begin{eqnarray}\label{strato132}
\mathbb{S}_{r^2=\frac{128}{3}}\bigcap\Lambda^\star_{hexag} & = &
O_1(\{6,1\})+O_2(\{12,3\})+O_3(\{12,2\})+O_4(\{12,2\})\nonumber\\
&&+O_5(\{12,2\})+O_6(\{12,2\}) +O_7(\{12,2\})+O_8(\{12,2\})
\end{eqnarray}
and yield a $90 \times 90 $-dimensional representation of the
Universal Classifying Group $\mathfrak{U}_{72}$ which admits the
following decomposition into irreps:
\begin{eqnarray}
\label{novantina} D\left[\mathfrak{U}_{72},90\right] &=&  2
D_1(\mathfrak{U}_{72},1)+2 D_3(\mathfrak{U}_{72},1)+3
D_5(\mathfrak{U}_{72},1)+3 D_7(\mathfrak{U}_{72},1)+5
   D_{10}(\mathfrak{U}_{72},2)\nonumber\\
   &&+5 D_{11}(\mathfrak{U}_{72},2)+5 D_{13}(\mathfrak{U}_{72},2)+5
   D_{15}(\mathfrak{U}_{72},2)+5 D_{17}(\mathfrak{U}_{72},2)\nonumber\\
   &&+5 D_{19}(\mathfrak{U}_{72},2)+5
   D_{21}(\mathfrak{U}_{72},2)+5 D_{24}(\mathfrak{U}_{72},2)
\end{eqnarray}
As one sees from eq.(\ref{novantina}) the 90-dimensional parameter
space contains a $2$-dimensional subspace invariant with respect to
the full group $\mathfrak{U}_{72}$, corresponding to the identity
representation. It is interesting to choose such an example and
consider its properties.
%%%%%%%%%%%%%%%
\subsubsection{Choice of the $\mathfrak{U}_{72}$ invariant subspace}
\label{singlechaos} Collecting respectively the coefficients of
$Y_{1,1}$ and $Y_{1,2}$, that parameterize the singlet 2-dimensional
subspace and using the hexagonal cell coordinates $u,v,r$ defined
by:
\begin{equation}\label{uvrcoordi}
    x\, = \,\frac{2 u-v}{\sqrt{2}} \quad ; \quad y \, = \,\sqrt{\frac{3}{2}} v\quad ;
    \quad z\, = \, \sqrt{2} \, r
\end{equation}
we obtain two explicit vector fields
$\mathbf{V}^{sing|1,2}\left(u,v,r\right)$ of which, due to the
massiveness of the formulae, we display only the first, in order to
give the reader some feeling of the result structure and quality.
Here it is:
\begin{eqnarray}
\label{sing1c1}
  \mathbf{V}^{sing|1}_1 &=& \frac{1}{8
   \sqrt{3}}\left\{11 \sin (2 \pi  (6 r+3 u-7 v))-28 \sin (2 \pi  (6 r+4 u-7 v))-17 \sin (2 \pi  (6
   r+7 u-4 v))\right.\nonumber\\
   &&\left.-17 \sin (2 \pi  (6 r+7 u-3 v))+17 \sin (2 \pi  (6 r-7 u+3 v))-28 \sin (2
   \pi  (-6 r+4 u+3 v))\right.\nonumber\\&&\left.-28 \sin (2 \pi  (6 r+4 u+3 v))+17 \sin (2 \pi  (6 r-7 u+4 v))  +11
   \sin (2 \pi  (-6 r+3 u+4 v))\right.\nonumber\\
   &&\left.+11 \sin (2 \pi  (6 r+3 u+4 v))+28 \sin (2 \pi  (6 r-4
   u+7 v))-11 \sin (2 \pi  (6 r-3 u+7 v))\right.\nonumber\\
   &&\left.-32 \cos (2 \pi  (6 r+3 u-7 v))-16 \cos (2 \pi
   (6 r+4 u-7 v))-16 \cos (2 \pi  (6 r+7 u-4 v))\right. \nonumber\\
   && \left.+16 \cos (2 \pi  (6 r+7 u-3 v))-16 \cos
   (2 \pi  (6 r-7 u+3 v))-16 \cos (2 \pi  (-6 r+4 u+3 v))\right.\nonumber\\
   &&\left. +16 \cos (2 \pi  (6 r+4 u+3
   v))+16 \cos (2 \pi  (6 r-7 u+4 v))-32 \cos (2 \pi  (-6 r+3 u+4 v))\right.\nonumber\\
   &&\left.+32 \cos (2 \pi  (6
   r+3 u+4 v))+16 \cos (2 \pi  (6 r-4 u+7 v))+32 \cos (2 \pi  (6 r-3 u+7 v))\right\}\nonumber\\
\end{eqnarray}
\begin{eqnarray}
\label{sing1c2} \mathbf{V}^{sing|1}_2 &=& \frac{1}{8} \left( 15 \sin
(2 \pi (6 r+3 u-7 v))+2 \sin (2 \pi  (6 r+4 u-7 v))+13 \sin (2 \pi
    (6 r+7 u-4 v))\right.\nonumber\\
&&\left.  -13 \sin (2 \pi  (6 r+7 u-3 v))+13 \sin (2 \pi  (6 r-7 u+3
v))-2 \sin
   (2 \pi  (-6 r+4 u+3 v))\right.\nonumber\\
   &&\left.-2 \sin (2 \pi  (6 r+4 u+3 v))-13 \sin (2 \pi  (6 r-7 u+4
   v)) \right.\nonumber\\
&&\left.  -15 \sin (2 \pi  (-6 r+3 u+4 v))-15 \sin (2 \pi  (6 r+3
u+4 v))-2 \sin (2 \pi  (6
   r-4 u+7 v)) \right.\nonumber\\
&&\left.  -15 \sin (2 \pi  (6 r-3 u+7 v))-16 \cos (2 \pi  (6 r+4 u-7
v))-16 \cos (2
   \pi  (6 r+7 u-4 v)) \right.\nonumber\\
&&\left.  -16 \cos (2 \pi  (6 r+7 u-3 v))+16 \cos (2 \pi  (6 r-7 u+3
v))+16
   \cos (2 \pi  (-6 r+4 u+3 v)) \right.\nonumber\\
&&\left.  -16 \cos (2 \pi  (6 r+4 u+3 v))+16 \cos (2 \pi  (6 r-7
   u+4 v))+16 \cos (2 \pi  (6 r-4 u+7 v)) \right.\nonumber\\
\end{eqnarray}
\begin{eqnarray}
\label{sing1c3} \mathbf{V}^{sing|1}_3 &=& \frac{1}{4
\sqrt{3}}\,\left\{11 \sin (2 \pi (6 r+3 u-7 v))+11 \sin (2 \pi  (6
r+4 u-7 v))+11 \sin (2 \pi  (6
   r+7 u-4 v)) \right.\nonumber\\
&&\left.  +11 \sin (2 \pi  (6 r+7 u-3 v))+11 \sin (2 \pi  (6 r-7 u+3
v))-11 \sin (2
   \pi  (-6 r+4 u+3 v))\right.\nonumber\\
   &&\left.+11 \sin (2 \pi  (6 r+4 u+3 v))-8 \cos (2 \pi  (6 r-4 u+7 v))+8 \cos (2 \pi  (6 r-3 u+7 v)) \right.\nonumber\\
&&\left.  +11 \sin (2 \pi  (6 r-7 u+4 v))-11
   \sin (2 \pi  (-6 r+3 u+4 v))+11 \sin (2 \pi  (6 r+3 u+4 v)) \right.\nonumber\\
&&\left.  +11 \sin (2 \pi  (6 r-4
   u+7 v))+11 \sin (2 \pi  (6 r-3 u+7 v))+8 \cos (2 \pi  (6 r+3 u-7 v))\right.\nonumber\\
   &&\left. -8 \cos (2 \pi
   (6 r+4 u-7 v)) +8 \cos (2 \pi  (-6 r+4 u+3 v))+8 \cos (2 \pi  (6 r+4 u+3 v)) \right.\nonumber\\
&&\left.  +8 \cos (2 \pi  (6 r+7 u-4 v))-8 \cos (2 \pi  (6 r+7 u-3
v))-8 \cos (2
   \pi  (6 r-7 u+3 v)) \right.\nonumber\\
&&\left.  +8
   \cos (2 \pi  (6 r-7 u+4 v))-8 \cos (2 \pi  (-6 r+3 u+4 v))-8 \cos (2 \pi  (6 r+3 u+4
   v)) \right\}\nonumber\\
\end{eqnarray}
In order to perceive what \textit{chaos from symmetry} really means
we focus on the above singlet Beltrami vector field  and we make a
vector plot of it inside the cubic shaped fundamental cell
$u\in[0,1],v\in[0,1],r\in[0,1]$ which can be smoothly mapped into
one of the three sectors of the hexagonal cell.
\par
\begin{figure}[!htb]
\begin{center}
\includegraphics[width=70mm]{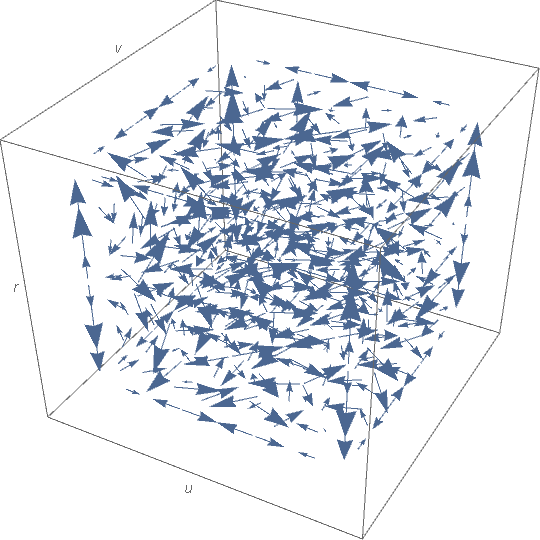}
\end{center}
\caption{\it Plot of the vector field displayed in eq.s
(\ref{sing1c1}-\ref{sing1c3}) which is invariant with respect to the
full group $\mathfrak{U}_{72}$. The high symmetry of this vector
field is visible at eight sight. The pattern repeats itself under
rotation and reflections but also under the $1/6$ translations in
the vertical direction $r$. \label{ordineesagonale}}
\end{figure}

The result for our singlet field is displayed in
fig.\ref{ordineesagonale}.
\par
Given the high symmetry of the vector plot the capricious chaotic
development of the stream-lines follows from the integration of the
first order equations. As an exemplification we begin with a single
stream line starting at a generic initial point of the hexagonal
fundamental cell.
\par
\begin{figure}[!htb]
\begin{flushleft}
\includegraphics[width=60mm]{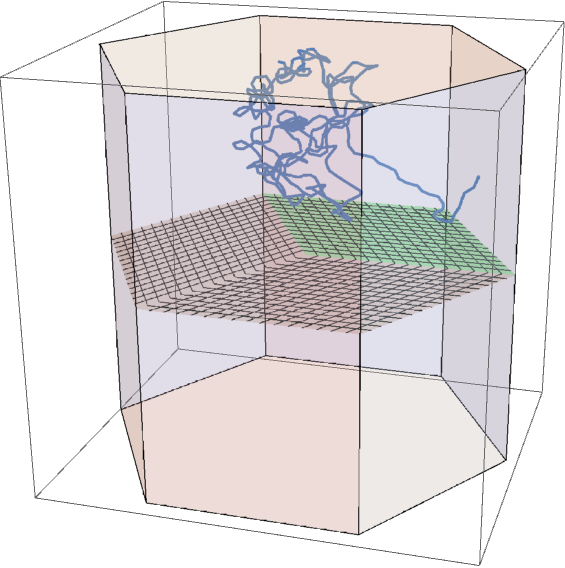}
\includegraphics[width=50mm]{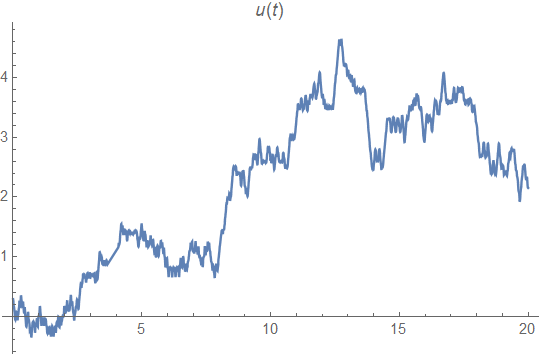}
\includegraphics[width=50mm]{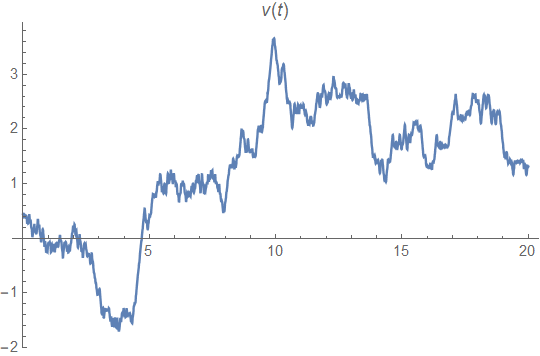}
\includegraphics[width=50mm]{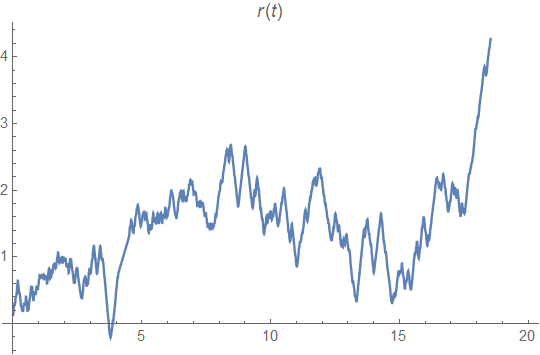}
\end{flushleft}
\caption{\it Plot of the single streamline starting at the point
(\ref{inicondahex})  for the vector field invariant under
$\mathfrak{U}_{72}$ defined in eq.s (\ref{sing1c1}-\ref{sing1c3})
and displayed in  fig. \ref{ordineesagonale}. In the first picture
we present the three-dimensional path of the fluid element within
the hexagonal cell, whose basis is divided in the three sectors,
related to each other by a $2\pi/3$ rotation. In green we have the
fundamental sector, image of the square $u\in[0,1]$,$v\in[0,1]$. The
other pictures display the time plots $u(t),v(t),r(t)$ of the three
hexagonal coordinates. The chaotic behavior is fully evident.
\label{vadoaspassoesag}}
\end{figure}

We choose:
\begin{equation}\label{inicondahex}
    \{u_0,v_0,r_0\} \, = \, \left\{\frac{2}{7},\frac{4}{9},\frac{2}{15} \right\}
\end{equation}
The response elaborated by the computer is the wandering path
presented in fig.\ref{vadoaspassoesag}.
\par
\begin{figure}[!htb]
\begin{center}
\includegraphics[width=70mm]{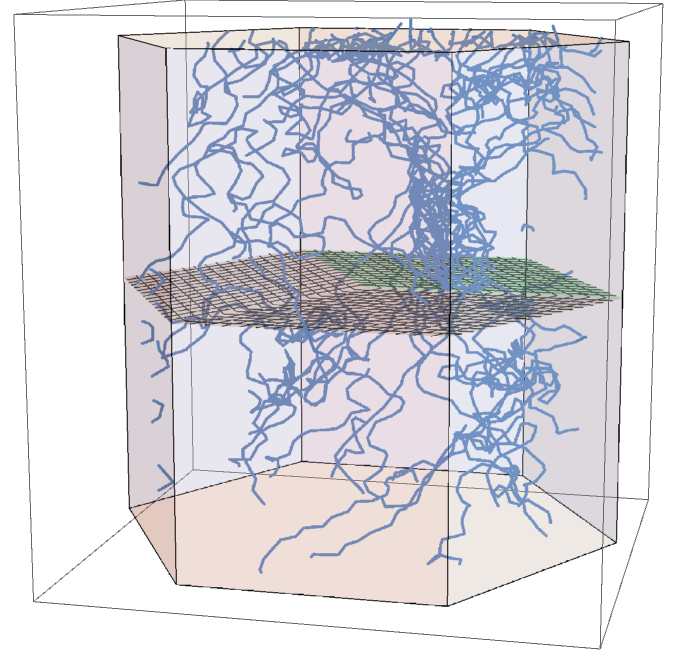}
\end{center}
\caption{\it Plot of 25 streamlines, all originating at equally
spaced points  in the fundamental sector (green parallelogram) of
the hexagonal basis  for the vector field invariant under
$\mathfrak{U}_{72}$ defined in eq.s (\ref{sing1c1}-\ref{sing1c3})
and displayed in  fig. \ref{ordineesagonale}. The chaotic behavior
after $t_{max}=20$ units of integration time are quite
evident.\label{casinonellesagono}}
\end{figure}

Next we proceeded to the calculation of 25 streamlines that have
equally spaced starting points in the fundamental planar cell
$u\in[0,1]$, $v\in[0,1]$ but after $30$ integration steps have
already  diffused capriciously and chaotically throughout the entire
hexagonal cell. The result is what you see in figure
\ref{casinonellesagono}.
\subsection{A vertical motion} 
\begin{figure}[!htb]
\begin{center}
\includegraphics[width=60mm]{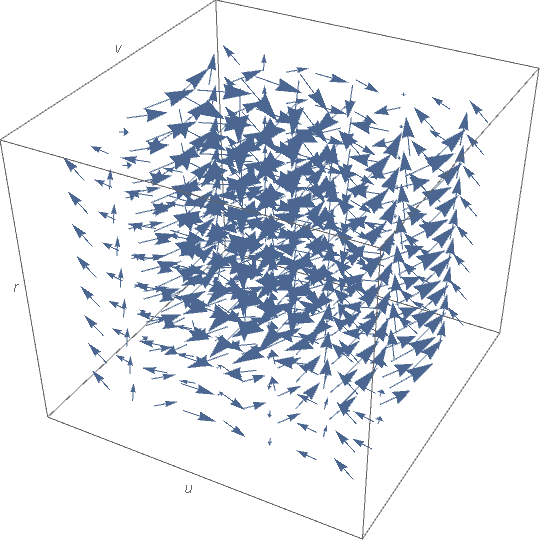}
\includegraphics[width=60mm]{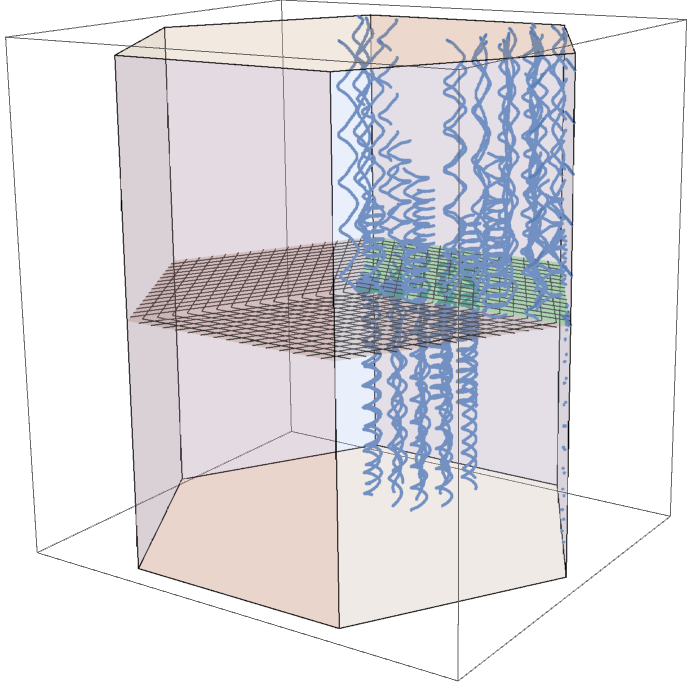}
\end{center}
\caption{\it Plot of a  Beltrami field from a planar 6 orbit in
sided by a plot of 25 of its streamlines, originating from the
fundamental planar cell. As one realizes some are uprising spirals,
some other descending spirals. \label{girinispaventati}}
\end{figure}

The main problem one meets in several
applications of hydrodynamics is, as we already stressed, that of
mixing a chaotic behavior at small scales with an approximate global
motion, at larger scales, in one definite direction that we can
conventionally assume to be the $z$-axis. The superposition is
intrinsically forbidden by the non linearity of the NS and Euler
equations, yet within the scope of the Beltrami fields and the
landscape approach there is a limited superposition freedom:
\textit{Beltrami flows having the same eigenvalue parameter
$\lambda$ can be linearly combined.} Hence it is interesting to
consider whether in the same spherical layer that contains highly
symmetric and hence chaotic flows like that described in the
previous section \ref{singlechaos} there are other orbits that
provide instead rather orderly flows uniformly directed. The answer
is yes and it is also of a general type. All orbits of the point
group $\mathrm{Dih_6}$ in the momentum lattice
$\Lambda_{exag}^\star$ that are of type $O({6,1})$ have the
following features:
\begin{description}
\item[a)] The orbits is planar at $z=0$
\item[b)] The Beltrami flow associated with the orbit has a 6 dimensional parameter space that decomposes with respect
to the $\mathfrak{U}_{72}$ group according to the following scheme:
\begin{equation}\label{localinosfizioso}
    D[\mathfrak{U}_{72},6]\, = \,
    D_\alpha[\mathfrak{U}_{72},1]+D_\beta[\mathfrak{U}_{72},1]+D_\gamma[\mathfrak{U}_{72},2]+D_\delta[\mathfrak{U}_{72},2]
\end{equation}
where $D_\alpha,D_\beta$ are two different one-dimensional and
$D_\gamma,D_\delta$ are two different two-dimensional
representations.
\item[c)] The restriction of the Beltrami field to the two
one-dimensional representations provides an integral model whose
streamlines are parallel spirals directed in the $z$ direction that
wind around their central vertical axis with wider or more tight
coils.
\end{description}
An example is shown in fig.\ref{girinispaventati}
\section{Conclusions}
In the previous sections of the present paper we have outlined and
presented the theoretical basis of the mechanism \textbf{Chaos from
symmetry}, emerging from the use of Beltrami fields as ingredients
of exact periodic solutions of the Navier-Stokes equations. When the
compact space in which they occur is a three torus, as introduced in
eq. (\ref{metricT3}), these exact solutions are governed quite
efficiently by Group Theory. This is the fundamental message that is
not widely and fully appreciated neither among the differential
geometers and dynamical system theorists that give important
contributions to the field of mathematical hydrodynamics, nor among
the applied scientists doing numerical simulations and working in
CFD.
\par
Notwithstanding their long life Navier-Stokes equations have few
exact solutions, the existing ones providing already a wide spectrum
of qualitatively different behaviors and enucleating the essential
point of difficulty that can be summarized as follows.
\par
As stressed in the introduction both conceptually and at the level
of applications, one would like to consider hydro flows that have at
least two scales, a \textit{macro scale} where we observe a
directional, reasonably ordered flow and a \textit{micro scale}
where the flow is instead chaotic. How to combine the two aspects
into exact solutions is the open unsolved problem.
\par
Vladimir Arnold unveiled since the years 70.s of the XXth century
the profound topological nature of chaotic behavior
\cite{arnoldus,ArnoldBook}. His theorem \ref{fundarnoltheo}
emphasizes the essential role of Beltrami vector fields that are
precisely what, after the work of one of us with A.Sorin of
2014-2015 \cite{Fre:2015mla}, can be precisely classified and
constructed in terms of Finite Group Theory. On the other hand
Beltrami fields have a natural relation, in the capacity of Reeb
fields, with the geometrical conception of \textit{contact
structures} on odd-dimensional manifolds. Contact Manifolds in
odd-dimensions  have symbiotic relations with symplectic manifolds
one dimension above and one dimension below and so does their
defining contact one-form; all that brings into the field of
hydrodynamics  the visions and the arguments of differential
geometry of symplectic related type. The group-theoretical
classification of Beltrami fields therefore reflects into a
group-theoretical classification of contact-structures and of their
allied even-manifolds. Contact structures are the deep root of
chaotic behavior being the geometric obstruction to the existence of
a foliation of the ambient manifold $\mathcal{M}$ in which the fluid
moves and foliations being, instead, the essential ingredient of
potential or laminar ordered flows.
\par
So the difficulty in reconciling the above mentioned two scale
regimes within one and the same exact solution of Navier-Stokes
equation is not an occasional one, rather it is a very much
conceptual antinomy.
\par
What are the possible strategic way out? Three have emerged that
might be combined together:
\begin{enumerate}
  \item Within the scope of the landscape approach to Beltrami fields one can
  superimpose motions that look like directional ones on larger
  scales,
  although they reveal themselves as winding spirals at smaller scales, with properly
  chaotic motions at short distances. From the point of view of
  contact structure Beltrami fields are necessary for chaos yet not
  viceversa. There are Beltrami fields that give rise to integral
  systems and cohexist on the same spherical layer with really
  chaos-generating Beltrami fields that typically are the most
  symmetric ones.
  \item Consider the new development of singular contact structures
  and singular Beltrami fields in so named $\mathfrak{b}$-manifolds.
  \item Reconsider the results of \cite{beltraspectra} where it was
  shown that solutions of Navier Sokes equations display the feature
  of weekly interacting Beltrami spectra.
\end{enumerate}
As we have shown in the present work the case 2) of the above list
that was initiated by the authors of
\cite{guglielminoconmiranda,unochiuseinteg,trasversozero,cardona2020universality,
cardona2021integrable,pollosingolare,cardone2019,miranda2021} has an
unsuspected strong relation with the group theoretical structure of
Beltrami fields that requires to be clarified in detail and is
potentially very powerful.
\par
Similarly the in depth analysis of the landscape properties in
search of algorithmic recipes for the optimal synthesis of  over all
directional flows with low scale chaotic flows requires appropriate
group theoretical investigations and also extensive surveys of the
landscape at large. For instance the classification of the 48
momentum classes of the cubic lattice achieved in \cite{Fre:2015mla}
has not yet been done for the hexagonal lattice. To this effect
implementation of the \textbf{AlmafluidaNSPsytem} on large powerful
computers would be quite appropriate.
\par
That of point 3) is anyhow the master direction to be explored. This
was already emphasized in section \ref{furetto}, the Beltrami
operator is a chiral one and a generic periodic solution of Navier
Stokes equations is layer by layer the superposition of a Beltrami
and an anti-Beltrami field. This leads to the concept of the
Beltrami spectral index. On the other hand the very fact that
Beltrami and anti-Beltrami fields have the same group-theoretical
structure combined with the fact that the same representations of
the Universal Classifying Group reappear on successive layers
provides the opportunity of constructing candidate solutions of the
Navier Stokes equations in the form of Fourier expansions with
prescribed hidden symmetries. Whether the free coefficients can be
determined in such a ways as to provide exact solutions of the
Navier Stokes  differential equations is something to be explored.
Reversely known exact solutions of the NS equations have to be
analyzed from the point of view of hidden symmetries. This is the
most promising direction for future work.

\subsection*{Aknowledgements} The present scientific investigation has
been conducted within the framework of the Project \textit{ALMA
FLUIDA} partially financed by the Toscana Region as part of the
Consultancy Agreement between  ITALMATIC Presse e Stampi  and the
DISAT of Politecnico di Torino. The origin of this investigation is
 traced back to a previous study of Arnold-Beltrami Flows
conducted in 2014-15 by one of us (P.F.), together with his long
time collaborator and close friend Alexander Sorin. P.F. would  like
to express his gratitude to A. Sorin for introducing him to this
topic that now finds new life. Last but not least we desire to
express our gratitude to our great friend Sauro Additati for
envisaging the whole scheme of the  project and to the Fredianis,
father Adriano and son Roberto, directing ITALMATIC Presse e Stampi,
who  made this scientific mission not only possible but also very
pleasant due to their warm and deep friendship. Last but not least
it is a pleasure to express our gratitude to Daniele Marchisio first
of all for our amicable relations that we  developed  during these
months since the time when the ALMA FLUIDA project was firstly
conceived and constructed together, secondly for his excellent
coordination of the Politecnico consultancy still going on.

\end{document}